\documentclass[prx,twocolumn,
preprintnumbers,superscriptaddress,amsmath,amssymb,longbibliography]{revtex4-1}
\usepackage{graphicx,braket}
\usepackage{subfigure}
\usepackage{mathrsfs}
\usepackage{amsfonts}
\usepackage{amsmath}
\usepackage{times}
\usepackage[colorlinks,linkcolor=blue,citecolor=blue]{hyperref}

\begin{document}
\title{Topological phases of non-Hermitian systems}
\author{Zongping Gong}
\email{gong@cat.phys.s.u-tokyo.ac.jp}
\affiliation{Department of Physics, University of Tokyo, 7-3-1 Hongo, Bunkyo-ku, Tokyo 113-0033, Japan}
\author{Yuto Ashida}
\email{ashida@cat.phys.s.u-tokyo.ac.jp}
\affiliation{Department of Physics, University of Tokyo, 7-3-1 Hongo, Bunkyo-ku, Tokyo 113-0033, Japan}
\author{Kohei Kawabata}
\affiliation{Department of Physics, University of Tokyo, 7-3-1 Hongo, Bunkyo-ku, Tokyo 113-0033, Japan}
\author{Kazuaki Takasan}
\affiliation{Department of Physics, Kyoto University, Kyoto 606-8502, Japan}
\author{Sho Higashikawa}
\affiliation{Department of Physics, University of Tokyo, 7-3-1 Hongo, Bunkyo-ku, Tokyo 113-0033, Japan}
\author{Masahito Ueda}
\affiliation{Department of Physics, University of Tokyo, 7-3-1 Hongo, Bunkyo-ku, Tokyo 113-0033, Japan}
\affiliation{RIKEN Center for Emergent Matter Science (CEMS), Wako, Saitama 351-0198, Japan}
\date{\today}
\date{\today}

\begin{abstract}
While Hermiticity lies at the heart of quantum mechanics, recent experimental advances in controlling dissipation have brought about unprecedented flexibility in engineering non-Hermitian Hamiltonians in open classical and quantum systems. Examples include parity-time-symmetric optical systems with gain and loss, dissipative Bose-Einstein condensates, exciton-polariton systems and biological networks. A particular interest centers on the topological properties of non-Hermitian systems, which exhibit unique phases with no Hermitian counterparts. However, no systematic understanding in analogy with the periodic table of topological insulators and superconductors has been achieved. In this paper, we develop a coherent framework of topological phases of non-Hermitian systems. After elucidating the physical meaning and the mathematical definition of non-Hermitian topological phases, we start with one-dimensional lattices, which exhibit topological phases with no Hermitian counterparts and are found to be characterized by an integer topological winding number even with no symmetry constraint, reminiscent of the quantum Hall insulator in Hermitian systems. A system with a nonzero winding number, which is experimentally measurable from the wave-packet dynamics, is shown to be robust against disorder, a phenomenon observed in the Hatano-Nelson model with asymmetric hopping amplitudes. We also unveil a novel bulk-edge correspondence that features an infinite number of (quasi-)edge modes. We then apply the K-theory to systematically classify all the non-Hermitian topological phases in the Altland-Zirnbauer (AZ) classes in all dimensions. The obtained periodic table unifies time-reversal and particle-hole symmetries, leading to highly nontrivial predictions such as the absence of non-Hermitian topological phases in two dimensions. We provide concrete examples for all the nontrivial non-Hermitian AZ classes in zero and one dimensions. In particular, we identify a $\mathbb{Z}_2$ topological index for arbitrary quantum channels (CPTP maps). Our work lays the cornerstone for a unified understanding of the role of topology in non-Hermitian systems.
\end{abstract}
\maketitle

\section{Introduction}
Topological phases of matter \cite{Thouless1982,Haldane1983,Haldane1988,Wen1995,Kane2005} have attracted growing interest over the last decade in many subfields of physics, including condensed matter physics \cite{Bernevig2006,Molenkamp2007,Kane2010,Qi2011,Beenakker2015,Ryu2016,Wen2017}, ultracold atomic gases \cite{Bloch2012,Monika2013b,Esslinger2014,Monika2015,Spielman2015,Inguscio2015,Pan2016,Goldman2016,Pan2017}, quantum information \cite{Kitaev2003,Nayak2008,Alicea2011,Martinis2014}, photonics \cite{Haldane2008,Hafezi2011,Fan2012,MacDonald2012,Ciuti2013,Hafezi2013,Segev2013,Soljacic2014,Lindner2015,Ozawa2018} and mechanics \cite{Kane2014,Marquardt2015,Huber2015,He2016}. Topological phase transitions lie outside the Ginzburg-Landau-Wilson paradigm of spontaneous symmetry breaking \cite{Sachdev2011}, can occur in noninteracting systems, and may require the existence of certain symmetries \cite{Bernevig2013}. Systematic classifications have been achieved for such symmetry-protected-topological (SPT) phases, ranging from the Altland-Zirnbauer (AZ) classes \cite{Altland1997,Ryu2008,Kitaev2009,Ryu2010,Teo2010} to crystalline insulators and superconductors \cite{Fu2011,Slager2013,Ryu2013,Morimoto2013,Shiozaki2014,Fu2015,Slager2017}. These SPT states of matter exhibit robust edge states (gapless or zero modes) localized at open boundaries \cite{Hatsugai1993,Kitaev2001} and novel entanglement spectra for subsystems \cite{Fidkowski2010}. The gapped bulk SPT phases are characterized by highly nonlocal topological indices, which can give rise to quantized transport phenomena immune to disorder \cite{Niu1985}. More recently, the notion of SPT phases has been generalized from equilibrium to periodically driven (Floquet) systems \cite{Else2016,Keyserlingk2016a,Potter2016,Roy2017}, which accommodate new topological phases with no static counterparts \cite{Kitagawa2010,Jiang2011,Lindner2013}. 

In recent years, considerable efforts have been devoted to explore topological phases in \emph{non-Hermitian} systems \cite{Hughes2011,Sato2011,Schomerus2013,Malzard2015,Jose2016,Tony2016,Nori2017,Duan2017,Kawabata2018a}, which are open and out of equilibrium. This burgeoning research arena is largely driven by the experimental progress on atomic, optical and optomechanical systems \cite{Ott2013,Aspelmeyer2014,Wiersig2015,Xiao2016,Xu2016,Yang2016,Yang2017,Zhou2018,Segev2018b}, where gain and loss can be introduced in a controllable manner. Controlled dissipation can be harnessed to engineer an effective non-Hermitian Hamiltonian $H\neq H^\dag$, represented by parity-time ($PT$)-symmetric systems \cite{Segev2010,Yang2014,Zhang2014,Konotop2016,Xue2017,Segev2017,Christodoulides2018}, which feature real spectra in the $PT$-unbroken phases \cite{Bender1998,Bender2007}. 
Unlike Hermitian systems, the eigenvalues of $H$ are generally complex, and its right eigenstates need not be orthogonal to each other and are not equivalent to the left eigenstates in general. Furthermore, the right eigenstates can coalesce and become orthogonal to the corresponding left ones at an exceptional point \cite{Heiss2012}, where $H$ cannot be diagonalized. Previous works have mostly focused on topological properties associated with the exceptional point. Some unique topological objects with no Hermitian counterparts are identified, such as anomalous edge modes characterized by half-integers \cite{Tony2016} and Weyl exceptional rings with both the quantized Chern number and the quantized Berry phase \cite{Duan2017}. Non-Hermitian systems emerge ubiquitously in a variety of situations including open quantum systems \cite{Breuer2002,Rudner2009,Tony2014a,Tony2014b,Ashida2016,Ashida2017a,Gong2017,Kawabata2017}, mesoscopic physics \cite{Esposito2009,Ren2010,Sagawa2011b}, biological physics \cite{Nelson1998,Amir2016,Vaikuntanathan2017} and chemistry \cite{Ren2013,Cao2015,Ouldridge2017}, where topology can play important roles \cite{Rudner2009,Gong2017,Vaikuntanathan2017,Ren2013}.

\begin{table*}[tbp]
\caption{Periodic table for non-Hermitian Hamiltonians. The Altland-Zirnbauer ten-fold classes \cite{Ryu2008,Kitaev2009,Ryu2010} are grouped into six such that classes A, DIII and CI, classes AI and D, and classes AII and C are unified. The Bott periodicity of classifying space $\mathcal{C}_1$ ($\mathcal{C}_1\times\mathcal{C}_1$) is $2$, and that of $\mathcal{R}_s$ ($\mathcal{R}_s\times\mathcal{R}_s$, $s=1,5$) is $8$. Note that all the classes are nontrivial (trivial) in $d=4n+1$ ($d=4n+2$) dimensions, where $n=0,1,2,\cdots$.}
\begin{center}
\begin{tabular}{cccccccccc}
\hline\hline
\;\;\;AZ class\;\;\; & \;\;\;Classifying space\;\;\; & \;\;\;\;$d=0$\;\;\;\; & \;\;\;\;\;\;1\;\;\;\;\;\; & \;\;\;\;\;\;\;2\;\;\;\;\;\;\; & \;\;\;\;\;\;3\;\;\;\;\;\; & \;\;\;\;\;\;4\;\;\;\;\;\; & \;\;\;\;\;\;5\;\;\;\;\;\; & \;\;\;\;\;\;6\;\;\;\;\;\; & \;\;\;\;\;\;7\;\;\;\;\;\; \\
\hline
A,\;DIII,\;CI & $\mathcal{C}_1$ & 0 & $\mathbb{Z}$ & 0 & $\mathbb{Z}$ & 0 & $\mathbb{Z}$ & 0 & $\mathbb{Z}$ \\
AIII & $\mathcal{C}_1\times\mathcal{C}_1$ & 0 & $\mathbb{Z}\oplus\mathbb{Z}$ & 0 & $\mathbb{Z}\oplus\mathbb{Z}$ & 0 & $\mathbb{Z}\oplus\mathbb{Z}$ & 0 & $\mathbb{Z}\oplus\mathbb{Z}$ \\
AI,\;D & $\mathcal{R}_1$ & $\mathbb{Z}_2$  & $\mathbb{Z}$ & 0 & 0 &0 & $2\mathbb{Z}$ & 0 & $\mathbb{Z}_2$ \\
BDI & $\mathcal{R}_1\times\mathcal{R}_1$ & $\mathbb{Z}_2\oplus\mathbb{Z}_2$ & $\mathbb{Z}\oplus\mathbb{Z}$ & 0 & 0 &0 & $2\mathbb{Z}\oplus2\mathbb{Z}$ & 0 & $\mathbb{Z}_2\oplus\mathbb{Z}_2$ \\
AII,\;C & $\mathcal{R}_5$ & 0 & $2\mathbb{Z}$ & 0 & $\mathbb{Z}_2$ & $\mathbb{Z}_2$ & $\mathbb{Z}$ & 0 & 0 \\
CII & $\mathcal{R}_5\times\mathcal{R}_5$ & 0 & $2\mathbb{Z}\oplus2\mathbb{Z}$ & 0 & $\mathbb{Z}_2\oplus\mathbb{Z}_2$ & $\mathbb{Z}_2\oplus\mathbb{Z}_2$ & $\mathbb{Z}\oplus\mathbb{Z}$ & 0 & 0 \\
\hline\hline
\end{tabular}
\end{center}
\label{table1}
\end{table*}

Nevertheless, a systematic understanding of topological phases of non-Hermitian systems is still elusive. Inspired by the periodic table for Hermitian topological insulators and superconductors \cite{Ryu2008,Kitaev2009,Ryu2010}, we are naturally led to the following questions:
\begin{quote}
(i) Can we classify non-Hermitian systems in analogy with the SPT phases in closed quantum systems? \newline
(ii) If yes, then what is the non-Hermitian counterparts of AZ classes? \newline
(iii) Is there a quantum-Hall-like non-Hermitian system which has no symmetry yet is topologically nontrivial? \newline
(iv) Is there a bulk-edge correspondence in non-Hermitian systems? 
\end{quote}

Regarding these fundamental questions, it seems that exceptional points, while unique to non-Hermitian systems and of great experimental importance, 
may not be a good starting point for a systematic classification, since they imply band touching in the bulk and seem incompatible with a non-Hermitian generalization of gap. We note that two very recent works \cite{Rudner2016,Fu2018} have made efforts to build a general framework following the methodology for gapped Hermitian systems. In particular, Ref.~\cite{Rudner2016} focuses on one-dimensional lattices with on-site loss and no dark states, and identifies a topological winding number relevant to particle displacement; Ref.~\cite{Fu2018} mainly discusses two-dimensional non-Hermitian lattices with separable bands in the complex-energy plane, and identifies a Chern number for individual bands. However, these results are rather specific in spatial dimensions and/or the structure of the Hamiltonian.

Here, we present a systematic framework for studying the topological phases of generic non-Hermitian systems. For the sake of comparison with SPT phases in Hermitian systems, we focus  primarily on lattice systems described by non-Hermitian Bloch (or Bogoliubov-de Gennes) Hamiltonians $H(\boldsymbol{k})$, but our formalism can also be applied to other setups like quantum channels \cite{Chuang2010} and full counting statistics \cite{Ren2013}, where non-Hermiticity appears in completely positive trace-preserving (CPTP) superoperators and generators for characteristic functions, respectively. We shall discuss the $\mathbb{Z}_2$ topological index for arbitrary quantum channels in Sec.~\ref{QC}. Our framework is based on two guiding principles: \newline
\begin{quote}
(I) Topological phases of non-Hermitian systems can be understood as \emph{dynamical} phases, where not only the eigenstates but also the \emph{full complex spectra} should be taken into account; \newline
(II) The non-Hermitian generalization of the concept of the band gap is the \emph{prohibition} of touching a base energy, which is typically zero but generally complex, in the spectrum.  
\end{quote}

We show that (I) and (II) are well justified both physically and mathematically. On the basis of these two guiding principles, we find that a one-dimensional lattice with \emph{asymmetric} hopping amplitudes turns out to be the most prototypical example comparable to the quantum Hall insulator, in the sense that an integer topological number can be defined without any symmetry protection. This result gives an interesting topological interpretation to the emergent Anderson transition \cite{Anderson1958} in the Hatano-Nelson model \cite{Hatano1996,Hatano1997,Hatano1998}, which should otherwise be absent in one-dimensional Hermitian systems \cite{Anderson1979}. We also unveil a bulk-edge correspondence which is qualitatively different from the Hermitian case: There is a \emph{continuum} of (quasi-)edge modes in the \emph{semi-infinite} space (open chain), with the winding number being the degeneracy at a given base energy. These findings answer the last two questions (iii) and (iv) raised in the last paragraph.

Our guiding principles also enable a systematic application of the \emph{K-theory} \cite{Karoubi2008}, a technique widely used in classifying Hermitian topological systems \cite{Kitaev2009,Teo2010,Shiozaki2014}, to the non-Hermitian AZ classes, leading to a complete classification in all spatial dimensions. We introduce a \emph{unitarization} procedure as a non-Hermitian generalization of band flattening, followed by a \emph{Hermitianization} procedure to represent the classifying space as a Clifford-algebra extension \cite{Roy2017}. The classification problem turns out to be mathematically equivalent to that of the Hermitian AZ classes with an additional chiral symmetry, leading to a dramatically different periodic table as shown in Table~\ref{table1}. We identify the underlying topological numbers implied by the K-theory classification for all the non-Hermitian AZ classes in one dimension. We also unveil a $\mathbb{Z}_2$ topological index for zero-dimensional (anti-)$PT$-symmetric systems and quantum channels. These results answer the first two questions (i) and (ii) raised above, and can further be generalized to, e.g., systems with crystalline symmetries and especially to $PT$-symmetric systems.

The remainder of the paper is organized as follows. In Sec.~\ref{DVTP}, we introduce the dynamical point of view regarding topological phases and justify the guiding principle (I). In Sec.~\ref{NH1D}, we first justify the guiding principle (II) and then discuss the topological properties of non-Hermitian lattices in one dimension, including the definition of the winding number, edge physics and experimentally observable signatures. In Sec.~\ref{NHAZ}, we employ the K-theory to achieve a complete classification of non-Hermitian AZ classes in all dimensions, as shown in Table~\ref{table1}. The identification of topological numbers and some topologically nontrivial examples in zero and one dimensions are given in Sec.~\ref{EXNH}. We conclude the paper with an outlook in Sec.~\ref{CO}. Several technical details and an experimental implimentation on asymmetric hopping are relegated to Appendices to avoid digressing from the main subjects.

\section{Dynamical viewpoint on the topological phases}
\label{DVTP}
\begin{figure}
\begin{center}
       \includegraphics[width=7cm, clip]{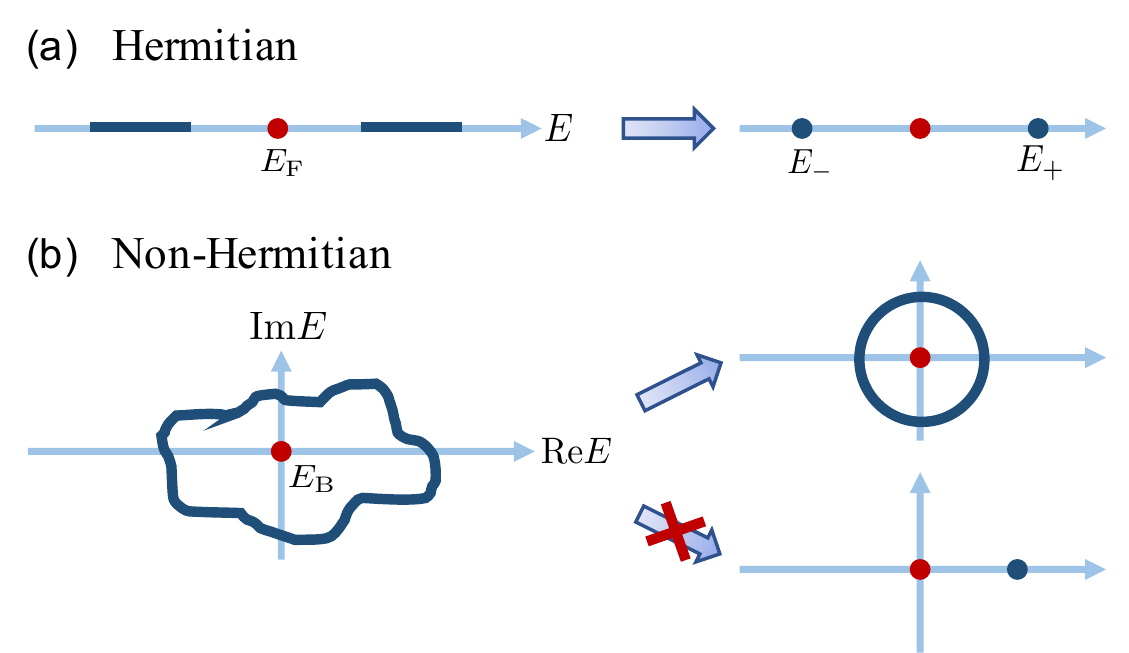}
       \end{center}
   \caption{(a) Energy spectrum (thick lines and dots) of a Hermitian insulator. We can always perform band flattening, i.e., continuously deform the spectrum into $\{E_-,E_+\}$ with $E_-<E_{\rm F}<E_+$, where $E_{\rm F}$ (red dot) is the Fermi energy. In particular, we can choose $E_\pm=\pm1$ for $E_{\rm F}=0$. (b) Energy spectrum of a non-Hermitian system forming a loop that encircles a base point $E_{\rm B}\in\mathbb{C}$. (In the figure we set $E_\mathrm{B}=0$ for simplicity.) While the shape can be deformed continuously, the loop can never shrink to a single point without crossing the base point.}
   \label{fig1}
\end{figure}

We begin by discussing how to define topological phases. In a Hermitian system, a topological phase can be analyzed from the many-body ground-state wave function $|\Psi\rangle$, 
which can be mapped through the projector 
\begin{equation}
P_-=\sum_{E_j<E_{\rm F}}|\varphi_j\rangle\langle\varphi_j|
\label{P-}
\end{equation}
onto all the single-particle eigenstates $|\varphi_j\rangle=f^\dag_j|{\rm vac}\rangle$ below the Fermi energy $E_{\rm F}$ for free fermions with $|\Psi\rangle=(\prod_{E_j<E_{\rm F}}f^\dag_j)|{\rm vac}\rangle$. Note that the spectrum plays no role here, since the Hamiltonian $H$ can be flattened by means of the projector (\ref{P-}) into $1-2P_-$ \cite{Ryu2008,Kitaev2009,Ryu2010} without closing the (band or many-body) energy gap, as schematically illustrated in Fig.~\ref{fig1} (a). Two gapped Hamiltonians $H$ and $H'$ differ topologically if and only if $|\Psi\rangle$ ($P_-$) cannot continuously be deformed into $|\Psi'\rangle$ ($P'_-$) under the constraint of the energy gap and certain symmetries. Such a topological distinction between wave functions accords with the ``states of matter" interpretation of phases. 

However, the very notion of the ground state, be it single- or many-body, breaks down for a non-Hermitian system, since its eigenenergy belongs to the complex-number field $\mathbb{C}$, where, unlike the real-number field $\mathbb{R}$, an order relation cannot be defined \cite{Ahlfors1979}. Indeed, from a physical point of view, non-Hermitian systems are intrinsically nonequilibrium and even unstable. According to the nonunitary Schr\"odinger equation
\begin{equation}
i\partial_t|\psi_t\rangle=H|\psi_t\rangle,
\label{Schro}
\end{equation}
where $H$ is non-Hermitian and the Planck constant is set to unity throughout this paper, only the single-particle eigenstate with the largest imaginary energy survives in the long-time limit, a phenomenon well known in photonics experiments \cite{Segev2017}. It thus cannot be justified to interpret non-Hermitian topological phases simply as topological states of matter.

In this paper we will show that the topological phases of non-Hermitian systems can be understood as topological \emph{dynamical} phases, for which not only the eigenstates but also the \emph{full complex spectra} play important roles. In fact, such a dynamical perspective has widely been adopted in the context of thermalization and many-body localization \cite{Nandkishore2015}, as well as Floquet systems \cite{Moessner2017}. Examples include the Wigner-Dyson (Poisson) level-spacing statistics in chaotic (integrable) systems \cite{Haake2010} and quasi-energy pairing in discrete time crystals \cite{Khemani2016,Keyserlingk2016b,Nayak2016}. As for non-Hermitian systems, we can immediately identify a unique topological object arising solely from the complex spectrum --- a \emph{loop} constituted from eigenvalues that encircles a prescribed base point (see Fig.~\ref{fig1} (b)). Here by unique we mean that the topological object discussed here never occurs in a Hermitian system with a real spectrum; by topological we mean that the loop can never be removed without crossing the base point at $E=E_{\rm B}$. If the base point is chosen to be zero, a loop ensures the existence of amplifying (${\rm Im} E>0$) and attenuating (${\rm Im} E<0$) modes. Such a topologically enforced dynamical instability (dynamical property) can be compared to topologically protected edge states (state property) in Hermitian systems. Note that the converse is not true, since instability or edge modes may not have a topological origin.

While only the complex spectrum is relevant in the above example, in general, however, both states and the spectrum are important in the complicated transient dynamics governed by Eq.~(\ref{Schro}). Since the full information of dynamical behavior is encoded in the non-Hermitian Hamiltonian $H$ in Eq.~(\ref{Schro}), we can generally define that \emph{two non-Hermitian systems differ topologically if and only if their Hamiltonians cannot continuously be deformed into each other under certain constraints}. Here the minimal constraint follows guiding principle (II), which will be justified in the next section.

Remarkably, by imposing the constraints of Hermiticity and a finite gap, we can reproduce the states-of-matter interpretation in Hermitian systems, at least for noninteracting SPT phases. Without loss of generality \footnote{In the absence of a particle-hole or chiral symmetry, the classification of all the Hermitian Hamiltonians with a given $E_{\rm F}$ is equivalent to that with $E_{\rm F}=0$, since we have a time-reversal-symmetry-preserved one-to-one map $H\to H-E_{\rm F}$ between two sets of Hamiltonians. In the presence of a particle-hole or/and chiral symmetry, although $E_{\rm F}$ has arbitrariness for a given system, the only choice of $E_{\rm F}$ is zero when considering the set of all such Hermitian Hamiltonians.}, assuming that $E_{\rm F}=0$ lies in the band gap, the real spectrum can always be trivialized into $\pm1$, leaving the only difference arising from $P_-$ given in Eq.~(\ref{P-}). In this sense, the dynamical viewpoint on topological phases is a generalization of the static one.

We would like to mention that Eq.~(\ref{Schro}) should not necessarily be interpreted as a nonunitary equation of motion for a wave function. Indeed, it can be any linear dynamics, such as a classical Markovian process, where $|\psi_t\rangle$ is a probability distribution \cite{Kampen2007}, or a quantum master equation, where $|\psi_t\rangle$ is a density operator or a supervector in the Liouville space \cite{Jiang2016}. In some cases we may consider a discrete version of Eq.~(\ref{Schro}):
\begin{equation}
|\psi_{t+T}\rangle=U_T|\psi_t\rangle,
\label{UT}
\end{equation}
which can be any linear stroboscopic dynamics or even a single input-output process, such as nonunitary quantum walk \cite{Xue2017,Obuse2016} or quantum channels \cite{Chuang2010}. A recent work \cite{Diehl2013} on classifying Gaussian nonequilibrium steady states $\rho_{\rm ss}$ can be regarded as a specific case of Eq.~(\ref{UT}) with $\mathcal{U}_\infty(\rho)=\rho_{\rm ss}$ for all $\rho$, where $\mathcal{U}_\infty=\lim_{t\to\infty}e^{\mathcal{L}t}$, and $\rho_{\rm ss}$ is the unique (under the periodic boundary condition) kernel of a quadratic Lindbladian $\mathcal{L}$ with a finite damping gap.

\section{Topological non-Hermitian lattices in one dimension with no symmetry}
\label{NH1D}
Before performing a general classification, it is instructive to start from the most illustrative case --- one-dimensional lattices without any symmetry requirements. These systems are found to be classified by a topological winding number, provided that a base energy $E_{\rm B}$ is not involved in the energy spectrum. We show that such a winding number corresponds to the number of edge states at $E_{\rm B}$ in a \emph{semi-infinite} space and is measurable from the wave-packet dynamics.

\subsection{Topological winding number}
\label{TWN}
Let us first clarify the allowed continuous deformation. Note that all the matrices $M$ can continuously be deformed to $0$ via the path $M_\lambda=(1-\lambda)M$, $\lambda\in[0,1]$ if there is no constraint. 
To avoid the case in which all non-Hermitian systems in all dimensions are trivial, we must impose at least one constraint. In the Hermitian case, such a constraint is the existence of an energy gap near the Fermi energy $E_{\rm F}$, which is equivalent to the condition that $E_{\rm F}$ does not belong to the energy spectrum of the Hamiltonian. As a possible generalization to the non-Hermitian case, we impose the condition that a base energy $E_{\rm B}\in\mathbb{C}$ does not belong to the energy spectrum of $H(k)$ for all $k\in[-\pi,\pi]$, where $k$ is the wave vector. In analogy with the Hermitian case where $E_{\rm F}$ is typically set to be zero, we assume without loss of generality $E_{\rm B}=0$ such that $H(k)\in{\rm GL}(V)$, where ${\rm GL}(V)$ is the general linear group on the Hilbert space $V$ at a given wave vector $k$. Such a minimal constraint is not only natural from a mathematical viewpoint, but also physically reasonable, since breaking the invertibility of a Hamiltonian usually requires fine-tuning of parameters. In other words, the constraint should easily be satisfied under random perturbations, as is typically the case with experimental imperfection. Indeed, as will be detailed from now on, our setup does bring fruitful physical insights into non-Hermitian systems.

Mathematically, our minimal constraint reads
\begin{equation}
{\rm det}H(k)\neq0,\;\;\;\;{\rm for\;all\;} k\in[-\pi,\pi],
\label{cneq0}
\end{equation}
which allows one to define a topological winding number:
\begin{equation}
w\equiv\int^\pi_{-\pi}\frac{dk}{2\pi i}\partial_k\ln {\rm det}H(k).
\label{wH1}
\end{equation}
We note that the generalization to the case of $E_{\rm B}\neq0$ can be achieved by simply replacing $H(k)$ by $H(k)-E_{\rm B}$ in Eqs.~(\ref{cneq0}) and (\ref{wH1}). Let $E_1(k),E_2(k),...,E_N(k)\in\mathbb{C}/\{0\}$ be the eigenenergies of $H(k)$, where $N={\rm dim}V$ is the total number of bands. Then the winding number (\ref{wH1}) can be expressed as
\begin{equation}
w=\sum^N_{n=1}\int^\pi_{-\pi}\frac{dk}{2\pi}\partial_k\arg E_n(k),
\label{wE}
\end{equation}
where $\arg E_n(k)$ is the argument of the complex energy $E_n(k)$. Note that $w$ vanishes identically for Hermitian Hamiltonians because the real energy spectrum implies ${\rm Arg} E_n(k)=0,\pi$, where ${\rm Arg}$ denotes the principle value of the argument belonging to $[0,2\pi)$. In this sense, a nontrivial winding number, which gives the number of times the complex eigenenergies encircle $E_{\rm B}$, is unique to non-Hermitian systems. Mathematically, the existence of this winding number is ensured by the fact that the fundamental group of ${\rm GL}(V)$ is isomorphic to $\mathbb{Z}$. In the next section, we will show that the K-theory approach also gives the same $\mathbb{Z}$ classification for one-dimensional systems belonging to class A, which imposes no symmetries. In contrast, class A is trivial in one-dimensional Hermitian systems \cite{Ryu2008}.

\begin{figure}
\begin{center}
       \includegraphics[width=6cm, clip]{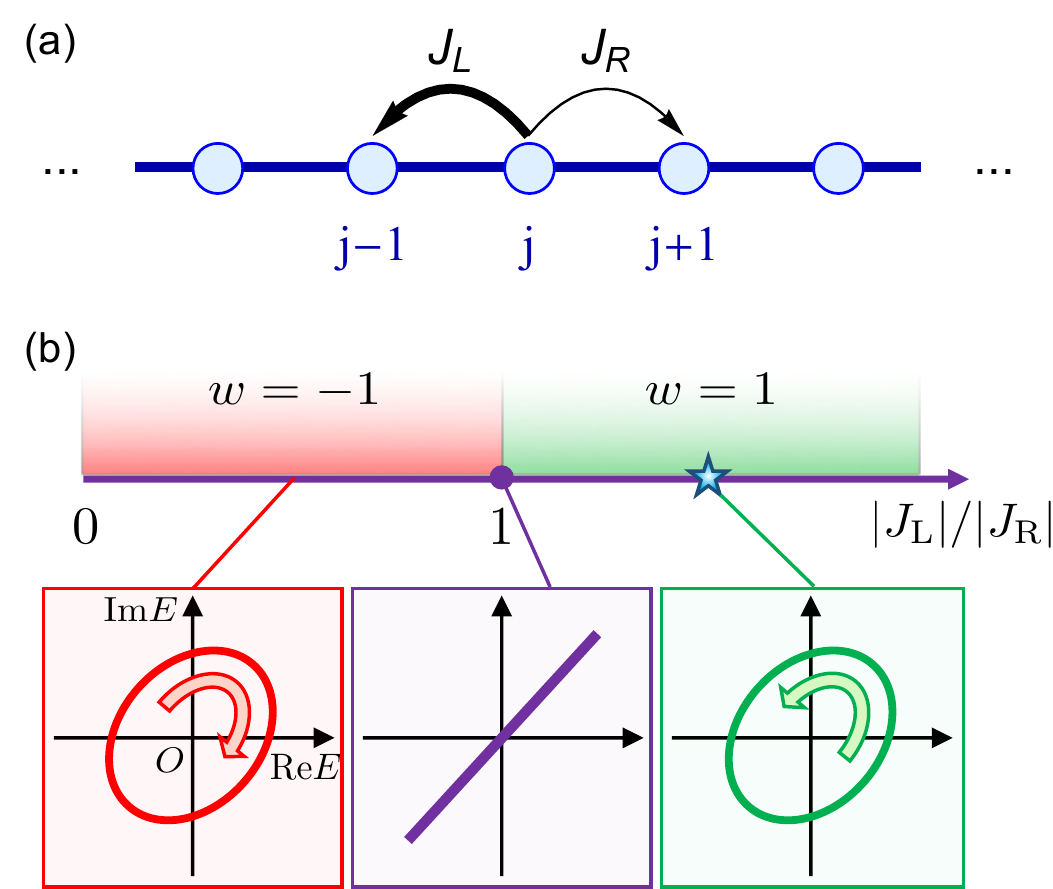}
       \end{center}
   \caption{(a) One-dimensional lattice with asymmetric hopping amplitudes $J_{\rm L}\neq J^*_{\rm R}$. Here, we show the case in which $|J_{\rm L}|>|J_{\rm R}|$, as indicated by the thickness of the arrows. (b) Phase diagram and typical complex energy spectra for the model in (a), where $w$ is the winding number. A topological phase transition occurs at $|J_{\rm L}|=|J_{\rm R}|$ (purple dot), where the spectrum touches the origin, while the specific case of (a) (blue star) belongs to the $w=1$ phase, where the energy spectrum forms a loop encircling the origin. 
   An arrow inside each loop 
   indicates the direction of increasing $k$ which corresponds to the sign of the winding number $w$.}
   \label{fig2}
\end{figure}

As a minimal setup to observe a topological phase transition, we consider a ring geometry with \emph{asymmetric} hopping amplitudes $J_{\rm R},J_{\rm L}\in\mathbb{C}$ (see Fig.~\ref{fig2}(a)):
\begin{equation}
H=\sum_j(J_{\rm R}c^\dag_{j+1}c_j+J_{\rm L}c^\dag_jc_{j+1}).
\label{HNH}
\end{equation}
Fourier transforming Eq. (\ref{HNH}) to moment space, we obtain the Bloch Hamiltonian as
\begin{equation}
H(k)=J_{\rm R}e^{-ik}+J_{\rm L}e^{ik},
\label{BHN}
\end{equation}
whose winding number is evaluated to give 
\begin{equation}
w=\left\{\begin{array}{ll} 1 & \;\;|J_{\rm R}|<|J_{\rm L}|; \\ -1 & \;\;|J_{\rm R}|>|J_{\rm L}|.\end{array}\right. 
\end{equation}
The topological phase-transition point thus locates at $|J_{\rm R}|=|J_{\rm L}|$ (see Fig.~\ref{fig2}(b)), where $H(k)=0$ for $k=[{\rm arg}(J_{\rm R}/J_{\rm L})\pm\pi]/2$ and thus $H(k)$ is not invertible.

\begin{figure*}
\begin{center}
       \includegraphics[width=17.6cm, clip]{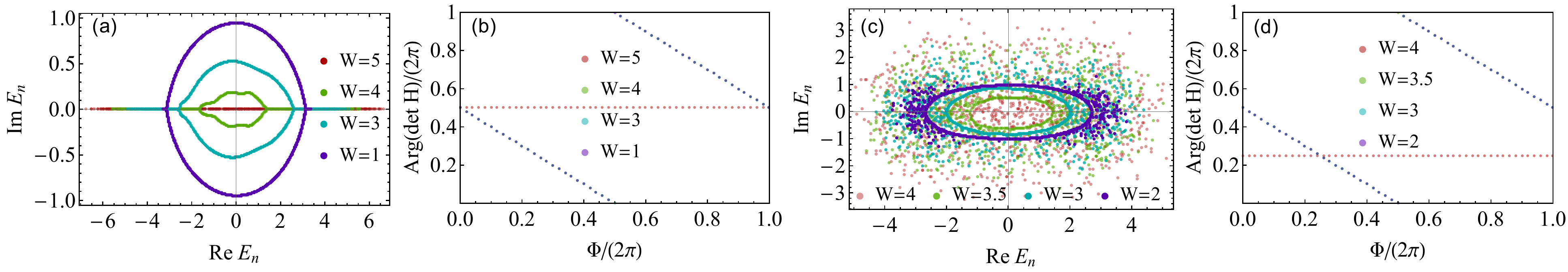}
       \end{center}
   \caption{(a) Complex-energy spectra and (b) flows of ${\rm Arg}({\rm det}H)$ with respect to the flux $\Phi$ for typical realizations of the Hatano-Nelson Hamiltonian (\ref{HNHD}) with $L=10^3$, $J_{\rm R}=2$, $J_{\rm L}=1$ and real on-site disorder $V_j\in[-W,W]$, where $W=1,3,4,5$. (c) and (d) correspond to (a) and (b), respectively, with the same set of parameters except for inclusion of a complex on-site disorder $V_j=|V_j|e^{i\phi_j}$, where $|V_j|\in[0,W]$ with $W=2,3,3.5,4$ and $\phi_j\in[0,2\pi]$. Note that the flows of ${\rm Arg}({\rm det}H)$ almost overlap in (b) for $W=1,3,4$ and in (d) for $W=2,3,3.5$, and that they also overlap with each other between (b) and (d). We see that the transition occurs between $W=4$ and $W=5$ in (a) and between $W=3.5$ and $W=4$ in (c). In the nontrivial phase ($W=1,3,4$ in (a) and $W=2,3, 3.5$ in (c)), the spectra encircle the base point at $E=0$, giving the winding number $w=-1$. In the trivial phase, the data points lie on the real axis in (a) and scatter in the complex energy plane without forming a closed loop in (c).}
   \label{fig3}
\end{figure*}

Note that Eq.~(\ref{BHN}) becomes $H(k)=e^{-ik}$ for the specific choice of $J_{\rm R}=1$ and $J_{\rm L}=0$. In this case, the non-Hermitian Hamiltonian becomes unitary. If we regard $H(k)$ as the Floquet operator $U_{\rm F}(k)$, we obtain a Thouless pump \cite{Thouless1983}, which is characterized by the winding number proposed in Ref.~\cite{Kitagawa2010}:
\begin{equation}
w=\int^\pi_{-\pi}\frac{dk}{2\pi i}{\rm Tr}[U^{-1}_{\rm F}(k)\partial_kU_{\rm F}(k)].
\label{wH2}
\end{equation}
In fact, Eq.~(\ref{wH2}) reduces to Eq.~(\ref{wH1}) if we replace $U_{\rm F}(k)$ by $H(k)$ (see Appendix~\ref{WNE}). The formal similarity and the essential difference between non-Hermitian Hamiltonians and Floquet operators will be clarified in the next section.

Remarkably, without symmetry constraints, non-Hermitian systems can support topological phases and transitions even for a single-band lattice like Eq.~(\ref{HNH}). Indeed, we can easily write down a single-band Bloch Hamiltonian $H(k)=e^{ink}$, which corresponds to an $|n|$-site (leftward when $n>0$ and rightward when $n<0$) unidirectional hopping and features an arbitrary winding number $n\in\mathbb{Z}$. This makes a sharp contrast with Hermitian systems which require at least two bands for observing topological phenomena \footnote{The phenomenon alone may be relevant to a single band, like the integer or anomalous quantum Hall effect. After all, at least another band is necessary to make the present band nontrivial, such as another band with the opposite Chern number.}, either with (in one dimension) or without (in two dimensions) additional symmetries. Such a sharp distinction can be understood as follows: According to Eq.~(\ref{wE}), the winding numbers in non-Hermitian systems are determined solely from complex energies. On the other hand, winding numbers (or Chern numbers) in Hermitian systems are usually related to the Berry phase, which automatically becomes trivial if there is only a single band. We will return to these crucial points in Sec.~\ref{NHBO}.

\subsection{Robustness against disorder --- revisiting the Hatano-Nelson model}
So far we have focused on the case with translation invariance and used the Bloch Hamiltonian. For Hermitian systems belonging to class A, we know that the integer quantum Hall states in two dimensions are robust against spatial disorder. As a consequence, while the Anderson transition is forbbiden in two dimensions \cite{Anderson1979} in the absence of spin-orbit interactions \cite{Asada2002}, mobility edges emerge in an integer quantum Hall state and the delocalized modes contribute to the quantized Hall conductivity $Ce^2/h$ \cite{Aoki1981}, with $C$ being the Chern number \cite{Thouless1982}. These well-established results naturally raise a question of whether or not a topological non-Hermitian system like Eq.~(\ref{HNH}) is robust against disorder and, if yes, in what sense. 

To address this question, we consider the following modification of Eq.~(\ref{HNH}):
\begin{equation}
H=\sum_j(J_{\rm R}c^\dag_{j+1}c_j+J_{\rm L}c^\dag_jc_{j+1}+V_jc^\dag_jc_j),
\label{HNHD}
\end{equation}
which describes a one-dimensional ring with asymmetric hopping amplitudes and on-site disorder $V_j$. This is a well studied model proposed by Hatano and Nelson \cite{Hatano1996,Hatano1997,Hatano1998}. While a one-dimensional Hermitian system is always localized in the presence of a random potential \cite{Anderson1979}, e.g., $V_j\in[-W,W]$ with a uniform probability, the Hatano-Nelson model (\ref{HNHD}) exhibits an Anderson transition \cite{Anderson1958}. Recalling the emergence of mobility edges in the quantum Hall state, we may conjecture that the Anderson transition is ensured by the nontrivial topological winding number, which is expected to be trivial \footnote{Here we tacitly assume a finite system size ($L$), and therefore the probability is zero for the disordered Hamiltonian to be not invertible.} if the system is fully localized.

To verify the conjecture, we have to first generalize the definition of the winding number to disordered systems. Following the idea of defining the Chern number for disordered quantum Hall states \cite{Niu1985}, we apply a magnetic flux $\Phi$ through a finite non-Hermitian ring with length $L$ such that the hopping amplitudes are multiplied by $e^{\mp i\Phi/L}$ under a specific choice of gauge. For the Hatano-Nelson model (\ref{HNHD}), we have
\begin{equation}
H(\Phi)=\sum^L_{j=1}(J_{\rm R}e^{-i\frac{\Phi}{L}}c^\dag_{j+1}c_j+J_{\rm L}e^{i\frac{\Phi}{L}}c^\dag_jc_{j+1}+V_jc^\dag_jc_j).
\end{equation}
While $H(\Phi)$ is not periodic in $\Phi$, there exists a large-gauge transformation $U_{\rm LG}=e^{\frac{2\pi i}{L}\sum_j jc^\dag_jc_j}$ such that
\begin{equation}
H(\Phi+2\pi)=U_{\rm LG}H(\Phi)U^\dag_{\rm LG}.
\end{equation}
Therefore, the gauge-independent quantity ${\rm det}H(\Phi)$ is periodic in $\Phi$ and the winding number can be defined as
\begin{equation}
w\equiv\int^{2\pi}_0\frac{d\Phi}{2\pi i}\partial_\Phi\ln {\rm det}H(\Phi).
\label{wPhi}
\end{equation}
We can show that Eq.~(\ref{wPhi}) reproduces Eq.~(\ref{wH1}) in the presence of translation invariance (see Appendix~\ref{WNE}). In general, $w$ counts the number of 
times the complex spectral trajectory encircles the base point $E_{\rm B}=0$ when the flux is increased from 0 to $2\pi$. Having in mind that a time-varying flux induces an electric field, we expect that both the eigenenergy and the wave function of a localized mode stay almost unchanged when changing $\Phi$. Accordingly, the winding number should vanish if the system is fully localized (see Appendix~\ref{SFL} for details). 

We perform an exact-diagonalization analysis of a Hatano-Nelson model with $L=10^3$, $J_{\rm R}=2$ and $J_{\rm L}=1$ subject to the periodic boundary condition. We present the numerical results in Fig.~\ref{fig3} for four different disorder strengths $W=1,3,4,5$. As $W$ increases, the fraction of localized modes (indicated by the points located on the real axis in Fig.~\ref{fig3}(a)) increases and the mobility edges (points encircling the origin) shrink to the origin. Nevertheless, even if the fraction of delocalized modes is small, the winding number (\ref{wPhi}) is always quantized at $w=-1$. Moreover, ${\rm arg\;det}H(\Phi)$ is approximately given by $\pi-\Phi$, as can be seen from the following explicit expression 
\begin{equation}
{\rm det}H(\Phi)=(-)^{L-1}(J^L_{\rm R}e^{-i\Phi}+J^L_{\rm L}e^{i\Phi})+P(\{V_j\}),
\label{detHPhi}
\end{equation}
where an overwhelming majority of the random magnitudes of the polynomial $P(\{V_j\})$ (see Appendix~\ref{PVj} for the detailed expression), which are independent of $\Phi$, should be much smaller than $J^L_{\rm R}$ before localization. With further increasing the disorder strength, an Anderson transition occurs at $W_{\rm c}\simeq4.3$ and all the states become localized, leading to a trivial topological number. 

In fact, the real parameters used in numerical calculations endows the Hatano-Nelson model with time-reversal symmetry $T=K$ ($K$: complex conjugate), which makes the spectra symmetric under reflection with respect to the real axis (see Fig.~\ref{fig3}(a)). To demonstrate that the time-reversal symmetry is irrelevant to the winding number discussed here, we also calculate the energy spectra for complex random potentials $V_j=|V_j|e^{i\phi_j}$, where the magnitude $|V_j|$ (phase $\phi_j$) is randomly sampled from a uniform distribution over $[0,W]$ ($[0,2\pi]$). Then the symmetry with respect to the real axis is lost, 
yet for disorder strength $W=2,3$ and $3.5$, we still find that the complex spectrum encircles the origin (see Fig.~\ref{fig3}(c)), as listed in 
a nontrivial winding number $w=-1$ (see Fig.~\ref{fig3}(d)). When the disorder is too strong (the critical value is about $W_{\rm c}\simeq3.9$), e.g., for $W=4$, the winding number becomes zero. Note that ${\rm Arg}({\rm det}H)$ in Fig.~\ref{fig3}(d) for $W=4$ does not take on special values like $0$ or $\pi$ unlike the Hermitian case. This is because the constant term $P(\{V_j\})$ in Eq.~(\ref{detHPhi}) now becomes complex due to $V_j\in\mathbb{C}$. 

It should be mentioned that while the topological transition and the localization transition coincide in the above two models, this may not be the case for other forms of disorder (see Appendix~\ref{SER}). On the other hand, one may conjecture that the system is fully localized if and only if the winding number with respect to an arbitrary base energy vanishes, provided that the eigenvalues of robust delocalized modes always form some loops. That is to say, a topological transition is certainly not sufficient but probably necessary for a localization transition.

While both the Hatano-Nelson model and the quantum Hall insulator are topologically nontrivial with no symmetry requirement, we would like to mention two crucial differences. First, due to the difference in spatial dimension, the former is characterized by a winding number, while the latter is characterized by a Chern number. Second, as indicated by Table~\ref{table1}, the topological winding number of the Hatano-Nelson model survives if the time-reversal symmetry is imposed. In stark contrast, a quantum Hall insulator (or Chern insulator) necessarily breaks the time-reversal symmetry.

\subsection{Bulk-edge correspondence}
As is well known in Hermitian systems, a nontrivial topological number in the bulk usually implies the existence of edge states, such as chiral edge modes in a quantum (anomalous) Hall state with open boundaries \cite{Hatsugai1993}. It is thus natural to ask whether the bulk-edge correspondence exists in topological non-Hermitian systems. We answer this question in the affirmative, at least for the single-band case. 
However, the correspondence turns out to be very different from that in Hermitian systems --- given a base energy $E_{\rm B}$, a positive (negative) winding number $w$ implies $w$ ($-w$) independent edge modes with energy $E=E_{\rm B}$ and localized at the left (right) boundary in the \emph{semi-infinite} space.

\begin{figure*}
\begin{center}
       \includegraphics[width=16cm, clip]{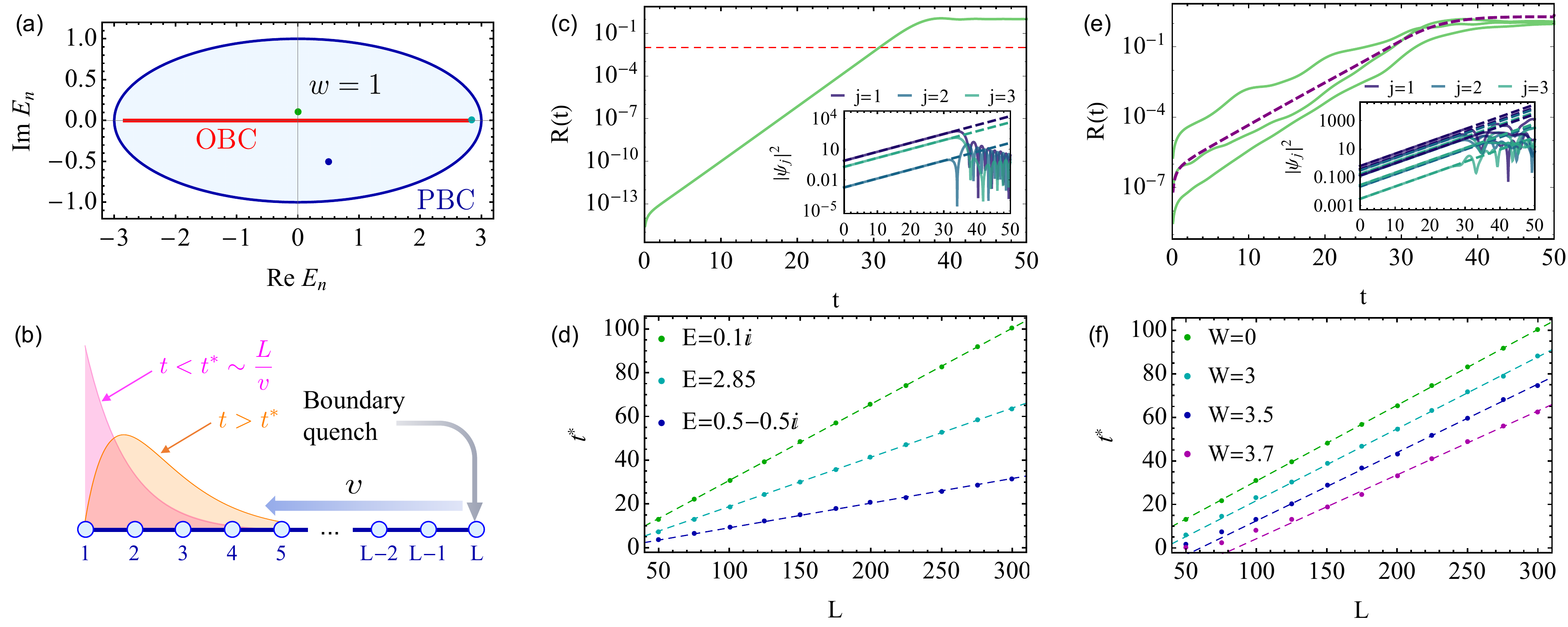}
       \end{center}
   \caption{(a) Energy spectrum of Eq.~(\ref{HNH}) with $J_{\rm L}=2$ and $J_{\rm R}=1$ under the periodic boundary condition (PBC, blue ellipse) and the open boundary condition (OBC, red line). For each energy $E$ inside the ellipse (light-blue region) there exists a $w=1$ edge state localized at the left boundary in the semi-infinite space. Three colored points show energies of the three quasi-edge modes in (d). (b) An edge state in the semi-infinite space (magenta wave packet) will eventually become unstable (orange wave packet) in a finite open chain with length $L$ after a time $t^*\sim\frac{L}{v}$, where $v$ is the Lieb-Robinson velocity. 
   (c) Time evolution of the relative deviation $R(t)\equiv\|[e^{-i(H-E)t}-1]|\psi\rangle\|$ for the edge state $|\psi\rangle$ with $E=0.1i$ in an open chain with $L=100$. Inset: Time evolution (solid curves) of $|\psi_j(t)|^2$ at the leftmost three sites ($j=1,2,3$) in comparison with that of $|\psi_j(t)|^2=e^{2{\rm Im} E t}|\psi_j(0)|^2$ (dashed lines). (d) Finite-size scaling of $t^*$ for three different quasi-edge states with energies $E=0.1i,2.85$ and $0.5-0.5i$ (marked in (a)). We define $t^*$ by $R(t^*)=10^{-2}$, as indicated by the dashed red line in (c). (e) The same as (c) but in the presence of real on-site disorder $V_j\in[-W,W]$ with $W=3$. The green curves and the inset show three typical realizations, and the dashed purple curve gives the disorder average over $10^3$ realizations. The average is taken for $\ln R(t)$, and thus gives the geometric mean for $R(t)$. (f) The same as (d) but for different disorder strengths $W=3,3.5,3.7$ (compared to $W=0$, the same as $E=0.1i$ in (d)) with fixed $E=0.1i$.}
   \label{fig4}
\end{figure*}

Let us first focus on the minimal model described by Eq.~(\ref{HNH}). By assuming $|J_{\rm L}|>|J_{\rm R}|$, we expect an edge state at the left boundary. Indeed, in the limiting case of $J_{\rm R}=0$, $\psi_j=\delta_{j,1}$ (localized at the first site) is an eigenstate with zero energy. More generally, by imposing the right-half-infinite boundary condition, a state localized at the left boundary can be obtained by solving
\begin{equation}
J_{\rm R}\psi_{j-1}+J_{\rm L}\psi_{j+1}=E\psi_j,\;\;j=1,2,...
\label{JRLP}
\end{equation}
subject to
\begin{equation}
\psi_0=0,\;\;\;\;\lim_{j\to\infty}\psi_j=0.
\label{BC}
\end{equation}
This is a standard problem on a recursive sequence. Denoting $z_1$ and $z_2$ as the roots of 
\begin{equation}
E=J_{\rm R}z^{-1}+J_{\rm L}z,
\label{Ez}
\end{equation}
which is the characteristic equation of Eq.~(\ref{JRLP}), the general form of the wave function can be written as \footnote{If $z_1=z_2$, we have $\psi_j=c_1z^j_1+c_2jz^{j-1}_1$.}
\begin{equation}
\psi_j=c_1z^j_1+c_2z^j_2.
\label{psij}
\end{equation}
Accordingly, the conditions in Eq.~(\ref{BC}) become
\begin{equation}
c_1+c_2=0,\;\;\;\;|z_1|<1,\;|z_2|<1.
\label{inic}
\end{equation}
These conditions lead to a \emph{continuum} of solutions $\psi_j\propto z^j_1-z^j_2$ with energies that fill the interior of the bulk energy spectrum --- a closed loop (see Fig.~\ref{fig4} (a)) specified by Eq.~(\ref{BHN}) or Eq.~(\ref{Ez}) with $|z|=|e^{ik}|=1$. Note that the winding number is $1$ for any base energy within this loop, including $E_{\rm B}=0$.

With the above concrete example in mind, we are ready to generalize the conclusion to arbitrary single bands with positive winding numbers. While the full proof is somewhat technical (see Appendix~\ref{BEC}), the key idea is simply the argument principle \cite{Ahlfors1979}
\begin{equation}
\oint_{|z|=1}\frac{dz}{2\pi i}\frac{f'(z)}{f(z)}=Z-P,
\label{AP}
\end{equation}
where $E=f(z)$ is the characteristic equation and $Z$ ($P$) denotes the number of zeros (poles) of $f(z)$ in the area $|z|<1$. Replacing $z$ with $e^{ik}$, we find that the left-hand side of Eq.~(\ref{AP}) gives nothing but the winding number $w$ introduced in Eq.~(\ref{wE}). 
A general form of the wave function can be written as $\psi_j=\sum^Z_{l=1}c_lz^j_l$, where $z_l$'s are the zeros and $c_l$'s are subject to $P$ different constraints stemming from the inhomogeneity at the edge. These are straightforward generalizations of Eqs.~(\ref{psij}) and (\ref{inic}). As a result, there are $Z-P=w$-fold degeneracies of edge states at $E=0$, or generally at $E=E_{\rm B}$ if we replace $f(z)$ with $f(z)-E_{\rm B}$ in Eq.~(\ref{AP}). Note that the same analysis applies to single bands with negative winding numbers by interchanging $z$ and $z^{-1}$.

In a realistic one-dimensional system, such as a photonic lattice \cite{Segev2017}, open boundaries always appear in pairs. In the presence of two edges, only a one-dimensional part is picked out from the edge-state continuum, making the topological degeneracy generally invisible for a given base energy. For example, the spectrum of an open chain described by Eq.~(\ref{HNH}) with length $L$ can be determined as $E_n=2\sqrt{J_{\rm L}J_{\rm R}}\cos\frac{n\pi}{L+1}$ ($n=1,2,...,L$) which distributes over an interval $(-2\sqrt{J_{\rm L}J_{\rm R}},2\sqrt{J_{\rm L}J_{\rm R}})$ on the real-energy axis in the thermodynamic limit (see the red line in Fig.~\ref{fig4} (a)). A sudden change in the spectrum under different boundary conditions has also been found in Ref.~\cite{Xiong2018}. Here, we can provide 
a topological understanding --- the winding number (\ref{wPhi}) should either vanish or become ill-defined in an open chain, since the flux can always be gauged out and thus ${\rm det}H(\Phi)$ is $\Phi$ independent. Therefore, the spectrum no longer encircles any base point inside the spectrum loop under the periodic boundary condition. Since the spectrum should change continuously when the boundary hopping 
is gradually switched on, the spectrum must be very sensitive to the boundary condition. Indeed, it is already shown in Ref.~\cite{Xiong2018} that an \emph{exponentially small} modification of the boundary condition 
can lead to an order-one change in the spectrum.

As stated above, an energy eigenstate localized at the edge of a semi-infinite space generally disappears if the system size is finite. 
Nevertheless, \emph{quasi-edge modes} may exist for finite-size systems. By quasi-edge modes, we mean that they are not genuine eigenstates, yet their dynamics look just like eigenstates up to a time scale that increases with the system size and diverges in the thermodynamic limit. To investigate them, suppose that an edge state with energy $E$ for the semi-infinite condition is prepared in a finite lattice with length $L$, whose spectrum does not include $E$. Then the time evolution can be obtained to a good approximation simply by multiplying $e^{-iEt}$ up to a time scale (at least) proportional to $L$ (see Figs.~\ref{fig4}(c) and (d)). Note that this quasi-eigenstate of a finite chain becomes exact in the semi-infinite limit $L\to\infty$. While a formal proof is available (see Appendix~\ref{LLQE}), we can intuitively interpret this linear scaling as a manifestation of the Lieb-Robinson bound \cite{Lieb1972} after a boundary-condition quench roughly $L$ sites away from the edge mode, as illustrated in Fig.~\ref{fig4} (b). In the presence of disorder, these quasi-edge modes stay robust, although they are irregularly modified depending on the disorder configuration. As for on-site disorder in Eq.~(\ref{HNH}), the wave function of a quasi-edge mode (if exist) at $E$ can iteratively be determined by $\psi_{j+1}=[(E-V_j)\psi_j-J_{\rm R}\psi_{j-1}]/J_{\rm L}$. The lifetime upon disorder average obeys the same linear scaling with respect to (sufficiently large) $L$ as the clean limit (see Fig.~\ref{fig4}(f)).

The dramatic changes in the spectra for different boundary conditions has already been investigated in a purely mathematical context regarding non-Hermitian Toeplitz matrices (i.e., the matrices satisfying $M_{jl}=M_{j-l}$) and operators \cite{Reichel1992}. A generalization of the conventional eigenvalues and eigenvectors, which is called the $\epsilon$-\emph{pseudo-eigenvalues and eigenvectors}, was made to explain the apparent inconsistency. The exact definition is as follows: Given a matrix or operator $H$, if there exists $V$ such that the operator norm satisfies $\|V\|\le\epsilon$ and $(H+V)\psi=E\psi$, then $E$ and $\psi$ constitute a pair of  $\epsilon$-pseudo-eigenvalue and eigenvector of $H$. In our language, Toeplitz matrices and operators correspond to finite and semi-infinite chains, respectively, and a pseudo-eigenvector is nothing but a quasi-edge mode. The spectrum of a Toeplitz operator must be obtained by first taking the thermodynamic limit $L\to\infty$ followed by $\epsilon\to0$, which is generally \emph{inequivalent} to the limit $\epsilon\rightarrow0$ followed by $L\rightarrow\infty$ \cite{Reichel1992}. 
This fact is reminiscent of quantum phase transitions \cite{Sachdev2011}, where spontaneous symmetry breaking occurs only by first taking the thermodynamic limit and then making the symmetry breaking perturbations vanish. Here, the noncommutativity of the limiting procedures stems from the topologically enforced sensitivity to the boundary condition, as already explained previously.

\subsection{Numerical and experimental schemes to extract the winding number}
\label{NHBO}
In Hermitian systems, the only direct signature of $w$ in one dimension seems to be the number of edge states. Due to the subtlety of the bulk-edge correspondence discussed above, we can hardly identify $w$ simply from the energy spectrum of a finite non-Hermitian system.

Nevertheless, we can numerically extract the winding number by counting the zero modes of the following enlarged \emph{Hermitian} Hamiltonian constructed from $H$:
\begin{equation}
H_{\rm H}\equiv\sigma_+\otimes H+\sigma_-\otimes H^\dag,
\label{HH}
\end{equation}
where $\sigma_\pm\equiv(\sigma_x\pm i\sigma_y)/2$, with $\sigma_x$ and $\sigma_y$ being the Pauli matrices. Such an idea of Hermitianization (\ref{HH}) actually lies at the heart of the K-theory classification discussed in the next section. Using the bulk-edge correspondence of $H$, we can show that the number of zero modes of Eq.~(\ref{HH}) equals to $2|w|$ (see Appendix~\ref{BEC}). This result is actually nothing but the bulk-edge correspondence for Hermitian systems with chiral symmetry alone (class AIII). If the chiral symmetry stems from the sublattice degrees of freedom, the sign of $w$ determines in which sublattice the edge state is localized. Note that the generalization to arbitrary base energies can be done through replacement of $H$ by $H-E_{\rm B}$ in Eq.~(\ref{HH}). 

In practice, we can measure the winding number from the wave-packet dynamics. For Hermitian lattice systems, the semiclassical equations of motion of a particle in a single band are given by \cite{Niu2010}
\begin{equation}
\frac{d\boldsymbol{k}}{dt}=\boldsymbol{F},\;\;\;\;
\frac{d\boldsymbol{r}}{dt}=\nabla_{\boldsymbol{k}}E(\boldsymbol{k})-\frac{d\boldsymbol{k}}{dt}\times\boldsymbol{\Omega}(\boldsymbol{k}),
\end{equation}
where $\boldsymbol{F}$ is the potential gradient, $E(\boldsymbol{k})$ is the band dispersion, and $\boldsymbol{\Omega}(\boldsymbol{k})=i\langle\nabla_{\boldsymbol{k}}u(\boldsymbol{k})|\times|\nabla_{\boldsymbol{k}}u(\boldsymbol{k})\rangle$ is the Berry curvature, which requires at least two dimensions and two bands (as mentioned in Sec.~\ref{TWN}) to be nonzero. In two dimensions, it suffices to determine the Chern number directly from the transverse motion of particles \cite{Monika2015}. However, in a one-dimensional lattice, rather sophisticated operations are needed to measure the winding number or the Zak phase \cite{Monika2013b}. That is, we have to isolate the geometric phase from the dynamical phase \cite{Abanin2013}. In a non-Hermitian one-dimensional system, however, the winding number (\ref{wE}) is determined solely from the eigenenergies, which are relevant to the \emph{dynamical} phase. It turns out that $w$ can be measured simply from the nonunitary Bloch oscillations \cite{Longhi2009,Longhi2015c}, whose semiclassical equation of motion is given by (see Appendix~\ref{DSM})
\begin{equation}
\frac{dk}{dt}=F,\;\;\;\;
\frac{dx}{dt}={\rm Re}\frac{dE(k)}{dk},\;\;\;\;
\frac{d\ln\mathcal{N}_t}{dt}=2{\rm Im}E(k),
\label{semicl}
\end{equation}
where $\mathcal{N}_t\equiv\langle\psi_t|\psi_t\rangle$ is not, in general, equal to unity due to the nonunitary nature of the dynamics. By simultaneously tracing the center of mass and the total weight of the wave packet, we can reconstruct the energy spectrum when the wave vector runs over the Brillouin zone. The winding number $w$ can thus be determined by counting how many times the complex-energy trajectory encircles a base point. Such a simple scenario can be implemented in photonic lattices \cite{Carusotto2017} with asymmetric backscattering \cite{Yang2016,Yang2017} or by using auxiliary microresonators with gain and loss \cite{Longhi2015a,Longhi2015b}. Here we propose another implementation based on ultracold atoms in optical lattices with engineered dissipation (see Appendix~\ref{AHOL} for details). Comparing with photonic lattices, ultracold atoms have the advantage in controlling interactions flexibly and thus are promising for exploring non-Hermitian quantum many-body physics \cite{Ashida2016,Ashida2017a}.

\begin{figure}
\begin{center}
       \includegraphics[width=8.5cm, clip]{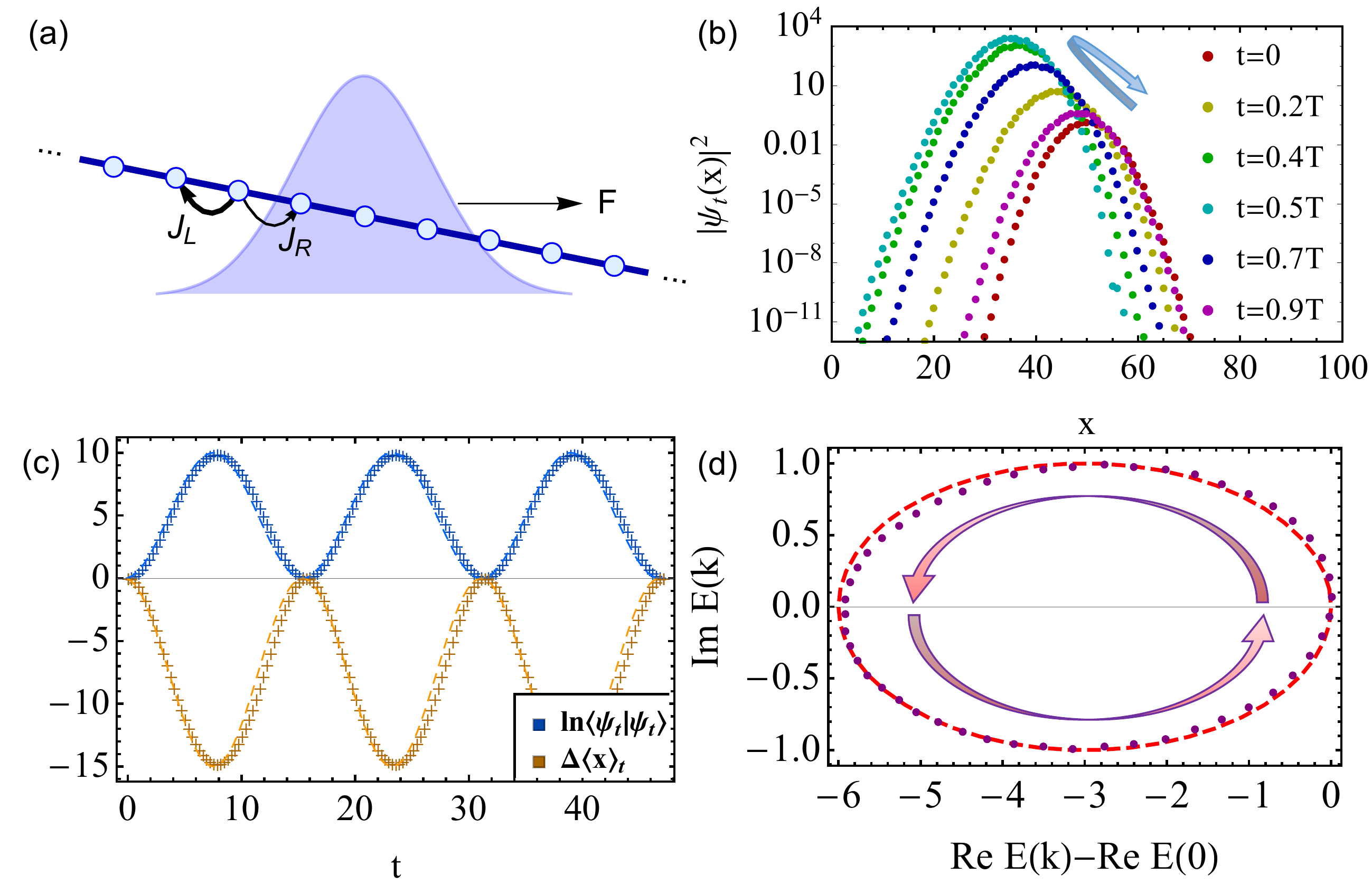}
       \end{center}
   \caption{(a) Gaussian wave packet in a lattice with asymmetric hopping amplitudes $J_{\rm L}=2$ and $J_{\rm R}=1$ and tilted by a potential gradient $F=0.4$. (b) Profiles of the wave packet in real space at $t=0,0.2T,0.4T,0.5T,0.7T$ and $0.9T$, with $T=\frac{2\pi}{F}$ for the lattice length $L=100$. (c) Numerical (``$+$" marks) and semiclassical (dashed curves, obtained from Eq.~(\ref{semicl})) results for the wave-packet dynamics in real space. Here $\Delta\langle x\rangle_t\equiv\langle x\rangle_t-\langle x\rangle_0$ denotes the center-of-mass displacement at time $t$. (d) Complex eigenenergies reconstructed from (c) (dots) in comparison with the theoretical results (dashed curve). The arrows in (b) and (d) show the direction of time. Since the data are taken stroboscopically, the imaginary energies ${\rm Im}E$ are estimated from $\ln(\langle\psi_{t+\Delta t}|\psi_{t+\Delta t}\rangle/\langle\psi_t|\psi_t\rangle)/(2\Delta t)$.} 
   \label{fig5}
\end{figure}

As a simple example, we consider the wave-packet dynamics in a disorder-free Hatano-Nelson lattice (\ref{HNH}) with $J_{\rm L}=2$, $J_{\rm R}=1$ and $L=100$. While the open-boundary condition is imposed, we have checked that the difference from the periodic-boundary condition is negligible. At the initial time, we prepare a Gaussian packet in the middle of the lattice with dispersion $\sigma_r=\sqrt{L/(4\pi)}$ and located at $k=0$ in the Brillouin zone (see Fig.~\ref{fig5}(a)). After applying a potential gradient $F=0.4$ in the positive $x$ (right) direction, both the center of mass and the intensity starts to oscillate. As shown in Fig.~\ref{fig5}(c), the numerical results (dots) agree quite well with the semiclassical predictions (dashed curves). Thus, the reconstructed complex energies based on Eq.~(\ref{semicl}) accurately reproduce those of the ideal dispersion relation (see Fig.~\ref{fig5}(d)). We have also plotted the wave-packet densities at several different times in Fig.~\ref{fig5}(b) and confirmed that the profile stays approximately Gaussian during the time evolution. Note that the initial direction of motion is opposite to $F$ due to the negative effective mass $m_{\rm eff}=-(J_{\rm L}+J_{\rm R})$ at $k=0$.

\section{Classification of non-Hermitian topological phases in the Altland-Zirnbauer classes}
\label{NHAZ}
The non-Hermitian systems discussed in the previous section are special in the sense that the spatial dimension is $d=1$ and no symmetry requirement is imposed. Such a non-Hermitian counterpart of class A in one dimension, however, exhibits an integer topological winding number (\ref{wH1}) reminiscent of Floquet systems \cite{Kitagawa2010} and Hermitian systems belonging to class AIII \cite{Ryu2008}. These observations suggest a connection between a non-Hermitian Hamiltonian and a unitary operator, the latter of which has a one-to-one correspondence to an involutory Hermitian Hamiltonian with a prescribed chiral symmetry \cite{Roy2017}. In this section, we establish such a connection, which enables a systematic classification of non-Hermitian Bloch Hamiltonians in all dimensions and in the presence of additional symmetries. In particular, we show that the topological classifications of non-Hermitian AZ classes differ significantly from those of Hermitian AZ classes \cite{Ryu2008,Kitaev2009,Ryu2010,Teo2010}.

\subsection{Unitarization under symmetry constraints}
\label{CNAZU}
In the previous sections we have already clarified that two Hamiltonians are topologically equivalent if they can continuously be deformed into each other under certain constraints. Without symmetries, the only constraint is that a base point $E_{\rm B}$ cannot be touched by the energy spectrum. Such a constraint is imposed to satisfy the condition of invertibility of the Hamiltonian for $E_{\rm B}=0$, which we primarily assume in the following discussions. For a given AZ class, we have to further impose symmetry constraints. We define that $H_0(\boldsymbol{k})$ and $H_1(\boldsymbol{k})$ are \emph{homotopically} equivalent, denoted as $H_0(\boldsymbol{k})\simeq H_1(\boldsymbol{k})$, if and only if there exists a path $H_\lambda(\boldsymbol{k})$ ($0\le\lambda\le1$) in the space of invertible matrices (i.e., the ${\rm GL}(V)$ group, where $V$ is the Hilbert space) such that 
\begin{equation}
AH_\lambda(\boldsymbol{k})=\eta_AH_\lambda(-\boldsymbol{k}) A,\;\;\;\;{\rm for\;all\;} \lambda\in[0,1],\label{homo}
\end{equation}
where $A=T$ (time-reversal operator) and $C$ (particle-hole operator) are anti-unitary operators, with $\eta_T=1$ and $\eta_C=-1$, respectively. We emphasize again that the condition of $H_\lambda(\boldsymbol{k})$ being invertible is equivalent to the condition that the system stays gapped in the Hermitian case, if we prescribe the Fermi energy to be $0$. When generalizing to non-Hermitian systems, the concepts of upper and lower bands disappear since we cannot establish an order relation for complex energies. 

\begin{figure}
\begin{center}
       \includegraphics[width=8cm, clip]{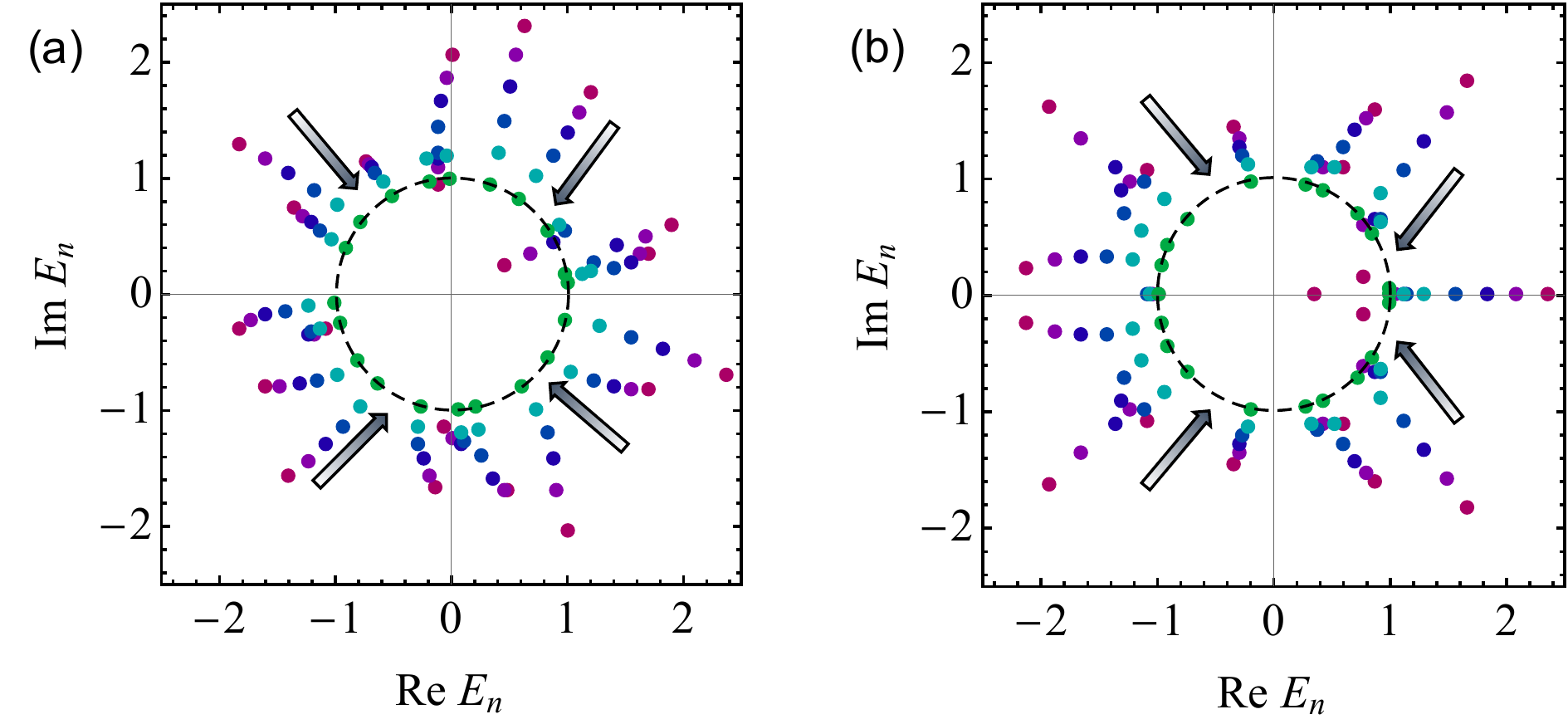}
       \end{center}
   \caption{(a) Spectral flow (from red to green, guided by the arrows) in the course of the unitarization process of an invertible complex matrix with size $20$. Note that the spectrum of the unitarized matrix locates on a unit circle (black dashed). (b) The same as in (a) but for a time-reversal-symmetric matrix. The time-reversal symmetry, which manifests itself as the mirror symmetry of the spectrum with respect to the real axis, is kept in the unitarization process.}
   \label{fig6}
\end{figure}

From now on, we may omit the variable $\boldsymbol{k}$ for simplicity. The definition of the homotopical equivalence based on Eq.~(\ref{homo}) implies the following theorem: 
\newtheorem{theorem}{Theorem}
\begin{theorem}
For an arbitrary invertible Hamiltonian $H$, which has a unique polar decomposition $H=UP$ with $U$ being unitary and $P=\sqrt{H^\dag H}$ being positive-definite and Hermitian, we have $H\simeq U$.
\label{HU}
\end{theorem}
This theorem is proved in Appendix \ref{Pf} and applicable also to crystalline symmetries. We provide two examples of unitarization from $H$ to $U$ in Fig.~\ref{fig6}. According to this theorem, it suffices to consider the classification of all the \emph{unitary} matrices. Note that this result is consistent with band flattening in the Hermitian case \cite{Ryu2008,Kitaev2009,Ryu2010}. By diagonalizing a Hermitian Hamiltonian as
\begin{equation}
H=V\begin{pmatrix} \Lambda^+_{p\times p} & 0 \\ 0 & \Lambda^-_{q\times q} \end{pmatrix}V^\dag,
\end{equation}
where $\Lambda^+_{p\times p}$ ($\Lambda^-_{q\times q}$) is the diagonal block of all the positive (negative) energies, we find the polar decomposition to be $H=UP$ with
\begin{equation}
\begin{split}
U&=V\begin{pmatrix} 1_{p\times p} & 0 \\ 0 & -1_{q\times q} \end{pmatrix}V^\dag,\\
P&=V\begin{pmatrix} \Lambda^+_{p\times p} & 0 \\ 0 & -\Lambda^-_{q\times q} \end{pmatrix}V^\dag,
\end{split}
\end{equation}
where $U$ is nothing but the flattened Hamiltonian.

\subsection{K-theory and Clifford-algebra extension}
\label{KTC}
The classification based on the homotopy equivalence is appropriate for a given Hilbert space, but is not so if the operations of inserting extra bands are also allowed. These operations are indeed possible in experiments of ultracold atoms, where we can, for example, couple two or more individual one-dimensional chains \cite{Schneider2016}. In this case, the correct classification should be carried out on the basis of the \emph{K-theory} \cite{Kitaev2009,Teo2010,Morimoto2013,Shiozaki2014,Shiozaki2017}, i.e., all we have to do is to figure out the K-group of the map from the Brillouin zone $M=T^d$ ($d$: spatial dimension) to a matrix space subject to specific symmetry requirements (but with no Hermiticity constraints). If we are only interested in the strong topological numbers \cite{Kitaev2009}, the manifold is $M=S^d$.

It is worthwhile to sketch the basics of the K-theory, so as to understand why it is compatible with band-inserting operations. The K-group is an Abelian group consisting of equivalence classes, denoted as $[H_0,H_1]$, of Hamiltonian \emph{pairs} $(H_0,H_1)$, where $H_0$ and $H_1$ act on the same Hilbert space. For $(H_0,H_1)$, we define an addition structure as
\begin{equation}
(H_0,H_1)+(H'_0,H'_1)=(H_0\oplus H'_0,H_1\oplus H'_1).
\end{equation}
We also impose $(H_0,H_1)=(H'_0,H'_1)$ if $H_0\simeq H'_0$ and $H_1\simeq H'_1$. To specify the equivalence classes, we require that $(H_0,H_1)$ should be identified as $(H_0\oplus H,H_1\oplus H)$ for all $H$, i.e., $[H_0\oplus H,H_1\oplus H]\equiv[H_0,H_1]$. By naturally defining the addition between equivalence classes
as
\begin{equation}
[H_0,H_1]+[H'_0,H'_1]=[H_0\oplus H'_0,H_1\oplus H'_1],
\end{equation}
we can deduce that they form an Abelian group, which is called the \emph{K-group} and denoted as $K(M)$, with zero element $[H,H]=0$ and the inverse of $[H_0,H_1]$ being $[H_1,H_0]$. We say that $H_0$ and $H_1$ belong to the same topological phase if and only if $[H_0,H_1]=0$. 

A crucial observation here is that although $H_0\simeq H_1$ implies $[H_0,H_1]=0$, the converse is \emph{not} true. A prototypical example is the Hopf insulator \cite{Moore2008} which is a two-band system in three dimensions and has no symmetry. While a Hopf insulator differs homotopically from a trivial insulator by a nonzero Hopf charge, it becomes trivial in the K-theory classification since we can insert additional bands into the system to trivialize the homotopy from $S^3$ to the entire Hilbert space. In other words, nontrivial topological phases emerge in class A in three dimensions only if there are two bands.

While it is generally difficult to calculate the K-group, well-developed techniques are available if the Hamiltonian space subjected to specific symmetry constraints is an extension of a Clifford algebra \cite{Kitaev2009}, which is generated by a set of \emph{anti-commutative} elements $\{e_j\}^n_{j=1}$, i.e., $e_je_{j'}=-e_{j'}e_j$ for all $j\neq j'$. If $e^2_j=1$ for all $j=1,2,...,n$, the algebra generated by $\{e_j\}^n_{j=1}$ over the complex-number field $\mathbb{C}$ is called a complex Clifford algebra ${\rm C}\ell_n$. If $e^2_j=-1$ for $j=1,2,...,p$ ($p\le n$) and $e^2_j=1$ for $j=p+1,p+2,...,n$, the algebra generated by $\{e_j\}^n_{j=1}$ over the real-number field $\mathbb{R}$ is called a real Clifford algebra ${\rm C}\ell_{p,q}$, where $q=n-p$. For a flattened Hermitian Hamiltonian $H$, we naturally have $H^2=1$, which can already be regarded as an element of a Clifford algebra ${\rm C}\ell_H$ generated by $H$ and its two-fold symmetry operators (as well as $i$, if there is an anti-unitary symmetry). Noting that the symmetry operators themselves generate another Clifford algebra ${\rm C}\ell_S$, we can thus represent the Hamiltonian space by the Clifford-algebra extension ${\rm C}\ell_S\to{\rm C}\ell_H$. In particular, we denote ${\rm C}\ell_s\to{\rm C}\ell_{s+1}$ and ${\rm C}\ell_{0,s}\to{\rm C}\ell_{0,s+1}$ as $\mathcal{C}_s$ and $\mathcal{R}_s$, respectively, which satisfy $\mathcal{C}_{s+2}=\mathcal{C}_s$ and $\mathcal{R}_{s+8}=\mathcal{R}_s$. It is well known for Hermitian systems that the two complex AZ classes correspond to $\mathcal{C}_s$ with $s=0,1$ and the eight real AZ classes correspond to $\mathcal{R}_s$ with $s=0,1,...,7$ \cite{Kitaev2009}. Denoting the K-group for a complex or real AZ class parametrized by $s$ and in $d$ dimensions as $K_{\mathbb{C}}(s;d)$ or $K_{\mathbb{R}}(s;d)$, we have
\begin{equation}
\begin{split}
K_{\mathbb{C}}(s;d)&=\pi_d(\mathcal{C}_s)=\pi_0(\mathcal{C}_{s-d}),\\
K_{\mathbb{R}}(s;d)&=\pi_d(\mathcal{R}_s)=\pi_0(\mathcal{R}_{s-d}),
\end{split}
\end{equation}
where $\pi_d$ is the $d$th homotopy group.

For a unitarized non-Hermitian Hamiltonian $U$, we do not have $U^2=\pm1$ in general. Nevertheless, we can introduce the corresponding Hermitian Hamiltonian
\begin{equation}
H_U\equiv\sigma_+\otimes U+\sigma_-\otimes U^\dag=\begin{bmatrix}
0 & U \\
U^\dag & 0
\end{bmatrix},
\label{HUU}
\end{equation}
which now satisfies $H^2_U=1$. Remarkably, by such construction, we naturally have a chiral symmetry $\Sigma\equiv\sigma_z\otimes 1$ which satisfies $\Sigma^2=1$ and
\begin{equation}
\Sigma H_U=-H_U\Sigma.
\label{HS}
\end{equation}
It has been proved (see, e.g., Appendix D in Ref.~\cite{Roy2017}) that $H_U$ must take the form of Eq.~(\ref{HUU}) if we impose Eq.~(\ref{HS}). Therefore, one can find properties of 
$U$ from those of $H_U$.

\subsection{Explicit classification}
Now let us study how the non-Hermiticity changes the topological classification for each AZ class. We start from the two complex AZ classes A and AIII, which correspond to $\mathcal{C}_0$ and $\mathcal{C}_1$ in the Hermitian case. Due to the emergent chiral symmetry (\ref{HS}), class A is shifted to class AIII, which is characterized by $\pi_d(\mathcal{C}_1)=\mathbb{Z}$ ($0$) for odd (even) $d$. As for class AIII with an intrinsic chiral symmetry $\Gamma$, due to $[\Sigma,
\sigma_0\otimes\Gamma]=0$ ($\sigma_0\equiv1_{2\times2}$), the topological number simply duplicates, i.e., it becomes $\pi_d(\mathcal{C}_1\times\mathcal{C}_1)=\mathbb{Z}\oplus\mathbb{Z}$ ($0$) for odd (even) $d$. 

Let us move on to the real AZ classes with only a single anti-unitary symmetry $A=U_AK$, including AI ($T^2=1$), D ($C^2=1$), AII ($T^2=-1$) and C ($C^2=-1$). By using the fact that
\begin{equation}
\begin{split}
AU=\eta_AUA\;\;\;\;&\Leftrightarrow\;\;\;\;AU^\dag=\eta_AU^\dag A\\
\Leftrightarrow
\begin{bmatrix}
A & 0 \\
0 & A
\end{bmatrix}
\begin{bmatrix}
0 & U \\
U^\dag & 0
\end{bmatrix}
&=\eta_A
\begin{bmatrix}
0 & U \\
U^\dag & 0
\end{bmatrix}
\begin{bmatrix}
A & 0 \\
0 & A
\end{bmatrix},
\end{split}
\end{equation}
we find that the action of an anti-unitary symmetry $\sigma_0
\otimes A$ on $H_U$ is the same as that on $U$. Since $[\sigma_0
\otimes A,\Sigma]=0$, such a chiral symmetry $\Sigma$ implies another anti-unitary symmetry whose square is the same as $A^2$. Therefore, classes AI and D (classes AII and C), which correspond to $\mathcal{R}_0$ and $\mathcal{R}_2$ ($\mathcal{R}_4$ and $\mathcal{R}_6$) in the Hermitian case, are unified into BDI (CII) described by $\mathcal{R}_1$ ($\mathcal{R}_5$).

Finally, let us discuss the AZ classes with two anti-unitary symmetries, including DIII ($T^2=-1,C^2=1$), CI ($T^2=1,C^2=-1$), BDI ($T^2=1,C^2=1$) and CII ($T^2=-1,C^2=-1$). For the former two classes, we can construct $i\Sigma(\sigma_0
\otimes\Gamma)=i\sigma_z\otimes\Gamma$; this operator gives $-1$ upon squaring and commutes with all the elements in the original Clifford algebra excluding $\Sigma$. This implies that DIII and CI, which correspond to $\mathcal{R}_3$ and $\mathcal{R}_7$ in the Hermitian case, are unified into AIII ($\mathcal{C}_1$), since $i\sigma_z\otimes\Gamma$ behaves like a complex unit that changes the real AZ classes into the complex ones \cite{Morimoto2013}. For the latter two classes, we can construct $\Sigma(\sigma_0
\otimes\Gamma)=\sigma_z\otimes\Gamma$; this operator gives $1$ upon squaring and commutes with all the elements in the original Clifford algebra excluding $\Sigma$. This implies that the topological number of classes BDI and CII simply gets doubled, since $\sigma_z\otimes\Gamma$ has two different subspaces of eigenstates with eigenvalues $\pm1$ \cite{Morimoto2013}.

We list all the results in Table~\ref{table1}. To summarize, the effect of non-Hermiticity is equivalent to adding a chiral symmetry that commutes with all the original symmetries. As a result, A, DIII and CI are unified into AIII, AI and D are unified into BDI, AII and C are unified into CII, and AIII, BDI and CII become duplicated.

\subsection{Discussions}
\label{Disc}
A few remarks are in order here. First, the unification of classes AI and D, AII and C as well as that of classes DIII and CI, can be understood as a consequence of the one-to-one mapping between a time-reversal symmetric Hamiltonian and a particle-hole symmetric Hamiltonian which are transformed to each other by simple multiplication of one or the other by $i$ \cite{Kawabata2018b}. Such a unification holds true for very general requirements of continuous deformation other than maintaining invertibility, such as the existence of a complex band gap \cite{Fu2018}. 

Second, despite the fact that the classification of non-Hermitian matrices is equivalent to that of unitary matrices, the periodic table (Table~\ref{table1}) differs significantly from that of Floquet systems \cite{Roy2017}. This is partly \footnote{Another reason is that the full information of $U(\boldsymbol{k},t)=\mathcal{T}e^{-i\int^t_0dt'H(t')}$ from $t=0$ to $t=T$ is important in a Floquet system. A good illustration is the anomalous edge states \cite{Lindner2013}, which exist in spite of a trivial $U(\boldsymbol{k},T)=1$. In contrast, we focus on time-independent non-Hermitian Hamiltonians, so that the base manifold for classification only contains $\boldsymbol{k}$ but not $t$.} due to the different meanings of time-reversal symmetric and particle-hole symmetric operators in the context of Hamiltonians and time-development operators. In the former case, we require $AHA^{-1}=\eta_AH$, while in the latter case we require $AUA^{-1}=U^{-\eta_A}$. 

\begin{figure}
\begin{center}
       \includegraphics[width=8.5cm, clip]{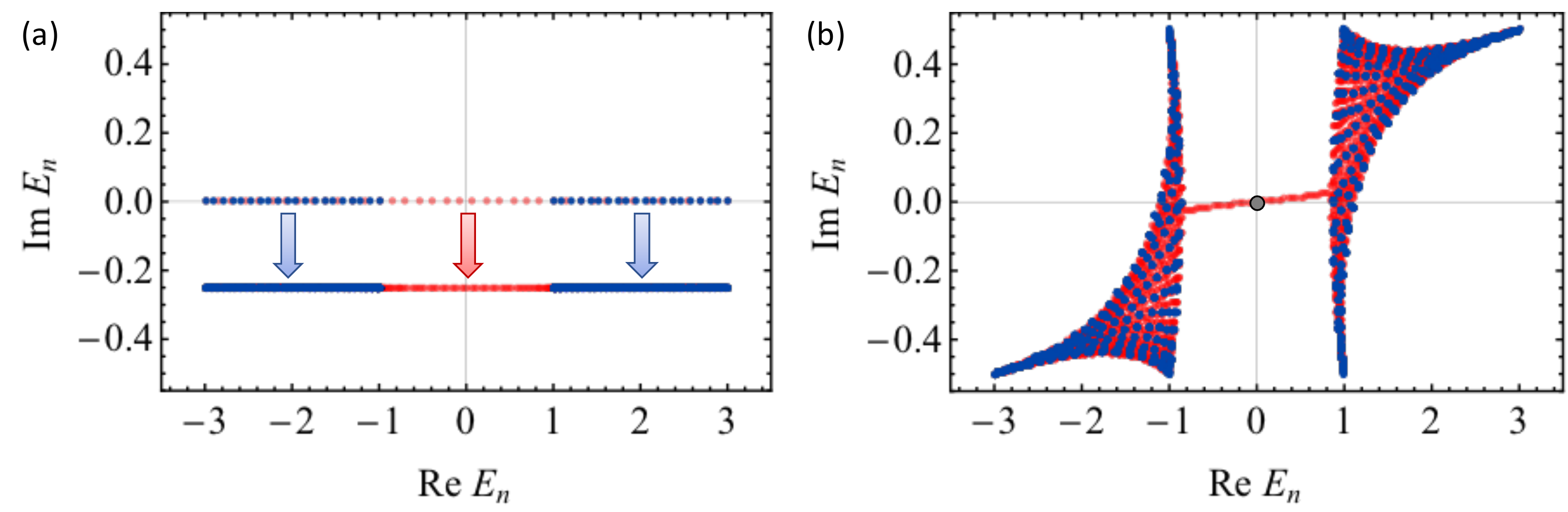}
       \end{center}
   \caption{(a) Spectrum of Eq.~(\ref{HBHZ}) with $(\gamma,m)=(0.25,1)$. The zero mode in the $\gamma=0$ limit (sparse dots) disappears due to the global spectrum shift along the imaginary energy axis (indicated by the arrows). (b) Same as (a) but with $(\gamma,m)=(0,1+0.5i)$. The symmetry constraint given in Eq.~(\ref{uncchiral}) enforces the spectrum to be inversion symmetric, leading to a robust zero mode (grey dot). In both (a) and (b), the blue (red) dots correspond to the periodic (open) boundary condition, and the system size is $40\times40$.}
   \label{fig7}
\end{figure}

Third, a two-dimensional non-Hermitian system turns out to be always trivial in our classification. This does not contradict a recently discovered Chern number for separable non-Hermitian bands \cite{Fu2018}, since all the bands can be deformed to touch each other without hitting a base energy. For example, let us show how to trivialize a Chern insulator without the spectrum touching at the origin (here, we assume $E_{\rm B}=0$). We consider a two-band system
\begin{equation}
\begin{split}
H(k_x,k_y)=&-i\gamma\sigma_0+\sin k_x\sigma_x+\sin k_y\sigma_y\\
&+(m-\cos k_x-\cos k_y)\sigma_z,
\end{split}
\label{HBHZ}
\end{equation}
 and start from $(\gamma,m)=(0,1)$, which describes a Hermitian Chern insulator \cite{Bernevig2013}. We can first gradually introduce a global loss up to, e.g., $\gamma=0.25$ (see Fig.~\ref{fig7}(a)), then  change $m$ into, e.g., $m=3$, and finally remove the global loss by reducing $\gamma$ to zero. It is clear that the origin is not touched by the spectrum of $H(k_x,k_y)$ during the whole process. Such kind of continuous deformation is, however, forbidden in Ref.~\cite{Fu2018}, because a band touching occurs at $m=2$.

 Although the AZ classes are always trivial in two dimensions in our framework, nontrivial topological phases \emph{do} exist in other symmetry classes. For example, by setting $\gamma=0$ in Eq.~(\ref{HBHZ}), we have  
 \begin{equation}
 \sigma_xH(k_x,k_y)\sigma_x=-H(-k_x,k_y)
 \label{uncchiral}
 \end{equation}
even for a complex $m$. With the symmetry constraint in Eq.~(\ref{uncchiral}) alone, we know that the Hermitianized Hamiltonian (\ref{HH}) exhibits not only a chiral symmetry $\Sigma$ but also a \emph{mirror} symmetry (with respect to the $y$ axis) $\sigma_z\otimes\sigma_x$ that commutes with $\Sigma$, leading to a $\mathbb{Z}$ classification \cite{Ryu2013}. In Fig.~\ref{fig7}(b), we plot the spectrum for $m=1+0.5i$ with a nontrivial mirror winding number $1$ \cite{Ryu2013}, and find a mode with \emph{zero} energy under the open boundary condition. Such a zero-energy mode should be robust due to the interplay of a nontrivial non-Hermitian Chern number \cite{Fu2018} and the inversion symmetry of the spectrum enforced by Eq.~(\ref{uncchiral}). This observation, together with the bulk-edge correspondence found in one dimension, suggests that a topologically nontrivial bulk with respect to a base energy $E_{\rm B}$ implies one or more robust edge modes \emph{at $E_{\rm B}$} (or crossing $E_{\rm B}$ upon the change of boundary condition). 
This is much stronger a requirement than the existence of robust edge modes (that may appear anywhere), which can be ensured by a nontrivial non-Hermitian Chern number as discussed in Ref.~\cite{Fu2018}. From this viewpoint, it may not be so incomprehensible that two-dimensional non-Hermitian systems in AZ classes are always trivial --- \emph{these systems may exhibit robust edge modes, but are not expected to exhibit an edge mode at the base energy in general.}

Finally, we again emphasize that weak topological numbers \cite{Kitaev2009} are not shown in Table~\ref{table1}. Indeed, we can define two winding numbers 
\begin{equation}
w_\mu\equiv\int^\pi_{-\pi}\frac{dk_\mu}{2\pi i}\partial_{k_\mu}\ln{\rm det}H(\boldsymbol{k}),\;\;\;\;\mu=x,y
\end{equation}
for any two-dimensional lattices, but they inherit from the lower dimension ($d=1$) and are not genuinely two-dimensional topological invariants. On the other hand, a nontrivial weak topological number can lead to a dramatic change in the spectrum under different boundary conditions, just like the one-dimensional case shown in Fig.~\ref{fig4}(a).

\section{Topological indices for non-Hermitian systems}
\label{EXNH}
In this section, we identify the topological indices and provide some concrete examples for all the nontrivial non-Hermitian AZ classes in zero and one dimensions.

\subsection{Zero dimension}
\label{ZD}
According to the K-theory classification (see Table \ref{table1}), if we impose either time-reversal or (and) particle-hole symmetry, we obtain two (four) types of topologically different matrices. Since a matrix of class BDI is made from two independent matrices of class AI (or D), it suffices to focus on a single $\mathbb{Z}_2$ topological number. Furthermore, class AI and class D can be mapped into each other by simply multiplying the imaginary unit $i$ \cite{Kawabata2018b}; therefore we will primarily discuss the case of class AI without loss of generality.

\begin{figure}
\begin{center}
       \includegraphics[width=8.5cm, clip]{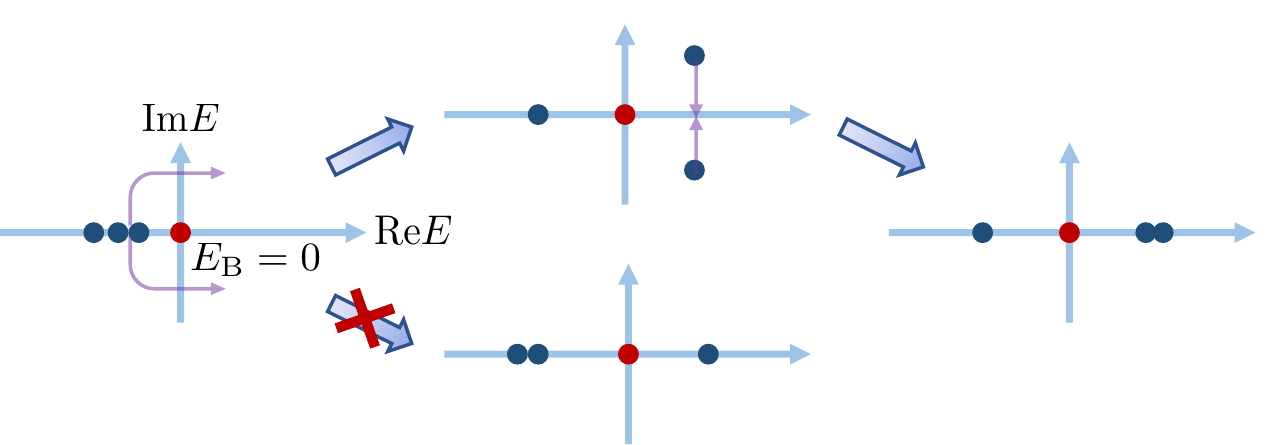}
       \end{center}
   \caption{Spectrum deformation in a class AI system described by a $3\times3$ matrix in zero dimension. The spectrum is always symmetric with respect to the real axis. Without touching $E_{\rm B}=0$, the number of eigenvalues on the negative or positive real axis can only change by an even number, so a $\mathbb{Z}_2$ index ($s=-1$) can be defined as in Eq.~(\ref{AIZ2}).}
   \label{fig8}
\end{figure}

Note that an involutory ($T^2=1$)  time-reversal symmetry can always be represented as $T=K$ in an appropriate basis \cite{Wigner1960}, under which all the time-reversal symmetric matrices are real. In this case, the polar decomposition becomes $H=OR$, where $O$ is orthogonal and $R$ is real, symmetric and positive-definite. Since $H\simeq O$, we conclude that the $\mathbb{Z}_2$ topological number characterizes the two disconnected sectors of an orthogonal group. In terms of $H$, this topological number can be defined as
\begin{equation}
s\equiv{\rm sgn}({\rm det}\;H),
\label{AIZ2}
\end{equation}
which takes on $1$ ($-1$) if there is an even (odd) number of eigenvalues on the negative real axis (see Fig.~\ref{fig8}). Using the correspondence between classes AI and D, the $\mathbb{Z}_2$ index of a particle-hole symmetric Hamiltonian can be defined as
\begin{equation}
s'\equiv{\rm sgn}({\rm det}\;iH),
\label{DZ2}
\end{equation}
which takes on $1$ ($-1$) if there is an even (odd) number of eigenvalues on the positive imaginary axis.

\subsubsection{$PT$-symmetric systems}
Remarkably, in the sense of Eq.~(\ref{AIZ2}) (Eq.~(\ref{DZ2})), a $PT$-symmetry-breaking (an anti-$PT$-symmetry-breaking \cite{Xiao2016}) transition across an exceptional point can be identified as a topological transition. While the $PT$ symmetry physically differs from the $T$ symmetry, as long as the symmetry operator is involutory and anti-unitary, the topological classification in zero dimension is the same as class AI. Note that the classification differs in higher dimensions (see Table~\ref{table2} and Appendix~\ref{CAPT}). As a minimal example, we consider a non-Hermitian two-level system \cite{Segev2010}
\begin{equation}
H=\Omega\sigma_x+i\gamma\sigma_z,\;\;\;\;
\Omega,\gamma\in\mathbb{R},
\label{TLSPT}
\end{equation}
which features a $PT$ symmetry $\sigma_xK$. It is easy to check that ${\rm det}H=\gamma^2-\Omega^2$ and thus $s=-1$ ($s=1$) in the $PT$-unbroken ($PT$-broken) phase. A topological transition with anti-$PT$-symmetry breaking (class D) can similarly be constructed by multiplying Eq.~(\ref{TLSPT}) by $i$. 

At first glance, the conclusion that a $PT$-symmetry breaking transition is topological seems rather odd, since in Hermitian systems the concept of SPT is complementary to spontaneous symmetry breaking. As for non-Hermitian systems, this is possible due to the conceptual difference in defining topological phases as dynamical phases instead of states of matter, so that the eigenstates do not necessarily respect the symmetry. In particular, the $\mathbb{Z}_2$ topological number (\ref{AIZ2}) for class AI in zero dimension is solely determined by the energy spectrum. The emergence of $E$ and $E^*$ is indeed topologically forbidden if they originate from two real energies with opposite signs. This is because in $PT$-symmetric systems a pair of complex conjugate eigenvalues emerges when two real eigenenergies coalesce; if these real eigenenergies have opposite signs, they have no alternative but to meet at the origin which, however, is forbidden by our assumption. Now the sign of the product of the two eigenvalues, which gives the $\mathbb{Z}_2$ index in Eq.~(\ref{AIZ2}), is negative before the $PT$-symmetry breaking and positive after it. Thus the $PT$ transition is topologically forbidden unless the origin is touched.

\subsubsection{Quantum channels}
\label{QC}
Another important example is \emph{quantum channels} or completely positive (CP) and trace-preserving (TP) maps. A CPTP map always has a Kraus representation \cite{Kraus1971}
\begin{equation}
\mathcal{E}(\rho)=\sum_\alpha K_\alpha\rho K^\dag_\alpha,
\label{Kraus}
\end{equation}
where the Kraus operators $K_\alpha$ satisfy $\sum_\alpha K^\dag_\alpha K_\alpha=I$. Alternatively, $\mathcal{E}$ can be represented as an enlarged non-Hermitian matrix $\mathcal{E}=\sum_\alpha K_\alpha\otimes K^*_\alpha$ on the Liouville space $\mathcal{V}\equiv V\otimes V^*$. Remarkably, defining $\mathcal{K}(\rho)\equiv\rho^\dag$ as the Hermitian-conjugate superoperator, which is anti-unitary \footnote{This should be understood with respect to the Hilbert-Schmidt inner product $(A,B)\equiv{\rm Tr}[A^\dag B]$. We can check that $(\mathcal{K}A,\mathcal{K}B)={\rm Tr}[AB^\dag]={\rm Tr}[B^\dag A]=(B,A)$.} and involutory ($\mathcal{K}^2(\rho)=\rho$), we have
\begin{equation}
\mathcal{E}\mathcal{K}(\rho)=\mathcal{K}\mathcal{E}(\rho)=\sum_\alpha K_\alpha\rho^\dag K^\dag_\alpha,
\label{EKKE}
\end{equation}
which is actually the Hermiticity-preserving property of $\mathcal{E}$ \cite{Haake2010}. Such an inherent symmetry is absolutely robust, unlike the $PT$ symmetry which can hardly be exact due to experimental imperfection. Therefore, a CPTP map $\mathcal{E}$ always belongs to the AI class and is classified by a $\mathbb{Z}_2$ topological index, determined by the sign of ${\rm det}\mathcal{E}\in\mathbb{R}$. We note that the same classification applies to a CP map, which can also be represented by Eq.~(\ref{Kraus}) with no constraints on $K_\alpha$'s. With the TP property imposed, the eigenvalues of $\mathcal{E}$ are enforced to be on or inside the unit circle in the complex plane \cite{Wolf2008}.

It is natural to define a trivial map if it is connected to the identity channel $\mathcal{I}$. It follows that $\mathcal{E}$ is trivial as long as ${\rm det}\mathcal{E}>0$. In this sense, each invertible quantum dynamical map $\Phi_t$ is trivial since $\Phi_t$ can continuously be deformed into $\Phi_0=\mathcal{I}$, irrespective of whether $\Phi_t$ is Markovian or not \cite{Breuer2016}. Conversely, we can conclude that a topologically nontrivial quantum channel with ${\rm det}\mathcal{E}<0$ can never be continuously generated by a Markovian dynamics. It is nevertheless easy to construct a nontrivial channel via random unitary circuits which take the form $\mathcal{E}(\rho)=\sum_j p_j U_j\rho U^\dag_j$ with $\sum_j p_j=1$. A prototypical example is the isotropic depolarization channel for a single qubit \cite{Chuang2010}:
\begin{equation}
\mathcal{E}_{\rm d}(\rho)=p\rho+\frac{1-p}{3}\sum_{\mu=x,y,z}\sigma_\mu\rho\sigma_\mu,
\end{equation}
whose extension $\mathcal{E}_{\rm d}\otimes\mathcal{I}$ has widely been used to introduce imperfection into a maximally entangled qubit pair \cite{Bennett1996}. We can check that ${\rm det}\mathcal{E}_{\rm d}=(\frac{4p-1}{3})^3$, so that a topological transition occurs at $p=\frac{1}{4}$, where the channel becomes a constant (fully depolarized) map $\mathcal{E}_{\rm d}(\rho)=\frac{\sigma_0}{2}$.

If the quantum channel plays a role of a Floquet superoperator for a periodically driven open system \cite{Gong2018}, the stroboscopic evolution is governed by $\rho_{(n+1)T}=\mathcal{E}(\rho_{nT})$, where $T$ is the driving period. If we look at the long-time dynamics, the topological index ${\rm sgn( det}\mathcal{E})$ might become meaningless since only the long-lived modes with eigenvalues with nearly unit norm are relevant. Denoting the superprojector onto such a metastable manifold $\mathcal{V}_{\rm ms}$ as $\mathcal{P}$, which can always be made Hermiticity-preserving \cite{Lesanovsky2016}, we expect the sign of ${\rm det}_{\mathcal{V}_{\rm ms}}\mathcal{PEP}$, denoted by $s_{\rm ms}$, to be important for the long-time dynamics. If $s_{\rm ms}=-1$, there must be an odd number of long-lived modes near $-1$. When the system is perturbed, we expect that at least one long-lived mode stays on the real axis near $-1$. This cannot be ensured by $s_{\rm ms}=1$, since all the long-lived modes near $-1$ can leave the real axis in a pairwise manner. The above discussion is parallel to the $\mathbb{Z}_2$ topological insulators \cite{Kane2005,Bernevig2006,Molenkamp2007}, on the surface of which at least one Dirac cone survives under time-reversal symmetric perturbations.

\begin{figure}
\begin{center}
       \includegraphics[width=8.5cm, clip]{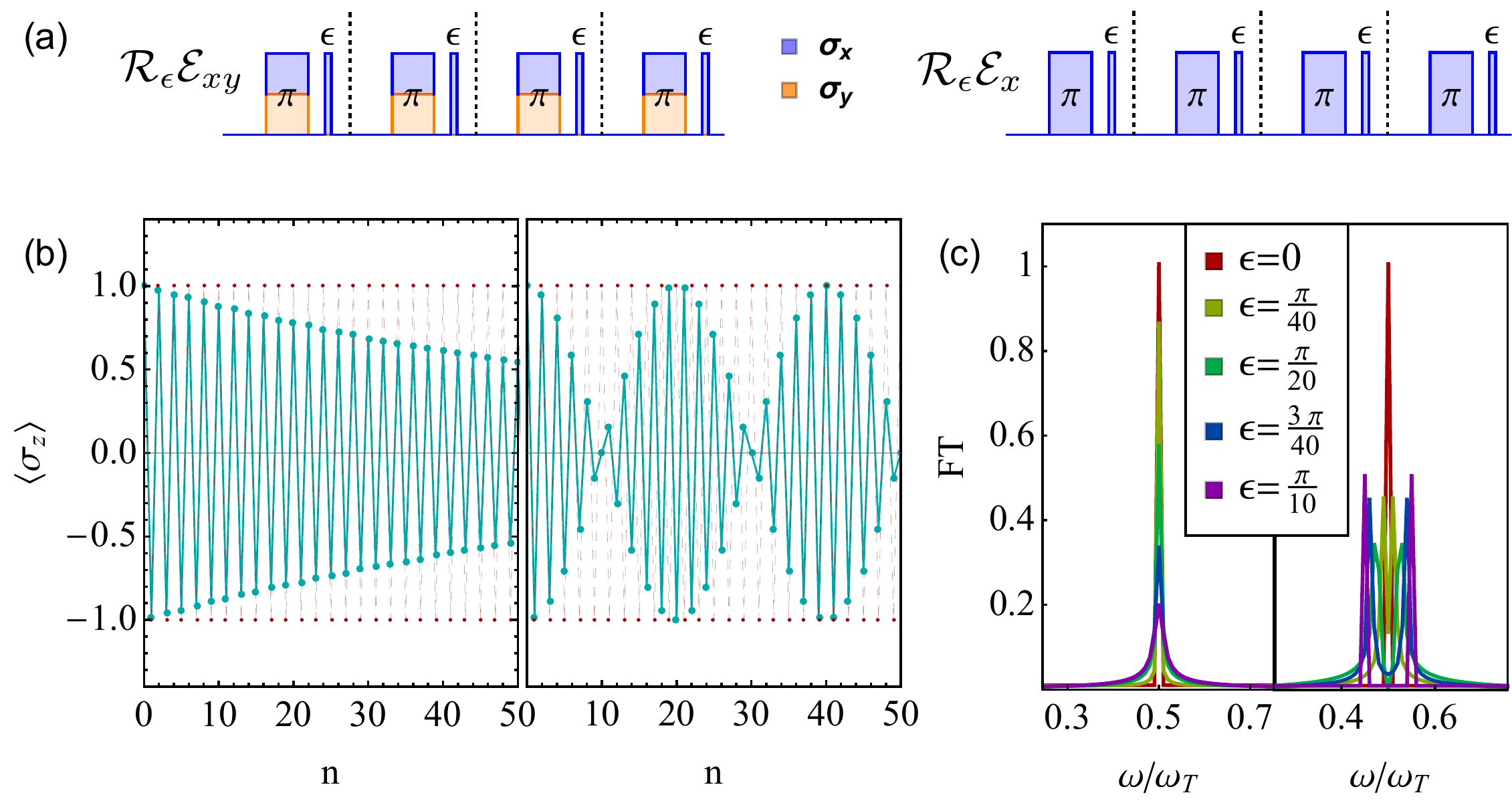}
       \end{center}
   \caption{(a) Pulse sequence of the stroboscopic qubit dynamics governed by two types of operations $\mathcal{R}_\epsilon\mathcal{E}_{xy}$ and $\mathcal{R}_\epsilon\mathcal{E}_x$. In the former case, $\pi$-pulses are applied randomly in the $x$ and $y$ directions with equal probability, leading to $s_{\rm ms}=-1$. In the latter case, $\pi$-pulses are applied in the $x$ direction, leading to $s_{\rm ms}=1$. (b) Starting from $\rho_0=\ket{\uparrow}\bra{\uparrow}$, the dynamics of $\langle\sigma_z\rangle$ for $\epsilon=0$ (red dots) are the same between the two cases. As for $\epsilon=0.05\pi$ (green dots), the dynamics governed by $\mathcal{R}_\epsilon\mathcal{E}_{xy}$ (left) exhibit a discrete time-crystalline-like behavior \cite{Yao2017,Zhang2017,Choi2017}, but the dynamics governed by $\mathcal{R}_\epsilon\mathcal{E}_x$ do not. (c) Fourier transform of $\langle\sigma_z\rangle_{t=nT}$ into the frequency domain. The single peak located at $\omega=0.5\omega_T$ ($\omega_T\equiv\frac{2\pi}{T}$) stays robust for $\mathcal{R}_\epsilon\mathcal{E}_{xy}$ (left), but splits into two peaks for $\mathcal{R}_\epsilon\mathcal{E}_x$ (right).}
   \label{fig9}
\end{figure}

As a minimal illustration, let us consider a critical (zero full determinant) quantum channel 
\begin{equation}
\mathcal{E}_{xy}(\rho)=\frac{1}{2}(\sigma_x\rho\sigma_x+\sigma_y\rho\sigma_y),
\label{Exy}
\end{equation}
which has a single long-lived mode $\sigma_z$ with eigenvalue $-1$ in addition to the steady state $\frac{\sigma_0}{2}$, so that $s_{\rm ms}=-1$. Starting from $\ket{\uparrow}$, we find an antiferromagnetic ($\uparrow\downarrow\uparrow\downarrow...$) stroboscopic dynamics (see red dots in Fig.~\ref{fig9}(b)). The same dynamics can be realized by unitary $\pi$-rotation along the $x$ axis, i.e., $\mathcal{E}_x(\rho)=\sigma_x\rho\sigma_x$, which has two modes with eigenvalues $-1$ so that $s_{\rm ms}=s=1$. Now let us disturb the temporal antiferromagnetic pattern by inserting a sudden pulse $\mathcal{R}_\epsilon(\rho)=e^{-i\pi\epsilon\sigma_x}\rho e^{i\pi\epsilon\sigma_x}$ at the end of each evolution period (see Fig.~\ref{fig9}(a)). As clearly shown by the Fourier transform of $\langle\sigma_z\rangle_{t=nT}$ in Fig.~\ref{fig9}(c), the antiferromagnetic pattern is robust against perturbation to $\mathcal{E}_{xy}$ 
with $s_{\rm ms}=-1$, but is fragile for $\mathcal{E}_x$ with $s_{\rm ms}=1$. This observation is reminiscent of discrete time crystals \cite{Khemani2016,Keyserlingk2016b,Nayak2016,Yao2017,Zhang2017,Choi2017}, which are Floquet systems that spontaneously break the discrete time-translation symmetry. Akin to intrinsic topological order \cite{Kitaev2003}, long-range entanglement has been identified as the origin of the rigidity of unitary discrete time crystals in one dimension \cite{Keyserlingk2016}. It would be interesting to study whether a nontrivial $\mathbb{Z}_2$ topological index, which emerges from the inherent time-reversal-like symmetry (\ref{EKKE}), can lead to the absolute rigidity of a dissipative discrete time crystal in zero dimension \cite{Gong2018}.

\subsection{One dimension}
We discuss the general structures of non-Hermitian Hamiltonians in one dimension and the corresponding topological numbers in addition to class A. 

For class DIII (CI), we can always find a basis under which $\Gamma=\sigma_z\otimes1$ and $T=\sigma_x\otimes i\sigma_y K$ ($C=\sigma_x\otimes i\sigma_y K$). The symmetry requirements $\Gamma H(k)=-H(k)\Gamma$ and $TH(k)=H(-k)T$ ($CH(k)=-H(-k)C$) lead to the following general form of the Hamiltonian:
\begin{equation}
H(k)=\begin{bmatrix} 0 & h(k) \\ \pm\sigma_yh^*(-k)\sigma_y & 0 \end{bmatrix},
\end{equation} 
where $h(k)$ can be an arbitrary invertible matrix, and $+$ and $-$ correspond to class DIII and CI, respectively. Due to the arbitrariness of $h(k)$, the topological classification coincides with class A and the topological number is determined by $w_h\in\mathbb{Z}$, i.e., the winding number of $h(k)$.

\begin{figure}
\begin{center}
       \includegraphics[width=7cm, clip]{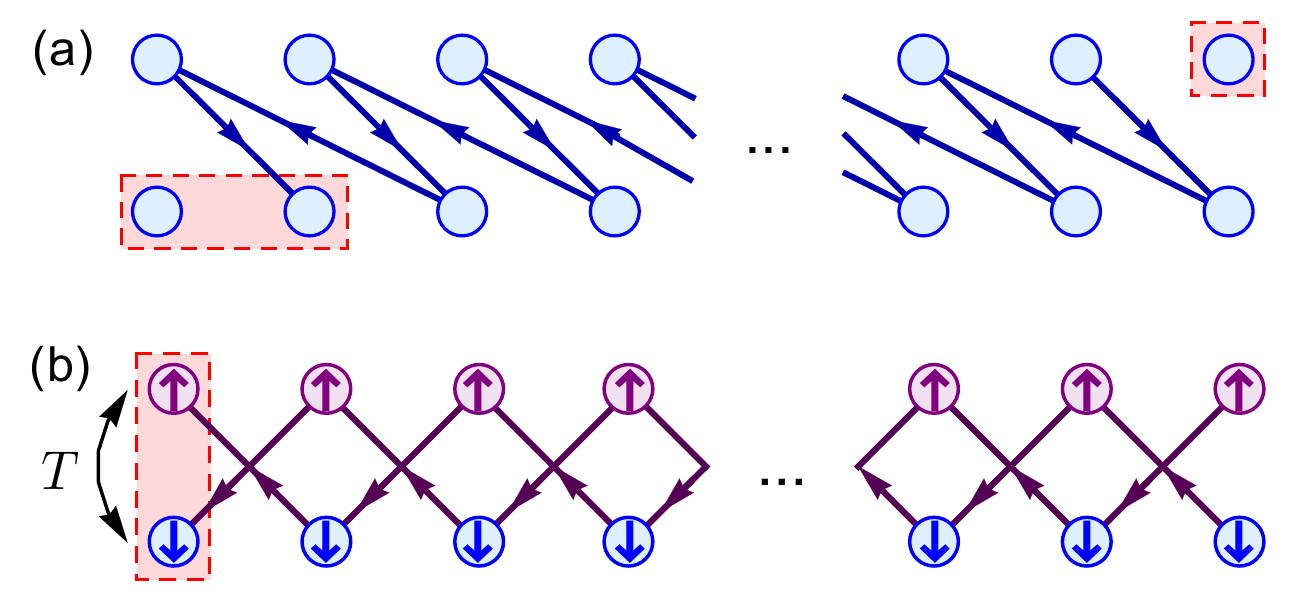}
       \end{center}
   \caption{Non-Hermitian open chains with unidirectional hoppings (indicated by the arrows) belonging to (a) class AIII and (b) class AII. In (a), the number of zero modes at the left boundary (enclosed by a red rectangle) is not the same as that on the right boundary. In (b), the zero modes form Kramers pairs, which interchange via the time-reversal operator $T$, and therefore the total number of the edge modes must be even.}
   \label{fig10}
\end{figure}

For class AIII, we can always find a basis under which $\Gamma=\sigma_z\otimes1$. The general form of the Hamiltonian reads
\begin{equation}
H(k)=\begin{bmatrix} 0 & h_1(k) \\ h_2(k) & 0 \end{bmatrix},
\label{HGa}
\end{equation} 
with $h_{1,2}(k)$ being arbitrary invertible matrices. Note that there are two independent winding numbers $w_{h_1}$ and $w_{h_2}$ in accordance with the classification $\mathbb{Z}\oplus\mathbb{Z}$. We can generally have $w_{h_2}\neq-w_{h_1}$, implying different numbers of (quasi-)zero modes localized at the two open boundaries. As shown in Fig.~\ref{fig10}(a), a two-band model with $h_1(k)=J_1e^{2ik}$ and $h_2(k)=J_2e^{-ik}$ in Eq.~(\ref{HGa}) has two and one zero modes at the left and right boundaries, respectively, as a consequence of asymmetric hopping amplitudes.  It is interesting to note that for the Hermitian case the non-Hermitian $\mathbb{Z}\oplus\mathbb{Z}$ group degenerates into its subset $\{(n,-n):n\in\mathbb{Z}\}$ due to the Hermitian constraint ($w_{h_2}=-w_{h_1}$), which is nothing but the $\mathbb{Z}$ classification of class AIII. It is worth mentioning that the Hamiltonian studied in Ref.~\cite{Tony2016}, which can be expressed as $H(k)=(v+r\cos k)\sigma_x+r(\sin k-i)\sigma_y$ ($v,r\in\mathbb{R}$), gives an example of the two generators of $\mathbb{Z}\oplus\mathbb{Z}$ by taking $0<\frac{v}{r}<2$ and $-2<\frac{v}{r}<0$. The $\frac{1}{2}\mathbb{Z}$ topological number identified therein turns out to be $\frac{1}{2}(w_{h_1}-w_{h_2})$, which can be a half-integer only if the system is non-Hermitian.

For class AI (D), we can always find a basis under which $T=K$ ($C=K$), so that $H^*(k)=H(-k)$ ($H^*(k)=-H(-k)$). This requirement enforces the matrix elements of $H(k)$ to be $\sum_{n\in\mathbb{Z}} c_ne^{ink}$, with $c_n$'s being real (purely imaginary) numbers, yet the winding number of $H(k)$ does run over $\mathbb{Z}$. All the different topological phases can be realized in a single-band model $H(k)=e^{ink}$ ($H(k)=ie^{ink}$) with $n\in\mathbb{Z}$.

For class BDI, we can always find a basis under which $\Gamma=\sigma_z$, $T=K$ and $C=\sigma_zK$. The general form of the Hamiltonian is again given by Eq.~(\ref{HGa}), but $h^*_{1,2}(k)=h_{1,2}(-k)$ is required. Similar to class AIII, we have two independent winding numbers $w_{h_1}$ and $w_{h_2}$ and the topological classification is $\mathbb{Z}\oplus\mathbb{Z}$. 

For class AII (C), we can always find a basis under which $T=i\sigma_y K$ ($C=i\sigma_y K$), so that $\sigma_yH^*(k)=H(-k)\sigma_y$ ($\sigma_yH^*(k)=-H(-k)\sigma_y$). This symmetry requirement restricts the form of the Hamiltonian to be
\begin{equation}
H(k)=\begin{bmatrix} h_1(k) & h_2(k) \\ \mp h^*_2(-k) & \pm h^*_1(-k) \end{bmatrix},
\label{HAC}
\end{equation}
where $h_1(k)$ and $h_2(k)$ can be arbitrary (but $H(k)$ should be invertible afterall) and the upper (lower) signs correspond to class AII (C). In this case, we can prove that the winding number of a Hamiltonian must be even (see Appendix~\ref{even}), as indicated by the $2\mathbb{Z}$ classification. An important physical implication is that there must be an even number of (quasi)-edge modes, which actually form Kramers pairs. In Fig.~\ref{fig10}(b), we present a minimal model of spin-$\frac{1}{2}$ fermions with $h_1(k)=0$ and $h_2(k)=Je^{ik}$ in Eq.~(\ref{HAC}).
 
For class CII, we can always find a basis under which $\Gamma=\sigma_z\otimes1$ and $T=\sigma_0\otimes i\sigma_y K$ ($C=\sigma_z\otimes i\sigma_y K$). The general form of the Hamiltonian in this case is again given by Eq.~(\ref{HGa}), but with $\sigma_yh^*_{1,2}(k)=h_{1,2}(-k)\sigma_y$, namely, both $h_1(k)$ and $h_2(k)$ belong to class AII. The topological characterization is thus given by two even integers $w_{h_1}$ and $w_{h_2}$, consistent with the $2\mathbb{Z}\oplus2\mathbb{Z}$ classification.


\section{Conclusion and outlook}
\label{CO}

In summary, we have established a fundamental framework for a systematic study of topological non-Hermitian systems. The two guiding principles are a dynamical viewpoint on topological systems and the constraint such that the energy spectrum neither touches nor crosses the base point. We have studied one-dimensional non-Hermitian lattices belonging to class A in details, identified the topological winding number, demonstrated the robustness against disorder, unveiled an exotic bulk-edge correspondence, and discussed the experimental relevance. We have given a systematic classification based on K-theory and obtained the periodic table (Table~\ref{table1}) for non-Hermitian AZ classes. All the nontrivial classes in zero dimension and one dimension have been exemplified. 

Our work opens up many possibilities for future studies. Even if we confine ourselves to non-Hermitian AZ classes, physical properties of topological phases in three dimensions are yet to be explored, though the formal classifications have been worked out. For class A, we expect the $\mathbb{Z}$ winding number to be given by
\begin{equation}
w_{\rm 3D}=\int_{\rm B.Z.}\frac{d^3\boldsymbol{k}}{24\pi^2}\epsilon_{\mu\nu\sigma}{\rm Tr}[Q_\mu (\boldsymbol{k})Q_\nu(\boldsymbol{k})Q_\sigma(\boldsymbol{k})],
\label{w3D}
\end{equation}
where $Q_\mu(\boldsymbol{k})=H^{-1}(\boldsymbol{k})\partial_{k_\mu}H(\boldsymbol{k})$. Such an expectation is based on the fact that Eq.~(\ref{w3D}) gives the winding number for a three-dimensional Hermitian system belonging to class AIII, if $H(\boldsymbol{k})$ is the off-diagonal block of the entire Hamiltonian \cite{Ryu2008}. It follows from Eq.~(\ref{w3D}) that, once two components of $Q_\mu(\boldsymbol{k})$ commute, $w_{\rm 3D}$ vanishes. This rules out the possibility for a nontrivial system with a single band, in stark contrast to the one-dimensional case. Since the noncommutativity between  $Q_\mu(\boldsymbol{k})$'s is essential for a nonzero $w_{\rm 3D}$, not only the spectrum but also the eigenstates become important. It would be interesting to explore the edge physics and dynamical response in such a system with nonzero $w_{\rm 3D}$. We also note that the topological phases in four dimensions can be realized by using the time direction \cite{Zilberberg2018,Bloch2018} or the synthetic dimension \cite{Goldman2015}; thus they are also physically relevant.

\begin{table}[tbp]
\caption{Topological classification of $PT$-symmetric systems without other symmetries.}
\begin{center}
\begin{tabular}{ccccccccc}
\hline\hline
\;\;$d$\;\; & \;\;\;0\;\;\; & \;\;\;1\;\;\; & \;\;\;2\;\;\; & \;\;\;3\;\;\; & \;\;\;4\;\;\; & \;\;\;5\;\;\; & \;\;\;6\;\;\; & \;\;\;7\;\;\; \\
\hline
$\pi_0(\mathcal{R}_{d+1})$ & $\mathbb{Z}_2$ & $\mathbb{Z}_2$ & 0 & $2\mathbb{Z}$ & 0 & 0 & 0 & $\mathbb{Z}$\\
\hline\hline
\end{tabular}
\end{center}
\label{table2}
\end{table}

Compatible with the K-theory, our framework can readily be extended to including crystalline symmetries \cite{Morimoto2013,Shiozaki2014}. An important class is $PT$-symmetric systems, whose Bloch Hamiltonians satisfy
\begin{equation}
PTH(\boldsymbol{k})=H(\boldsymbol{k})PT,
\label{PTH}
\end{equation}
with $PT$ being anti-unitary and involutory. Unless the spatial dimension is zero (as discussed in Sec.~\ref{ZD}), Eq.~(\ref{PTH}) differs from the time-reversal symmetry $TH(\boldsymbol{k})=H(-\boldsymbol{k})T$ in the sense that the sign of $\boldsymbol{k}$ is not inverted. As shown in Table~\ref{table2}, we have obtained the complete classification for $PT$-symmetric systems without any other symmetries (see Appendix~\ref{CAPT} for details). In particular, we have a $\mathbb{Z}_2$ classification in one dimension. Dramatic changes in classification are expected when additional symmetries are imposed. We also recall that crystalline symmetries open up the possibilities for exploring topological phases of non-Hermitian systems in two dimensions. Indeed, we have already provided such an example in Sec.~\ref{Disc}.

We can also modify the setup to perform a systematic classification for nonunitary quantum walks, as mentioned in Sec.~\ref{KTC}. Moreover, in analogy with Hermitian systems for which the K-theory approach has been applied to classify bulk-gapless topological phases \cite{Shiozaki2017}, our framework has a potential to be generalized to non-Hermitian systems with exceptional points in the bulk \cite{Malzard2015,Tony2016,Nori2017,Duan2017}. We can even go beyond the K-theory classification to seek for homotopically distinguishable (like the Hopf insulator \cite{Moore2008}) non-Hermitian topological phases with a definite Hilbert-space dimension as exemplified in Appendix~\ref{CAPT}. Last but not the least, it could be an intriguing theoretical issue to consider the topological characterization for interacting many-body non-Hermitian systems \cite{Ashida2016,Ashida2017a}, which are expected to be accessible in near-future atomic, molecular and optical experiments in light of the rapid development in reservoir engineering \cite{Muller2012}.

\acknowledgements
We thank M. Nakagawa, R. Hamazaki and S. Furukawa for valuable discussions. This work was supported by KAKENHI Grant No. JP26287088 from the Japan Society for the Promotion of Science, a Grant-in-Aid for Scientic Research on Innovative Areas ``Topological Materials Science" (KAKENHI Grant No. JP15H05855), and the Photon Frontier Network Program from MEXT of Japan, and the Mitsubishi Foundation. Z. G. was supported by MEXT.  Y. A., K. K. and S. H. were supported by the JSPS through Program for Leading Graduate Schools (ALPS). Y. A., K.T. and S. H. acknowledge support from JSPS (Grants No. JP16J03613, No. JP16J05078 and No. JP16J03619).

\appendix

\section{Consistency between the winding-number expressions}
\label{WNE}
We first show that Eq.~(\ref{wH2}) is equivalent to Eq.~(\ref{wH1}). For this purpose, it suffices to show the following identity for an invertible matrix with a single parameter:
\begin{equation}
\partial_k\ln{\;\rm det}H(k)={\rm Tr}[H^{-1}(k)\partial_kH(k)].
\label{HdetT}
\end{equation}
By definition, the left-hand side of Eq.~(\ref{HdetT}) reads
\begin{equation}
\partial_k\ln{\;\rm det}H(k)\equiv\lim_{\epsilon\to0}\frac{\ln{\;\rm det}H(k+\epsilon)-\ln{\;\rm det}H(k)}{\epsilon}.
\label{detHk}
\end{equation}
Since only the leading-order term ($O(\epsilon)$) survives in the numerator, we can approximate ${\rm det}H(k+\epsilon)$ as
\begin{equation}
\begin{split}
&\;\;\;\;\;{\rm det}[H(k)+\epsilon\partial_kH(k)]+O(\epsilon^2)\\
&={\rm det}H(k){\rm det}[I+\epsilon H^{-1}(k)\partial_kH(k)]+O(\epsilon^2)\\
&={\rm det}H(k)(1+\epsilon{\rm Tr}[H^{-1}(k)\partial_kH(k)])+O(\epsilon^2).
\end{split}
\label{Hke}
\end{equation}
Substituting the last expression in Eq.~(\ref{Hke}) into Eq.~(\ref{detHk}) and using $\ln(1+x)=x+O(x^2)$, we obtain Eq.~(\ref{HdetT}).

We then show that Eq.~(\ref{wPhi}) reproduces Eq.~(\ref{wH1}) if the translation invariance is imposed. In the quasi-momentum representation, the entire Hamiltonian $H_{\rm tot}$ with flux $\Phi$ can be block-diagonalized as
\begin{equation}
H_{\rm tot}(\Phi)=\bigoplus_{k=\frac{2j\pi}{L}-\pi} H\left(k+\frac{\Phi}{L}\right),
\end{equation}
which leads to 
\begin{equation}
\ln{\rm det}H_{\rm tot}(\Phi)=\sum_{k=\frac{2j\pi}{L}-\pi}\ln{\rm det}H\left(k+\frac{\Phi}{L}\right).
\end{equation}
Therefore, we have
\begin{equation}
\begin{split}
&\;\;\;\;\;\int^{2\pi}_0\frac{d\Phi}{2\pi}\partial_\Phi\ln{\rm det}H_{\rm tot}(\Phi)\\
&=\sum_{k=\frac{2j\pi}{L}-\pi}\int^{2\pi}_0\frac{d\Phi}{2\pi L}\partial_k\ln{\rm det}H\left(k+\frac{\Phi}{L}\right)\\
&=\sum^{L-1}_{j=0}\int^{\frac{2(j+1)\pi}{L}-\pi}_{\frac{2j\pi}{L}-\pi}\frac{d\phi}{2\pi}\partial_k\ln{\rm det}H(k+\phi)\\
&=\int^\pi_{-\pi}\frac{dk}{2\pi}\partial_k\ln{\rm det}H(k).
\end{split}
\end{equation}

\begin{figure}
\begin{center}
       \includegraphics[width=8cm, clip]{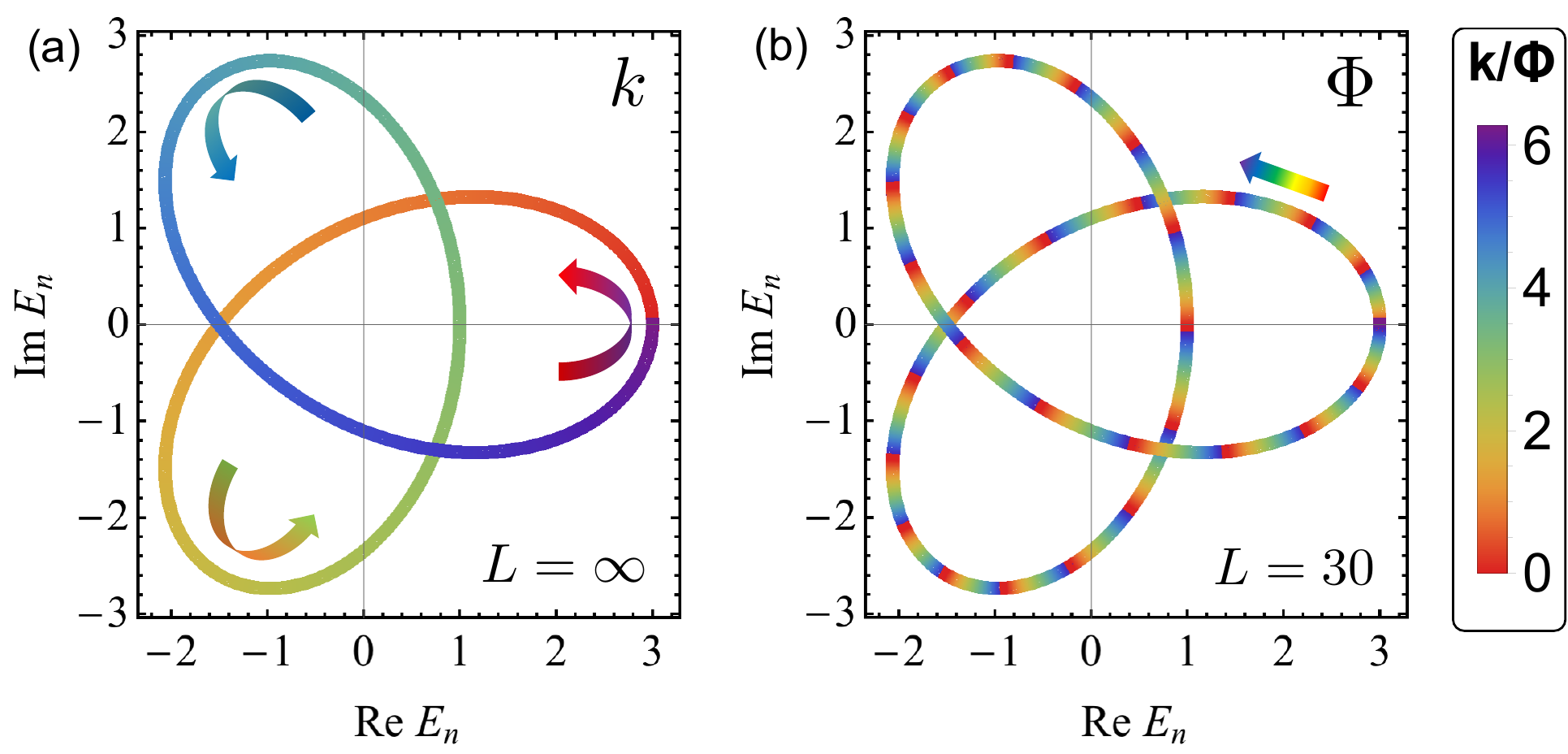}
       \end{center}
   \caption{(a) Energy spectrum of an infinite translation-invariant lattice described by Eq.~(\ref{J12}). The arrows indicate the flow of eigenenergy as the wave vector $k$ increases from $0$ to $2\pi$. (b) The same as in (a) but for a finite ($L=30$) ring subjected to a flux $\Phi$. The arrow indicates the spectral flow as $\Phi$ changes from $0$ to $2\pi$.}
   \label{fig11}
\end{figure}

It is instructive to illustrate the equivalence between the $k$-based and $\Phi$-based winding numbers in a concrete model, such as
\begin{equation}
H=\sum_j(J_1c^\dag_{j+1}c_j+J_2c^\dag_{j-1}c_{j+1})
\label{J12}
\end{equation}
with $J_1=1$ and $J_2=2$. According to the dispersion relation $H(k)=J_1e^{-ik}+J_2e^{2ik}$, it is easy to know that ${\rm det}H(k)$ encircles the origin twice when $k$ runs over the Brillouin zone, as shown in Fig.~\ref{fig11}(a). Note that a given $k$ corresponds to a single eigenenergy since there is only a single band. On the other hand, for a finite ring with length $L$ and subjected to a flux $\Phi$, the Hamiltonian becomes $H_{\rm tot}(\Phi)=\sum_j(e^{-i\frac{\Phi}{L}}J_1c^\dag_{j+1}c_j+e^{2i\frac{\Phi}{L}}J_2c^\dag_{j-1}c_{j+1})$, where a given $\Phi$ corresponds to $L$ eigenenergies that form a discretized configuration of the continuous curve $H(k)$ (see Fig.~\ref{fig11}(b)). When $\Phi$ increases from $0$ to $2\pi$, the spectrum of $H_{\rm tot}(\Phi)$ returns to itself and the trajectory exactly generates the energy spectrum in the thermodynamic limit in a counterclockwise manner, leading to the same winding number $w=2$.

\section{Further details on the Hatano-Nelson model}
In this appendix, we explain in details how the topological transition is related to the Anderson transition, and provide some quantitative results.

\subsection{Spectral flow and localization}
\label{SFL}
The topological interpretation of the Anderson transition in the Hatano-Nelson model is based on the intuition that a fully localized system is topologically trivial. Here, we justify this statement from the viewpoint of the potential-gradient response of wave functions.

To judge whether an eigenstate is localized, we can either look at its static properties such as the real-space profile, or the dynamical properties such as the response to a potential gradient. Here we apply the latter which turns out to work well even in small systems. For an open chain with length $L$ and described by the Hamiltonian $H=\sum_{j,l}J_{jl}c^\dag_jc_l$ subject to a perturbation $\delta H=-\frac{V}{L}\sum_j jc^\dag_jc_j$, starting from an eigenstate $|\varphi_0\rangle$ of $H$ and assuming the adiabaticity, the normalized wave function $|\psi_t\rangle$ at time $t$ can well be approximated by $e^{-i\delta H t}|\varphi_t\rangle$, with $|\varphi_t\rangle$ being the eigenstate of $H(t)\equiv e^{i\delta H t}He^{-i\delta H t}=\sum_{j,l}J_{jl}e^{-i\frac{Vt}{L}(j-l)}c^\dag_jc_l$. Note that $|\psi_t\rangle\simeq e^{-i\delta H t}|\varphi_t\rangle$ shares almost the same real-space profile as $|\varphi_t\rangle$. When the system becomes a ring, by replacing $Vt$ with $\Phi$ in $H(t)$, the obtained Hamiltonian $H(\Phi)$ is equivalent to that of a ring with a flux $\Phi$ inside. This correspondence can be understood from the fact that a temporally changing magnetic flux will induce an electromotive force. If $|\varphi_0\rangle$ is localized, then by definition the wave function should be rigid against the induced electric field. In contrast, a delocalized state should be flexible in response to a change of $\Phi$, giving rise to transport phenomena. Recalling that the spectra of $H(\Phi)$ and $H(\Phi+2\pi)$ coincide, we expect the complex energy of a localized (delocalized) state to almost stay unchanged (flow to another eigenvalue) when varying $\Phi$ from $0$ to $2\pi$. Accordingly, the spectral trajectory of $H(\Phi)$ cannot form any loop and is topologically trivial for a fully localized system.

\begin{figure}
\begin{center}
       \includegraphics[width=8cm, clip]{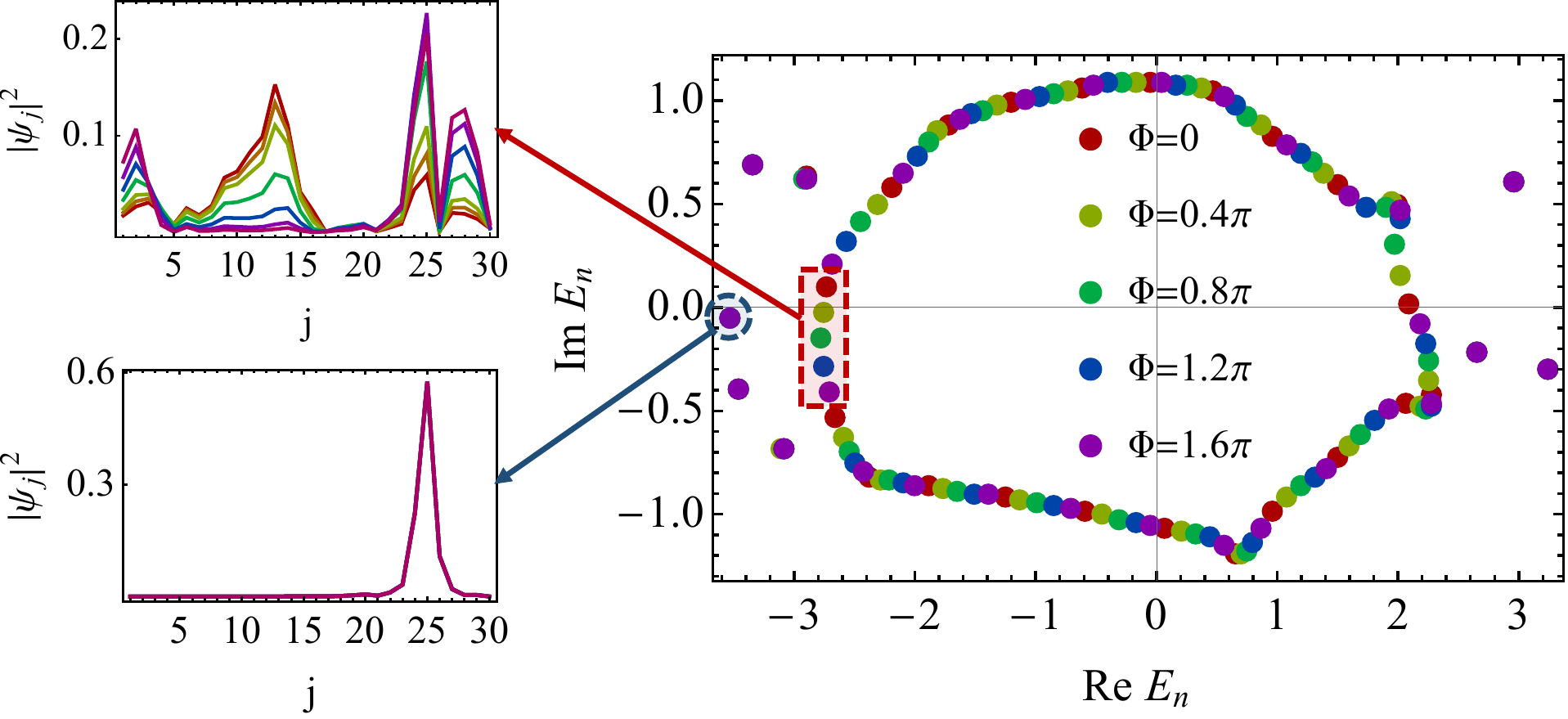}
       \end{center}
   \caption{Spectral flow (right) and two representative eigen wave functions (left) of a Hatano-Nelson ring with complex disorder $W=2.5$, $L=30$, $J_{\rm L}=2$, $J_{\rm R}=1$ and threaded by a varying flux $\Phi$. A delocalized wave function (left-upper panel) behaves flexibly, while a localized wave function (left-lower panel) exhibits rigidity.}
   \label{fig12}
\end{figure}

We illustrate the above argument for a Hatano-Nelson ring (\ref{HNHD}) with a complex on-site random potential and $L=30$. As shown in Fig.~\ref{fig12}, when changing $\Phi$ from $0$ to $2\pi$, $8$ of the $30$ eigenvalues almost stay unchanged, while the rest $22$ eigenvalues flow clockwise to their nearest neighbors, forming a loop. We also show the $\Phi$-dependence of two representative wave functions on and outside the loop. The former wave function (left-upper panel) is relatively extensive in real space and changes dramatically with respect to $\Phi$, while the latter one is localized and exhibits rigidity against a change in $\Phi$. Given a base point (e.g., $E_{\rm B}=0$) inside the loop, the spectral flow of the delocalized modes contributes to the winding number of $w=1$.

\subsection{Derivation of Eq.~(\ref{detHPhi})}
\label{PVj}
For convenience, we choose the gauge for which only the hopping between the $L$th site and the first site is multiplied by $e^{\mp i\Phi}$, such that
\begin{equation}
H(\Phi)=
\begin{bmatrix} 
V_1 & J_{\rm L} & 0 & \cdots & 0 & J_{\rm R}e^{-i\Phi} \\ 
J_{\rm R} & V_2 & J_{\rm L} & \cdots & 0 & 0 \\  
0 & J_{\rm R} & V_3 & \cdots & 0 & 0 \\
\vdots & \vdots & \vdots & \ddots & \vdots & \vdots \\ 
0 & 0 & 0  & \cdots & V_{L-1} & J_{\rm L} \\
J_{\rm L}e^{i\Phi} & 0 & 0  & \cdots & J_{\rm R} & V_L \\
\end{bmatrix}.
\end{equation}
Expanding the determinant of $H(\Phi)$ in terms of the first column, we obtain
\begin{align*}
{\rm det}H(\Phi)=V_1{\rm det}
\begin{bmatrix} 
V_2 & J_{\rm L} & \cdots & 0 & 0 \\  
J_{\rm R} & V_3 & \cdots & 0 & 0 \\
\vdots & \vdots & \ddots & \vdots & \vdots \\ 
0 & 0  & \cdots & V_{L-1} & J_{\rm L} \\
0 & 0  & \cdots & J_{\rm R} & V_L \\
\end{bmatrix}
\end{align*}
\begin{equation}
\;\;\;\;\;\;\;\;\;\;\;\;\;\;\;-J_{\rm R}
{\rm det}\begin{bmatrix} 
J_{\rm L} & 0 & \cdots & 0 & J_{\rm R}e^{-i\Phi} \\ 
J_{\rm R} & V_3 & \cdots & 0 & 0 \\
\vdots & \vdots & \ddots & \vdots & \vdots \\ 
0 & 0  & \cdots & V_{L-1} & J_{\rm L} \\
0 & 0  & \cdots & J_{\rm R} & V_L \\
\end{bmatrix}\\
\end{equation}
\begin{align*}
+(-)^{L-1}J_{\rm L}e^{i\Phi}{\rm det}
\begin{bmatrix} 
J_{\rm L} & 0 & \cdots & 0 & J_{\rm R}e^{-i\Phi} \\ 
V_2 & J_{\rm L} & \cdots & 0 & 0 \\  
\vdots & \vdots & \ddots & \vdots & \vdots \\ 
0 & 0  & \cdots & J_{\rm L} & 0 \\
0 & 0  & \cdots & V_{L-1} & J_{\rm L} \\
\end{bmatrix}.
\end{align*}
Denoting $Q_{m,n}$ as the determinant of the truncated Hatano-Nelson Hamiltonian (always subjected to the open-boundary condition) from site $m$ to $n$, we have
\begin{equation}
\begin{split}
{\rm det}H(\Phi)&=V_1Q_{2,L}\\
&-J_{\rm R}J_{\rm L}Q_{3,L}+(-)^{L-1}J^L_{\rm R}e^{-i\Phi}\\
&+(-)^{L-1}J^L_{\rm L}e^{i\Phi}-J_{\rm R}J_{\rm L}Q_{2,L-1},
\end{split}
\end{equation}
which can be rewritten in the form of Eq.~(\ref{detHPhi}) with
\begin{equation}
P(\{V_j\})=Q_{1,L}-J_{\rm R}J_{\rm L}Q_{2,L-1}.
\end{equation}
Here we have used the recursion relation
\begin{equation}
\begin{split}
Q_{m,n}&=V_mQ_{m+1,n}-J_{\rm R}J_{\rm L}Q_{m+2,n}\\
&=V_nQ_{m,n-1}-J_{\rm R}J_{\rm L}Q_{m,n-2},
\end{split}
\end{equation}
from which we can explicitly write down 
\begin{equation}
P(\{V_j\})=V_1V_2...V_L\sum^{\lfloor\frac{L}{2}\rfloor}_{|S|=0}\sum_{\substack{S\subset\mathbb{Z}_L:|n-n'|>1\\ \forall n\neq n',\;n,n'\in S}}\prod_{n\in S}\frac{-J_{\rm R}J_{\rm L}}{V_nV_{n+1}}.
\label{PVE}
\end{equation}
The condition $|n-n'|>1$ in Eq.~(\ref{PVE}) should be imposed on $\mathbb{Z}_L$, where $|L-1|$ is identified as $1$.

\subsection{Some exact results}
\label{SER}

\begin{figure}
\begin{center}
       \includegraphics[width=8.5cm, clip]{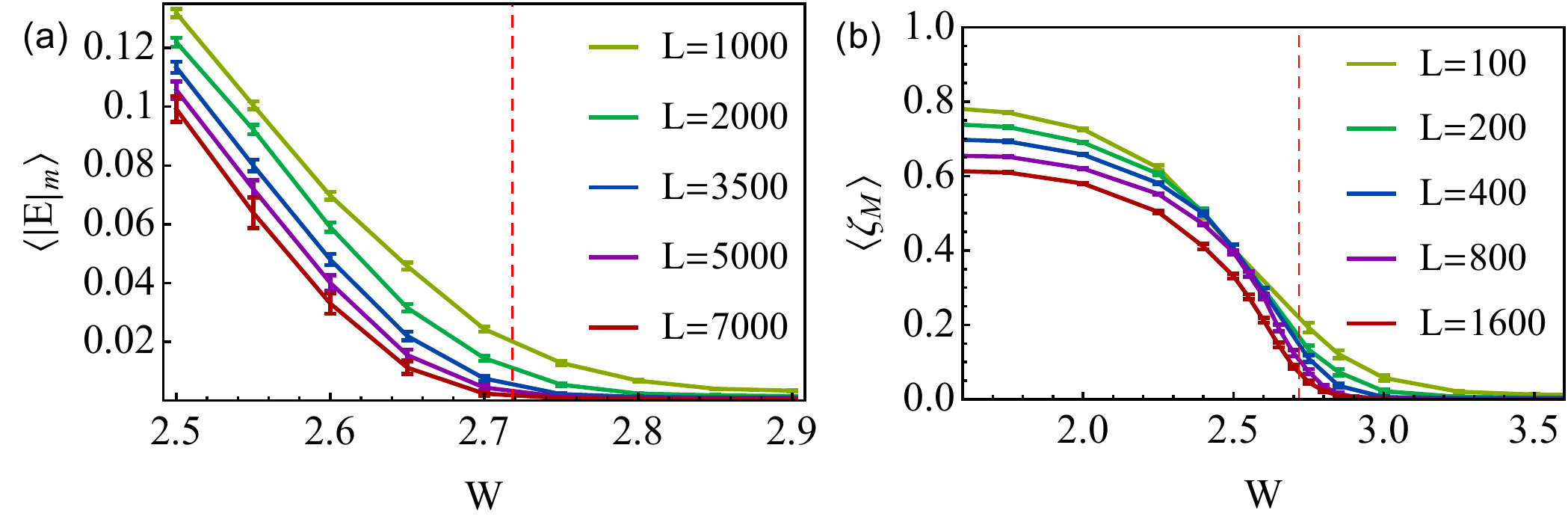}
       \end{center}
   \caption{(a) Disorder-averaged minimum absolute value of energy $\langle |E|_{\rm m}\rangle$ for the Hatano-Nelson model (\ref{HNHD}) with $J_{\rm L}=1$, $J_{\rm R}=0$, real on-site disorder $V_j\in[-W,W]$ and different system sizes ranging from $L=1000$ to $7000$. (b) Disorder-averaged maximum $\zeta$ (defined in Eq.~(\ref{RSIPR})) for the same model but with complex disorder $V_j=|V_j|e^{i\phi_j}$, where $|V_j|\in[0,W]$ and $\phi_j\in[0,2\pi]$, and different system sizes ranging from $L=100$ to $1600$. In both (a) and (b), the red dashed line indicates the theoretical transition point $W_{\rm c}=e=2.718...$. The number of disorder realizations ranges from thousands to hundreds, depending on the system size. The error bars denote twice the standard deviations of the mean.}
   \label{fig13}
\end{figure}

While it is difficult to obtain the distribution of $P(\{V_j\})$, analytical results are available under specific choices of parameters, e.g., $J_{\rm L}J_{\rm R}=0$ (unidirectional hopping) and $|V_j|$ obeys a uniform distribution over $[0,W]$. In this case, $P(\{V_j\})=\prod^L_{j=1}V_j$ and the distribution of $\Xi_L\equiv-\ln(|P(\{V_j\})|/W^L)\in[0,\infty)$ can explicitly be obtained as follows. Defining $\xi_j\equiv-\ln(|V_j|/W)\in[0,\infty)$, we find that $\xi_j$ obeys the standard exponential distribution, i.e., ${\rm Prob}(\xi_j=\xi)=e^{-\xi}\theta(\xi)$, where $\theta(\xi)$ is the Heaviside step function. Since $\Xi_L=\sum^L_{j=1}\xi_j$ with $\xi_j$'s being independent, $\Xi_L$ obeys the Gamma distribution
\begin{equation}
{\rm Prob}(\Xi_L=\Xi)=\frac{\Xi^{L-1}}{(L-1)!}e^{-\Xi}.
\end{equation}
For $L\gg1$, we can check that $\Xi_L/L$ approximately obeys the Gaussian distribution with mean $1$ and variance $L^{-1}$, and thus it approaches the delta distribution at $1$ in the thermodynamic limit. Recalling that the topological transition occurs at $|P(\{V_j\})|=J^L$ with $J\equiv\max\{|J_{\rm R}|,|J_{\rm L}|\}$, or equivalently $\Xi_L/L=-\ln(J/W)$; we thus obtain the critical disorder strength to be $W_{\rm c}=eJ$. Note that this critical value does not depend on whether $V_j$ is complex or not. However, this property should be unique to the unidirectional hopping. 

In Fig.~\ref{fig13}, we provide numerical evidence that supports the above prediction. For real disorder, we calculate the disorder average of $|E|_{\rm m}\equiv\min\{|E|:{\rm det}(E-H)=0,E\in\mathbb{C}\}$, which is the minimum absolute value of the complex eigenenergies. In the thermodynamic limit, we expect a nonzero (zero) $\langle |E|_{\rm m}\rangle$ in the delocalized (localized) phase. For a finite system, as shown in Fig.~\ref{fig13}(a), we find a sharper and sharper crossover near $W_{\rm c}$ when increasing the system size. For complex disorder, we use the inverse participation ratio, which is defined as ${\rm IPR}(\{\rho_j\})=\sum^L_{j=1}\rho^2_j$ for a normalized distribution $\sum^L_{j=1}\rho_j=1$, where $\rho_j\propto|\varphi_j\psi_j|$ (this quantity has been demonstrated to be a better indicator than $|\psi_j|^2$ and $|\varphi_j|^2$ \cite{Hatano1998}), $\psi_j$ is a right eigen-wave function of $H$ and $\varphi_j$ is the corresponding left eigen-wave function. We calculate the disorder average of the maximum of a rescaled quantity 
\begin{equation}
\zeta\equiv\frac{1}{L\times{\rm IPR}(\{\rho_j\})}\in(0,1] 
\label{RSIPR}
\end{equation}
for individual realizations. In the thermodynamic limit, we have 
$\zeta\neq0$ if $\rho_j$ decays no faster than the square-root power law and $\zeta=0$ otherwise, especially for an exponentially localized $\rho_j$. As shown in Fig.~\ref{fig13}(b), we find a similar crossover for $\langle\zeta_{\rm M}\rangle$ from finite to zero near $W_{\rm c}$, and the crossover becomes sharper for larger $L$.  

More generally, even if the analytic expression of ${\rm Prob}(\Xi_L=\Xi)$ is not available, the distribution of $\Xi_L/L$ asymptotically approaches the Gaussian distribution with mean ${\rm E}(\xi_j)$ and variance ${\rm Var}[\xi_j]/L$ as long as the central limit theorem is applicable. For example, when $|V_j|$ obeys the Lorentz distribution ${\rm Prob}(|V_j|=V)=\frac{2W}{\pi(V^2+W^2)}\theta(V)$, the rescaled variable $\xi_j\equiv-\ln(|V_j|/W)$ obeys the hyperbolic secant distribution ${\rm Prob}(\xi_j=\xi)=(\pi\cosh\xi)^{-1}$ with mean 0 and variance $\pi^2/4$. Therefore, the critical disorder strength for the Lorentz distribution is $W_{\rm c}=J$, which is consistent with that obtained by the Green's function method \cite{Zee1999}. 

\begin{figure}
\begin{center}
       \includegraphics[width=8.5cm, clip]{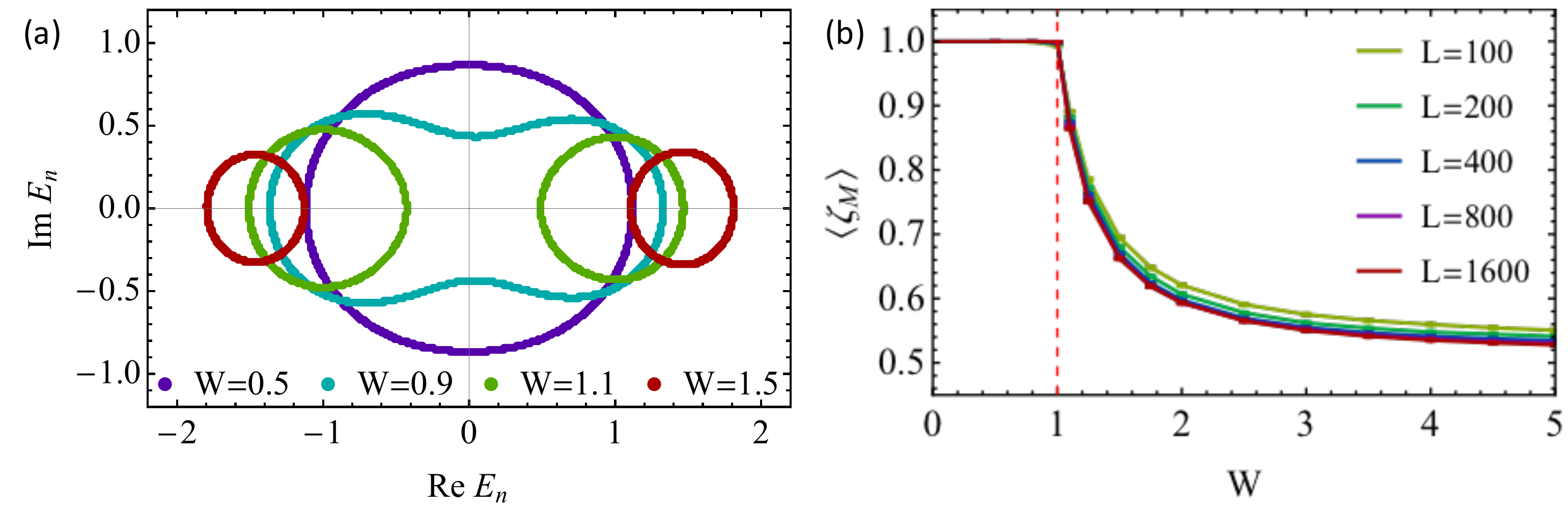}
       \end{center}
   \caption{(a) Complex-energy spectra of Eq.~(\ref{HNHD}) with $L=10^3$, $J_{\rm R}=0$, $J_{\rm L}=1$ and binary on-site disorder $V_j=\pm W$ with equal probability of occurrence for $W$ and $-W$, where $W=0.5,0.9,1.1,1.5$. (b) Disorder-averaged maximum $\zeta$ (see Eq.~(\ref{RSIPR})) for the same model but with different system sizes ranging from $L=100$ to $1600$. The red dashed line indicates the theoretical topological transition point $W_{\rm c}=1$.}
   \label{fig14}
\end{figure}

Finally, we provide an example which demonstrates a topological transition without a localization transition. We consider a \emph{binary} disorder $V_j=\pm W$ with equal probability of occurrence for $W$ and $-W$. In this case, $|P(\{V_j\})|=W^L$ in an arbitrary disorder realization, so the critical disorder strength for the topological transition is given by $W_{\rm c}=J$. On the other hand, the winding number with respect to $E_{\rm B}=\pm W$ is always one in the thermodynamic limit, no matter how large $W$ is. This implies that there are always some delocalized modes and the system never undergoes a localization transition. Nevertheless, there is indeed a qualitative change in the spectrum when $W$ exceeds $W_{\rm c}$ --- a single loop splits into two loops (see Fig.~\ref{fig14}(a)). As shown in Fig.~\ref{fig14}(b), such a transition is accompanied 
by the onset of the deviation of $\zeta_{\rm M}$ from one.

\section{Proof of the bulk-edge correspondence} 
\label{BEC}
To be specific, we focus on a single-band lattice with finite-range hopping amplitudes $J_j$'s. That is, we have at most $p$-site ($q$-site) hopping towards the right (left) direction. Hence, denoting $z=e^{ik}$, the dispersion relation, or the characteristic equation of the Schr\"odinger equation, can be written as
\begin{equation}
E=f(z)=\sum^q_{j=-p}J_jz^j,
\label{Efz}
\end{equation}
with $J_{-p},J_q\neq0$. Assuming that the winding number $w$ is non-negative, we impose the right semi-infinite condition, so the general solution of an edge state takes the form 
\begin{equation}
\psi_j=\sum^S_{l=1}\sum^{n_l}_{m=1}c_{l,m}\left.\frac{d^{m-1}}{dz^{m-1}}z^j\right |_{z=z_l},
\label{C2}
\end{equation}
where $z_l$ ($l=1,2,...,S$) is the $n_l$th-order zero of $f(z)=0$ given in Eq.~(\ref{Efz}) and inside of the unit circle $|z|=1$, i.e., $|z_l|<1$. Using the argument principle (\ref{AP}) and the assumption $w\ge0$, we have $\sum^S_{l=1}n_l=Z=p+w\ge p$, with $p$ being the effective number of poles for $|z|<1$. Indeed, there is a single $p$th-order pole at $z=0$, implying $z_l\neq0$ for all $l=1,2,...,S$. The initial condition reads
\begin{equation}
\psi_0=\psi_{-1}=...=\psi_{-p+1}=0,
\end{equation}
which, together with Eq.~(\ref{C2}), leads to a set of homogeneous linear equations 
\begin{equation}
M\boldsymbol{c}=\boldsymbol{0}, 
\label{Mc0}
\end{equation}
where the elements of the generalized Vandermonde matrix \cite{Meyer2000} $M=[M_{uv}]_{p\times Z}$ and the coefficient vector $\boldsymbol{c}=(c_1,c_2,...,c_Z)^{\rm T}$ are given by
\begin{equation}
\begin{split}
M_{j,\sigma(l,m)}&=\left.\frac{d^{m-1}}{dz^{m-1}}z^{-j+1}\right |_{z=z_l},\\
c_{\sigma(l,m)}&=c_{l,m},
\end{split}
\label{Mcele}
\end{equation}
with $\sigma(l,m)\equiv\sum^{l-1}_{r=1}n_r+m$, $1\le l\le S$ and $1\le m\le n_l$. To see how many degrees of freedom survive under the condition imposed by Eq.~(\ref{Mc0}), we have to determine the rank of $M$, which equals that of $M^{\rm T}$. Suppose that the rank of $M^{\rm T}$ does not saturate the maximum $p$, there must exist a nonzero vector $\boldsymbol{a}=(a_1,a_2,...,a_p)^{\rm T}$ that satisfies
\begin{equation}
M^{\rm T}\boldsymbol{a}=\boldsymbol{0}.
\label{MTa0}
\end{equation}
Defining a polynominal $g(z)\equiv\sum^p_{j=1}a_jz^{j-1}$ with $0<{\rm deg}\;g(z)\le p-1$ due to the fundamental theorem of algebra, Eq.~(\ref{MTa0}) can explicitly be written down as
\begin{equation}
\left.\frac{d^{m-1}}{dz^{m-1}}g(z^{-1})\right |_{z=z_l}=0,
\end{equation}
implying that $g(z)$ contains a polynomial factor $\prod^S_{l=1}(z-z^{-1}_l)^{n_l}$ and thus ${\rm deg}\;g(z)\ge\sum^S_{l=1}n_l=Z$. Recalling that $Z\ge p$, ${\rm deg}\;g(z)\ge Z$ contradicts ${\rm deg}\;g(z)\le p-1$, so the original assumption that ${\rm rank}(M^{\rm T})<p$ must be wrong. In other words, both the rank of $M^{\rm T}$ and that of $M$ saturate the maximum $p$.  Therefore, the number of independent $c_j$'s satisfying Eq.~(\ref{Mc0}), or the degeneracy of zero modes localized at the left edge, turns out to be $Z-p=w$. As an example with two-fold degeneracy, we can examine the model given in Eq.~(\ref{J12}) and check that
\begin{equation}
\begin{split}
\psi^{(1)}_j=&(-)^j\sqrt{\frac{1}{3}(1+\beta+\beta^2)\beta^{j-1}}(1-e^{\frac{2\pi i}{3}j}),\\
\psi^{(2)}_j=&\frac{(-)^j}{3}\sqrt{\frac{(1-\beta^3)\beta^{j-2}}{1+\beta}}[1+\beta e^{\frac{2\pi i}{3}}+\\
&(\beta+e^{\frac{2\pi i}{3}})e^{\frac{2\pi i}{3}j}+(\beta+1)e^{-\frac{2\pi i}{3}(j+1)}],
\end{split}
\end{equation}
span the zero-mode space, where $\beta=(\frac{J_1}{J_2})^{\frac{2}{3}}$.

Now let us next discuss the case of $w<0$. If we use the same boundary condition as above, we will again obtain Eq.~(\ref{Mc0}), but there are more rows than columns in $M$ since $p=Z-w>Z$. We can thus pick out the first $Z$ rows of $M$ to construct a square matrix $\tilde M$, such that
\begin{equation}
\tilde M\boldsymbol{c}=\boldsymbol{0}
\label{Mtc0}
\end{equation}
is necessarily satisfied. Straightforward calculations give
\begin{equation}
{\rm det}\tilde M=C\prod_{1\le r<s\le l}(z^{-1}_s-z^{-1}_r)^{n_rn_s}\neq0,
\end{equation}
where the factor $C=\prod^S_{l=1}(-)^{n_l-1}z^{-n_l(n_l-1)}_l\prod^{n_l}_{m=1}(m-1)!$. Therefore, as a necessary condition of Eq.~(\ref{Mc0}), Eq.~(\ref{Mtc0}) is sufficient to enforce $\boldsymbol{c}$ to be $\boldsymbol{0}$, implying no edge modes localized at the left boundary. On the other hand, if we change the boundary condition to be left semi-infinite, 
\begin{equation}
\psi_{-j}=\sum^R_{l=1}\sum^{m_l}_{n=1}c_{l,n}\left.\frac{d^{n-1}}{dz^{n-1}}z^{-j}\right |_{z=\zeta_l},
\end{equation}
where $\zeta_l$ ($l=1,2,...,R$) is the $m_l$th zero of $f(z)$ outside $|z|=1$. Recalling that $z^pf(z)$ is a polynomial with degree $p+q$, we have $Z'\equiv\sum^R_{l=1}m_l=p+q-Z=q-w$. This result is consistent with directly applying the argument principle to $f(z^{-1})$, which has a single $q$th-order pole $z=0$ inside the circle of $|z|=1$, leading to
\begin{equation}
\oint_{|z|=1}\frac{dz}{2\pi i}\frac{\frac{d}{dz}f(z^{-1})}{f(z^{-1})}=
Z'-q.
\label{AP2}
\end{equation}
Here we have used the fact that $\zeta^{-1}_l$'s are the zeros of $f(z^{-1})$ inside the unit circle $|z|=1$. Noting that the left-hand side in Eq.~(\ref{AP2}) can be shown to be the minus of that in Eq.~(\ref{AP}) via a change of the integration variable, we obtain $Z'=q-w$. The initial condition
\begin{equation}
\psi_0=\psi_1=...=\psi_{q-1}=0
\end{equation}
can again be written in the form of Eq.~(\ref{Mc0}), but the elements of the generalized Vandermonde matrix $M=[M_{uv}]_{q\times Z'}$ and the coefficient vector $\boldsymbol{c}=(c_1,c_2,...,c_{Z'})^{\rm T}$ become
\begin{equation}
\begin{split}
M_{j,\mu(l,n)}
&=\left.\frac{d^{n-1}}{dz^{n-1}}z^{j-1}\right |_{z=\zeta_l},\\
c_{\mu(l,n)}
&=c_{l,n},
\end{split}
\label{Mcele2}
\end{equation}
where $\mu(l,n)\equiv\sum^{l-1}_{r=1}m_r+n$, $1\le l\le R$ and $1\le n\le m_r$. Using the same technique as in the previous paragraph, we can prove that $M$ takes the maximum rank $q$, so the number of independent degrees of freedom, or the degeneracy of the zero modes localized at the right boundary, turns out to be $Z'-q=-w$.

As an application of the bulk-edge correspondence for non-Hermitian Hamiltonians, we can demonstrate the bulk-edge correspondence in Hermitian systems with a chiral symmetry (class AIII), whose Hamiltonian is given by Eq.~(\ref{HH}). Such a Hamiltonian can be unitarily transformed into $\sigma_x\otimes\sqrt{H^\dag H}$, so the full spectrum reads $\{\pm E_1,\pm E_2,...\}$, with $\{E_1,E_2,...\}$ being the eigenvalues of $\sqrt{H^\dag H}$, which is semi-positive-definite. Therefore the statement that there are $2|w|$ zero modes of Eq.~(\ref{HH}) is equivalent to the fact that there are $|w|$ zero modes of $\sqrt{H^\dag H}$. We have already known that $|w|$ gives the number of edge states of $H$ at $E=0$ in a semi-infinite space, but generally does not 
for an open chain. However, it gives the number of 
quasi-eigenstates at $E=0$, which almost vanish after being acted on by $H$. Using this property, we can show that $|w|$ does give the number of zero modes for the \emph{Hermitian} operator $\sqrt{H^\dag H}$.

To this end, we first prove the following theorem: 
\begin{theorem}
Given $D$ different wave functions $|\psi_n\rangle$ \emph{(}$n=1,2,...,D$\emph{)} satisfying $\|H|\psi_n\rangle\|<\epsilon_1$ and $|\langle\psi_m|\psi_n\rangle|<\epsilon_2\ll D^{-1}$ for all $m\neq n$, there must be at least $D$ different eigenstates of $\sqrt{H^\dag H}$ with energies less than $E_{\rm b}=\frac{D\epsilon_1}{\sqrt{1-(D-1)\epsilon_2}}$. 
\end{theorem}
\emph{Proof.---} We note that $|\psi_n\rangle$'s are linearly independent. Otherwise, we can find $c_j$'s ($j=1,2,...,D$) such that $\max_{1\le j\le D}|c_j|=|c_{j_0}|>0$ and $\sum^D_{j=1}c_j|\psi_j\rangle=0$, leading to the contradiction 
\begin{equation}
\begin{split}
&|c_{j_0}|=|c_{j_0}\langle\psi_{j_0}|\psi_{j_0}\rangle|=\left|\sum_{j\neq j_0}c_j\langle\psi_{j_0}|\psi_j\rangle\right|\\
&\le\sum_{j\neq j_0}|c_j||\langle\psi_{j_0}|\psi_j\rangle|
<\epsilon_2(D-1)|c_{j_0}|\ll|c_{j_0}|.
\end{split}
\end{equation}
Therefore, denoting $V_0\equiv{\rm span}\{|\psi_j\rangle:j=1,2,...,D\}$, we have ${\rm dim}V_0=D$. For an arbitrary $|\psi\rangle\in V_0$, which can always be expressed as $|\psi\rangle=\sum^D_{j=1}c_j|\psi_j\rangle/\|\sum^D_{j=1}c_j|\psi_j\rangle\|$, we can bound $\|H|\psi\rangle\|$ from above as
\begin{equation}
\begin{split}
\|H|\psi\rangle\|&\le\frac{\sum^D_{j=1}|c_j|\|H|\psi_j\rangle\|}{\|\sum^D_{j=1}c_j|\psi_j\rangle\|}\\
&<\frac{\epsilon_1\sum^D_{j=1}|c_j|}{\sqrt{\sum^D_{j=1}|c_j|^2-\sum_{m\neq n}|c^*_mc_n\langle\psi_m|\psi_n\rangle|}}\\
&<\frac{\epsilon_1}{\sqrt{1-(D-1)\epsilon_2}}\frac{\sum^D_{j=1}|c_j|}{\sqrt{\sum^D_{j=1}|c_j|^2}}\\
&\le\frac{\sqrt{D}\epsilon_1}{\sqrt{1-(D-1)\epsilon_2}}=\frac{E_{\rm b}}{\sqrt{D}}.
\end{split}
\end{equation}
Consequently, we have
\begin{equation}
{\rm Tr}_{V_0}[H^\dag H]<E^2_{\rm b},
\end{equation}
where ${\rm Tr}_{V_0}[...]$ denotes the trace over the subspace $V_0$. Denoting $P_{\rm g}$ as the projector onto the Hilbert subspace $V_{\rm g}$ spanned by all the eigenstates of $\sqrt{H^\dag H}$ with energies less than $E_{\rm b}$, we can construct $H'\equiv E^2_{\rm b}(1-P_{\rm g})\le H^\dag H$, leading to
\begin{equation}
\begin{split}
{\rm Tr}_{V_0}[H']=E^2_{\rm b}(D-{\rm Tr}_{V_0}[P_{\rm g}])<E^2_{\rm b}\\
\Leftrightarrow\;\;\;\;{\rm Tr}_{V_0}[P_{\rm g}]={\rm Tr}_{V_{\rm g}}[P_0]>D-1,
\end{split}
\end{equation}
where $P_0$ is the projector onto $V_0$. Since ${\rm Tr}_{V_{\rm g}}[P_0]\le{\rm Tr}_{V_{\rm g}}[1]={\rm dim}V_{\rm g}$, which should be an integer, we finally obtain ${\rm dim}V_{\rm g}\ge D$. $\blacksquare$

Now let us come back to the eigenvalue problem of $\sqrt{H^\dag H}$ for an open chain with length $L$. We can first work in the semi-infinite limit to determine a set of orthonormal zero modes $|\phi_j\rangle$'s ($j=1,2,...,|w|$) of $H$, and then truncate and normalize them on a finite chain, obtaining $|\psi_j\rangle$'s. Note that $|\psi_j\rangle$'s are now not exact eigenstates of $H$, but the conditions of the theorem proved above are satisfied, with $\epsilon_1$ and $\epsilon_2$ exponentially small in $L$, since the deviations stem from the exponential tail. According to the theorem, we can find at least $|w|$ eigenstates with exponentially small energies. We should furthermore mention the impossibility to find the $(|w|+1)$th eigenstate with a small energy that eventually vanishes in the thermodynamic limit; otherwise we will have at least $|w|+1$ zero modes of $H$ in a semi-infinite space, leading to a contradiction.

It is worthwhile to mention that the bulk-edge correspondence for class AIII (or BDI) alone can alternatively be proved using the Callias index theorem \cite{Callias1978} following Ref.~\cite{Kane2014}. However, it seems rather nontrivial whether a similar method can be applied to a single off-diagonal block in a class AIII Hamiltonian. 

\section{Long-lived quasi-eigenstates and their absence in Hermitian systems}
\label{LLQE}
According to the bulk-edge correspondence proved in the last appendix, we know that for an open chain and a given base energy $E_{\rm B}=E$ with respect to which the winding number is nonzero ($w\neq0$) for the corresponding periodic ring, there exists $w$ independent quasi-edge modes satisfying 
\begin{equation}
\|(H-E)|\psi\rangle\|<A_Ee^{-\alpha_E L},
\label{HEAa}
\end{equation}
where the constants $A_E$ and $\alpha_E$ depend on $E$ but not on $L$. In the following we show that such a quasi-eigenstate is long-lived to a time scale at least proportional to $L$, and thus becomes an exact eigenstate in the limit of $L\to\infty$.

\begin{figure}
\begin{center}
       \includegraphics[width=6cm, clip]{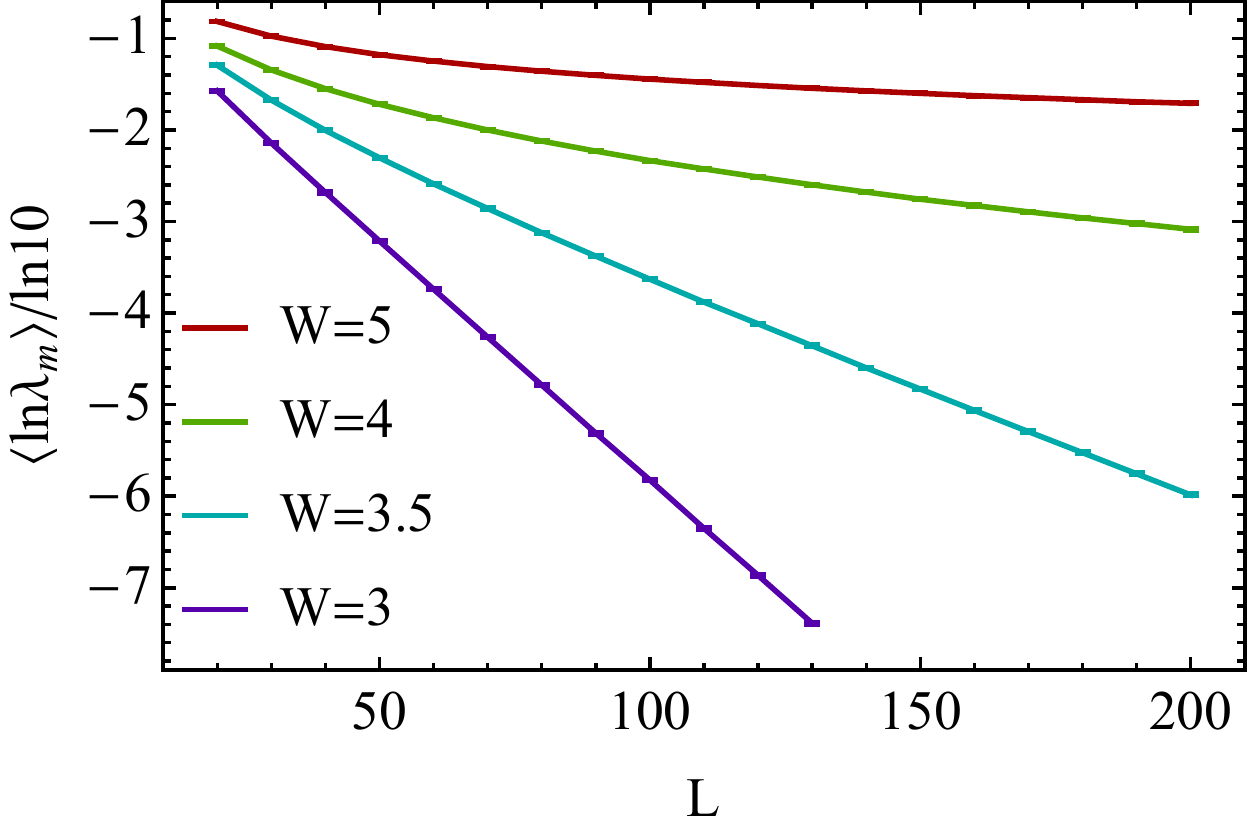}
       \end{center}
   \caption{Finite-size scaling for the logarithmic-disorder-averaged smallest singular value $\langle\ln\lambda_{\rm m}\rangle$ of the Hatano-Nelson Hamiltonian (\ref{HNHD}) with $J_{\rm L}=2$, $J_{\rm R}=1$ and complex disorder. Each point is obtained from $2.5\times10^5$ disorder realizations.}
   \label{fig15}
\end{figure}

We first clarify that ``long-lived" means that the free evolution $e^{-iHt}|\psi\rangle$ can be well approximated by $e^{-iEt}|\psi\rangle$ up to a long time. To quantify how close $|\psi\rangle$ is to an eigenstate, we examine the relative deviation 
\begin{equation}
R(t)\equiv\frac{\|e^{-iHt}|\psi\rangle-e^{-iEt}|\psi\rangle\|}{\|e^{-iEt}|\psi\rangle\|},
\end{equation}
which cancels the effect of amplification or decay due to a large imaginary part in $E$. We can thus bound $R(t)$ as
\begin{equation}
\begin{split}
R(t)&=
\|[e^{-i(H-E)t}-1]|\psi\rangle\|\\
&\le
\sum^\infty_{n=1}\frac{t^n}{n!}\|(H-E)^n|\psi\rangle\|\\
&\le 
\sum^\infty_{n=1}\frac{t^n\|H-E\|^{n-1}}{n!}\|(H-E)|\psi\rangle\|\\
&<A_E\|H-E\|^{-1}e^{-\alpha_E L+\|H-E\|
t}\\
&=\tilde A_Ee^{-\alpha_E(L-v_E t)},
\end{split}
\label{Rtb}
\end{equation}
where $\tilde A_E=A_E\|H-E\|^{-1}$ and $v_E=\|H-E\|/\alpha_E$, with $\|O\|\equiv\max_{\||\psi\rangle\|=1}\|O|\psi\rangle\|$ being the operator norm. Here we have iteratively used the inequality $\|O_1O_2|\psi\rangle\|\le\|O_1\|\|O_2|\psi\rangle\|$. Since $v_E\le\|H\|+|E|$ can be bounded by an $L$-independent quantity, Eq.~(\ref{Rtb}) implies that up to a time $t^*\sim O(\frac{L}{v_E})$ the relative deviation is exponentially small, i.e., $-\ln R(t)\sim O(L)$, which is consistent with a na\"{i}ve expectation from the Lieb-Robinson bound \cite{Lieb1972}. Nevertheless, we should mention that the Lieb-Robinson picture may break down in some many-particle non-Hermitian systems after a global quench \cite{Ashida2018}.

Remarkably, the above analysis does not depend on the translation invariance. That is to say, as long as Eq.~(\ref{HEAa}) holds true, we can claim the existence of long-lived quasi-eigenstates even for a disordered system. Equation (\ref{HEAa}) can numerically be justified by calculating the \emph{singular values} of $H-E$ followed by finite-size scaling. To be concrete, we focus on $E=0$ and the smallest singular value $\lambda_{\rm m}\equiv\min_{\||\psi\rangle\|=1}\|H|\psi\rangle\|$ in the Hatano-Nelson model with $J_{\rm L}=2$, $J_{\rm R}=1$ and a complex on-site random potential. As for the disorder average, we consider $\langle\ln \lambda_{\rm m}\rangle$ for up to $2.5\times10^5$ realizations. We choose $\langle\ln \lambda_{\rm m}\rangle$ rather than $\ln\langle\lambda_{\rm m}\rangle$ because the latter is sensitive to rare events while the former is not. As shown in Fig.~\ref{fig15}, $\langle\ln\lambda_{\rm m}\rangle$ scales linearly with respect to $L$ for a not-too-strong disorder strength ($W=3,3.5$), implying the robustness of (quasi-)edge modes. When the disorder is strong enough ($W=4,5$), however, the scaling seems to deviate from a linear one. We also note that the above analysis implies a similar Lieb-Robinson behavior for the quench from a finite chain to the semi-infinite boundary condition, since Eq.~(\ref{HEAa}) holds true also in this situation.

Let us move on to disprove the existence of a quasi-eigenstate in a general Hermitian system at any energy $E$ separated from the spectrum of $H=\sum_j E_j|\psi_j\rangle\langle\psi_j|$. We first prove that ${\rm Im}E=0$ is necessary for the existence of a quasi-eigenstate. Otherwise, by using the inequality $\||\psi_1\rangle-|\psi_2\rangle\|\ge|\||\psi_1\rangle\|-\||\psi_2\rangle\||$, we have
\begin{equation}
R(t)=\|e^{-i(H-E)t}|\psi\rangle-|\psi\rangle\|\ge|e^{-2{\rm Im}Et}-1|,
\end{equation}
where the right-hand side can exceed any small threshold after a time interval independent of the system size. We thus focus on $E\in\mathbb{R}/\{E_j\}$ from now on. Defining $d_E\equiv\min_j|E_j-E|$ and $D_E\equiv\max_j|E_j-E|$, which are finite even when $L\to\infty$, we have
\begin{equation}
\sin\frac{|E_j-E|t}{2}>\frac{|E_j-E|t}{\pi}
\end{equation}
for all $t<\pi/D_E$. Expanding the initial state as $|\psi\rangle=\sum_j c_j|\psi_j\rangle$, we have 
\begin{equation}
\begin{split}
R(t)&=\|\sum_j c_j [e^{-i(E_j-E)t}-1]|\psi_j\rangle\|\\
&=2\sqrt{\sum_j|c_j|^2\sin^2\frac{(E_j-E)t}{2}}\\
&>\frac{2t}{\pi}\sqrt{\sum_j|c_j|^2(E_j-E)^2}\\
&\ge \frac{2d_E}{\pi}t,
\end{split}
\label{Rtl}
\end{equation}
at least for $t<\pi/D_E$. This result (\ref{Rtl}) implies a finite time interval during which the deviation of $e^{-iHt}|\psi\rangle$ from $e^{-iEt}|\psi\rangle$ grows faster than a finite speed $2d_E/\pi$ for all $|\psi\rangle$, no matter how large the system size $L$ is. Therefore, no quasi-eigenstate whose lifetime increases with respect to $L$ exists in a Hermitian system.

\begin{figure}
\begin{center}
       \includegraphics[width=8.5cm, clip]{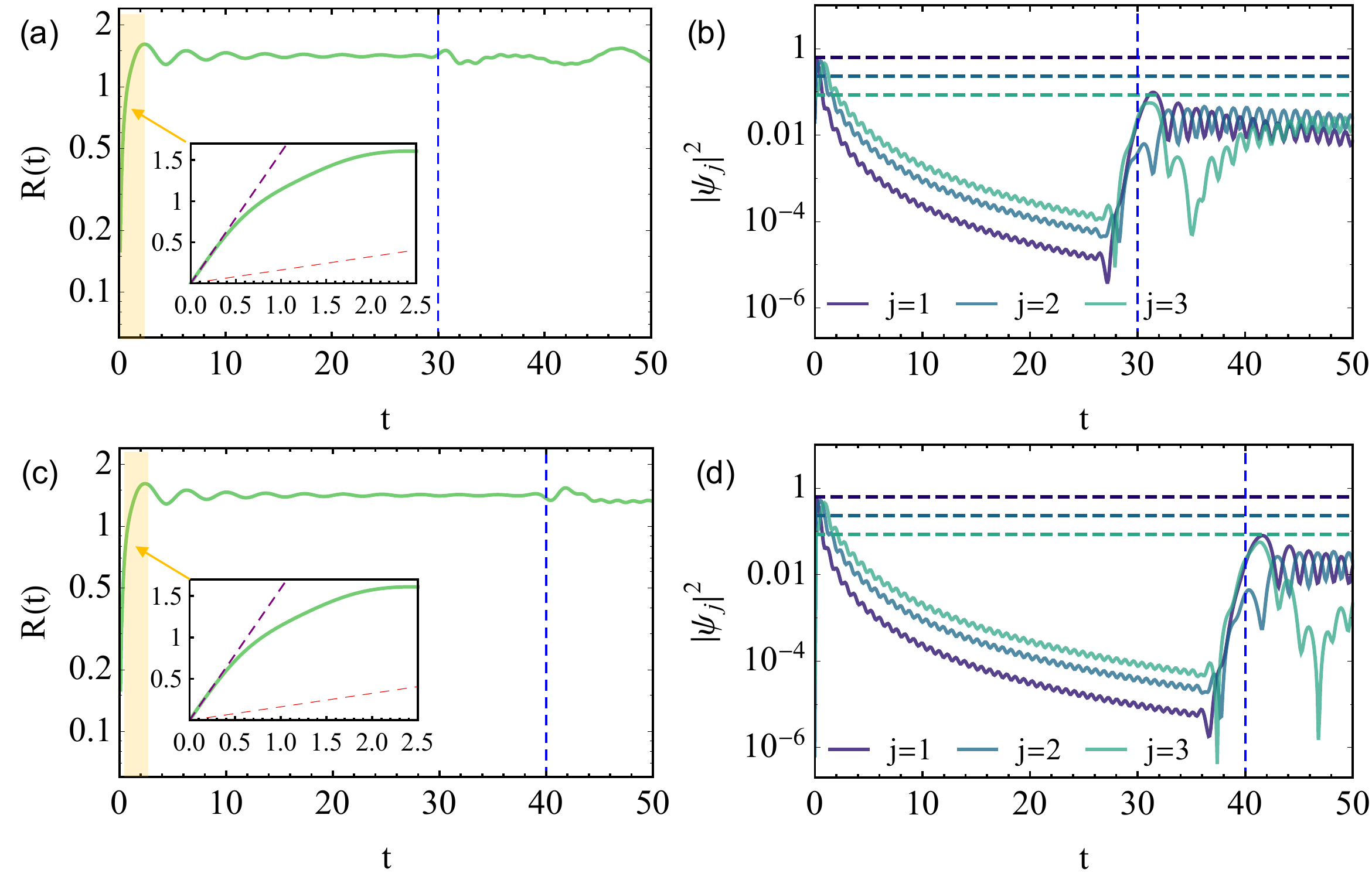}
       \end{center}
   \caption{Dynamics in a Hermitian single-band lattice with $J=1$ and starting from a wave function $\psi_j\propto e^{-j/\ell}$ with $\ell=2$. (a) Relative deviation $R(t)$. Inset: enlarged view of an initial behavior up to $t=2.5$ (orange region). The purple (red) line shows a linear fit (the lower bound in Eq.~(\ref{Rtl})). (b) Density profile $|\psi_j(t)|^2$ at the left-most three sites ($j=1,2,3$) in comparison with the initial values $|\psi_j(0)|^2$ (dashed lines). The system size is $L=60$ in (a) and (b). (c) and (d) are the same as (a) and (b), except for $L=80$. In (a)-(d), the blue dashed line denotes the revival time $t_{\rm r}=L/(2J)$.}
   \label{fig16}
\end{figure}

As a simple example, we consider the Hermitian limit of a Hatano-Nelson lattice: 
\begin{equation}
H=J\sum_j(c^\dag_{j+1}c_j+c^\dag_jc_{j+1}),
\end{equation}
whose band dispersion reads $E(k)=2J\cos k$, with a maximum group velocity $2J$. A naive extension to imaginary wave vector $k=i\ell^{-1}$ may suggest that a localized wave function $\psi_j\propto e^{ikj}=e^{-j/\ell}$ corresponds to an energy $E=2\cosh\ell^{-1}$ outside the spectrum of $H$. According to Eq.~(\ref{Rtl}), however, such a wave function can never be a quasi-eigenstate. To confirm this, we explicitly calculate the dynamics for $\ell=2$ and two different system sizes $L=60$ and $80$ (see Fig.~\ref{fig16}). In stark contrast to the Lieb-Robinson picture, we find a quick saturation of $R(t)$ independent of the system size. The bound in Eq.~(\ref{Rtl}) (dashed red lines in Fig.~\ref{fig16}(a) and (c)), although not tight, correctly predicts the linear growth at the initial stage (fitted by dashed purple lines). 
On the other hand, the finite system size sets a time scale after which the wave-packet dynamics fails to be approximated by free propagation, leading to revival in $R(t)$. As shown in Figs.~\ref{fig16}(a)-(d), such a revival time turns out to be well approximated by $t_{\rm r}=L/(2J)$.

\section{Derivation of the semiclassical equation of motion for nonunitary Bloch oscillations}
\label{DSM}
In the continuous limit, the Schr\"odinger equation (generally nonunitary) in momentum space is given by
\begin{equation}
i\partial_t\psi_t(k)=[E(k)-iF\partial_k]\psi_t(k),
\label{Schrok}
\end{equation}
where $E(k)$ is the dispersion relation of the band and $F$ is a potential gradient. Starting from an arbitrary initial state $\psi_0(k)$, we can write down a formal solution to Eq.~(\ref{Schrok}) as 
\begin{equation}
\psi_t(k)=e^{-i\int^t_0dt'E(k-F(t-t'))}\psi_0(k-Ft),
\label{gensol}
\end{equation}
which satisfies the quasi-periodicity $\psi_{t+\frac{2\pi}{F}}(k)=e^{-i\frac{2\pi}{F}\bar E}\psi_t(k)$ with $\bar E=\int^\pi_{-\pi}\frac{dk}{2\pi}E(k)$. Note that no approximation is made so far except for the continuous limit. We mention that a similar semi-classical analysis on nonunitary wave-packet dynamics is made in Ref.~\cite{Rush2016}.

If we focus on the semiclassical regime, $\psi_0(k)$ should be highly localized in the Brillouin zone, such as a Gaussian packet $\psi_0(k)=(\sqrt{2\pi}\sigma_k)^{-\frac{1}{2}}e^{-k^2/(2\sigma^2_k)}$ near $k=0$, with a small dispersion $\sigma_k\ll 1$. In this case, we can expand $E(k-F(t-t'))$ in Eq.~(\ref{gensol}) near $Ft'$ (in terms of $k-Ft\sim\sigma_k$) up to $(k-Ft)^2$, such that the wave packet $\psi_t(k)$ stays (approximately) Gaussian:
\begin{equation}
\begin{split}
&\;\;\;\;\psi_t(k)\simeq(\sqrt{2\pi}\sigma_k)^{-\frac{1}{2}}e^{-i\int^t_0dt'E(Ft')}\\
&\times e^{-i\frac{E(Ft)-E(0)}{F}(k-Ft)-\left[\frac{1}{4\sigma^2_k}+i\frac{E'(Ft)-E'(0)}{2F}\right](k-Ft)^2},
\end{split}
\label{psiGauss}
\end{equation}
where $E'(k)$ is the simplified notation for $\frac{dE(k)}{dk}$. We can thus calculate the normalization $\mathcal{N}_t\equiv\langle\psi_t|\psi_t\rangle$ as $\int^\pi_{-\pi} dk|\psi_t(k)|^2$, which turns out to be
\begin{equation}
\mathcal{N}_t=\frac{e^{2\int^t_0dt'{\rm Im}E(Ft')+\frac{2[{\rm Im}E(Ft)-{\rm Im}E(0)]^2\sigma^2_k}{F^2-2\sigma^2_kF[{\rm Im}E'(Ft)-{\rm Im}E'(0)]}}}{\sqrt{1-\frac{2\sigma^2_k}{F}[{\rm Im}E'(Ft)-{\rm Im}E'(0)]}}.
\label{Nt}
\end{equation}
By taking the limit $\sigma_k\to0$, we obtain the rightmost equation in Eq.~(\ref{semicl}). The center of mass in the Brillouin zone can also be read out from Eq.~(\ref{psiGauss}) as
\begin{equation}
\langle k \rangle_t=Ft+\frac{2[{\rm Im}E(Ft)-{\rm Im}E(0)]\sigma^2_k}{F-2\sigma^2_k[{\rm Im}E'(Ft)-{\rm Im}E'(0)]},
\label{kt}
\end{equation}
which reduces to $Ft$ in the $\sigma_k\to0$ limit. After the Fourier transform $\psi_t(x)=\int^\pi_{-\pi}\frac{dk}{\sqrt{2\pi}}\psi_t(k)e^{ikx}$, we can obtain the real-space wave function and determine the center of mass in real space as
\begin{equation}
\begin{split}
&\langle x \rangle_t=\frac{{\rm Re}E(Ft)-{\rm Re}E(0)}{F}\\
&-\frac{2\sigma^2_k}{F^2}{\rm Im}[(E^*(Ft)-E^*(0))(E'(Ft)-E'(0))],
\end{split}
\label{xt}
\end{equation}
which reduces to the middle equation in Eq.~(\ref{semicl}) in the limit of $\sigma_k\to0$.

It is worthwhile to consider the specific case of free diffusion with $F=0$. Taking the limit of $F\to0$ in Eqs.~(\ref{Nt}), (\ref{kt}) and (\ref{xt}), we obtain
\begin{equation}
\begin{split}
\mathcal{N}_t&=\frac{e^{2{\rm Im}E(0)t+\frac{2[\sigma_k{\rm Im}E'(0)t]^2}{1-2\sigma^2_k{\rm Im}E''(0)t}}}{\sqrt{1-2\sigma^2_k{\rm Im}E''(0)t}},\\
\langle k\rangle_t&=\frac{2\sigma^2_k{\rm Im}E'(0)t}{1-2\sigma^2_k{\rm Im}E''(0)t},\\
\langle x\rangle_t&={\rm Re}E'(0)t-2\sigma^2_k{\rm Im}[E'^*(0)E''(0)]t^2.
\end{split}
\end{equation}
Applying the last equation to a wave packet with the momentum-space spread of $\sigma^2_k=\pi/L$ in the clean Hatano-Nelson model (\ref{HNH}) with $J_{\rm L},J_{\rm R}\in\mathbb{R}$, we have
\begin{equation}
\langle x\rangle_t=-\frac{2\pi}{L}(J^2_{\rm L}-J^2_{\rm R})t^2.
\end{equation}
This result implies that the shift of center of mass due to asymmetric hopping amplitudes is a finite-size effect. In other words, a wave packet in the classical limit does not move in spite of the asymmetry in hopping amplitudes.

\section{Implementation of asymmetric hopping amplitudes with ultracold atoms in optical lattices}
\label{AHOL}
We here show that it is possible to realize a non-Hermitian system on the basis of reservoir engineering \cite{Zoller1996,Buchler2008,Diehl2008,Diehl2011,Muller2012}. Generally speaking, by engineering a Lindblad master equation \cite{Lindblad1976}
\begin{equation}
\dot\rho_t=-i[H,\rho_t]+\sum_j
\mathcal{D}[L_j]\rho_t,
\label{LE}
\end{equation}
where $\mathcal{D}[L]\rho\equiv L\rho L^\dag-\{L^\dag L,\rho\}/2$, we can obtain an effective non-Hermitian Hamiltonian
\begin{equation}
H_{\rm eff}=H-\frac{i}{2}\sum_jL^\dag_jL_j
\end{equation}
under postselection \cite{Ashida2016,Ashida2017a,Ashida2018} or for loss processes of a coherent condensate \cite{Ott2013,Duan2017,Gong2017}. In particular, if we choose
\begin{equation}
H=-J\sum_j(c^\dag_{j+1}c_j+{\rm H.c.}),\;\;\;\;
L_j=\sqrt{\kappa}(c_j\pm ic_{j+1}),
\label{HLj}
\end{equation}
where $L_j$'s describe a \emph{collective} one-body loss \cite{Gong2017}, the  effective non-Hermitian Hamiltonian involves asymmetric hopping amplitudes: 
\begin{equation}
H_{\rm eff}=\sum_j(J_{\rm R}c^\dag_{j+1}c_j+J_{\rm L}c^\dag_jc_{j+1})-i\kappa N,
\end{equation}
where $J_{\rm R}=-J\mp\frac{\kappa}{2}$ differs from $J_{\rm L}=-J\pm\frac{\kappa}{2}$ and $N=\sum_jc^\dag_jc_j$ is the total particle-number operator, so that the last term corresponds to a background loss. Unlike Fig.~\ref{fig5}(d), the energy spectrum is now below the real axis due to atom loss, and the imaginary part of its center is located at $-i\kappa$. 

It is not straightforward to engineer a nonlocal one-body loss like $L_j$'s in Eq.~(\ref{HLj}), since the usual loss process occurs locally \cite{Ott2013}. However, we can effectively engineer such a novel nonlocal loss by using a nonlocal Rabi coupling to some auxiliary degrees of freedom which undergo rapid local loss. After adiabatically eliminating the fast decay modes \cite{Sorensen2012}, we end up with an effective dynamics with target degrees of freedom alone, which now effectively undergo nonlocal loss. 

\begin{figure}
\begin{center}
       \includegraphics[width=7cm, clip]{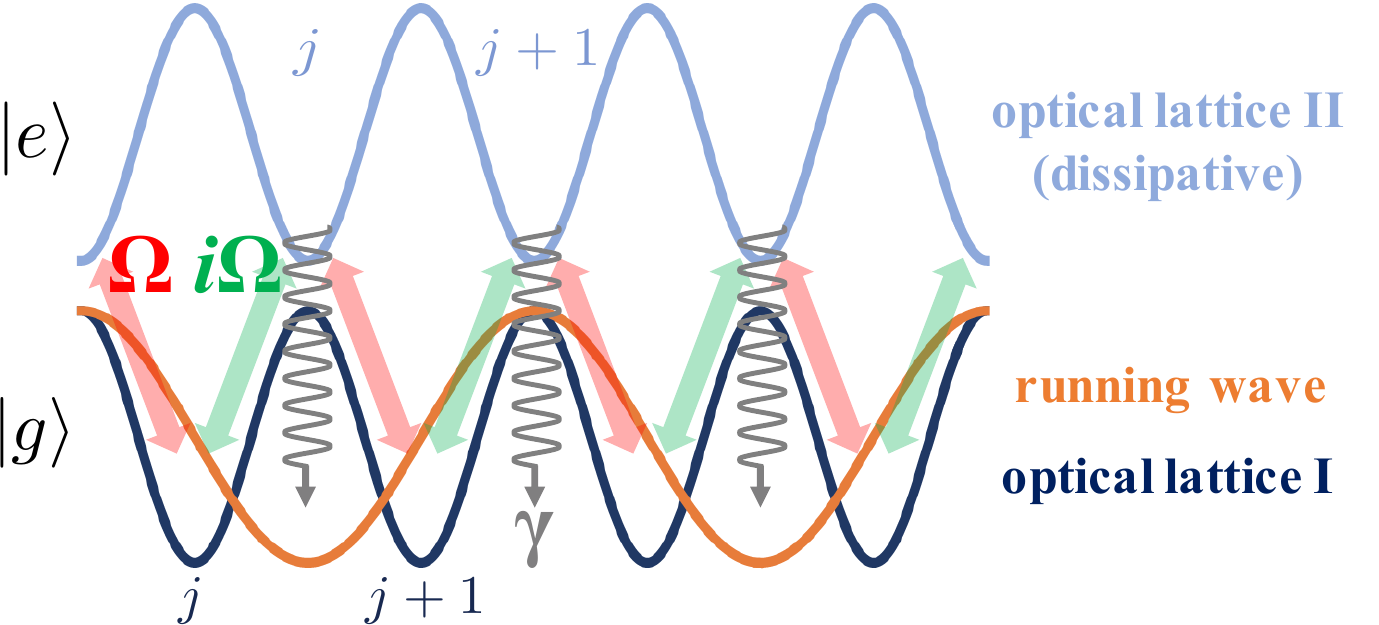}
       \end{center}
   \caption{Implementation of asymmetric hopping amplitudes in optical lattices. A stable (dissipative) optical lattice is applied to the ground (excited) state $|g\rangle$ ($|e\rangle$). A running wave parallel to an optical lattice couples $|g\rangle$ to $|e\rangle$, which undergoes rapid on-site loss at a rate $\gamma$. By making the wavelength of the running wave equal to that of the lattice constant, the phases of Rabi couplings can be adjusted to change by $\pi/2$ compared with the left nearest ones.}
   \label{fig17}
\end{figure}

As illustrated in Fig.~\ref{fig17}, we consider a system of two-level atoms with internal states $|g\rangle$ and $|e\rangle$ in a one-dimensional optical lattice with lattice constant $a$. Due to an opposite Stark shift, the potential minima for $|e\rangle$ locate in the middles of each of those for $|g\rangle$. The excited state $|e\rangle$ is assumed to be unstable and rapidly escape from the lattice at a rate $\gamma$. Parallel to the optical lattices, we further apply a running-wave laser with frequency $\omega_{\rm R}$, which is detuned from the atomic frequency $\omega_{eg}$ by $\Delta=\omega_{\rm R}-\omega_{eg}$. The strength of the laser-atom dipole coupling is characterized by a Rabi frequency $\Omega_{\rm R}$. Within the tight binding approximation and neglecting the interactions between atoms, we can write down the master equation in the rotating frame of reference as
\begin{equation}
\dot\rho_t=-i\sum_j[H_0+V,\rho_t]+
\mathcal{D}[c_{je}]\rho_t,
\label{BME}
\end{equation}
where $H_0$ includes the bare tunneling and $V$ couples different internal states:
\begin{equation}
\begin{split}
&H_0=-J\sum_{j,\alpha=g,e}(c^\dag_{j+1,\alpha}c_{j\alpha}+{\rm H.c.})-\sum_j\Delta c^\dag_{je}c_{je},\\
&V=\frac{1}{2}\sum_j[\Omega e^{\pm ik_{\rm R}ja}c^\dag_{je}(c_{jg}+e^{\pm\frac{i}{2}k_{\rm R}a}c_{j+1,g})+{\rm H.c.}].
\end{split}
\end{equation}
Here $+$ ($-$) corresponds to the right (left) propagating wave and the Rabi coupling $\Omega$ can be determined from $\Omega=\Omega_{\rm R}\int dx e^{\pm ik_{\rm R}x}W(x)W(x-\frac{a}{2})$, with $W(x)$ being the Wannier function. In the regime of $\max\{\Delta,\gamma\}\gg J,\Omega$, we can adiabatically eliminate $c_{je}$ in Eq.~(\ref{BME}) to obtain Eq.~(\ref{LE}) with the same $H$ as in Eq.~(\ref{HLj}) and a more general $L_j$: 
\begin{equation}
L_j=\frac{\sqrt{\gamma}|\Omega|}{\sqrt{\gamma^2+4\Delta^2}}(c_{jg}+e^{\pm\frac{i}{2}k_{\rm R}a}c_{j+1,g}),
\end{equation}
which gives the second equation in Eq.~(\ref{HLj}) if $k_{\rm R}a=\pi$ or $\lambda_{\rm R}=2a$. Note that even if $\lambda_{\rm R}$ differs from $\lambda_{\rm L}=2a$, which is the wavelength of the optical lattice laser, we can still obtain Eq.~(\ref{HLj}) by tilting the running wave from the optical lattice by an angle $\theta_{\rm R}=\arccos(\lambda_{\rm R}/\lambda_{\rm L})$ as long as $\lambda_{\rm R}<\lambda_{\rm L}$.

In a realistic experiment, we can, for example, use $^{174}$Yb atoms and $1117$ nm-wavelength lasers to create the anti-magic optical lattice with opposite Stark shifts for $g={}^1{\rm S}_0$ and $e={}^3{\rm P}_0$ \cite{Dalibard2010}. We choose a relatively shallow (yet the tight-binding approximation still works well) lattice depth $V_0=5E_{\rm r}$, with $E_{\rm r}\equiv h^2/(2m\lambda^2_{\rm L})=2\pi\times0.92$ kHz being the recoil energy. The bare hopping amplitude is thus estimated to be $J=0.066E_{\rm r}=2\pi\times60$ Hz \cite{Bloch2008}. The on-site loss rate $\gamma$ of $|e\rangle$ can be controlled by a $1285$ nm laser that couples $^3{\rm P}_0$ to $^1{\rm P}_1$ resonantly, and we can still make $\kappa=\gamma|\Omega|^2/(\gamma^2+4\Delta^2)$ as small as, e.g., $0.2J=2\pi\times 12$ Hz by tuning $\gamma$, $|\Omega|$ or/and $\Delta$. Here, we should make $\gamma$ much less than the band gap $4.6E_{\rm r}$ of the optical lattice to justify the tight-binding approximation for $|e\rangle$ (e.g., we can choose $\Delta=0$ and $\gamma=5\Omega=0.33E_{\rm r}=2\pi\times0.30$ kHz). The wavelength of the running-wave laser is fixed at $578.42$ nm (clock transition \cite{Barber2006}) and the tilting angle should be $\theta_{\rm R}=58.8^{\circ}$. The potential gradient can be made from an optical dipole force via an additional laser beam \cite{Monika2015}, and may be chosen to be, e.g., $F=4\kappa=2\pi\times 48$ Hz, which is much smaller than the band gap and thus justifies the single-band treatment. The maximum displacement can thus be evaluated to be $2J/F=5$ lattice sites, which is enough to be measured by single-site resolved quantum gas microscopy \cite{Kozuma2015}. The period of Bloch oscillations is $T_{\rm B}=2\pi/F=21$ ms, after which the survival fraction of atoms is given by $e^{-2\kappa T_{\rm B}}=4.3\%$, which should be sufficiently large for reconstructing the complex energy spectrum if there are at least thousands of atoms at the initial time.

\section{Proof of Theorem \ref{HU}}
\label{Pf}
To prove $H(\boldsymbol{k})\simeq U(\boldsymbol{k})$, we have to first confirm that $U(\boldsymbol{k})$ belongs to the same \emph{symmetry class} of $H(\boldsymbol{k})$. For an arbitrary anti-unitary symmetry or anti-symmetry $A=U_AK$ ($U_A$ is unitary and $K$ denotes complex conjugation), if $AH(\boldsymbol{k})=\eta_AH(-\boldsymbol{k})A$ ($\eta_A=\pm1$), by performing the polar decomposition $H(\boldsymbol{k})=U(\boldsymbol{k})P(\boldsymbol{k})$ ($P(\boldsymbol{k})=\sqrt{H^\dag(\boldsymbol{k})H(\boldsymbol{k})}$), we obtain
\begin{equation}
\begin{split}
&\;\;\;\;\;\;U_AU^*(\boldsymbol{k})P^*(\boldsymbol{k})=\eta_AU(-\boldsymbol{k})P(-\boldsymbol{k})U_A\\
&\Rightarrow U_AP^{*2}(\boldsymbol{k})U^\dag_A=P^2(-\boldsymbol{k})\\
&\Rightarrow [P(-\boldsymbol{k})+U_AP^*(\boldsymbol{k})U^\dag_A][P(-\boldsymbol{k})-U_AP^*(\boldsymbol{k})U^\dag_A]=0,
\end{split}
\end{equation}
where the unitarity of $U_A$ and $U(\boldsymbol{k})$ ($U^*(\boldsymbol{k})$) and the Hermiticity of $P(\boldsymbol{k})$ ($P^*(\boldsymbol{k})$) are used. Recalling that $P(\boldsymbol{k})$ ($P^*(\boldsymbol{k})$) is positive-definite, we can infer that $P(-\boldsymbol{k})+U_AP^*(\boldsymbol{k})U^\dag_A$ is also positive-definite and thus invertible. This fact implies 
\begin{equation}
\begin{split}
&\;\;\;\;\;\;\;\;P(-\boldsymbol{k})=U_AP^*(\boldsymbol{k})U^\dag_A\\
&\Rightarrow\;\;U_AU^*(\boldsymbol{k})=\eta_AU(-\boldsymbol{k})U_A.
\end{split}
\label{UAU}
\end{equation}
Following a similar procedure, we can prove that $U(\boldsymbol{k})$ and $H(\boldsymbol{k})$ share the same unitary symmetry or anti-symmetry, irrespective of the fact that $\boldsymbol{k}$ is flipped or not. This is why we use the term \emph{symmetry class} in the beginning, which is much wider than the AZ class (for example, we can consider crystalline symmetries).

We can now construct the following path
\begin{equation}
\begin{split}
H_\lambda(\boldsymbol{k})&=(1-\lambda)H(\boldsymbol{k})+\lambda U(\boldsymbol{k})\\
&=U(\boldsymbol{k})[(1-\lambda)P(\boldsymbol{k})+\lambda],
\end{split}
\end{equation}
which satisfy $H_0(\boldsymbol{k})=H(\boldsymbol{k})$ and $H_1(\boldsymbol{k})=U(\boldsymbol{k})$. Furthermore, $H_\lambda(\boldsymbol{k})$ shares the same symmetry as $H(\boldsymbol{k})$ and $U(\boldsymbol{k})$ and is indeed invertible due to the fact that $(1-\lambda)P(\boldsymbol{k})+\lambda$ is positive-definite.

\section{Evenness of the winding numbers for classes AII and C}
\label{even}
Since the winding number is a topological invariant, it can be calculated from the unitarized Hamiltonian $U(k)$. Let us first show that ${\rm Tr}[U^\dag(k)\partial_k U(k)]$ is an even function of $k$. From Eq.~(\ref{UAU}) we know that
\begin{equation}
\begin{split}
&\;\;\;\;-{\rm Tr}[U^\dag(-k)\partial_k U(-k)]\\
&=-\eta^2_A{\rm Tr}[U_AU^{{\rm T}}(k)U^\dag_A\partial_k(U_AU^*(k)U^\dag_A)]\\
&=-{\rm Tr}[U^{{\rm T}}(k)\partial_kU^*(k)]=-{\rm Tr}[(\partial_kU^\dag(k))U(k)]\\
&=-\partial_k{\rm Tr}[U^\dag(k)U(k)]+{\rm Tr}[U^\dag(k)\partial_kU(k)]\\
&={\rm Tr}[U^\dag(k)\partial_k U(k)],
\end{split}
\end{equation}
where we have used ${\rm Tr}[A^{\rm T}]={\rm Tr}[A]$, $\partial_k(AB)=(\partial_kA)B+A\partial_kB$ and $U^\dag(k)U(k)=1$. Using the fact that ${\rm Tr}[U^\dag(k)\partial_k U(k)]$ is even in terms of $k$, the winding number can be expressed as
\begin{equation}
\begin{split}
w&=\int^\pi_{-\pi}\frac{dk}{2\pi i}{\rm Tr}[U^\dag(k)\partial_k U(k)]\\
&=2\int^\pi_0\frac{dk}{2\pi i}{\rm Tr}[U^\dag(k)\partial_k U(k)]\\
&=2\int^\pi_0\frac{dk}{2\pi i}\partial_k\ln{\rm det}U(k).
\end{split}
\label{w2}
\end{equation}
However, this is not sufficient to ensure $w\in2\mathbb{Z}$ since $\int^\pi_0\frac{dk}{2\pi i}\partial_k\ln{\rm det}U(k)$ may be a half-integer. Ineed, Eq.~(\ref{w2}) is applicable also to classes AI and D. To rule out this possibility, we should show that ${\rm det}U(0)$ and ${\rm det}U(\pi)$ share the same argument. 

To this end, we first write down the explicit form of $U(\Gamma)$ ($\Gamma=0,\pi$):
\begin{equation}
U(\Gamma)=\begin{bmatrix} u_1 & u_2 \\ \mp u^*_2 & \pm u^*_1 \end{bmatrix},
\label{UGamma}
\end{equation}
where, due to $U(\Gamma)U^\dag(\Gamma)=1$, the blocks $u_{1,2}$ satisfy
\begin{equation}
u_1u^\dag_1+u_2u^\dag_2=1,\;\;\;\;
u_1u^{\rm T}_2=u_2u^{\rm T}_1.
\label{u12}
\end{equation}
If $u_1$ is invertible, the second identity in Eq.~(\ref{u12}) is equivalent to $u^{-1}_1u_2=(u^{-1}_1u_2)^{\rm T}$ and we can apply the  determinant formula for block matrices \cite{Meyer2000}
\begin{equation}
{\rm det}\begin{bmatrix} A & B \\ C & D \end{bmatrix}={\rm det} A\;{\rm det}(D-CA^{-1}B)
\end{equation}
to Eq.~(\ref{UGamma}), obtaining
\begin{equation}
\begin{split}
{\rm det} U(\Gamma)&={\rm det}u_1\;{\rm det}(\pm u^*_1\pm u^*_2u^{-1}_1u_2)\\
&=(\pm1)^m{\rm det}u_1\;{\rm det}[u^*_1+u^*_2(u^{-1}_1u_2)^{\rm T}]\\
&=(\pm1)^m{\rm det}u_1\;{\rm det}(u^\dag_1+u^{-1}_1u_2u^\dag_2)\\
&=(\pm1)^m{\rm det}(u_1u^\dag_1+u_2u^\dag_2)=(\pm1)^m.\\
\end{split}
\end{equation}
Here $m$ is the size of $u_{1,2}$ in Eq.~(\ref{UGamma}) (or that of $h_{1,2}$ in Eq.~(\ref{HAC})) and the properties ${\rm det} A={\rm det} A^{\rm T}$ and ${\rm det}(AB)={\rm det}A{\rm det}B$ have been used. If $u_1$ is not invertible, we expect $U(\Gamma)$ to be connected to some nearby time-reversal or particle-hole symmetric (with $T^2=-1$ or $C^2=-1$) unitary matrices with invertible $u_1$ and we arrive at the same result due to the fact that ${\rm det} U(\Gamma)=\pm1$ cannot change suddenly during continuous deformation.

In fact, we can easily obtain the same result by looking at the individual eigenvalues. Note that $TU(\Gamma)=U(\Gamma)T$ ($CU(\Gamma)=-U(\Gamma)C$) with $\Gamma=0,\pi$, and $T^2=-1$ ($C^2=-1$) enforce the eigenvalues to appear in pairs like $e^{\pm i\theta_\alpha}$ ($\pm e^{\pm i\theta_\alpha}$), leading to ${\rm det}\;U(0)={\rm det}\;U(\pi)=1$ (${\rm det}\;U(0)={\rm det}\;U(\pi)=(-1)^m$). This fact ensures that $w=2\int^\pi_0\frac{dk}{2\pi i}\partial_k\ln{\rm det}\;U(k)$ is quantized as an even integer.

\section{Class A with $PT$ symmetry}
\label{CAPT}

The set of all the $PT$-symmetric systems without any other symmetry requirements can be obtained by imposing $PT$ symmetry into the non-Hermitian class A, which is equivalent to the Hermitian class AIII. Since the $PT$ symmetry does not invert the sign of the wave vector $\boldsymbol{k}$, we have to use the formula developed in Ref.~\cite{Shiozaki2014}:
\begin{equation}
K^A_{\mathbb{C}}(s;d,d_\parallel)=\pi_0(\mathcal{R}_{s-d+2d_\parallel}),
\end{equation}
where $s$ is determined by the properties of the anti-unitary symmetry $A$ and $d_\parallel$ is the number of $\boldsymbol{k}$ components that do not change their signs under $A$. It is clear that $d_\parallel=d$ for $A=PT$, and $s$ should be 1 (BDI-like) since $PT$ is involutory and commutes with the virtual chiral symmetry. We can thus obtain the classification in all dimensions shown in Table~\ref{table2}. Note that this classification does not rule out the possibilities of other topological numbers in $PT$-symmetric systems with exceptional points in the bulk \cite{Malzard2015,Tony2016,Nori2017,Duan2017}.

A remarkable result in Table~\ref{table2} is that a one-dimensional $PT$-symmetric system $H(k)$ is characterized by a $\mathbb{Z}_2$ topological index rather than a $\mathbb{Z}$ winding number (see Table~\ref{table1}). Such a $\mathbb{Z}_2$ index should be different from ${\rm sgn [det}H(k)]$, which is like a weak topological index inheriting from zero dimension. Instead, this result should be understood from the fact that the fundamental group of ${\rm GL}^+_n(\mathbb{R})$ is $\mathbb{Z}_2$ for $n>2$, where ${\rm GL}^+_n(\mathbb{R})$ denotes the general linear group of all the $n\times n$ real matrices with positive determinant. This is because $PT$ can always be represented as $K$ under a properly chosen basis and the sign of ${\rm det}H(k)$ determines the branch of ${\rm GL}_n(\mathbb{R})$ to which $H(k)$'s belong. 

We should mention that if the Hilber-space dimension is fixed to be 2, which is the case in a recent experiment \cite{Segev2017}, we will obtain a different classification as $\mathbb{Z}$. This is because each matrix in ${\rm GL}^+_2(\mathbb{R})$ can continuously be deformed into that in ${\rm SO}(2)$, which is isomorphic to $S^1$, giving $\pi_1({\rm GL}^+_2(\mathbb{R}))=\pi_1(S^1)=\mathbb{Z}$. This example, similar to the Hopf insulator in Hermitian systems \cite{Moore2008}, illustrates the fact that the homotopy classification does not always coincide with the K-theory classification, since the latter allows the operation of band insertion and thus contains more general operations of continuous deformation. 

\begin{figure}
\begin{center}
       \includegraphics[width=8.5cm, clip]{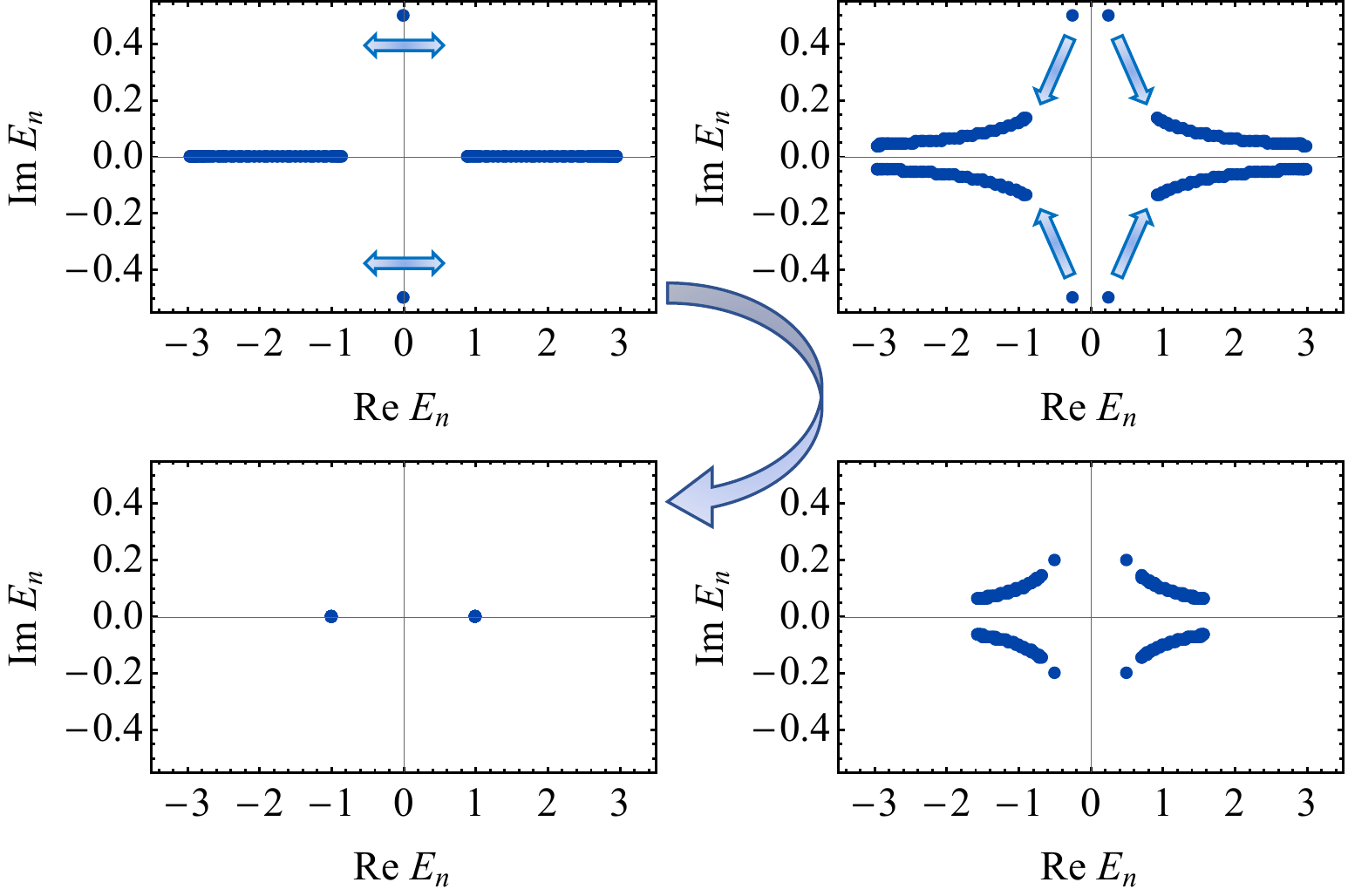}
       \end{center}
   \caption{Snapshots of the energy spectra during the trivialization process for two coupled $PT$ symmetry Su-Schrieffer-Heeger chains (\ref{H12}) under the open-boundary condition. The arrows indicate the direction of the spectral flow. The parameters $(J,J',\gamma,J_{\rm c})$ are given by $(1,2,0.5,0)$ and $(0,0,0,1)$ at the initial and final times, respectively.}
   \label{fig18}
\end{figure}

An important implication of the $\mathbb{Z}_2$ classification is that any $PT$-symmetric two-band lattice can be trivialized if we combine it 
with its copy. Let us demonstrate such a trivialization process for two copies of $PT$-symmetric Su-Schrieffer-Heeger chains realized in Ref.~\cite{,Segev2017}:
\begin{equation}
\begin{split}
H_\alpha&=\sum_j(Jb^\dag_{j\alpha}a_{j\alpha}+J'a^\dag_{j+1,\alpha}b_{j\alpha}+{\rm H.c.})\\
&+i\gamma\sum_j(a^\dag_{j\alpha}a_{j\alpha}-b^\dag_{j\alpha}b_{j\alpha}),
\end{split}
\label{PTSSH}
\end{equation}
where $a_{j\alpha}$ and $b_{j\alpha}$ correspond to two sublattice degrees of freedom, $j\alpha$ labels the $j$th unit cell in the $\alpha$th chain ($\alpha=1,2$) and $J,J',\gamma\in\mathbb{R}^+$. In terms of Pauli matrices, the Bloch Hamiltonian can be written as $H_\alpha(k)=(J+J'\cos k)\sigma_x+J'\sin k\sigma_y+i\gamma\sigma_z$, with $\max_k\{{\rm det}H_\alpha(k)\}=\gamma^2-(J-J')^2$. Assuming $J'-J>\gamma$, after unitarization and changing the basis such that $PT=K$, we obtain the ${\rm O}(2)$ matrix $[J'\sin k\sigma_z+(J'\cos k+J)\sigma_x]/|q(k)|$ with $q(k)=J+J'e^{-ik}$, which has a nontrivial winding number $w=1$. While $H_\alpha(k)$ alone is nontrivial, we can trivialize the combined system of the two chains ($\alpha=1,2$) via a phase-staggered coupling:
\begin{equation}
H=H_1+H_2+J_{\rm c}\sum_j(ia^\dag_{j1}a_{j2}-ib^\dag_{j1}b_{j2}+{\rm H.c.}),
\label{H12}
\end{equation}
which respects the $PT$ symmetry for  $J_{\rm c}\in\mathbb{R}$. The determinant of the four-band Bloch Hamiltonian of Eq.~(\ref{H12}) can be evaluated to be
\begin{equation}
{\rm det}H(k)=[J^2_{\rm c}+(\gamma+|q(k)|)^2][J^2_{\rm c}+(\gamma-|q(k)|)^2],
\end{equation}
which does not vanish as long as $J_{\rm c}\neq0$. Therefore, after introducing a finite phase-staggered coupling $J_{\rm c}$, we can safely change $J,J',\gamma$ to zero to obtain a trivial band insulator. 

It is worthwhile to trace the spectral flow in the above trivialization process. As shown in Fig.~\ref{fig18}, under the open-boundary condition, two pairs of $PT$-broken edge modes of the coupled $PT$ symmetry Su-Schrieffer-Heeger chains are gradually absorbed into the bulk spectrum. Such a process is impossible for a single pair of $PT$-broken edge modes without touching or crossing the origin and retrieving the $PT$ symmetry. Note that the reflection symmetry with respect to the imaginary axis arises from a particle-hole symmetry $C=\sigma_zK$, which anti-commutes with $PT=\sigma_xK$ and also leads to a $\mathbb{Z}_2$ classification in one dimension \cite{Shiozaki2014}: 
\begin{equation}
\begin{split}
&K^A_{\mathbb{R}}(s,t;d,d_\parallel)|_{s=1,t=0,d=1,d_\parallel=1}\\
=&K^A_{\mathbb{R}}(s-d,t-d_\parallel;0,0)|_{s=1,t=0,d=1,d_\parallel=1}\\
=&K^A_{\mathbb{R}}(0,-1;0,0)=\pi_0(\mathcal{R}_1)=\mathbb{Z}_2.
\end{split}
\end{equation}

Finally, 
we make a conjecture that the $\mathbb{Z}_2$ index manifests itself as the number of potentially 
$PT$-broken edge-mode pairs, and can thus be computed as
\begin{equation}
s={\rm sgn}({\rm det}(H_{\rm o}+\epsilon)){\rm sgn}({\rm det}H_{\rm p}),
\label{sop}
\end{equation}
where $H_{\rm o}$ ($H_{\rm p}$) is the full Hamiltonian under the open-boundary (periodic-boundary) condition, and $\epsilon$ is an arbitrarily small \emph{real} number that is necessary for avoiding an ill definition in the presence of zero modes, which are counted as potentially $PT$-broken pairs. Note that unlike the chiral symmetry, a $PT$-symmetric Hamiltonian maintains the $PT$-symmetry under the translation $H\to H+E$ for all $E\in\mathbb{R}$. 
The topological index given by Eq.~(\ref{sop}) can be interpreted as whether a topological transition occurs at the edge that changes the zero-dimensional $\mathbb{Z}_2$ index (discussed in Sec.~\ref{ZD}) when the boundary condition changes. Similar to a $\mathbb{Z}_2$ topological insulator \cite{Kane2005,Bernevig2006}, which has an odd number of helical modes at the edge, a nontrivial $PT$-symmetric system in one dimension should exhibt an odd number of edge-mode pairs, leading to $s=-1$. If there is additional particle-hole symmetry, we can conclude that a system with $s=-1$ must have an odd number of pairs of $PT$-broken edge modes with purely imaginary eigenenergies, and thus the system possesses at least one pair.


\bibliography{GZP_references}

\begin{thebibliography}{182}%
\makeatletter
\providecommand \@ifxundefined [1]{%
 \@ifx{#1\undefined}
}%
\providecommand \@ifnum [1]{%
 \ifnum #1\expandafter \@firstoftwo
 \else \expandafter \@secondoftwo
 \fi
}%
\providecommand \@ifx [1]{%
 \ifx #1\expandafter \@firstoftwo
 \else \expandafter \@secondoftwo
 \fi
}%
\providecommand \natexlab [1]{#1}%
\providecommand \enquote  [1]{``#1''}%
\providecommand \bibnamefont  [1]{#1}%
\providecommand \bibfnamefont [1]{#1}%
\providecommand \citenamefont [1]{#1}%
\providecommand \href@noop [0]{\@secondoftwo}%
\providecommand \href [0]{\begingroup \@sanitize@url \@href}%
\providecommand \@href[1]{\@@startlink{#1}\@@href}%
\providecommand \@@href[1]{\endgroup#1\@@endlink}%
\providecommand \@sanitize@url [0]{\catcode `\\12\catcode `\$12\catcode
  `\&12\catcode `\#12\catcode `\^12\catcode `\_12\catcode `\%12\relax}%
\providecommand \@@startlink[1]{}%
\providecommand \@@endlink[0]{}%
\providecommand \url  [0]{\begingroup\@sanitize@url \@url }%
\providecommand \@url [1]{\endgroup\@href {#1}{\urlprefix }}%
\providecommand \urlprefix  [0]{URL }%
\providecommand \Eprint [0]{\href }%
\providecommand \doibase [0]{http://dx.doi.org/}%
\providecommand \selectlanguage [0]{\@gobble}%
\providecommand \bibinfo  [0]{\@secondoftwo}%
\providecommand \bibfield  [0]{\@secondoftwo}%
\providecommand \translation [1]{[#1]}%
\providecommand \BibitemOpen [0]{}%
\providecommand \bibitemStop [0]{}%
\providecommand \bibitemNoStop [0]{.\EOS\space}%
\providecommand \EOS [0]{\spacefactor3000\relax}%
\providecommand \BibitemShut  [1]{\csname bibitem#1\endcsname}%
\let\auto@bib@innerbib\@empty
\bibitem [{\citenamefont {Thouless}\ \emph {et~al.}(1982)\citenamefont
  {Thouless}, \citenamefont {Kohmoto}, \citenamefont {Nightingale},\ and\
  \citenamefont {den Nijs}}]{Thouless1982}%
  \BibitemOpen
  \bibfield  {author} {\bibinfo {author} {\bibfnamefont {D.~J.}\ \bibnamefont
  {Thouless}}, \bibinfo {author} {\bibfnamefont {M.}~\bibnamefont {Kohmoto}},
  \bibinfo {author} {\bibfnamefont {M.~P.}\ \bibnamefont {Nightingale}}, \ and\
  \bibinfo {author} {\bibfnamefont {M.}~\bibnamefont {den Nijs}},\ }\bibfield
  {title} {\enquote {\bibinfo {title} {Quantized hall conductance in a
  two-dimensional periodic potential},}\ }\href {\doibase
  10.1103/PhysRevLett.49.405} {\bibfield  {journal} {\bibinfo  {journal} {Phys.
  Rev. Lett.}\ }\textbf {\bibinfo {volume} {49}},\ \bibinfo {pages} {405}
  (\bibinfo {year} {1982})}\BibitemShut {NoStop}%
\bibitem [{\citenamefont {Haldane}(1983)}]{Haldane1983}%
  \BibitemOpen
  \bibfield  {author} {\bibinfo {author} {\bibfnamefont {F.~D.~M.}\
  \bibnamefont {Haldane}},\ }\bibfield  {title} {\enquote {\bibinfo {title}
  {Nonlinear field theory of large-spin heisenberg antiferromagnets:
  Semiclassically quantized solitons of the one-dimensional easy-axis n\'eel
  state},}\ }\href {\doibase 10.1103/PhysRevLett.50.1153} {\bibfield  {journal}
  {\bibinfo  {journal} {Phys. Rev. Lett.}\ }\textbf {\bibinfo {volume} {50}},\
  \bibinfo {pages} {1153} (\bibinfo {year} {1983})}\BibitemShut {NoStop}%
\bibitem [{\citenamefont {Haldane}(1988)}]{Haldane1988}%
  \BibitemOpen
  \bibfield  {author} {\bibinfo {author} {\bibfnamefont {F.~D.~M.}\
  \bibnamefont {Haldane}},\ }\bibfield  {title} {\enquote {\bibinfo {title}
  {Model for a quantum hall effect without landau levels: Condensed-matter
  realization of the "parity anomaly"},}\ }\href {\doibase
  10.1103/PhysRevLett.61.2015} {\bibfield  {journal} {\bibinfo  {journal}
  {Phys. Rev. Lett.}\ }\textbf {\bibinfo {volume} {61}},\ \bibinfo {pages}
  {2015} (\bibinfo {year} {1988})}\BibitemShut {NoStop}%
\bibitem [{\citenamefont {Wen}(1995)}]{Wen1995}%
  \BibitemOpen
  \bibfield  {author} {\bibinfo {author} {\bibfnamefont {Xiao-Gang}\
  \bibnamefont {Wen}},\ }\bibfield  {title} {\enquote {\bibinfo {title}
  {Topological orders and edge excitations in fractional quantum hall
  states},}\ }\href {\doibase 10.1080/00018739500101566} {\bibfield  {journal}
  {\bibinfo  {journal} {Adv. Phys.}\ }\textbf {\bibinfo {volume} {44}},\
  \bibinfo {pages} {405} (\bibinfo {year} {1995})}\BibitemShut {NoStop}%
\bibitem [{\citenamefont {Kane}\ and\ \citenamefont {Mele}(2005)}]{Kane2005}%
  \BibitemOpen
  \bibfield  {author} {\bibinfo {author} {\bibfnamefont {C.~L.}\ \bibnamefont
  {Kane}}\ and\ \bibinfo {author} {\bibfnamefont {E.~J.}\ \bibnamefont
  {Mele}},\ }\bibfield  {title} {\enquote {\bibinfo {title} {Quantum spin hall
  effect in graphene},}\ }\href {\doibase 10.1103/PhysRevLett.95.226801}
  {\bibfield  {journal} {\bibinfo  {journal} {Phys. Rev. Lett.}\ }\textbf
  {\bibinfo {volume} {95}},\ \bibinfo {pages} {226801} (\bibinfo {year}
  {2005})}\BibitemShut {NoStop}%
\bibitem [{\citenamefont {Bernevig}\ \emph {et~al.}(2006)\citenamefont
  {Bernevig}, \citenamefont {Hughes},\ and\ \citenamefont
  {Zhang}}]{Bernevig2006}%
  \BibitemOpen
  \bibfield  {author} {\bibinfo {author} {\bibfnamefont {B.~Andrei}\
  \bibnamefont {Bernevig}}, \bibinfo {author} {\bibfnamefont {Taylor~L.}\
  \bibnamefont {Hughes}}, \ and\ \bibinfo {author} {\bibfnamefont {Shou-Cheng}\
  \bibnamefont {Zhang}},\ }\bibfield  {title} {\enquote {\bibinfo {title}
  {Quantum spin hall effect and topological phase transition in hgte quantum
  wells},}\ }\href {\doibase 10.1126/science.1133734} {\bibfield  {journal}
  {\bibinfo  {journal} {Science}\ }\textbf {\bibinfo {volume} {314}},\ \bibinfo
  {pages} {1757} (\bibinfo {year} {2006})}\BibitemShut {NoStop}%
\bibitem [{\citenamefont {K{\"o}nig}\ \emph {et~al.}(2007)\citenamefont
  {K{\"o}nig}, \citenamefont {Wiedmann}, \citenamefont {Br{\"u}ne},
  \citenamefont {Roth}, \citenamefont {Buhmann}, \citenamefont {Molenkamp},
  \citenamefont {Qi},\ and\ \citenamefont {Zhang}}]{Molenkamp2007}%
  \BibitemOpen
  \bibfield  {author} {\bibinfo {author} {\bibfnamefont {Markus}\ \bibnamefont
  {K{\"o}nig}}, \bibinfo {author} {\bibfnamefont {Steffen}\ \bibnamefont
  {Wiedmann}}, \bibinfo {author} {\bibfnamefont {Christoph}\ \bibnamefont
  {Br{\"u}ne}}, \bibinfo {author} {\bibfnamefont {Andreas}\ \bibnamefont
  {Roth}}, \bibinfo {author} {\bibfnamefont {Hartmut}\ \bibnamefont {Buhmann}},
  \bibinfo {author} {\bibfnamefont {Laurens~W.}\ \bibnamefont {Molenkamp}},
  \bibinfo {author} {\bibfnamefont {Xiao-Liang}\ \bibnamefont {Qi}}, \ and\
  \bibinfo {author} {\bibfnamefont {Shou-Cheng}\ \bibnamefont {Zhang}},\
  }\bibfield  {title} {\enquote {\bibinfo {title} {Quantum spin hall insulator
  state in hgte quantum wells},}\ }\href {\doibase 10.1126/science.1148047}
  {\bibfield  {journal} {\bibinfo  {journal} {Science}\ }\textbf {\bibinfo
  {volume} {318}},\ \bibinfo {pages} {766} (\bibinfo {year}
  {2007})}\BibitemShut {NoStop}%
\bibitem [{\citenamefont {Hasan}\ and\ \citenamefont {Kane}(2010)}]{Kane2010}%
  \BibitemOpen
  \bibfield  {author} {\bibinfo {author} {\bibfnamefont {M.~Z.}\ \bibnamefont
  {Hasan}}\ and\ \bibinfo {author} {\bibfnamefont {C.~L.}\ \bibnamefont
  {Kane}},\ }\bibfield  {title} {\enquote {\bibinfo {title} {Colloquium:
  Topological insulators},}\ }\href {\doibase 10.1103/RevModPhys.82.3045}
  {\bibfield  {journal} {\bibinfo  {journal} {Rev. Mod. Phys.}\ }\textbf
  {\bibinfo {volume} {82}},\ \bibinfo {pages} {3045} (\bibinfo {year}
  {2010})}\BibitemShut {NoStop}%
\bibitem [{\citenamefont {Qi}\ and\ \citenamefont {Zhang}(2011)}]{Qi2011}%
  \BibitemOpen
  \bibfield  {author} {\bibinfo {author} {\bibfnamefont {Xiao-Liang}\
  \bibnamefont {Qi}}\ and\ \bibinfo {author} {\bibfnamefont {Shou-Cheng}\
  \bibnamefont {Zhang}},\ }\bibfield  {title} {\enquote {\bibinfo {title}
  {Topological insulators and superconductors},}\ }\href {\doibase
  10.1103/RevModPhys.83.1057} {\bibfield  {journal} {\bibinfo  {journal} {Rev.
  Mod. Phys.}\ }\textbf {\bibinfo {volume} {83}},\ \bibinfo {pages} {1057}
  (\bibinfo {year} {2011})}\BibitemShut {NoStop}%
\bibitem [{\citenamefont {Beenakker}(2015)}]{Beenakker2015}%
  \BibitemOpen
  \bibfield  {author} {\bibinfo {author} {\bibfnamefont {C.~W.~J.}\
  \bibnamefont {Beenakker}},\ }\bibfield  {title} {\enquote {\bibinfo {title}
  {Random-matrix theory of majorana fermions and topological
  superconductors},}\ }\href {\doibase 10.1103/RevModPhys.87.1037} {\bibfield
  {journal} {\bibinfo  {journal} {Rev. Mod. Phys.}\ }\textbf {\bibinfo {volume}
  {87}},\ \bibinfo {pages} {1037} (\bibinfo {year} {2015})}\BibitemShut
  {NoStop}%
\bibitem [{\citenamefont {Chiu}\ \emph {et~al.}(2016)\citenamefont {Chiu},
  \citenamefont {Teo}, \citenamefont {Schnyder},\ and\ \citenamefont
  {Ryu}}]{Ryu2016}%
  \BibitemOpen
  \bibfield  {author} {\bibinfo {author} {\bibfnamefont {Ching-Kai}\
  \bibnamefont {Chiu}}, \bibinfo {author} {\bibfnamefont {Jeffrey C.~Y.}\
  \bibnamefont {Teo}}, \bibinfo {author} {\bibfnamefont {Andreas~P.}\
  \bibnamefont {Schnyder}}, \ and\ \bibinfo {author} {\bibfnamefont {Shinsei}\
  \bibnamefont {Ryu}},\ }\bibfield  {title} {\enquote {\bibinfo {title}
  {Classification of topological quantum matter with symmetries},}\ }\href
  {\doibase 10.1103/RevModPhys.88.035005} {\bibfield  {journal} {\bibinfo
  {journal} {Rev. Mod. Phys.}\ }\textbf {\bibinfo {volume} {88}},\ \bibinfo
  {pages} {035005} (\bibinfo {year} {2016})}\BibitemShut {NoStop}%
\bibitem [{\citenamefont {Wen}(2017)}]{Wen2017}%
  \BibitemOpen
  \bibfield  {author} {\bibinfo {author} {\bibfnamefont {Xiao-Gang}\
  \bibnamefont {Wen}},\ }\bibfield  {title} {\enquote {\bibinfo {title}
  {Colloquium: Zoo of quantum-topological phases of matter},}\ }\href {\doibase
  10.1103/RevModPhys.89.041004} {\bibfield  {journal} {\bibinfo  {journal}
  {Rev. Mod. Phys.}\ }\textbf {\bibinfo {volume} {89}},\ \bibinfo {pages}
  {041004} (\bibinfo {year} {2017})}\BibitemShut {NoStop}%
\bibitem [{\citenamefont {Bloch}\ \emph {et~al.}(2012)\citenamefont {Bloch},
  \citenamefont {Dalibard},\ and\ \citenamefont {Nascimb\`ene}}]{Bloch2012}%
  \BibitemOpen
  \bibfield  {author} {\bibinfo {author} {\bibfnamefont {Immanuel}\
  \bibnamefont {Bloch}}, \bibinfo {author} {\bibfnamefont {Jean}\ \bibnamefont
  {Dalibard}}, \ and\ \bibinfo {author} {\bibfnamefont {Sylvain}\ \bibnamefont
  {Nascimb\`ene}},\ }\bibfield  {title} {\enquote {\bibinfo {title} {Quantum
  simulations with ultracold quantum gases},}\ }\href
  {http://dx.doi.org/10.1038/nphys2259} {\bibfield  {journal} {\bibinfo
  {journal} {Nat. Phys.}\ }\textbf {\bibinfo {volume} {8}},\ \bibinfo {pages}
  {267} (\bibinfo {year} {2012})}\BibitemShut {NoStop}%
\bibitem [{\citenamefont {Atala}\ \emph {et~al.}(2013)\citenamefont {Atala},
  \citenamefont {Aidelsburger}, \citenamefont {Barreiro}, \citenamefont
  {Abanin}, \citenamefont {Kitagawa}, \citenamefont {Demler},\ and\
  \citenamefont {Bloch}}]{Monika2013b}%
  \BibitemOpen
  \bibfield  {author} {\bibinfo {author} {\bibfnamefont {Marcos}\ \bibnamefont
  {Atala}}, \bibinfo {author} {\bibfnamefont {Monika}\ \bibnamefont
  {Aidelsburger}}, \bibinfo {author} {\bibfnamefont {Julio~T.}\ \bibnamefont
  {Barreiro}}, \bibinfo {author} {\bibfnamefont {Dmitry}\ \bibnamefont
  {Abanin}}, \bibinfo {author} {\bibfnamefont {Takuya}\ \bibnamefont
  {Kitagawa}}, \bibinfo {author} {\bibfnamefont {Eugene}\ \bibnamefont
  {Demler}}, \ and\ \bibinfo {author} {\bibfnamefont {Immanuel}\ \bibnamefont
  {Bloch}},\ }\bibfield  {title} {\enquote {\bibinfo {title} {Direct
  measurement of the zak phase in topological bloch bands},}\ }\href
  {http://dx.doi.org/10.1038/nphys2790} {\bibfield  {journal} {\bibinfo
  {journal} {Nat. Phys.}\ }\textbf {\bibinfo {volume} {9}},\ \bibinfo {pages}
  {795} (\bibinfo {year} {2013})}\BibitemShut {NoStop}%
\bibitem [{\citenamefont {Jotzu}\ \emph {et~al.}(2014)\citenamefont {Jotzu},
  \citenamefont {Messer}, \citenamefont {Desbuquois}, \citenamefont {Lebrat},
  \citenamefont {Uehlinger}, \citenamefont {Greif},\ and\ \citenamefont
  {Esslinger}}]{Esslinger2014}%
  \BibitemOpen
  \bibfield  {author} {\bibinfo {author} {\bibfnamefont {Gregor}\ \bibnamefont
  {Jotzu}}, \bibinfo {author} {\bibfnamefont {Michael}\ \bibnamefont {Messer}},
  \bibinfo {author} {\bibfnamefont {R\'{e}mi}\ \bibnamefont {Desbuquois}},
  \bibinfo {author} {\bibfnamefont {Martin}\ \bibnamefont {Lebrat}}, \bibinfo
  {author} {\bibfnamefont {Thomas}\ \bibnamefont {Uehlinger}}, \bibinfo
  {author} {\bibfnamefont {Daniel}\ \bibnamefont {Greif}}, \ and\ \bibinfo
  {author} {\bibfnamefont {Tilman}\ \bibnamefont {Esslinger}},\ }\bibfield
  {title} {\enquote {\bibinfo {title} {Experimental realization of the
  topological haldane model with ultracold fermions},}\ }\href
  {http://dx.doi.org/10.1038/nature13915} {\bibfield  {journal} {\bibinfo
  {journal} {Nature}\ }\textbf {\bibinfo {volume} {515}},\ \bibinfo {pages}
  {237} (\bibinfo {year} {2014})}\BibitemShut {NoStop}%
\bibitem [{\citenamefont {Aidelsburger}\ \emph {et~al.}(2015)\citenamefont
  {Aidelsburger}, \citenamefont {Lohse}, \citenamefont {Schweizer},
  \citenamefont {Atala}, \citenamefont {Barreiro}, \citenamefont
  {Nascimb\`ene}, \citenamefont {Cooper}, \citenamefont {Bloch},\ and\
  \citenamefont {Goldman}}]{Monika2015}%
  \BibitemOpen
  \bibfield  {author} {\bibinfo {author} {\bibfnamefont {M.}~\bibnamefont
  {Aidelsburger}}, \bibinfo {author} {\bibfnamefont {M.}~\bibnamefont {Lohse}},
  \bibinfo {author} {\bibfnamefont {C.}~\bibnamefont {Schweizer}}, \bibinfo
  {author} {\bibfnamefont {M.}~\bibnamefont {Atala}}, \bibinfo {author}
  {\bibfnamefont {J.~T.}\ \bibnamefont {Barreiro}}, \bibinfo {author}
  {\bibfnamefont {S.}~\bibnamefont {Nascimb\`ene}}, \bibinfo {author}
  {\bibfnamefont {N.~R.}\ \bibnamefont {Cooper}}, \bibinfo {author}
  {\bibfnamefont {I.}~\bibnamefont {Bloch}}, \ and\ \bibinfo {author}
  {\bibfnamefont {N.}~\bibnamefont {Goldman}},\ }\bibfield  {title} {\enquote
  {\bibinfo {title} {Measuring the chern number of hofstadter bands with
  ultracold bosonic atoms},}\ }\href {http://dx.doi.org/10.1038/nphys3171}
  {\bibfield  {journal} {\bibinfo  {journal} {Nat. Phys.}\ }\textbf {\bibinfo
  {volume} {11}},\ \bibinfo {pages} {162} (\bibinfo {year} {2015})}\BibitemShut
  {NoStop}%
\bibitem [{\citenamefont {Stuhl}\ \emph {et~al.}(2015)\citenamefont {Stuhl},
  \citenamefont {Lu}, \citenamefont {Aycock}, \citenamefont {Genkina},\ and\
  \citenamefont {Spielman}}]{Spielman2015}%
  \BibitemOpen
  \bibfield  {author} {\bibinfo {author} {\bibfnamefont {B.~K.}\ \bibnamefont
  {Stuhl}}, \bibinfo {author} {\bibfnamefont {H.-I.}\ \bibnamefont {Lu}},
  \bibinfo {author} {\bibfnamefont {L.~M.}\ \bibnamefont {Aycock}}, \bibinfo
  {author} {\bibfnamefont {D.}~\bibnamefont {Genkina}}, \ and\ \bibinfo
  {author} {\bibfnamefont {I.~B.}\ \bibnamefont {Spielman}},\ }\bibfield
  {title} {\enquote {\bibinfo {title} {Visualizing edge states with an atomic
  bose gas in the quantum hall regime},}\ }\href {\doibase
  10.1126/science.aaa8515} {\bibfield  {journal} {\bibinfo  {journal}
  {Science}\ }\textbf {\bibinfo {volume} {349}},\ \bibinfo {pages} {1514}
  (\bibinfo {year} {2015})}\BibitemShut {NoStop}%
\bibitem [{\citenamefont {Mancini}\ \emph {et~al.}(2015)\citenamefont
  {Mancini}, \citenamefont {Pagano}, \citenamefont {Cappellini}, \citenamefont
  {Livi}, \citenamefont {Rider}, \citenamefont {Catani}, \citenamefont {Sias},
  \citenamefont {Zoller}, \citenamefont {Inguscio}, \citenamefont {Dalmonte},\
  and\ \citenamefont {Fallani}}]{Inguscio2015}%
  \BibitemOpen
  \bibfield  {author} {\bibinfo {author} {\bibfnamefont {M.}~\bibnamefont
  {Mancini}}, \bibinfo {author} {\bibfnamefont {G.}~\bibnamefont {Pagano}},
  \bibinfo {author} {\bibfnamefont {G.}~\bibnamefont {Cappellini}}, \bibinfo
  {author} {\bibfnamefont {L.}~\bibnamefont {Livi}}, \bibinfo {author}
  {\bibfnamefont {M.}~\bibnamefont {Rider}}, \bibinfo {author} {\bibfnamefont
  {J.}~\bibnamefont {Catani}}, \bibinfo {author} {\bibfnamefont
  {C.}~\bibnamefont {Sias}}, \bibinfo {author} {\bibfnamefont {P.}~\bibnamefont
  {Zoller}}, \bibinfo {author} {\bibfnamefont {M.}~\bibnamefont {Inguscio}},
  \bibinfo {author} {\bibfnamefont {M.}~\bibnamefont {Dalmonte}}, \ and\
  \bibinfo {author} {\bibfnamefont {L.}~\bibnamefont {Fallani}},\ }\bibfield
  {title} {\enquote {\bibinfo {title} {Observation of chiral edge states with
  neutral fermions in synthetic hall ribbons},}\ }\href {\doibase
  10.1126/science.aaa8736} {\bibfield  {journal} {\bibinfo  {journal}
  {Science}\ }\textbf {\bibinfo {volume} {349}},\ \bibinfo {pages} {1510}
  (\bibinfo {year} {2015})}\BibitemShut {NoStop}%
\bibitem [{\citenamefont {Wu}\ \emph {et~al.}(2016)\citenamefont {Wu},
  \citenamefont {Zhang}, \citenamefont {Sun}, \citenamefont {Xu}, \citenamefont
  {Wang}, \citenamefont {Ji}, \citenamefont {Deng}, \citenamefont {Chen},
  \citenamefont {Liu},\ and\ \citenamefont {Pan}}]{Pan2016}%
  \BibitemOpen
  \bibfield  {author} {\bibinfo {author} {\bibfnamefont {Zhan}\ \bibnamefont
  {Wu}}, \bibinfo {author} {\bibfnamefont {Long}\ \bibnamefont {Zhang}},
  \bibinfo {author} {\bibfnamefont {Wei}\ \bibnamefont {Sun}}, \bibinfo
  {author} {\bibfnamefont {Xiao-Tian}\ \bibnamefont {Xu}}, \bibinfo {author}
  {\bibfnamefont {Bao-Zong}\ \bibnamefont {Wang}}, \bibinfo {author}
  {\bibfnamefont {Si-Cong}\ \bibnamefont {Ji}}, \bibinfo {author}
  {\bibfnamefont {Youjin}\ \bibnamefont {Deng}}, \bibinfo {author}
  {\bibfnamefont {Shuai}\ \bibnamefont {Chen}}, \bibinfo {author}
  {\bibfnamefont {Xiong-Jun}\ \bibnamefont {Liu}}, \ and\ \bibinfo {author}
  {\bibfnamefont {Jian-Wei}\ \bibnamefont {Pan}},\ }\bibfield  {title}
  {\enquote {\bibinfo {title} {Realization of two-dimensional spin-orbit
  coupling for bose-einstein condensates},}\ }\href
  {http://science.sciencemag.org/content/354/6308/83} {\bibfield  {journal}
  {\bibinfo  {journal} {Science}\ }\textbf {\bibinfo {volume} {354}},\ \bibinfo
  {pages} {83} (\bibinfo {year} {2016})}\BibitemShut {NoStop}%
\bibitem [{\citenamefont {Goldman}\ \emph {et~al.}(2016)\citenamefont
  {Goldman}, \citenamefont {Budich},\ and\ \citenamefont
  {Zoller}}]{Goldman2016}%
  \BibitemOpen
  \bibfield  {author} {\bibinfo {author} {\bibfnamefont {N.}~\bibnamefont
  {Goldman}}, \bibinfo {author} {\bibfnamefont {J.~C.}\ \bibnamefont {Budich}},
  \ and\ \bibinfo {author} {\bibfnamefont {P.}~\bibnamefont {Zoller}},\
  }\bibfield  {title} {\enquote {\bibinfo {title} {Topological quantum matter
  with ultracold gases in optical lattices},}\ }\href
  {http://dx.doi.org/10.1038/nphys3803} {\bibfield  {journal} {\bibinfo
  {journal} {Nat. Phys.}\ }\textbf {\bibinfo {volume} {12}},\ \bibinfo {pages}
  {639} (\bibinfo {year} {2016})}\BibitemShut {NoStop}%
\bibitem [{\citenamefont {Dai}\ \emph {et~al.}(2017)\citenamefont {Dai},
  \citenamefont {Yang}, \citenamefont {Reingruber}, \citenamefont {Sun},
  \citenamefont {Xu}, \citenamefont {Chen}, \citenamefont {Yuan},\ and\
  \citenamefont {Pan}}]{Pan2017}%
  \BibitemOpen
  \bibfield  {author} {\bibinfo {author} {\bibfnamefont {Han-Ning}\
  \bibnamefont {Dai}}, \bibinfo {author} {\bibfnamefont {Bing}\ \bibnamefont
  {Yang}}, \bibinfo {author} {\bibfnamefont {Andreas}\ \bibnamefont
  {Reingruber}}, \bibinfo {author} {\bibfnamefont {Hui}\ \bibnamefont {Sun}},
  \bibinfo {author} {\bibfnamefont {Xiao-Fan}\ \bibnamefont {Xu}}, \bibinfo
  {author} {\bibfnamefont {Yu-Ao}\ \bibnamefont {Chen}}, \bibinfo {author}
  {\bibfnamefont {Zhen-Sheng}\ \bibnamefont {Yuan}}, \ and\ \bibinfo {author}
  {\bibfnamefont {Jian-Wei}\ \bibnamefont {Pan}},\ }\bibfield  {title}
  {\enquote {\bibinfo {title} {Four-body ring-exchange interactions and anyonic
  statistics within a minimal toric-code hamiltonian},}\ }\href
  {http://dx.doi.org/10.1038/nphys4243} {\bibfield  {journal} {\bibinfo
  {journal} {Nat. Phys.}\ }\textbf {\bibinfo {volume} {13}},\ \bibinfo {pages}
  {1195} (\bibinfo {year} {2017})}\BibitemShut {NoStop}%
\bibitem [{\citenamefont {Kitaev}(2003)}]{Kitaev2003}%
  \BibitemOpen
  \bibfield  {author} {\bibinfo {author} {\bibfnamefont {A.~Y.}\ \bibnamefont
  {Kitaev}},\ }\bibfield  {title} {\enquote {\bibinfo {title} {Fault-tolerant
  quantum computation by anyons},}\ }\href
  {https://doi.org/10.1016/S0003-4916(02)00018-0} {\bibfield  {journal}
  {\bibinfo  {journal} {Ann. Phys.}\ }\textbf {\bibinfo {volume} {303}},\
  \bibinfo {pages} {2} (\bibinfo {year} {2003})}\BibitemShut {NoStop}%
\bibitem [{\citenamefont {Nayak}\ \emph {et~al.}(2008)\citenamefont {Nayak},
  \citenamefont {Simon}, \citenamefont {Stern}, \citenamefont {Freedman},\ and\
  \citenamefont {Das~Sarma}}]{Nayak2008}%
  \BibitemOpen
  \bibfield  {author} {\bibinfo {author} {\bibfnamefont {Chetan}\ \bibnamefont
  {Nayak}}, \bibinfo {author} {\bibfnamefont {Steven~H.}\ \bibnamefont
  {Simon}}, \bibinfo {author} {\bibfnamefont {Ady}\ \bibnamefont {Stern}},
  \bibinfo {author} {\bibfnamefont {Michael}\ \bibnamefont {Freedman}}, \ and\
  \bibinfo {author} {\bibfnamefont {Sankar}\ \bibnamefont {Das~Sarma}},\
  }\bibfield  {title} {\enquote {\bibinfo {title} {Non-abelian anyons and
  topological quantum computation},}\ }\href {\doibase
  10.1103/RevModPhys.80.1083} {\bibfield  {journal} {\bibinfo  {journal} {Rev.
  Mod. Phys.}\ }\textbf {\bibinfo {volume} {80}},\ \bibinfo {pages} {1083}
  (\bibinfo {year} {2008})}\BibitemShut {NoStop}%
\bibitem [{\citenamefont {Alicea}\ \emph {et~al.}(2011)\citenamefont {Alicea},
  \citenamefont {Oreg}, \citenamefont {Refael}, \citenamefont {von Oppen},\
  and\ \citenamefont {Fisher}}]{Alicea2011}%
  \BibitemOpen
  \bibfield  {author} {\bibinfo {author} {\bibfnamefont {Jason}\ \bibnamefont
  {Alicea}}, \bibinfo {author} {\bibfnamefont {Yuval}\ \bibnamefont {Oreg}},
  \bibinfo {author} {\bibfnamefont {Gil}\ \bibnamefont {Refael}}, \bibinfo
  {author} {\bibfnamefont {Felix}\ \bibnamefont {von Oppen}}, \ and\ \bibinfo
  {author} {\bibfnamefont {Matthew P.~A.}\ \bibnamefont {Fisher}},\ }\bibfield
  {title} {\enquote {\bibinfo {title} {Non-abelian statistics and topological
  quantum information processing in 1d wire networks},}\ }\href
  {http://dx.doi.org/10.1038/nphys1915} {\bibfield  {journal} {\bibinfo
  {journal} {Nat. Phys.}\ }\textbf {\bibinfo {volume} {7}},\ \bibinfo {pages}
  {412} (\bibinfo {year} {2011})}\BibitemShut {NoStop}%
\bibitem [{\citenamefont {Barends}\ \emph {et~al.}(2014)\citenamefont
  {Barends}, \citenamefont {Kelly}, \citenamefont {Megrant}, \citenamefont
  {Veitia}, \citenamefont {Sank}, \citenamefont {Jeffrey}, \citenamefont
  {White}, \citenamefont {Mutus}, \citenamefont {Fowler}, \citenamefont
  {Campbell}, \citenamefont {Chen}, \citenamefont {Chen}, \citenamefont
  {Chiaro}, \citenamefont {Dunsworth}, \citenamefont {Neill}, \citenamefont
  {O'Malley}, \citenamefont {Roushan}, \citenamefont {Vainsencher},
  \citenamefont {Wenner}, \citenamefont {Korotkov}, \citenamefont {Cleland},\
  and\ \citenamefont {Martinis}}]{Martinis2014}%
  \BibitemOpen
  \bibfield  {author} {\bibinfo {author} {\bibfnamefont {R.}~\bibnamefont
  {Barends}}, \bibinfo {author} {\bibfnamefont {J.}~\bibnamefont {Kelly}},
  \bibinfo {author} {\bibfnamefont {A.}~\bibnamefont {Megrant}}, \bibinfo
  {author} {\bibfnamefont {A.}~\bibnamefont {Veitia}}, \bibinfo {author}
  {\bibfnamefont {D.}~\bibnamefont {Sank}}, \bibinfo {author} {\bibfnamefont
  {E.}~\bibnamefont {Jeffrey}}, \bibinfo {author} {\bibfnamefont {T.~C.}\
  \bibnamefont {White}}, \bibinfo {author} {\bibfnamefont {J.}~\bibnamefont
  {Mutus}}, \bibinfo {author} {\bibfnamefont {A.~G.}\ \bibnamefont {Fowler}},
  \bibinfo {author} {\bibfnamefont {B.}~\bibnamefont {Campbell}}, \bibinfo
  {author} {\bibfnamefont {Y.}~\bibnamefont {Chen}}, \bibinfo {author}
  {\bibfnamefont {Z.}~\bibnamefont {Chen}}, \bibinfo {author} {\bibfnamefont
  {B.}~\bibnamefont {Chiaro}}, \bibinfo {author} {\bibfnamefont
  {A.}~\bibnamefont {Dunsworth}}, \bibinfo {author} {\bibfnamefont
  {C.}~\bibnamefont {Neill}}, \bibinfo {author} {\bibfnamefont
  {P.}~\bibnamefont {O'Malley}}, \bibinfo {author} {\bibfnamefont
  {P.}~\bibnamefont {Roushan}}, \bibinfo {author} {\bibfnamefont
  {A.}~\bibnamefont {Vainsencher}}, \bibinfo {author} {\bibfnamefont
  {J.}~\bibnamefont {Wenner}}, \bibinfo {author} {\bibfnamefont {A.~N.}\
  \bibnamefont {Korotkov}}, \bibinfo {author} {\bibfnamefont {A.~N.}\
  \bibnamefont {Cleland}}, \ and\ \bibinfo {author} {\bibfnamefont {John~M.}\
  \bibnamefont {Martinis}},\ }\bibfield  {title} {\enquote {\bibinfo {title}
  {Superconducting quantum circuits at the surface code threshold for fault
  tolerance},}\ }\href {http://dx.doi.org/10.1038/nature13171} {\bibfield
  {journal} {\bibinfo  {journal} {Nature}\ }\textbf {\bibinfo {volume} {508}},\
  \bibinfo {pages} {500} (\bibinfo {year} {2014})}\BibitemShut {NoStop}%
\bibitem [{\citenamefont {Haldane}\ and\ \citenamefont
  {Raghu}(2008)}]{Haldane2008}%
  \BibitemOpen
  \bibfield  {author} {\bibinfo {author} {\bibfnamefont {F.~D.~M.}\
  \bibnamefont {Haldane}}\ and\ \bibinfo {author} {\bibfnamefont
  {S.}~\bibnamefont {Raghu}},\ }\bibfield  {title} {\enquote {\bibinfo {title}
  {Possible realization of directional optical waveguides in photonic crystals
  with broken time-reversal symmetry},}\ }\href {\doibase
  10.1103/PhysRevLett.100.013904} {\bibfield  {journal} {\bibinfo  {journal}
  {Phys. Rev. Lett.}\ }\textbf {\bibinfo {volume} {100}},\ \bibinfo {pages}
  {013904} (\bibinfo {year} {2008})}\BibitemShut {NoStop}%
\bibitem [{\citenamefont {Hafezi}\ \emph {et~al.}(2011)\citenamefont {Hafezi},
  \citenamefont {Demler}, \citenamefont {Lukin},\ and\ \citenamefont
  {Taylor}}]{Hafezi2011}%
  \BibitemOpen
  \bibfield  {author} {\bibinfo {author} {\bibfnamefont {Mohammad}\
  \bibnamefont {Hafezi}}, \bibinfo {author} {\bibfnamefont {Eugene~A.}\
  \bibnamefont {Demler}}, \bibinfo {author} {\bibfnamefont {Mikhail~D.}\
  \bibnamefont {Lukin}}, \ and\ \bibinfo {author} {\bibfnamefont {Jacob~M.}\
  \bibnamefont {Taylor}},\ }\bibfield  {title} {\enquote {\bibinfo {title}
  {Robust optical delay lines with topological protection},}\ }\href
  {http://dx.doi.org/10.1038/nphys2063} {\bibfield  {journal} {\bibinfo
  {journal} {Nat. Phys.}\ }\textbf {\bibinfo {volume} {7}},\ \bibinfo {pages}
  {907} (\bibinfo {year} {2011})}\BibitemShut {NoStop}%
\bibitem [{\citenamefont {Fang}\ \emph {et~al.}(2012)\citenamefont {Fang},
  \citenamefont {Yu},\ and\ \citenamefont {Fan}}]{Fan2012}%
  \BibitemOpen
  \bibfield  {author} {\bibinfo {author} {\bibfnamefont {Kejie}\ \bibnamefont
  {Fang}}, \bibinfo {author} {\bibfnamefont {Zongfu}\ \bibnamefont {Yu}}, \
  and\ \bibinfo {author} {\bibfnamefont {Shanhui}\ \bibnamefont {Fan}},\
  }\bibfield  {title} {\enquote {\bibinfo {title} {Realizing effective magnetic
  field for photons by controlling the phase of dynamic modulation},}\ }\href
  {\doibase 10.1038/nphoton.2012.236} {\bibfield  {journal} {\bibinfo
  {journal} {Nat. Photonics}\ }\textbf {\bibinfo {volume} {6}},\ \bibinfo
  {pages} {782} (\bibinfo {year} {2012})}\BibitemShut {NoStop}%
\bibitem [{\citenamefont {Khanikaev}\ \emph {et~al.}(2013)\citenamefont
  {Khanikaev}, \citenamefont {Mousavi}, \citenamefont {Tse}, \citenamefont
  {Kargarian}, \citenamefont {MacDonald},\ and\ \citenamefont
  {Shvets}}]{MacDonald2012}%
  \BibitemOpen
  \bibfield  {author} {\bibinfo {author} {\bibfnamefont {Alexander~B.}\
  \bibnamefont {Khanikaev}}, \bibinfo {author} {\bibfnamefont {S.~Hossein}\
  \bibnamefont {Mousavi}}, \bibinfo {author} {\bibfnamefont {Wang-Kong}\
  \bibnamefont {Tse}}, \bibinfo {author} {\bibfnamefont {Mehdi}\ \bibnamefont
  {Kargarian}}, \bibinfo {author} {\bibfnamefont {Allan~H.}\ \bibnamefont
  {MacDonald}}, \ and\ \bibinfo {author} {\bibfnamefont {Gennady}\ \bibnamefont
  {Shvets}},\ }\bibfield  {title} {\enquote {\bibinfo {title} {Photonic
  topological insulators},}\ }\href {http://dx.doi.org/10.1038/nmat3520}
  {\bibfield  {journal} {\bibinfo  {journal} {Nat. Mater.}\ }\textbf {\bibinfo
  {volume} {12}},\ \bibinfo {pages} {233} (\bibinfo {year} {2013})}\BibitemShut
  {NoStop}%
\bibitem [{\citenamefont {Carusotto}\ and\ \citenamefont
  {Ciuti}(2013)}]{Ciuti2013}%
  \BibitemOpen
  \bibfield  {author} {\bibinfo {author} {\bibfnamefont {Iacopo}\ \bibnamefont
  {Carusotto}}\ and\ \bibinfo {author} {\bibfnamefont {Cristiano}\ \bibnamefont
  {Ciuti}},\ }\bibfield  {title} {\enquote {\bibinfo {title} {Quantum fluids of
  light},}\ }\href {\doibase 10.1103/RevModPhys.85.299} {\bibfield  {journal}
  {\bibinfo  {journal} {Rev. Mod. Phys.}\ }\textbf {\bibinfo {volume} {85}},\
  \bibinfo {pages} {299} (\bibinfo {year} {2013})}\BibitemShut {NoStop}%
\bibitem [{\citenamefont {Hafezi}\ \emph {et~al.}(2013)\citenamefont {Hafezi},
  \citenamefont {Mittal}, \citenamefont {Fan}, \citenamefont {Migdall},\ and\
  \citenamefont {Taylor}}]{Hafezi2013}%
  \BibitemOpen
  \bibfield  {author} {\bibinfo {author} {\bibfnamefont {M.}~\bibnamefont
  {Hafezi}}, \bibinfo {author} {\bibfnamefont {S.}~\bibnamefont {Mittal}},
  \bibinfo {author} {\bibfnamefont {J.}~\bibnamefont {Fan}}, \bibinfo {author}
  {\bibfnamefont {A.}~\bibnamefont {Migdall}}, \ and\ \bibinfo {author}
  {\bibfnamefont {J.~M.}\ \bibnamefont {Taylor}},\ }\bibfield  {title}
  {\enquote {\bibinfo {title} {Imaging topological edge states in silicon
  photonics},}\ }\href {http://dx.doi.org/10.1038/nphoton.2013.274} {\bibfield
  {journal} {\bibinfo  {journal} {Nat. Photonics}\ }\textbf {\bibinfo {volume}
  {7}},\ \bibinfo {pages} {1001} (\bibinfo {year} {2013})}\BibitemShut
  {NoStop}%
\bibitem [{\citenamefont {Rechtsman}\ \emph {et~al.}(2013)\citenamefont
  {Rechtsman}, \citenamefont {Zeuner}, \citenamefont {Plotnik}, \citenamefont
  {Lumer}, \citenamefont {Podolsky}, \citenamefont {Dreisow}, \citenamefont
  {Nolte}, \citenamefont {Segev},\ and\ \citenamefont {Szameit}}]{Segev2013}%
  \BibitemOpen
  \bibfield  {author} {\bibinfo {author} {\bibfnamefont {Mikael~C.}\
  \bibnamefont {Rechtsman}}, \bibinfo {author} {\bibfnamefont {Julia~M.}\
  \bibnamefont {Zeuner}}, \bibinfo {author} {\bibfnamefont {Yonatan}\
  \bibnamefont {Plotnik}}, \bibinfo {author} {\bibfnamefont {Yaakov}\
  \bibnamefont {Lumer}}, \bibinfo {author} {\bibfnamefont {Daniel}\
  \bibnamefont {Podolsky}}, \bibinfo {author} {\bibfnamefont {Felix}\
  \bibnamefont {Dreisow}}, \bibinfo {author} {\bibfnamefont {Stefan}\
  \bibnamefont {Nolte}}, \bibinfo {author} {\bibfnamefont {Mordechai}\
  \bibnamefont {Segev}}, \ and\ \bibinfo {author} {\bibfnamefont {Alexander}\
  \bibnamefont {Szameit}},\ }\bibfield  {title} {\enquote {\bibinfo {title}
  {Photonic floquet topological insulators},}\ }\href
  {http://dx.doi.org/10.1038/nature12066} {\bibfield  {journal} {\bibinfo
  {journal} {Nature}\ }\textbf {\bibinfo {volume} {496}},\ \bibinfo {pages}
  {196} (\bibinfo {year} {2013})}\BibitemShut {NoStop}%
\bibitem [{\citenamefont {Lu}\ \emph {et~al.}(2014)\citenamefont {Lu},
  \citenamefont {Joannopoulos},\ and\ \citenamefont
  {Solja\v{c}i\'c}}]{Soljacic2014}%
  \BibitemOpen
  \bibfield  {author} {\bibinfo {author} {\bibfnamefont {Ling}\ \bibnamefont
  {Lu}}, \bibinfo {author} {\bibfnamefont {John~D.}\ \bibnamefont
  {Joannopoulos}}, \ and\ \bibinfo {author} {\bibfnamefont {Marin}\
  \bibnamefont {Solja\v{c}i\'c}},\ }\bibfield  {title} {\enquote {\bibinfo
  {title} {Topological photonics},}\ }\href
  {http://dx.doi.org/10.1038/nphoton.2014.248} {\bibfield  {journal} {\bibinfo
  {journal} {Nat. Photonics}\ }\textbf {\bibinfo {volume} {8}},\ \bibinfo
  {pages} {821} (\bibinfo {year} {2014})}\BibitemShut {NoStop}%
\bibitem [{\citenamefont {Karzig}\ \emph {et~al.}(2015)\citenamefont {Karzig},
  \citenamefont {Bardyn}, \citenamefont {Lindner},\ and\ \citenamefont
  {Refael}}]{Lindner2015}%
  \BibitemOpen
  \bibfield  {author} {\bibinfo {author} {\bibfnamefont {Torsten}\ \bibnamefont
  {Karzig}}, \bibinfo {author} {\bibfnamefont {Charles-Edouard}\ \bibnamefont
  {Bardyn}}, \bibinfo {author} {\bibfnamefont {Netanel~H.}\ \bibnamefont
  {Lindner}}, \ and\ \bibinfo {author} {\bibfnamefont {Gil}\ \bibnamefont
  {Refael}},\ }\bibfield  {title} {\enquote {\bibinfo {title} {Topological
  polaritons},}\ }\href {\doibase 10.1103/PhysRevX.5.031001} {\bibfield
  {journal} {\bibinfo  {journal} {Phys. Rev. X}\ }\textbf {\bibinfo {volume}
  {5}},\ \bibinfo {pages} {031001} (\bibinfo {year} {2015})}\BibitemShut
  {NoStop}%
\bibitem [{\citenamefont {Ozawa}\ \emph {et~al.}(2018)\citenamefont {Ozawa},
  \citenamefont {Price}, \citenamefont {Amo}, \citenamefont {Goldman},
  \citenamefont {Hafezi}, \citenamefont {Lu}, \citenamefont {Rechtsman},
  \citenamefont {Schuster}, \citenamefont {Simon}, \citenamefont {Zilberberg},\
  and\ \citenamefont {Carusotto}}]{Ozawa2018}%
  \BibitemOpen
  \bibfield  {author} {\bibinfo {author} {\bibfnamefont {Tomoki}\ \bibnamefont
  {Ozawa}}, \bibinfo {author} {\bibfnamefont {Hannah~M}\ \bibnamefont {Price}},
  \bibinfo {author} {\bibfnamefont {Alberto}\ \bibnamefont {Amo}}, \bibinfo
  {author} {\bibfnamefont {Nathan}\ \bibnamefont {Goldman}}, \bibinfo {author}
  {\bibfnamefont {Mohammad}\ \bibnamefont {Hafezi}}, \bibinfo {author}
  {\bibfnamefont {Ling}\ \bibnamefont {Lu}}, \bibinfo {author} {\bibfnamefont
  {Mikael}\ \bibnamefont {Rechtsman}}, \bibinfo {author} {\bibfnamefont
  {David}\ \bibnamefont {Schuster}}, \bibinfo {author} {\bibfnamefont
  {Jonathan}\ \bibnamefont {Simon}}, \bibinfo {author} {\bibfnamefont {Oded}\
  \bibnamefont {Zilberberg}}, \ and\ \bibinfo {author} {\bibfnamefont {Iacopo}\
  \bibnamefont {Carusotto}},\ }\href@noop {} {\enquote {\bibinfo {title}
  {Topological photonics},}\ } (\bibinfo {year} {2018}),\ \bibinfo {note}
  {arXiv:1802.04173}\BibitemShut {NoStop}%
\bibitem [{\citenamefont {Kane}\ and\ \citenamefont
  {Lubensky}(2014)}]{Kane2014}%
  \BibitemOpen
  \bibfield  {author} {\bibinfo {author} {\bibfnamefont {C.~L.}\ \bibnamefont
  {Kane}}\ and\ \bibinfo {author} {\bibfnamefont {T.~C.}\ \bibnamefont
  {Lubensky}},\ }\bibfield  {title} {\enquote {\bibinfo {title} {Topological
  boundary modes in isostatic lattices},}\ }\href
  {http://dx.doi.org/10.1038/nphys2835} {\bibfield  {journal} {\bibinfo
  {journal} {Nat. Phys.}\ }\textbf {\bibinfo {volume} {10}},\ \bibinfo {pages}
  {39} (\bibinfo {year} {2014})}\BibitemShut {NoStop}%
\bibitem [{\citenamefont {Para\"{\i}so}\ \emph {et~al.}(2015)\citenamefont
  {Para\"{\i}so}, \citenamefont {Kalaee}, \citenamefont {Zang}, \citenamefont
  {Pfeifer}, \citenamefont {Marquardt},\ and\ \citenamefont
  {Painter}}]{Marquardt2015}%
  \BibitemOpen
  \bibfield  {author} {\bibinfo {author} {\bibfnamefont {Taofiq~K.}\
  \bibnamefont {Para\"{\i}so}}, \bibinfo {author} {\bibfnamefont {Mahmoud}\
  \bibnamefont {Kalaee}}, \bibinfo {author} {\bibfnamefont {Leyun}\
  \bibnamefont {Zang}}, \bibinfo {author} {\bibfnamefont {Hannes}\ \bibnamefont
  {Pfeifer}}, \bibinfo {author} {\bibfnamefont {Florian}\ \bibnamefont
  {Marquardt}}, \ and\ \bibinfo {author} {\bibfnamefont {Oskar}\ \bibnamefont
  {Painter}},\ }\bibfield  {title} {\enquote {\bibinfo {title}
  {Position-squared coupling in a tunable photonic crystal optomechanical
  cavity},}\ }\href {\doibase 10.1103/PhysRevX.5.041024} {\bibfield  {journal}
  {\bibinfo  {journal} {Phys. Rev. X}\ }\textbf {\bibinfo {volume} {5}},\
  \bibinfo {pages} {041024} (\bibinfo {year} {2015})}\BibitemShut {NoStop}%
\bibitem [{\citenamefont {S{\"u}sstrunk}\ and\ \citenamefont
  {Huber}(2015)}]{Huber2015}%
  \BibitemOpen
  \bibfield  {author} {\bibinfo {author} {\bibfnamefont {Roman}\ \bibnamefont
  {S{\"u}sstrunk}}\ and\ \bibinfo {author} {\bibfnamefont {Sebastian~D.}\
  \bibnamefont {Huber}},\ }\bibfield  {title} {\enquote {\bibinfo {title}
  {Observation of phononic helical edge states in a mechanical topological
  insulator},}\ }\href {\doibase 10.1126/science.aab0239} {\bibfield  {journal}
  {\bibinfo  {journal} {Science}\ }\textbf {\bibinfo {volume} {349}},\ \bibinfo
  {pages} {47} (\bibinfo {year} {2015})}\BibitemShut {NoStop}%
\bibitem [{\citenamefont {He}\ \emph {et~al.}(2016)\citenamefont {He},
  \citenamefont {Ni}, \citenamefont {Ge}, \citenamefont {Sun}, \citenamefont
  {Chen}, \citenamefont {Lu}, \citenamefont {Liu},\ and\ \citenamefont
  {Chen}}]{He2016}%
  \BibitemOpen
  \bibfield  {author} {\bibinfo {author} {\bibfnamefont {Cheng}\ \bibnamefont
  {He}}, \bibinfo {author} {\bibfnamefont {Xu}~\bibnamefont {Ni}}, \bibinfo
  {author} {\bibfnamefont {Hao}\ \bibnamefont {Ge}}, \bibinfo {author}
  {\bibfnamefont {Xiao-Chen}\ \bibnamefont {Sun}}, \bibinfo {author}
  {\bibfnamefont {Yan-Bin}\ \bibnamefont {Chen}}, \bibinfo {author}
  {\bibfnamefont {Ming-Hui}\ \bibnamefont {Lu}}, \bibinfo {author}
  {\bibfnamefont {Xiao-Ping}\ \bibnamefont {Liu}}, \ and\ \bibinfo {author}
  {\bibfnamefont {Yan-Feng}\ \bibnamefont {Chen}},\ }\bibfield  {title}
  {\enquote {\bibinfo {title} {Acoustic topological insulator and robust
  one-way sound transport},}\ }\href {http://dx.doi.org/10.1038/nphys3867}
  {\bibfield  {journal} {\bibinfo  {journal} {Nat. Phys.}\ }\textbf {\bibinfo
  {volume} {12}},\ \bibinfo {pages} {1124} (\bibinfo {year}
  {2016})}\BibitemShut {NoStop}%
\bibitem [{\citenamefont {Sachdev}(2011)}]{Sachdev2011}%
  \BibitemOpen
  \bibfield  {author} {\bibinfo {author} {\bibfnamefont {Subir}\ \bibnamefont
  {Sachdev}},\ }\href@noop {} {\emph {\bibinfo {title} {Quantum Phase
  Transitions}}}\ (\bibinfo  {publisher} {Cambridge University Press,
  Cambridge},\ \bibinfo {year} {2011})\BibitemShut {NoStop}%
\bibitem [{\citenamefont {Bernevig}\ and\ \citenamefont
  {Hughes}(2013)}]{Bernevig2013}%
  \BibitemOpen
  \bibfield  {author} {\bibinfo {author} {\bibfnamefont {B.~Andrei}\
  \bibnamefont {Bernevig}}\ and\ \bibinfo {author} {\bibfnamefont {Taylor~L.}\
  \bibnamefont {Hughes}},\ }\href@noop {} {\emph {\bibinfo {title} {Topological
  insulators and topological superconductors}}}\ (\bibinfo  {publisher}
  {Princeton University Press, Princeton},\ \bibinfo {year} {2013})\BibitemShut
  {NoStop}%
\bibitem [{\citenamefont {Altland}\ and\ \citenamefont
  {Zirnbauer}(1997)}]{Altland1997}%
  \BibitemOpen
  \bibfield  {author} {\bibinfo {author} {\bibfnamefont {Alexander}\
  \bibnamefont {Altland}}\ and\ \bibinfo {author} {\bibfnamefont {Martin~R.}\
  \bibnamefont {Zirnbauer}},\ }\bibfield  {title} {\enquote {\bibinfo {title}
  {Nonstandard symmetry classes in mesoscopic normal-superconducting hybrid
  structures},}\ }\href {\doibase 10.1103/PhysRevB.55.1142} {\bibfield
  {journal} {\bibinfo  {journal} {Phys. Rev. B}\ }\textbf {\bibinfo {volume}
  {55}},\ \bibinfo {pages} {1142} (\bibinfo {year} {1997})}\BibitemShut
  {NoStop}%
\bibitem [{\citenamefont {Schnyder}\ \emph {et~al.}(2008)\citenamefont
  {Schnyder}, \citenamefont {Ryu}, \citenamefont {Furusaki},\ and\
  \citenamefont {Ludwig}}]{Ryu2008}%
  \BibitemOpen
  \bibfield  {author} {\bibinfo {author} {\bibfnamefont {Andreas~P.}\
  \bibnamefont {Schnyder}}, \bibinfo {author} {\bibfnamefont {Shinsei}\
  \bibnamefont {Ryu}}, \bibinfo {author} {\bibfnamefont {Akira}\ \bibnamefont
  {Furusaki}}, \ and\ \bibinfo {author} {\bibfnamefont {Andreas W.~W.}\
  \bibnamefont {Ludwig}},\ }\bibfield  {title} {\enquote {\bibinfo {title}
  {Classification of topological insulators and superconductors in three
  spatial dimensions},}\ }\href {\doibase 10.1103/PhysRevB.78.195125}
  {\bibfield  {journal} {\bibinfo  {journal} {Phys. Rev. B}\ }\textbf {\bibinfo
  {volume} {78}},\ \bibinfo {pages} {195125} (\bibinfo {year}
  {2008})}\BibitemShut {NoStop}%
\bibitem [{\citenamefont {Kitaev}(2009)}]{Kitaev2009}%
  \BibitemOpen
  \bibfield  {author} {\bibinfo {author} {\bibfnamefont {A.}~\bibnamefont
  {Kitaev}},\ }\bibfield  {title} {\enquote {\bibinfo {title} {Periodic table
  for topological insulators and superconductors},}\ }\href
  {https://doi.org/10.1063/1.3149495} {\bibfield  {journal} {\bibinfo
  {journal} {AIP Conf. Proc.}\ }\textbf {\bibinfo {volume} {1134}},\ \bibinfo
  {pages} {22} (\bibinfo {year} {2009})}\BibitemShut {NoStop}%
\bibitem [{\citenamefont {Ryu}\ \emph {et~al.}(2010)\citenamefont {Ryu},
  \citenamefont {Schnyder}, \citenamefont {Furusaki},\ and\ \citenamefont
  {Ludwig}}]{Ryu2010}%
  \BibitemOpen
  \bibfield  {author} {\bibinfo {author} {\bibfnamefont {Shinsei}\ \bibnamefont
  {Ryu}}, \bibinfo {author} {\bibfnamefont {Andreas~P.}\ \bibnamefont
  {Schnyder}}, \bibinfo {author} {\bibfnamefont {Akira}\ \bibnamefont
  {Furusaki}}, \ and\ \bibinfo {author} {\bibfnamefont {Andreas W.~W.}\
  \bibnamefont {Ludwig}},\ }\bibfield  {title} {\enquote {\bibinfo {title}
  {Topological insulators and superconductors: tenfold way and dimensional
  hierarchy},}\ }\href {http://stacks.iop.org/1367-2630/12/i=6/a=065010}
  {\bibfield  {journal} {\bibinfo  {journal} {New J. Phys.}\ }\textbf {\bibinfo
  {volume} {12}},\ \bibinfo {pages} {065010} (\bibinfo {year}
  {2010})}\BibitemShut {NoStop}%
\bibitem [{\citenamefont {Teo}\ and\ \citenamefont {Kane}(2010)}]{Teo2010}%
  \BibitemOpen
  \bibfield  {author} {\bibinfo {author} {\bibfnamefont {Jeffrey C.~Y.}\
  \bibnamefont {Teo}}\ and\ \bibinfo {author} {\bibfnamefont {C.~L.}\
  \bibnamefont {Kane}},\ }\bibfield  {title} {\enquote {\bibinfo {title}
  {Topological defects and gapless modes in insulators and superconductors},}\
  }\href {\doibase 10.1103/PhysRevB.82.115120} {\bibfield  {journal} {\bibinfo
  {journal} {Phys. Rev. B}\ }\textbf {\bibinfo {volume} {82}},\ \bibinfo
  {pages} {115120} (\bibinfo {year} {2010})}\BibitemShut {NoStop}%
\bibitem [{\citenamefont {Fu}(2011)}]{Fu2011}%
  \BibitemOpen
  \bibfield  {author} {\bibinfo {author} {\bibfnamefont {Liang}\ \bibnamefont
  {Fu}},\ }\bibfield  {title} {\enquote {\bibinfo {title} {Topological
  crystalline insulators},}\ }\href {\doibase 10.1103/PhysRevLett.106.106802}
  {\bibfield  {journal} {\bibinfo  {journal} {Phys. Rev. Lett.}\ }\textbf
  {\bibinfo {volume} {106}},\ \bibinfo {pages} {106802} (\bibinfo {year}
  {2011})}\BibitemShut {NoStop}%
\bibitem [{\citenamefont {Slager}\ \emph {et~al.}(2013)\citenamefont {Slager},
  \citenamefont {Mesaros}, \citenamefont {Juri\v{c}i\'c},\ and\ \citenamefont
  {Zaanen}}]{Slager2013}%
  \BibitemOpen
  \bibfield  {author} {\bibinfo {author} {\bibfnamefont {Robert-Jan}\
  \bibnamefont {Slager}}, \bibinfo {author} {\bibfnamefont {Andrej}\
  \bibnamefont {Mesaros}}, \bibinfo {author} {\bibfnamefont {Vladimir}\
  \bibnamefont {Juri\v{c}i\'c}}, \ and\ \bibinfo {author} {\bibfnamefont {Jan}\
  \bibnamefont {Zaanen}},\ }\bibfield  {title} {\enquote {\bibinfo {title} {The
  space group classification of topological band-insulators},}\ }\href
  {http://dx.doi.org/10.1038/nphys2513} {\bibfield  {journal} {\bibinfo
  {journal} {Nat. Phys.}\ }\textbf {\bibinfo {volume} {9}},\ \bibinfo {pages}
  {98} (\bibinfo {year} {2013})}\BibitemShut {NoStop}%
\bibitem [{\citenamefont {Chiu}\ \emph {et~al.}(2013)\citenamefont {Chiu},
  \citenamefont {Yao},\ and\ \citenamefont {Ryu}}]{Ryu2013}%
  \BibitemOpen
  \bibfield  {author} {\bibinfo {author} {\bibfnamefont {Ching-Kai}\
  \bibnamefont {Chiu}}, \bibinfo {author} {\bibfnamefont {Hong}\ \bibnamefont
  {Yao}}, \ and\ \bibinfo {author} {\bibfnamefont {Shinsei}\ \bibnamefont
  {Ryu}},\ }\bibfield  {title} {\enquote {\bibinfo {title} {Classification of
  topological insulators and superconductors in the presence of reflection
  symmetry},}\ }\href {\doibase 10.1103/PhysRevB.88.075142} {\bibfield
  {journal} {\bibinfo  {journal} {Phys. Rev. B}\ }\textbf {\bibinfo {volume}
  {88}},\ \bibinfo {pages} {075142} (\bibinfo {year} {2013})}\BibitemShut
  {NoStop}%
\bibitem [{\citenamefont {Morimoto}\ and\ \citenamefont
  {Furusaki}(2013)}]{Morimoto2013}%
  \BibitemOpen
  \bibfield  {author} {\bibinfo {author} {\bibfnamefont {Takahiro}\
  \bibnamefont {Morimoto}}\ and\ \bibinfo {author} {\bibfnamefont {Akira}\
  \bibnamefont {Furusaki}},\ }\bibfield  {title} {\enquote {\bibinfo {title}
  {Topological classification with additional symmetries from clifford
  algebras},}\ }\href {\doibase 10.1103/PhysRevB.88.125129} {\bibfield
  {journal} {\bibinfo  {journal} {Phys. Rev. B}\ }\textbf {\bibinfo {volume}
  {88}},\ \bibinfo {pages} {125129} (\bibinfo {year} {2013})}\BibitemShut
  {NoStop}%
\bibitem [{\citenamefont {Shiozaki}\ and\ \citenamefont
  {Sato}(2014)}]{Shiozaki2014}%
  \BibitemOpen
  \bibfield  {author} {\bibinfo {author} {\bibfnamefont {Ken}\ \bibnamefont
  {Shiozaki}}\ and\ \bibinfo {author} {\bibfnamefont {Masatoshi}\ \bibnamefont
  {Sato}},\ }\bibfield  {title} {\enquote {\bibinfo {title} {Topology of
  crystalline insulators and superconductors},}\ }\href {\doibase
  10.1103/PhysRevB.90.165114} {\bibfield  {journal} {\bibinfo  {journal} {Phys.
  Rev. B}\ }\textbf {\bibinfo {volume} {90}},\ \bibinfo {pages} {165114}
  (\bibinfo {year} {2014})}\BibitemShut {NoStop}%
\bibitem [{\citenamefont {Ando}\ and\ \citenamefont {Fu}(2015)}]{Fu2015}%
  \BibitemOpen
  \bibfield  {author} {\bibinfo {author} {\bibfnamefont {Yoichi}\ \bibnamefont
  {Ando}}\ and\ \bibinfo {author} {\bibfnamefont {Liang}\ \bibnamefont {Fu}},\
  }\bibfield  {title} {\enquote {\bibinfo {title} {Topological crystalline
  insulators and topological superconductors: From concepts to materials},}\
  }\href {\doibase 10.1146/annurev-conmatphys-031214-014501} {\bibfield
  {journal} {\bibinfo  {journal} {Annu. Rev. Cond. Matt. Phys.}\ }\textbf
  {\bibinfo {volume} {6}},\ \bibinfo {pages} {361} (\bibinfo {year}
  {2015})}\BibitemShut {NoStop}%
\bibitem [{\citenamefont {Kruthoff}\ \emph {et~al.}(2017)\citenamefont
  {Kruthoff}, \citenamefont {de~Boer}, \citenamefont {van Wezel}, \citenamefont
  {Kane},\ and\ \citenamefont {Slager}}]{Slager2017}%
  \BibitemOpen
  \bibfield  {author} {\bibinfo {author} {\bibfnamefont {Jorrit}\ \bibnamefont
  {Kruthoff}}, \bibinfo {author} {\bibfnamefont {Jan}\ \bibnamefont {de~Boer}},
  \bibinfo {author} {\bibfnamefont {Jasper}\ \bibnamefont {van Wezel}},
  \bibinfo {author} {\bibfnamefont {Charles~L.}\ \bibnamefont {Kane}}, \ and\
  \bibinfo {author} {\bibfnamefont {Robert-Jan}\ \bibnamefont {Slager}},\
  }\bibfield  {title} {\enquote {\bibinfo {title} {Topological classification
  of crystalline insulators through band structure combinatorics},}\ }\href
  {\doibase 10.1103/PhysRevX.7.041069} {\bibfield  {journal} {\bibinfo
  {journal} {Phys. Rev. X}\ }\textbf {\bibinfo {volume} {7}},\ \bibinfo {pages}
  {041069} (\bibinfo {year} {2017})}\BibitemShut {NoStop}%
\bibitem [{\citenamefont {Hatsugai}(1993)}]{Hatsugai1993}%
  \BibitemOpen
  \bibfield  {author} {\bibinfo {author} {\bibfnamefont {Yasuhiro}\
  \bibnamefont {Hatsugai}},\ }\bibfield  {title} {\enquote {\bibinfo {title}
  {Chern number and edge states in the integer quantum hall effect},}\ }\href
  {\doibase 10.1103/PhysRevLett.71.3697} {\bibfield  {journal} {\bibinfo
  {journal} {Phys. Rev. Lett.}\ }\textbf {\bibinfo {volume} {71}},\ \bibinfo
  {pages} {3697--3700} (\bibinfo {year} {1993})}\BibitemShut {NoStop}%
\bibitem [{\citenamefont {Kitaev}(2001)}]{Kitaev2001}%
  \BibitemOpen
  \bibfield  {author} {\bibinfo {author} {\bibfnamefont {A.~Y.}\ \bibnamefont
  {Kitaev}},\ }\bibfield  {title} {\enquote {\bibinfo {title} {Unpaired
  majorana fermions in quantum wires},}\ }\href
  {https://doi.org/10.1070/1063-7869/44/10S/S29} {\bibfield  {journal}
  {\bibinfo  {journal} {Phys. Usp.}\ }\textbf {\bibinfo {volume} {44}},\
  \bibinfo {pages} {131} (\bibinfo {year} {2001})}\BibitemShut {NoStop}%
\bibitem [{\citenamefont {Fidkowski}(2010)}]{Fidkowski2010}%
  \BibitemOpen
  \bibfield  {author} {\bibinfo {author} {\bibfnamefont {Lukasz}\ \bibnamefont
  {Fidkowski}},\ }\bibfield  {title} {\enquote {\bibinfo {title} {Entanglement
  spectrum of topological insulators and superconductors},}\ }\href {\doibase
  10.1103/PhysRevLett.104.130502} {\bibfield  {journal} {\bibinfo  {journal}
  {Phys. Rev. Lett.}\ }\textbf {\bibinfo {volume} {104}},\ \bibinfo {pages}
  {130502} (\bibinfo {year} {2010})}\BibitemShut {NoStop}%
\bibitem [{\citenamefont {Niu}\ \emph {et~al.}(1985)\citenamefont {Niu},
  \citenamefont {Thouless},\ and\ \citenamefont {Wu}}]{Niu1985}%
  \BibitemOpen
  \bibfield  {author} {\bibinfo {author} {\bibfnamefont {Qian}\ \bibnamefont
  {Niu}}, \bibinfo {author} {\bibfnamefont {D.~J.}\ \bibnamefont {Thouless}}, \
  and\ \bibinfo {author} {\bibfnamefont {Yong-Shi}\ \bibnamefont {Wu}},\
  }\bibfield  {title} {\enquote {\bibinfo {title} {Quantized hall conductance
  as a topological invariant},}\ }\href {\doibase 10.1103/PhysRevB.31.3372}
  {\bibfield  {journal} {\bibinfo  {journal} {Phys. Rev. B}\ }\textbf {\bibinfo
  {volume} {31}},\ \bibinfo {pages} {3372--3377} (\bibinfo {year}
  {1985})}\BibitemShut {NoStop}%
\bibitem [{\citenamefont {Else}\ and\ \citenamefont {Nayak}(2016)}]{Else2016}%
  \BibitemOpen
  \bibfield  {author} {\bibinfo {author} {\bibfnamefont {Dominic~V.}\
  \bibnamefont {Else}}\ and\ \bibinfo {author} {\bibfnamefont {Chetan}\
  \bibnamefont {Nayak}},\ }\bibfield  {title} {\enquote {\bibinfo {title}
  {Classification of topological phases in periodically driven interacting
  systems},}\ }\href {\doibase 10.1103/PhysRevB.93.201103} {\bibfield
  {journal} {\bibinfo  {journal} {Phys. Rev. B}\ }\textbf {\bibinfo {volume}
  {93}},\ \bibinfo {pages} {201103} (\bibinfo {year} {2016})}\BibitemShut
  {NoStop}%
\bibitem [{\citenamefont {von Keyserlingk}\ and\ \citenamefont
  {Sondhi}(2016{\natexlab{a}})}]{Keyserlingk2016a}%
  \BibitemOpen
  \bibfield  {author} {\bibinfo {author} {\bibfnamefont {C.~W.}\ \bibnamefont
  {von Keyserlingk}}\ and\ \bibinfo {author} {\bibfnamefont {S.~L.}\
  \bibnamefont {Sondhi}},\ }\bibfield  {title} {\enquote {\bibinfo {title}
  {Phase structure of one-dimensional interacting floquet systems. i. abelian
  symmetry-protected topological phases},}\ }\href {\doibase
  10.1103/PhysRevB.93.245145} {\bibfield  {journal} {\bibinfo  {journal} {Phys.
  Rev. B}\ }\textbf {\bibinfo {volume} {93}},\ \bibinfo {pages} {245145}
  (\bibinfo {year} {2016}{\natexlab{a}})}\BibitemShut {NoStop}%
\bibitem [{\citenamefont {Potter}\ \emph {et~al.}(2016)\citenamefont {Potter},
  \citenamefont {Morimoto},\ and\ \citenamefont {Vishwanath}}]{Potter2016}%
  \BibitemOpen
  \bibfield  {author} {\bibinfo {author} {\bibfnamefont {Andrew~C.}\
  \bibnamefont {Potter}}, \bibinfo {author} {\bibfnamefont {Takahiro}\
  \bibnamefont {Morimoto}}, \ and\ \bibinfo {author} {\bibfnamefont {Ashvin}\
  \bibnamefont {Vishwanath}},\ }\bibfield  {title} {\enquote {\bibinfo {title}
  {Classification of interacting topological floquet phases in one
  dimension},}\ }\href {\doibase 10.1103/PhysRevX.6.041001} {\bibfield
  {journal} {\bibinfo  {journal} {Phys. Rev. X}\ }\textbf {\bibinfo {volume}
  {6}},\ \bibinfo {pages} {041001} (\bibinfo {year} {2016})}\BibitemShut
  {NoStop}%
\bibitem [{\citenamefont {Roy}\ and\ \citenamefont {Harper}(2017)}]{Roy2017}%
  \BibitemOpen
  \bibfield  {author} {\bibinfo {author} {\bibfnamefont {Rahul}\ \bibnamefont
  {Roy}}\ and\ \bibinfo {author} {\bibfnamefont {Fenner}\ \bibnamefont
  {Harper}},\ }\bibfield  {title} {\enquote {\bibinfo {title} {Periodic table
  for floquet topological insulators},}\ }\href {\doibase
  10.1103/PhysRevB.96.155118} {\bibfield  {journal} {\bibinfo  {journal} {Phys.
  Rev. B}\ }\textbf {\bibinfo {volume} {96}},\ \bibinfo {pages} {155118}
  (\bibinfo {year} {2017})}\BibitemShut {NoStop}%
\bibitem [{\citenamefont {Kitagawa}\ \emph {et~al.}(2010)\citenamefont
  {Kitagawa}, \citenamefont {Berg}, \citenamefont {Rudner},\ and\ \citenamefont
  {Demler}}]{Kitagawa2010}%
  \BibitemOpen
  \bibfield  {author} {\bibinfo {author} {\bibfnamefont {Takuya}\ \bibnamefont
  {Kitagawa}}, \bibinfo {author} {\bibfnamefont {Erez}\ \bibnamefont {Berg}},
  \bibinfo {author} {\bibfnamefont {Mark}\ \bibnamefont {Rudner}}, \ and\
  \bibinfo {author} {\bibfnamefont {Eugene}\ \bibnamefont {Demler}},\
  }\bibfield  {title} {\enquote {\bibinfo {title} {Topological characterization
  of periodically driven quantum systems},}\ }\href {\doibase
  10.1103/PhysRevB.82.235114} {\bibfield  {journal} {\bibinfo  {journal} {Phys.
  Rev. B}\ }\textbf {\bibinfo {volume} {82}},\ \bibinfo {pages} {235114}
  (\bibinfo {year} {2010})}\BibitemShut {NoStop}%
\bibitem [{\citenamefont {Jiang}\ \emph {et~al.}(2011)\citenamefont {Jiang},
  \citenamefont {Kitagawa}, \citenamefont {Alicea}, \citenamefont {Akhmerov},
  \citenamefont {Pekker}, \citenamefont {Refael}, \citenamefont {Cirac},
  \citenamefont {Demler}, \citenamefont {Lukin},\ and\ \citenamefont
  {Zoller}}]{Jiang2011}%
  \BibitemOpen
  \bibfield  {author} {\bibinfo {author} {\bibfnamefont {Liang}\ \bibnamefont
  {Jiang}}, \bibinfo {author} {\bibfnamefont {Takuya}\ \bibnamefont
  {Kitagawa}}, \bibinfo {author} {\bibfnamefont {Jason}\ \bibnamefont
  {Alicea}}, \bibinfo {author} {\bibfnamefont {A.~R.}\ \bibnamefont
  {Akhmerov}}, \bibinfo {author} {\bibfnamefont {David}\ \bibnamefont
  {Pekker}}, \bibinfo {author} {\bibfnamefont {Gil}\ \bibnamefont {Refael}},
  \bibinfo {author} {\bibfnamefont {J.~Ignacio}\ \bibnamefont {Cirac}},
  \bibinfo {author} {\bibfnamefont {Eugene}\ \bibnamefont {Demler}}, \bibinfo
  {author} {\bibfnamefont {Mikhail~D.}\ \bibnamefont {Lukin}}, \ and\ \bibinfo
  {author} {\bibfnamefont {Peter}\ \bibnamefont {Zoller}},\ }\bibfield  {title}
  {\enquote {\bibinfo {title} {Majorana fermions in equilibrium and in driven
  cold-atom quantum wires},}\ }\href {\doibase 10.1103/PhysRevLett.106.220402}
  {\bibfield  {journal} {\bibinfo  {journal} {Phys. Rev. Lett.}\ }\textbf
  {\bibinfo {volume} {106}},\ \bibinfo {pages} {220402} (\bibinfo {year}
  {2011})}\BibitemShut {NoStop}%
\bibitem [{\citenamefont {Rudner}\ \emph {et~al.}(2013)\citenamefont {Rudner},
  \citenamefont {Lindner}, \citenamefont {Berg},\ and\ \citenamefont
  {Levin}}]{Lindner2013}%
  \BibitemOpen
  \bibfield  {author} {\bibinfo {author} {\bibfnamefont {Mark~S.}\ \bibnamefont
  {Rudner}}, \bibinfo {author} {\bibfnamefont {Netanel~H.}\ \bibnamefont
  {Lindner}}, \bibinfo {author} {\bibfnamefont {Erez}\ \bibnamefont {Berg}}, \
  and\ \bibinfo {author} {\bibfnamefont {Michael}\ \bibnamefont {Levin}},\
  }\bibfield  {title} {\enquote {\bibinfo {title} {Anomalous edge states and
  the bulk-edge correspondence for periodically driven two-dimensional
  systems},}\ }\href {\doibase 10.1103/PhysRevX.3.031005} {\bibfield  {journal}
  {\bibinfo  {journal} {Phys. Rev. X}\ }\textbf {\bibinfo {volume} {3}},\
  \bibinfo {pages} {031005} (\bibinfo {year} {2013})}\BibitemShut {NoStop}%
\bibitem [{\citenamefont {Hu}\ and\ \citenamefont {Hughes}(2011)}]{Hughes2011}%
  \BibitemOpen
  \bibfield  {author} {\bibinfo {author} {\bibfnamefont {Yi~Chen}\ \bibnamefont
  {Hu}}\ and\ \bibinfo {author} {\bibfnamefont {Taylor~L.}\ \bibnamefont
  {Hughes}},\ }\bibfield  {title} {\enquote {\bibinfo {title} {Absence of
  topological insulator phases in non-hermitian $pt$-symmetric hamiltonians},}\
  }\href {\doibase 10.1103/PhysRevB.84.153101} {\bibfield  {journal} {\bibinfo
  {journal} {Phys. Rev. B}\ }\textbf {\bibinfo {volume} {84}},\ \bibinfo
  {pages} {153101} (\bibinfo {year} {2011})}\BibitemShut {NoStop}%
\bibitem [{\citenamefont {Esaki}\ \emph {et~al.}(2011)\citenamefont {Esaki},
  \citenamefont {Sato}, \citenamefont {Hasebe},\ and\ \citenamefont
  {Kohmoto}}]{Sato2011}%
  \BibitemOpen
  \bibfield  {author} {\bibinfo {author} {\bibfnamefont {Kenta}\ \bibnamefont
  {Esaki}}, \bibinfo {author} {\bibfnamefont {Masatoshi}\ \bibnamefont {Sato}},
  \bibinfo {author} {\bibfnamefont {Kazuki}\ \bibnamefont {Hasebe}}, \ and\
  \bibinfo {author} {\bibfnamefont {Mahito}\ \bibnamefont {Kohmoto}},\
  }\bibfield  {title} {\enquote {\bibinfo {title} {Edge states and topological
  phases in non-hermitian systems},}\ }\href {\doibase
  10.1103/PhysRevB.84.205128} {\bibfield  {journal} {\bibinfo  {journal} {Phys.
  Rev. B}\ }\textbf {\bibinfo {volume} {84}},\ \bibinfo {pages} {205128}
  (\bibinfo {year} {2011})}\BibitemShut {NoStop}%
\bibitem [{\citenamefont {Schomerus}(2013)}]{Schomerus2013}%
  \BibitemOpen
  \bibfield  {author} {\bibinfo {author} {\bibfnamefont {Henning}\ \bibnamefont
  {Schomerus}},\ }\bibfield  {title} {\enquote {\bibinfo {title} {Topologically
  protected midgap states in complex photonic lattices},}\ }\href
  {http://ol.osa.org/abstract.cfm?URI=ol-38-11-1912} {\bibfield  {journal}
  {\bibinfo  {journal} {Opt. Lett.}\ }\textbf {\bibinfo {volume} {38}},\
  \bibinfo {pages} {1912} (\bibinfo {year} {2013})}\BibitemShut {NoStop}%
\bibitem [{\citenamefont {Malzard}\ \emph {et~al.}(2015)\citenamefont
  {Malzard}, \citenamefont {Poli},\ and\ \citenamefont
  {Schomerus}}]{Malzard2015}%
  \BibitemOpen
  \bibfield  {author} {\bibinfo {author} {\bibfnamefont {Simon}\ \bibnamefont
  {Malzard}}, \bibinfo {author} {\bibfnamefont {Charles}\ \bibnamefont {Poli}},
  \ and\ \bibinfo {author} {\bibfnamefont {Henning}\ \bibnamefont
  {Schomerus}},\ }\bibfield  {title} {\enquote {\bibinfo {title} {Topologically
  protected defect states in open photonic systems with non-hermitian
  charge-conjugation and parity-time symmetry},}\ }\href {\doibase
  10.1103/PhysRevLett.115.200402} {\bibfield  {journal} {\bibinfo  {journal}
  {Phys. Rev. Lett.}\ }\textbf {\bibinfo {volume} {115}},\ \bibinfo {pages}
  {200402} (\bibinfo {year} {2015})}\BibitemShut {NoStop}%
\bibitem [{\citenamefont {San-Jose}\ \emph {et~al.}(2016)\citenamefont
  {San-Jose}, \citenamefont {Cayao}, \citenamefont {Prada},\ and\ \citenamefont
  {Aguado}}]{Jose2016}%
  \BibitemOpen
  \bibfield  {author} {\bibinfo {author} {\bibfnamefont {Pablo}\ \bibnamefont
  {San-Jose}}, \bibinfo {author} {\bibfnamefont {Jorge}\ \bibnamefont {Cayao}},
  \bibinfo {author} {\bibfnamefont {Elsa}\ \bibnamefont {Prada}}, \ and\
  \bibinfo {author} {\bibfnamefont {Ram\'on}\ \bibnamefont {Aguado}},\
  }\bibfield  {title} {\enquote {\bibinfo {title} {Majorana bound states from
  exceptional points in non-topological superconductors},}\ }\href
  {http://dx.doi.org/10.1038/srep21427} {\bibfield  {journal} {\bibinfo
  {journal} {Sci. Rep.}\ }\textbf {\bibinfo {volume} {6}},\ \bibinfo {pages}
  {21427} (\bibinfo {year} {2016})}\BibitemShut {NoStop}%
\bibitem [{\citenamefont {Lee}(2016)}]{Tony2016}%
  \BibitemOpen
  \bibfield  {author} {\bibinfo {author} {\bibfnamefont {Tony~E.}\ \bibnamefont
  {Lee}},\ }\bibfield  {title} {\enquote {\bibinfo {title} {Anomalous edge
  state in a non-hermitian lattice},}\ }\href {\doibase
  10.1103/PhysRevLett.116.133903} {\bibfield  {journal} {\bibinfo  {journal}
  {Phys. Rev. Lett.}\ }\textbf {\bibinfo {volume} {116}},\ \bibinfo {pages}
  {133903} (\bibinfo {year} {2016})}\BibitemShut {NoStop}%
\bibitem [{\citenamefont {Leykam}\ \emph {et~al.}(2017)\citenamefont {Leykam},
  \citenamefont {Bliokh}, \citenamefont {Huang}, \citenamefont {Chong},\ and\
  \citenamefont {Nori}}]{Nori2017}%
  \BibitemOpen
  \bibfield  {author} {\bibinfo {author} {\bibfnamefont {Daniel}\ \bibnamefont
  {Leykam}}, \bibinfo {author} {\bibfnamefont {Konstantin~Y.}\ \bibnamefont
  {Bliokh}}, \bibinfo {author} {\bibfnamefont {Chunli}\ \bibnamefont {Huang}},
  \bibinfo {author} {\bibfnamefont {Y.~D.}\ \bibnamefont {Chong}}, \ and\
  \bibinfo {author} {\bibfnamefont {Franco}\ \bibnamefont {Nori}},\ }\bibfield
  {title} {\enquote {\bibinfo {title} {Edge modes, degeneracies, and
  topological numbers in non-hermitian systems},}\ }\href {\doibase
  10.1103/PhysRevLett.118.040401} {\bibfield  {journal} {\bibinfo  {journal}
  {Phys. Rev. Lett.}\ }\textbf {\bibinfo {volume} {118}},\ \bibinfo {pages}
  {040401} (\bibinfo {year} {2017})}\BibitemShut {NoStop}%
\bibitem [{\citenamefont {Xu}\ \emph {et~al.}(2017)\citenamefont {Xu},
  \citenamefont {Wang},\ and\ \citenamefont {Duan}}]{Duan2017}%
  \BibitemOpen
  \bibfield  {author} {\bibinfo {author} {\bibfnamefont {Yong}\ \bibnamefont
  {Xu}}, \bibinfo {author} {\bibfnamefont {Sheng-Tao}\ \bibnamefont {Wang}}, \
  and\ \bibinfo {author} {\bibfnamefont {L.-M.}\ \bibnamefont {Duan}},\
  }\bibfield  {title} {\enquote {\bibinfo {title} {Weyl exceptional rings in a
  three-dimensional dissipative cold atomic gas},}\ }\href {\doibase
  10.1103/PhysRevLett.118.045701} {\bibfield  {journal} {\bibinfo  {journal}
  {Phys. Rev. Lett.}\ }\textbf {\bibinfo {volume} {118}},\ \bibinfo {pages}
  {045701} (\bibinfo {year} {2017})}\BibitemShut {NoStop}%
\bibitem [{\citenamefont {Kawabata}\ \emph
  {et~al.}(2018{\natexlab{a}})\citenamefont {Kawabata}, \citenamefont {Ashida},
  \citenamefont {Katsura},\ and\ \citenamefont {Ueda}}]{Kawabata2018a}%
  \BibitemOpen
  \bibfield  {author} {\bibinfo {author} {\bibfnamefont {Kohei}\ \bibnamefont
  {Kawabata}}, \bibinfo {author} {\bibfnamefont {Yuto}\ \bibnamefont {Ashida}},
  \bibinfo {author} {\bibfnamefont {Hosho}\ \bibnamefont {Katsura}}, \ and\
  \bibinfo {author} {\bibfnamefont {Masahito}\ \bibnamefont {Ueda}},\
  }\href@noop {} {\enquote {\bibinfo {title} {Parity-time-symmetric topological
  superconductor},}\ } (\bibinfo {year} {2018}{\natexlab{a}}),\ \bibinfo {note}
  {arXiv:1801.00499}\BibitemShut {NoStop}%
\bibitem [{\citenamefont {Barontini}\ \emph {et~al.}(2013)\citenamefont
  {Barontini}, \citenamefont {Labouvie}, \citenamefont {Stubenrauch},
  \citenamefont {Vogler}, \citenamefont {Guarrera},\ and\ \citenamefont
  {Ott}}]{Ott2013}%
  \BibitemOpen
  \bibfield  {author} {\bibinfo {author} {\bibfnamefont {G.}~\bibnamefont
  {Barontini}}, \bibinfo {author} {\bibfnamefont {R.}~\bibnamefont {Labouvie}},
  \bibinfo {author} {\bibfnamefont {F.}~\bibnamefont {Stubenrauch}}, \bibinfo
  {author} {\bibfnamefont {A.}~\bibnamefont {Vogler}}, \bibinfo {author}
  {\bibfnamefont {V.}~\bibnamefont {Guarrera}}, \ and\ \bibinfo {author}
  {\bibfnamefont {H.}~\bibnamefont {Ott}},\ }\bibfield  {title} {\enquote
  {\bibinfo {title} {Controlling the dynamics of an open many-body quantum
  system with localized dissipation},}\ }\href
  {https://link.aps.org/doi/10.1103/PhysRevLett.110.035302} {\bibfield
  {journal} {\bibinfo  {journal} {Phys. Rev. Lett.}\ }\textbf {\bibinfo
  {volume} {110}},\ \bibinfo {pages} {035302} (\bibinfo {year}
  {2013})}\BibitemShut {NoStop}%
\bibitem [{\citenamefont {Aspelmeyer}\ \emph {et~al.}(2014)\citenamefont
  {Aspelmeyer}, \citenamefont {Kippenberg},\ and\ \citenamefont
  {Marquardt}}]{Aspelmeyer2014}%
  \BibitemOpen
  \bibfield  {author} {\bibinfo {author} {\bibfnamefont {Markus}\ \bibnamefont
  {Aspelmeyer}}, \bibinfo {author} {\bibfnamefont {Tobias~J.}\ \bibnamefont
  {Kippenberg}}, \ and\ \bibinfo {author} {\bibfnamefont {Florian}\
  \bibnamefont {Marquardt}},\ }\bibfield  {title} {\enquote {\bibinfo {title}
  {Cavity optomechanics},}\ }\href {\doibase 10.1103/RevModPhys.86.1391}
  {\bibfield  {journal} {\bibinfo  {journal} {Rev. Mod. Phys.}\ }\textbf
  {\bibinfo {volume} {86}},\ \bibinfo {pages} {1391--1452} (\bibinfo {year}
  {2014})}\BibitemShut {NoStop}%
\bibitem [{\citenamefont {Cao}\ and\ \citenamefont
  {Wiersig}(2015)}]{Wiersig2015}%
  \BibitemOpen
  \bibfield  {author} {\bibinfo {author} {\bibfnamefont {Hui}\ \bibnamefont
  {Cao}}\ and\ \bibinfo {author} {\bibfnamefont {Jan}\ \bibnamefont
  {Wiersig}},\ }\bibfield  {title} {\enquote {\bibinfo {title} {Dielectric
  microcavities: Model systems for wave chaos and non-hermitian physics},}\
  }\href {\doibase 10.1103/RevModPhys.87.61} {\bibfield  {journal} {\bibinfo
  {journal} {Rev. Mod. Phys.}\ }\textbf {\bibinfo {volume} {87}},\ \bibinfo
  {pages} {61} (\bibinfo {year} {2015})}\BibitemShut {NoStop}%
\bibitem [{\citenamefont {Peng}\ \emph
  {et~al.}(2016{\natexlab{a}})\citenamefont {Peng}, \citenamefont {Cao},
  \citenamefont {Shen}, \citenamefont {Qu}, \citenamefont {Wen}, \citenamefont
  {Jiang},\ and\ \citenamefont {Xiao}}]{Xiao2016}%
  \BibitemOpen
  \bibfield  {author} {\bibinfo {author} {\bibfnamefont {Peng}\ \bibnamefont
  {Peng}}, \bibinfo {author} {\bibfnamefont {Wanxia}\ \bibnamefont {Cao}},
  \bibinfo {author} {\bibfnamefont {Ce}~\bibnamefont {Shen}}, \bibinfo {author}
  {\bibfnamefont {Weizhi}\ \bibnamefont {Qu}}, \bibinfo {author} {\bibfnamefont
  {Jianming}\ \bibnamefont {Wen}}, \bibinfo {author} {\bibfnamefont {Liang}\
  \bibnamefont {Jiang}}, \ and\ \bibinfo {author} {\bibfnamefont {Yanhong}\
  \bibnamefont {Xiao}},\ }\bibfield  {title} {\enquote {\bibinfo {title}
  {Anti-parity-time symmetry with flying atoms},}\ }\href
  {http://dx.doi.org/10.1038/nphys3842} {\bibfield  {journal} {\bibinfo
  {journal} {Nat. Phys.}\ }\textbf {\bibinfo {volume} {12}},\ \bibinfo {pages}
  {1139} (\bibinfo {year} {2016}{\natexlab{a}})}\BibitemShut {NoStop}%
\bibitem [{\citenamefont {Xu}\ \emph {et~al.}(2016)\citenamefont {Xu},
  \citenamefont {Mason}, \citenamefont {Jiang},\ and\ \citenamefont
  {Harris}}]{Xu2016}%
  \BibitemOpen
  \bibfield  {author} {\bibinfo {author} {\bibfnamefont {H.}~\bibnamefont
  {Xu}}, \bibinfo {author} {\bibfnamefont {D.}~\bibnamefont {Mason}}, \bibinfo
  {author} {\bibfnamefont {Luyao}\ \bibnamefont {Jiang}}, \ and\ \bibinfo
  {author} {\bibfnamefont {J.~G.~E.}\ \bibnamefont {Harris}},\ }\bibfield
  {title} {\enquote {\bibinfo {title} {Topological energy transfer in an
  optomechanical system with exceptional points},}\ }\href
  {http://dx.doi.org/10.1038/nature18604} {\bibfield  {journal} {\bibinfo
  {journal} {Nature}\ }\textbf {\bibinfo {volume} {537}},\ \bibinfo {pages}
  {80} (\bibinfo {year} {2016})}\BibitemShut {NoStop}%
\bibitem [{\citenamefont {Peng}\ \emph
  {et~al.}(2016{\natexlab{b}})\citenamefont {Peng}, \citenamefont
  {{\"O}zdemir}, \citenamefont {Liertzer}, \citenamefont {Chen}, \citenamefont
  {Kramer}, \citenamefont {Y{\i}lmaz}, \citenamefont {Wiersig}, \citenamefont
  {Rotter},\ and\ \citenamefont {Yang}}]{Yang2016}%
  \BibitemOpen
  \bibfield  {author} {\bibinfo {author} {\bibfnamefont {Bo}~\bibnamefont
  {Peng}}, \bibinfo {author} {\bibfnamefont {{\c S}ahin~Kaya}\ \bibnamefont
  {{\"O}zdemir}}, \bibinfo {author} {\bibfnamefont {Matthias}\ \bibnamefont
  {Liertzer}}, \bibinfo {author} {\bibfnamefont {Weijian}\ \bibnamefont
  {Chen}}, \bibinfo {author} {\bibfnamefont {Johannes}\ \bibnamefont {Kramer}},
  \bibinfo {author} {\bibfnamefont {Huzeyfe}\ \bibnamefont {Y{\i}lmaz}},
  \bibinfo {author} {\bibfnamefont {Jan}\ \bibnamefont {Wiersig}}, \bibinfo
  {author} {\bibfnamefont {Stefan}\ \bibnamefont {Rotter}}, \ and\ \bibinfo
  {author} {\bibfnamefont {Lan}\ \bibnamefont {Yang}},\ }\bibfield  {title}
  {\enquote {\bibinfo {title} {Chiral modes and directional lasing at
  exceptional points},}\ }\href {\doibase 10.1073/pnas.1603318113} {\bibfield
  {journal} {\bibinfo  {journal} {Proc. Natl. Acad. Sci. USA}\ }\textbf
  {\bibinfo {volume} {113}},\ \bibinfo {pages} {6845} (\bibinfo {year}
  {2016}{\natexlab{b}})}\BibitemShut {NoStop}%
\bibitem [{\citenamefont {Chen}\ \emph {et~al.}(2017)\citenamefont {Chen},
  \citenamefont {\"Ozdemir}, \citenamefont {Zhao}, \citenamefont {Wiersig},\
  and\ \citenamefont {Yang}}]{Yang2017}%
  \BibitemOpen
  \bibfield  {author} {\bibinfo {author} {\bibfnamefont {Weijian}\ \bibnamefont
  {Chen}}, \bibinfo {author} {\bibfnamefont {{\c S}ahin~Kaya}\ \bibnamefont
  {\"Ozdemir}}, \bibinfo {author} {\bibfnamefont {Guangming}\ \bibnamefont
  {Zhao}}, \bibinfo {author} {\bibfnamefont {Jan}\ \bibnamefont {Wiersig}}, \
  and\ \bibinfo {author} {\bibfnamefont {Lan}\ \bibnamefont {Yang}},\
  }\bibfield  {title} {\enquote {\bibinfo {title} {Exceptional points enhance
  sensing in an optical microcavity},}\ }\href
  {http://dx.doi.org/10.1038/nature23281} {\bibfield  {journal} {\bibinfo
  {journal} {Nature}\ }\textbf {\bibinfo {volume} {548}},\ \bibinfo {pages}
  {192} (\bibinfo {year} {2017})}\BibitemShut {NoStop}%
\bibitem [{\citenamefont {Zhou}\ \emph {et~al.}(2018)\citenamefont {Zhou},
  \citenamefont {Peng}, \citenamefont {Yoon}, \citenamefont {Hsu},
  \citenamefont {Nelson}, \citenamefont {Fu}, \citenamefont {Joannopoulos},
  \citenamefont {Solja{\v c}i{\'c}},\ and\ \citenamefont {Zhen}}]{Zhou2018}%
  \BibitemOpen
  \bibfield  {author} {\bibinfo {author} {\bibfnamefont {Hengyun}\ \bibnamefont
  {Zhou}}, \bibinfo {author} {\bibfnamefont {Chao}\ \bibnamefont {Peng}},
  \bibinfo {author} {\bibfnamefont {Yoseob}\ \bibnamefont {Yoon}}, \bibinfo
  {author} {\bibfnamefont {Chia~Wei}\ \bibnamefont {Hsu}}, \bibinfo {author}
  {\bibfnamefont {Keith~A.}\ \bibnamefont {Nelson}}, \bibinfo {author}
  {\bibfnamefont {Liang}\ \bibnamefont {Fu}}, \bibinfo {author} {\bibfnamefont
  {John~D.}\ \bibnamefont {Joannopoulos}}, \bibinfo {author} {\bibfnamefont
  {Marin}\ \bibnamefont {Solja{\v c}i{\'c}}}, \ and\ \bibinfo {author}
  {\bibfnamefont {Bo}~\bibnamefont {Zhen}},\ }\bibfield  {title} {\enquote
  {\bibinfo {title} {Observation of bulk fermi arc and polarization half charge
  from paired exceptional points},}\ }\href {\doibase 10.1126/science.aap9859}
  {\bibfield  {journal} {\bibinfo  {journal} {Science}\ }\textbf {\bibinfo
  {volume} {359}},\ \bibinfo {pages} {1009} (\bibinfo {year}
  {2018})}\BibitemShut {NoStop}%
\bibitem [{\citenamefont {Bandres}\ \emph {et~al.}(2018)\citenamefont
  {Bandres}, \citenamefont {Wittek}, \citenamefont {Harari}, \citenamefont
  {Parto}, \citenamefont {Ren}, \citenamefont {Segev}, \citenamefont
  {Christodoulides},\ and\ \citenamefont {Khajavikhan}}]{Segev2018b}%
  \BibitemOpen
  \bibfield  {author} {\bibinfo {author} {\bibfnamefont {Miguel~A.}\
  \bibnamefont {Bandres}}, \bibinfo {author} {\bibfnamefont {Steffen}\
  \bibnamefont {Wittek}}, \bibinfo {author} {\bibfnamefont {Gal}\ \bibnamefont
  {Harari}}, \bibinfo {author} {\bibfnamefont {Midya}\ \bibnamefont {Parto}},
  \bibinfo {author} {\bibfnamefont {Jinhan}\ \bibnamefont {Ren}}, \bibinfo
  {author} {\bibfnamefont {Mordechai}\ \bibnamefont {Segev}}, \bibinfo {author}
  {\bibfnamefont {Demetrios~N.}\ \bibnamefont {Christodoulides}}, \ and\
  \bibinfo {author} {\bibfnamefont {Mercedeh}\ \bibnamefont {Khajavikhan}},\
  }\bibfield  {title} {\enquote {\bibinfo {title} {Topological insulator laser:
  Experiments},}\ }\href {\doibase 10.1126/science.aar4005} {\bibfield
  {journal} {\bibinfo  {journal} {Science}\ }\textbf {\bibinfo {volume} {359}}
  (\bibinfo {year} {2018}),\ 10.1126/science.aar4005}\BibitemShut {NoStop}%
\bibitem [{\citenamefont {R\"uter}\ \emph {et~al.}(2010)\citenamefont
  {R\"uter}, \citenamefont {Makris}, \citenamefont {El-Ganainy}, \citenamefont
  {Christodoulides}, \citenamefont {Segev},\ and\ \citenamefont
  {Kip}}]{Segev2010}%
  \BibitemOpen
  \bibfield  {author} {\bibinfo {author} {\bibfnamefont {Christian~E.}\
  \bibnamefont {R\"uter}}, \bibinfo {author} {\bibfnamefont {Konstantinos~G.}\
  \bibnamefont {Makris}}, \bibinfo {author} {\bibfnamefont {Ramy}\ \bibnamefont
  {El-Ganainy}}, \bibinfo {author} {\bibfnamefont {Demetrios~N.}\ \bibnamefont
  {Christodoulides}}, \bibinfo {author} {\bibfnamefont {Mordechai}\
  \bibnamefont {Segev}}, \ and\ \bibinfo {author} {\bibfnamefont {Detlef}\
  \bibnamefont {Kip}},\ }\bibfield  {title} {\enquote {\bibinfo {title}
  {Observation of parity-time symmetry in optics},}\ }\href
  {http://dx.doi.org/10.1038/nphys1515} {\bibfield  {journal} {\bibinfo
  {journal} {Nat. Phys.}\ }\textbf {\bibinfo {volume} {6}},\ \bibinfo {pages}
  {192} (\bibinfo {year} {2010})}\BibitemShut {NoStop}%
\bibitem [{\citenamefont {Peng}\ \emph {et~al.}(2014)\citenamefont {Peng},
  \citenamefont {\"Ozdemir}, \citenamefont {Lei}, \citenamefont {Monifi},
  \citenamefont {Gianfreda}, \citenamefont {Long}, \citenamefont {Fan},
  \citenamefont {Nori}, \citenamefont {Bender},\ and\ \citenamefont
  {Yang}}]{Yang2014}%
  \BibitemOpen
  \bibfield  {author} {\bibinfo {author} {\bibfnamefont {Bo}~\bibnamefont
  {Peng}}, \bibinfo {author} {\bibfnamefont {{\c S}ahin~Kaya}\ \bibnamefont
  {\"Ozdemir}}, \bibinfo {author} {\bibfnamefont {Fuchuan}\ \bibnamefont
  {Lei}}, \bibinfo {author} {\bibfnamefont {Faraz}\ \bibnamefont {Monifi}},
  \bibinfo {author} {\bibfnamefont {Mariagiovanna}\ \bibnamefont {Gianfreda}},
  \bibinfo {author} {\bibfnamefont {Gui~Lu}\ \bibnamefont {Long}}, \bibinfo
  {author} {\bibfnamefont {Shanhui}\ \bibnamefont {Fan}}, \bibinfo {author}
  {\bibfnamefont {Franco}\ \bibnamefont {Nori}}, \bibinfo {author}
  {\bibfnamefont {Carl~M.}\ \bibnamefont {Bender}}, \ and\ \bibinfo {author}
  {\bibfnamefont {Lan}\ \bibnamefont {Yang}},\ }\bibfield  {title} {\enquote
  {\bibinfo {title} {Parity-time-symmetric whispering-gallery microcavities},}\
  }\href {http://dx.doi.org/10.1038/nphys2927} {\bibfield  {journal} {\bibinfo
  {journal} {Nat. Phys.}\ }\textbf {\bibinfo {volume} {10}},\ \bibinfo {pages}
  {394} (\bibinfo {year} {2014})}\BibitemShut {NoStop}%
\bibitem [{\citenamefont {Feng}\ \emph {et~al.}(2014)\citenamefont {Feng},
  \citenamefont {Wong}, \citenamefont {Ma}, \citenamefont {Wang},\ and\
  \citenamefont {Zhang}}]{Zhang2014}%
  \BibitemOpen
  \bibfield  {author} {\bibinfo {author} {\bibfnamefont {Liang}\ \bibnamefont
  {Feng}}, \bibinfo {author} {\bibfnamefont {Zi~Jing}\ \bibnamefont {Wong}},
  \bibinfo {author} {\bibfnamefont {Ren-Min}\ \bibnamefont {Ma}}, \bibinfo
  {author} {\bibfnamefont {Yuan}\ \bibnamefont {Wang}}, \ and\ \bibinfo
  {author} {\bibfnamefont {Xiang}\ \bibnamefont {Zhang}},\ }\bibfield  {title}
  {\enquote {\bibinfo {title} {Single-mode laser by parity-time symmetry
  breaking},}\ }\href {\doibase 10.1126/science.1258479} {\bibfield  {journal}
  {\bibinfo  {journal} {Science}\ }\textbf {\bibinfo {volume} {346}},\ \bibinfo
  {pages} {972} (\bibinfo {year} {2014})}\BibitemShut {NoStop}%
\bibitem [{\citenamefont {Konotop}\ \emph {et~al.}(2016)\citenamefont
  {Konotop}, \citenamefont {Yang},\ and\ \citenamefont
  {Zezyulin}}]{Konotop2016}%
  \BibitemOpen
  \bibfield  {author} {\bibinfo {author} {\bibfnamefont {Vladimir~V.}\
  \bibnamefont {Konotop}}, \bibinfo {author} {\bibfnamefont {Jianke}\
  \bibnamefont {Yang}}, \ and\ \bibinfo {author} {\bibfnamefont {Dmitry~A.}\
  \bibnamefont {Zezyulin}},\ }\bibfield  {title} {\enquote {\bibinfo {title}
  {Nonlinear waves in $\mathcal{PT}$-symmetric systems},}\ }\href {\doibase
  10.1103/RevModPhys.88.035002} {\bibfield  {journal} {\bibinfo  {journal}
  {Rev. Mod. Phys.}\ }\textbf {\bibinfo {volume} {88}},\ \bibinfo {pages}
  {035002} (\bibinfo {year} {2016})}\BibitemShut {NoStop}%
\bibitem [{\citenamefont {Xiao}\ \emph {et~al.}(2017)\citenamefont {Xiao},
  \citenamefont {Zhan}, \citenamefont {Bian}, \citenamefont {Wang},
  \citenamefont {Zhang}, \citenamefont {Wang}, \citenamefont {Li},
  \citenamefont {Mochizuki}, \citenamefont {Kim}, \citenamefont {Kawakami},
  \citenamefont {Yi}, \citenamefont {Obuse}, \citenamefont {Sanders},\ and\
  \citenamefont {Xue}}]{Xue2017}%
  \BibitemOpen
  \bibfield  {author} {\bibinfo {author} {\bibfnamefont {L.}~\bibnamefont
  {Xiao}}, \bibinfo {author} {\bibfnamefont {X.}~\bibnamefont {Zhan}}, \bibinfo
  {author} {\bibfnamefont {Z.~H.}\ \bibnamefont {Bian}}, \bibinfo {author}
  {\bibfnamefont {K.~K.}\ \bibnamefont {Wang}}, \bibinfo {author}
  {\bibfnamefont {X.}~\bibnamefont {Zhang}}, \bibinfo {author} {\bibfnamefont
  {X.~P.}\ \bibnamefont {Wang}}, \bibinfo {author} {\bibfnamefont
  {J.}~\bibnamefont {Li}}, \bibinfo {author} {\bibfnamefont {K.}~\bibnamefont
  {Mochizuki}}, \bibinfo {author} {\bibfnamefont {D.}~\bibnamefont {Kim}},
  \bibinfo {author} {\bibfnamefont {N.}~\bibnamefont {Kawakami}}, \bibinfo
  {author} {\bibfnamefont {W.}~\bibnamefont {Yi}}, \bibinfo {author}
  {\bibfnamefont {H.}~\bibnamefont {Obuse}}, \bibinfo {author} {\bibfnamefont
  {B.~C.}\ \bibnamefont {Sanders}}, \ and\ \bibinfo {author} {\bibfnamefont
  {P.}~\bibnamefont {Xue}},\ }\bibfield  {title} {\enquote {\bibinfo {title}
  {Observation of topological edge states in parity-time-symmetric quantum
  walks},}\ }\href {http://dx.doi.org/10.1038/nphys4204} {\bibfield  {journal}
  {\bibinfo  {journal} {Nat. Phys.}\ }\textbf {\bibinfo {volume} {13}},\
  \bibinfo {pages} {1117} (\bibinfo {year} {2017})}\BibitemShut {NoStop}%
\bibitem [{\citenamefont {Weimann}\ \emph {et~al.}(2017)\citenamefont
  {Weimann}, \citenamefont {Kremer}, \citenamefont {Plotnik}, \citenamefont
  {Lumer}, \citenamefont {Nolte}, \citenamefont {Makris}, \citenamefont
  {Segev}, \citenamefont {Rechtsman},\ and\ \citenamefont
  {Szameit}}]{Segev2017}%
  \BibitemOpen
  \bibfield  {author} {\bibinfo {author} {\bibfnamefont {S.}~\bibnamefont
  {Weimann}}, \bibinfo {author} {\bibfnamefont {M.}~\bibnamefont {Kremer}},
  \bibinfo {author} {\bibfnamefont {Y.}~\bibnamefont {Plotnik}}, \bibinfo
  {author} {\bibfnamefont {Y.}~\bibnamefont {Lumer}}, \bibinfo {author}
  {\bibfnamefont {S.}~\bibnamefont {Nolte}}, \bibinfo {author} {\bibfnamefont
  {K.~G.}\ \bibnamefont {Makris}}, \bibinfo {author} {\bibfnamefont
  {M.}~\bibnamefont {Segev}}, \bibinfo {author} {\bibfnamefont {M.~C.}\
  \bibnamefont {Rechtsman}}, \ and\ \bibinfo {author} {\bibfnamefont
  {A.}~\bibnamefont {Szameit}},\ }\bibfield  {title} {\enquote {\bibinfo
  {title} {Topologically protected bound states in photonic
  parity-time-symmetric crystals},}\ }\href
  {http://dx.doi.org/10.1038/nmat4811} {\bibfield  {journal} {\bibinfo
  {journal} {Nat. Mater.}\ }\textbf {\bibinfo {volume} {16}},\ \bibinfo {pages}
  {433} (\bibinfo {year} {2017})}\BibitemShut {NoStop}%
\bibitem [{\citenamefont {El-Ganainy}\ \emph {et~al.}(2018)\citenamefont
  {El-Ganainy}, \citenamefont {Makris}, \citenamefont {Khajavikhan},
  \citenamefont {Musslimani}, \citenamefont {Rotter},\ and\ \citenamefont
  {Christodoulides}}]{Christodoulides2018}%
  \BibitemOpen
  \bibfield  {author} {\bibinfo {author} {\bibfnamefont {Ramy}\ \bibnamefont
  {El-Ganainy}}, \bibinfo {author} {\bibfnamefont {Konstantinos~G.}\
  \bibnamefont {Makris}}, \bibinfo {author} {\bibfnamefont {Mercedeh}\
  \bibnamefont {Khajavikhan}}, \bibinfo {author} {\bibfnamefont {Ziad~H.}\
  \bibnamefont {Musslimani}}, \bibinfo {author} {\bibfnamefont {Stefan}\
  \bibnamefont {Rotter}}, \ and\ \bibinfo {author} {\bibfnamefont
  {Demetrios~N.}\ \bibnamefont {Christodoulides}},\ }\bibfield  {title}
  {\enquote {\bibinfo {title} {Non-hermitian physics and pt symmetry},}\ }\href
  {http://dx.doi.org/10.1038/nphys4323} {\bibfield  {journal} {\bibinfo
  {journal} {Nat. Phys.}\ }\textbf {\bibinfo {volume} {14}},\ \bibinfo {pages}
  {11} (\bibinfo {year} {2018})}\BibitemShut {NoStop}%
\bibitem [{\citenamefont {Bender}\ and\ \citenamefont
  {Boettcher}(1998)}]{Bender1998}%
  \BibitemOpen
  \bibfield  {author} {\bibinfo {author} {\bibfnamefont {Carl~M.}\ \bibnamefont
  {Bender}}\ and\ \bibinfo {author} {\bibfnamefont {Stefan}\ \bibnamefont
  {Boettcher}},\ }\bibfield  {title} {\enquote {\bibinfo {title} {Real spectra
  in non-hermitian hamiltonians having $\mathscr{P}\mathscr{T}$ symmetry},}\
  }\href {\doibase 10.1103/PhysRevLett.80.5243} {\bibfield  {journal} {\bibinfo
   {journal} {Phys. Rev. Lett.}\ }\textbf {\bibinfo {volume} {80}},\ \bibinfo
  {pages} {5243} (\bibinfo {year} {1998})}\BibitemShut {NoStop}%
\bibitem [{\citenamefont {Bender}(2007)}]{Bender2007}%
  \BibitemOpen
  \bibfield  {author} {\bibinfo {author} {\bibfnamefont {Carl~M.}\ \bibnamefont
  {Bender}},\ }\bibfield  {title} {\enquote {\bibinfo {title} {Making sense of
  non-hermitian hamiltonians},}\ }\href
  {https://doi.org/10.1088/0034-4885/70/6/R03} {\bibfield  {journal} {\bibinfo
  {journal} {Rep. Prog. Phys.}\ }\textbf {\bibinfo {volume} {70}},\ \bibinfo
  {pages} {947} (\bibinfo {year} {2007})}\BibitemShut {NoStop}%
\bibitem [{\citenamefont {Heiss}(2012)}]{Heiss2012}%
  \BibitemOpen
  \bibfield  {author} {\bibinfo {author} {\bibfnamefont {W.~D.}\ \bibnamefont
  {Heiss}},\ }\bibfield  {title} {\enquote {\bibinfo {title} {The physics of
  exceptional points},}\ }\href
  {https://doi.org/10.1088/1751-8113/45/44/444016} {\bibfield  {journal}
  {\bibinfo  {journal} {J. Phys. A: Math. Theor.}\ }\textbf {\bibinfo {volume}
  {45}},\ \bibinfo {pages} {444016} (\bibinfo {year} {2012})}\BibitemShut
  {NoStop}%
\bibitem [{\citenamefont {Petruccione}\ and\ \citenamefont
  {Breuer}(2002)}]{Breuer2002}%
  \BibitemOpen
  \bibfield  {author} {\bibinfo {author} {\bibfnamefont {F.}~\bibnamefont
  {Petruccione}}\ and\ \bibinfo {author} {\bibfnamefont {H.~P.}\ \bibnamefont
  {Breuer}},\ }\href@noop {} {\emph {\bibinfo {title} {The theory of open
  quantum systems}}}\ (\bibinfo  {publisher} {Oxford University Press,
  London},\ \bibinfo {year} {2002})\BibitemShut {NoStop}%
\bibitem [{\citenamefont {Rudner}\ and\ \citenamefont
  {Levitov}(2009)}]{Rudner2009}%
  \BibitemOpen
  \bibfield  {author} {\bibinfo {author} {\bibfnamefont {M.~S.}\ \bibnamefont
  {Rudner}}\ and\ \bibinfo {author} {\bibfnamefont {L.~S.}\ \bibnamefont
  {Levitov}},\ }\bibfield  {title} {\enquote {\bibinfo {title} {Topological
  transition in a non-hermitian quantum walk},}\ }\href {\doibase
  10.1103/PhysRevLett.102.065703} {\bibfield  {journal} {\bibinfo  {journal}
  {Phys. Rev. Lett.}\ }\textbf {\bibinfo {volume} {102}},\ \bibinfo {pages}
  {065703} (\bibinfo {year} {2009})}\BibitemShut {NoStop}%
\bibitem [{\citenamefont {Lee}\ and\ \citenamefont {Chan}(2014)}]{Tony2014a}%
  \BibitemOpen
  \bibfield  {author} {\bibinfo {author} {\bibfnamefont {Tony~E.}\ \bibnamefont
  {Lee}}\ and\ \bibinfo {author} {\bibfnamefont {Ching-Kit}\ \bibnamefont
  {Chan}},\ }\bibfield  {title} {\enquote {\bibinfo {title} {Heralded magnetism
  in non-hermitian atomic systems},}\ }\href {\doibase
  10.1103/PhysRevX.4.041001} {\bibfield  {journal} {\bibinfo  {journal} {Phys.
  Rev. X}\ }\textbf {\bibinfo {volume} {4}},\ \bibinfo {pages} {041001}
  (\bibinfo {year} {2014})}\BibitemShut {NoStop}%
\bibitem [{\citenamefont {Lee}\ \emph {et~al.}(2014)\citenamefont {Lee},
  \citenamefont {Reiter},\ and\ \citenamefont {Moiseyev}}]{Tony2014b}%
  \BibitemOpen
  \bibfield  {author} {\bibinfo {author} {\bibfnamefont {Tony~E.}\ \bibnamefont
  {Lee}}, \bibinfo {author} {\bibfnamefont {Florentin}\ \bibnamefont {Reiter}},
  \ and\ \bibinfo {author} {\bibfnamefont {Nimrod}\ \bibnamefont {Moiseyev}},\
  }\bibfield  {title} {\enquote {\bibinfo {title} {Entanglement and spin
  squeezing in non-hermitian phase transitions},}\ }\href {\doibase
  10.1103/PhysRevLett.113.250401} {\bibfield  {journal} {\bibinfo  {journal}
  {Phys. Rev. Lett.}\ }\textbf {\bibinfo {volume} {113}},\ \bibinfo {pages}
  {250401} (\bibinfo {year} {2014})}\BibitemShut {NoStop}%
\bibitem [{\citenamefont {Ashida}\ \emph {et~al.}(2016)\citenamefont {Ashida},
  \citenamefont {Furukawa},\ and\ \citenamefont {Ueda}}]{Ashida2016}%
  \BibitemOpen
  \bibfield  {author} {\bibinfo {author} {\bibfnamefont {Yuto}\ \bibnamefont
  {Ashida}}, \bibinfo {author} {\bibfnamefont {Shunsuke}\ \bibnamefont
  {Furukawa}}, \ and\ \bibinfo {author} {\bibfnamefont {Masahito}\ \bibnamefont
  {Ueda}},\ }\bibfield  {title} {\enquote {\bibinfo {title} {Quantum critical
  behavior influenced by measurement backaction in ultracold gases},}\ }\href
  {\doibase 10.1103/PhysRevA.94.053615} {\bibfield  {journal} {\bibinfo
  {journal} {Phys. Rev. A}\ }\textbf {\bibinfo {volume} {94}},\ \bibinfo
  {pages} {053615} (\bibinfo {year} {2016})}\BibitemShut {NoStop}%
\bibitem [{\citenamefont {Ashida}\ \emph {et~al.}(2017)\citenamefont {Ashida},
  \citenamefont {Furukawa},\ and\ \citenamefont {Ueda}}]{Ashida2017a}%
  \BibitemOpen
  \bibfield  {author} {\bibinfo {author} {\bibfnamefont {Yuto}\ \bibnamefont
  {Ashida}}, \bibinfo {author} {\bibfnamefont {Shunsuke}\ \bibnamefont
  {Furukawa}}, \ and\ \bibinfo {author} {\bibfnamefont {Masahito}\ \bibnamefont
  {Ueda}},\ }\bibfield  {title} {\enquote {\bibinfo {title}
  {Parity-time-symmetric quantum critical phenomena},}\ }\href
  {http://dx.doi.org/10.1038/ncomms15791} {\bibfield  {journal} {\bibinfo
  {journal} {Nat. Commun.}\ }\textbf {\bibinfo {volume} {8}},\ \bibinfo {pages}
  {15791} (\bibinfo {year} {2017})}\BibitemShut {NoStop}%
\bibitem [{\citenamefont {Gong}\ \emph {et~al.}(2017)\citenamefont {Gong},
  \citenamefont {Higashikawa},\ and\ \citenamefont {Ueda}}]{Gong2017}%
  \BibitemOpen
  \bibfield  {author} {\bibinfo {author} {\bibfnamefont {Zongping}\
  \bibnamefont {Gong}}, \bibinfo {author} {\bibfnamefont {Sho}\ \bibnamefont
  {Higashikawa}}, \ and\ \bibinfo {author} {\bibfnamefont {Masahito}\
  \bibnamefont {Ueda}},\ }\bibfield  {title} {\enquote {\bibinfo {title} {Zeno
  hall effect},}\ }\href {\doibase 10.1103/PhysRevLett.118.200401} {\bibfield
  {journal} {\bibinfo  {journal} {Phys. Rev. Lett.}\ }\textbf {\bibinfo
  {volume} {118}},\ \bibinfo {pages} {200401} (\bibinfo {year}
  {2017})}\BibitemShut {NoStop}%
\bibitem [{\citenamefont {Kawabata}\ \emph {et~al.}(2017)\citenamefont
  {Kawabata}, \citenamefont {Ashida},\ and\ \citenamefont
  {Ueda}}]{Kawabata2017}%
  \BibitemOpen
  \bibfield  {author} {\bibinfo {author} {\bibfnamefont {Kohei}\ \bibnamefont
  {Kawabata}}, \bibinfo {author} {\bibfnamefont {Yuto}\ \bibnamefont {Ashida}},
  \ and\ \bibinfo {author} {\bibfnamefont {Masahito}\ \bibnamefont {Ueda}},\
  }\bibfield  {title} {\enquote {\bibinfo {title} {Information retrieval and
  criticality in parity-time-symmetric systems},}\ }\href {\doibase
  10.1103/PhysRevLett.119.190401} {\bibfield  {journal} {\bibinfo  {journal}
  {Phys. Rev. Lett.}\ }\textbf {\bibinfo {volume} {119}},\ \bibinfo {pages}
  {190401} (\bibinfo {year} {2017})}\BibitemShut {NoStop}%
\bibitem [{\citenamefont {Esposito}\ \emph {et~al.}(2009)\citenamefont
  {Esposito}, \citenamefont {Harbola},\ and\ \citenamefont
  {Mukamel}}]{Esposito2009}%
  \BibitemOpen
  \bibfield  {author} {\bibinfo {author} {\bibfnamefont {M.}~\bibnamefont
  {Esposito}}, \bibinfo {author} {\bibfnamefont {U.}~\bibnamefont {Harbola}}, \
  and\ \bibinfo {author} {\bibfnamefont {S.}~\bibnamefont {Mukamel}},\
  }\bibfield  {title} {\enquote {\bibinfo {title} {Nonequilibrium fluctuations,
  fluctuation theorems, and counting statistics in quantum systems},}\ }\href
  {https://link.aps.org/doi/10.1103/RevModPhys.81.1665} {\bibfield  {journal}
  {\bibinfo  {journal} {Rev. Mod. Phys.}\ }\textbf {\bibinfo {volume} {81}},\
  \bibinfo {pages} {1665} (\bibinfo {year} {2009})}\BibitemShut {NoStop}%
\bibitem [{\citenamefont {Ren}\ \emph {et~al.}(2010)\citenamefont {Ren},
  \citenamefont {H\"anggi},\ and\ \citenamefont {Li}}]{Ren2010}%
  \BibitemOpen
  \bibfield  {author} {\bibinfo {author} {\bibfnamefont {Jie}\ \bibnamefont
  {Ren}}, \bibinfo {author} {\bibfnamefont {Peter}\ \bibnamefont {H\"anggi}}, \
  and\ \bibinfo {author} {\bibfnamefont {Baowen}\ \bibnamefont {Li}},\
  }\bibfield  {title} {\enquote {\bibinfo {title} {Berry-phase-induced heat
  pumping and its impact on the fluctuation theorem},}\ }\href {\doibase
  10.1103/PhysRevLett.104.170601} {\bibfield  {journal} {\bibinfo  {journal}
  {Phys. Rev. Lett.}\ }\textbf {\bibinfo {volume} {104}},\ \bibinfo {pages}
  {170601} (\bibinfo {year} {2010})}\BibitemShut {NoStop}%
\bibitem [{\citenamefont {Sagawa}\ and\ \citenamefont
  {Hayakawa}(2011)}]{Sagawa2011b}%
  \BibitemOpen
  \bibfield  {author} {\bibinfo {author} {\bibfnamefont {Takahiro}\
  \bibnamefont {Sagawa}}\ and\ \bibinfo {author} {\bibfnamefont {Hisao}\
  \bibnamefont {Hayakawa}},\ }\bibfield  {title} {\enquote {\bibinfo {title}
  {Geometrical expression of excess entropy production},}\ }\href {\doibase
  10.1103/PhysRevE.84.051110} {\bibfield  {journal} {\bibinfo  {journal} {Phys.
  Rev. E}\ }\textbf {\bibinfo {volume} {84}},\ \bibinfo {pages} {051110}
  (\bibinfo {year} {2011})}\BibitemShut {NoStop}%
\bibitem [{\citenamefont {Nelson}\ and\ \citenamefont
  {Shnerb}(1998)}]{Nelson1998}%
  \BibitemOpen
  \bibfield  {author} {\bibinfo {author} {\bibfnamefont {David~R.}\
  \bibnamefont {Nelson}}\ and\ \bibinfo {author} {\bibfnamefont {Nadav~M.}\
  \bibnamefont {Shnerb}},\ }\bibfield  {title} {\enquote {\bibinfo {title}
  {Non-hermitian localization and population biology},}\ }\href {\doibase
  10.1103/PhysRevE.58.1383} {\bibfield  {journal} {\bibinfo  {journal} {Phys.
  Rev. E}\ }\textbf {\bibinfo {volume} {58}},\ \bibinfo {pages} {1383}
  (\bibinfo {year} {1998})}\BibitemShut {NoStop}%
\bibitem [{\citenamefont {Amir}\ \emph {et~al.}(2016)\citenamefont {Amir},
  \citenamefont {Hatano},\ and\ \citenamefont {Nelson}}]{Amir2016}%
  \BibitemOpen
  \bibfield  {author} {\bibinfo {author} {\bibfnamefont {Ariel}\ \bibnamefont
  {Amir}}, \bibinfo {author} {\bibfnamefont {Naomichi}\ \bibnamefont {Hatano}},
  \ and\ \bibinfo {author} {\bibfnamefont {David~R.}\ \bibnamefont {Nelson}},\
  }\bibfield  {title} {\enquote {\bibinfo {title} {Non-hermitian localization
  in biological networks},}\ }\href {\doibase 10.1103/PhysRevE.93.042310}
  {\bibfield  {journal} {\bibinfo  {journal} {Phys. Rev. E}\ }\textbf {\bibinfo
  {volume} {93}},\ \bibinfo {pages} {042310} (\bibinfo {year}
  {2016})}\BibitemShut {NoStop}%
\bibitem [{\citenamefont {Murugan}\ and\ \citenamefont
  {Vaikuntanathan}(2017)}]{Vaikuntanathan2017}%
  \BibitemOpen
  \bibfield  {author} {\bibinfo {author} {\bibfnamefont {Arvind}\ \bibnamefont
  {Murugan}}\ and\ \bibinfo {author} {\bibfnamefont {Suriyanarayanan}\
  \bibnamefont {Vaikuntanathan}},\ }\bibfield  {title} {\enquote {\bibinfo
  {title} {Topologically protected modes in non-equilibrium stochastic
  systems},}\ }\href {http://dx.doi.org/10.1038/ncomms13881} {\bibfield
  {journal} {\bibinfo  {journal} {Nat. Commun.}\ }\textbf {\bibinfo {volume}
  {8}},\ \bibinfo {pages} {13881} (\bibinfo {year} {2017})}\BibitemShut
  {NoStop}%
\bibitem [{\citenamefont {Ren}\ and\ \citenamefont {Sinitsyn}(2013)}]{Ren2013}%
  \BibitemOpen
  \bibfield  {author} {\bibinfo {author} {\bibfnamefont {Jie}\ \bibnamefont
  {Ren}}\ and\ \bibinfo {author} {\bibfnamefont {N.~A.}\ \bibnamefont
  {Sinitsyn}},\ }\bibfield  {title} {\enquote {\bibinfo {title} {Braid group
  and topological phase transitions in nonequilibrium stochastic dynamics},}\
  }\href {\doibase 10.1103/PhysRevE.87.050101} {\bibfield  {journal} {\bibinfo
  {journal} {Phys. Rev. E}\ }\textbf {\bibinfo {volume} {87}},\ \bibinfo
  {pages} {050101(R)} (\bibinfo {year} {2013})}\BibitemShut {NoStop}%
\bibitem [{\citenamefont {Cao}\ \emph {et~al.}(2015)\citenamefont {Cao},
  \citenamefont {Gong},\ and\ \citenamefont {Quan}}]{Cao2015}%
  \BibitemOpen
  \bibfield  {author} {\bibinfo {author} {\bibfnamefont {Yuansheng}\
  \bibnamefont {Cao}}, \bibinfo {author} {\bibfnamefont {Zongping}\
  \bibnamefont {Gong}}, \ and\ \bibinfo {author} {\bibfnamefont {H.~T.}\
  \bibnamefont {Quan}},\ }\bibfield  {title} {\enquote {\bibinfo {title}
  {Thermodynamics of information processing based on enzyme kinetics: An
  exactly solvable model of an information pump},}\ }\href {\doibase
  10.1103/PhysRevE.91.062117} {\bibfield  {journal} {\bibinfo  {journal} {Phys.
  Rev. E}\ }\textbf {\bibinfo {volume} {91}},\ \bibinfo {pages} {062117}
  (\bibinfo {year} {2015})}\BibitemShut {NoStop}%
\bibitem [{\citenamefont {McGrath}\ \emph {et~al.}(2017)\citenamefont
  {McGrath}, \citenamefont {Jones}, \citenamefont {ten Wolde},\ and\
  \citenamefont {Ouldridge}}]{Ouldridge2017}%
  \BibitemOpen
  \bibfield  {author} {\bibinfo {author} {\bibfnamefont {Thomas}\ \bibnamefont
  {McGrath}}, \bibinfo {author} {\bibfnamefont {Nick~S.}\ \bibnamefont
  {Jones}}, \bibinfo {author} {\bibfnamefont {Pieter~Rein}\ \bibnamefont {ten
  Wolde}}, \ and\ \bibinfo {author} {\bibfnamefont {Thomas~E.}\ \bibnamefont
  {Ouldridge}},\ }\bibfield  {title} {\enquote {\bibinfo {title} {Biochemical
  machines for the interconversion of mutual information and work},}\ }\href
  {\doibase 10.1103/PhysRevLett.118.028101} {\bibfield  {journal} {\bibinfo
  {journal} {Phys. Rev. Lett.}\ }\textbf {\bibinfo {volume} {118}},\ \bibinfo
  {pages} {028101} (\bibinfo {year} {2017})}\BibitemShut {NoStop}%
\bibitem [{\citenamefont {Rudner}\ \emph {et~al.}(2016)\citenamefont {Rudner},
  \citenamefont {Levin},\ and\ \citenamefont {Levitov}}]{Rudner2016}%
  \BibitemOpen
  \bibfield  {author} {\bibinfo {author} {\bibfnamefont {Mark~S.}\ \bibnamefont
  {Rudner}}, \bibinfo {author} {\bibfnamefont {Michael}\ \bibnamefont {Levin}},
  \ and\ \bibinfo {author} {\bibfnamefont {Leonid~S.}\ \bibnamefont
  {Levitov}},\ }\href@noop {} {\enquote {\bibinfo {title} {Survival, decay, and
  topological protection in non-hermitian quantum transport},}\ } (\bibinfo
  {year} {2016}),\ \bibinfo {note} {arXiv:1605.07652}\BibitemShut {NoStop}%
\bibitem [{\citenamefont {Shen}\ \emph {et~al.}(2018)\citenamefont {Shen},
  \citenamefont {Zhen},\ and\ \citenamefont {Fu}}]{Fu2018}%
  \BibitemOpen
  \bibfield  {author} {\bibinfo {author} {\bibfnamefont {Huitao}\ \bibnamefont
  {Shen}}, \bibinfo {author} {\bibfnamefont {Bo}~\bibnamefont {Zhen}}, \ and\
  \bibinfo {author} {\bibfnamefont {Liang}\ \bibnamefont {Fu}},\ }\bibfield
  {title} {\enquote {\bibinfo {title} {Topological band theory for
  non-hermitian hamiltonians},}\ }\href {\doibase
  10.1103/PhysRevLett.120.146402} {\bibfield  {journal} {\bibinfo  {journal}
  {Phys. Rev. Lett.}\ }\textbf {\bibinfo {volume} {120}},\ \bibinfo {pages}
  {146402} (\bibinfo {year} {2018})}\BibitemShut {NoStop}%
\bibitem [{\citenamefont {Nielsen}\ and\ \citenamefont
  {Chuang}(2010)}]{Chuang2010}%
  \BibitemOpen
  \bibfield  {author} {\bibinfo {author} {\bibfnamefont {M.~A.}\ \bibnamefont
  {Nielsen}}\ and\ \bibinfo {author} {\bibfnamefont {I.~L.}\ \bibnamefont
  {Chuang}},\ }\href@noop {} {\emph {\bibinfo {title} {Quantum Computation and
  Information}}}\ (\bibinfo  {publisher} {Cambridge University Press,
  Cambridge},\ \bibinfo {year} {2010})\BibitemShut {NoStop}%
\bibitem [{\citenamefont {Anderson}(1958)}]{Anderson1958}%
  \BibitemOpen
  \bibfield  {author} {\bibinfo {author} {\bibfnamefont {P.~W.}\ \bibnamefont
  {Anderson}},\ }\bibfield  {title} {\enquote {\bibinfo {title} {Absence of
  diffusion in certain random lattices},}\ }\href {\doibase
  10.1103/PhysRev.109.1492} {\bibfield  {journal} {\bibinfo  {journal} {Phys.
  Rev.}\ }\textbf {\bibinfo {volume} {109}},\ \bibinfo {pages} {1492} (\bibinfo
  {year} {1958})}\BibitemShut {NoStop}%
\bibitem [{\citenamefont {Hatano}\ and\ \citenamefont
  {Nelson}(1996)}]{Hatano1996}%
  \BibitemOpen
  \bibfield  {author} {\bibinfo {author} {\bibfnamefont {Naomichi}\
  \bibnamefont {Hatano}}\ and\ \bibinfo {author} {\bibfnamefont {David~R.}\
  \bibnamefont {Nelson}},\ }\bibfield  {title} {\enquote {\bibinfo {title}
  {Localization transitions in non-hermitian quantum mechanics},}\ }\href
  {\doibase 10.1103/PhysRevLett.77.570} {\bibfield  {journal} {\bibinfo
  {journal} {Phys. Rev. Lett.}\ }\textbf {\bibinfo {volume} {77}},\ \bibinfo
  {pages} {570} (\bibinfo {year} {1996})}\BibitemShut {NoStop}%
\bibitem [{\citenamefont {Hatano}\ and\ \citenamefont
  {Nelson}(1997)}]{Hatano1997}%
  \BibitemOpen
  \bibfield  {author} {\bibinfo {author} {\bibfnamefont {Naomichi}\
  \bibnamefont {Hatano}}\ and\ \bibinfo {author} {\bibfnamefont {David~R.}\
  \bibnamefont {Nelson}},\ }\bibfield  {title} {\enquote {\bibinfo {title}
  {Vortex pinning and non-hermitian quantum mechanics},}\ }\href {\doibase
  10.1103/PhysRevB.56.8651} {\bibfield  {journal} {\bibinfo  {journal} {Phys.
  Rev. B}\ }\textbf {\bibinfo {volume} {56}},\ \bibinfo {pages} {8651}
  (\bibinfo {year} {1997})}\BibitemShut {NoStop}%
\bibitem [{\citenamefont {Hatano}\ and\ \citenamefont
  {Nelson}(1998)}]{Hatano1998}%
  \BibitemOpen
  \bibfield  {author} {\bibinfo {author} {\bibfnamefont {Naomichi}\
  \bibnamefont {Hatano}}\ and\ \bibinfo {author} {\bibfnamefont {David~R.}\
  \bibnamefont {Nelson}},\ }\bibfield  {title} {\enquote {\bibinfo {title}
  {Non-hermitian delocalization and eigenfunctions},}\ }\href {\doibase
  10.1103/PhysRevB.58.8384} {\bibfield  {journal} {\bibinfo  {journal} {Phys.
  Rev. B}\ }\textbf {\bibinfo {volume} {58}},\ \bibinfo {pages} {8384}
  (\bibinfo {year} {1998})}\BibitemShut {NoStop}%
\bibitem [{\citenamefont {Abrahams}\ \emph {et~al.}(1979)\citenamefont
  {Abrahams}, \citenamefont {Anderson}, \citenamefont {Licciardello},\ and\
  \citenamefont {Ramakrishnan}}]{Anderson1979}%
  \BibitemOpen
  \bibfield  {author} {\bibinfo {author} {\bibfnamefont {E.}~\bibnamefont
  {Abrahams}}, \bibinfo {author} {\bibfnamefont {P.~W.}\ \bibnamefont
  {Anderson}}, \bibinfo {author} {\bibfnamefont {D.~C.}\ \bibnamefont
  {Licciardello}}, \ and\ \bibinfo {author} {\bibfnamefont {T.~V.}\
  \bibnamefont {Ramakrishnan}},\ }\bibfield  {title} {\enquote {\bibinfo
  {title} {Scaling theory of localization: Absence of quantum diffusion in two
  dimensions},}\ }\href {\doibase 10.1103/PhysRevLett.42.673} {\bibfield
  {journal} {\bibinfo  {journal} {Phys. Rev. Lett.}\ }\textbf {\bibinfo
  {volume} {42}},\ \bibinfo {pages} {673} (\bibinfo {year} {1979})}\BibitemShut
  {NoStop}%
\bibitem [{\citenamefont {Karoubi}(2008)}]{Karoubi2008}%
  \BibitemOpen
  \bibfield  {author} {\bibinfo {author} {\bibfnamefont {Max}\ \bibnamefont
  {Karoubi}},\ }\href@noop {} {\emph {\bibinfo {title} {K-theory: An
  Introduction}}}\ (\bibinfo  {publisher} {Springer, Berlin},\ \bibinfo {year}
  {2008})\BibitemShut {NoStop}%
\bibitem [{\citenamefont {Ahlfors}(1979)}]{Ahlfors1979}%
  \BibitemOpen
  \bibfield  {author} {\bibinfo {author} {\bibfnamefont {Lars~V.}\ \bibnamefont
  {Ahlfors}},\ }\href@noop {} {\emph {\bibinfo {title} {Complex Analysis}}}\
  (\bibinfo  {publisher} {McGraw-Hill, New York},\ \bibinfo {year}
  {1979})\BibitemShut {NoStop}%
\bibitem [{\citenamefont {Nandkishore}\ and\ \citenamefont
  {Huse}(2015)}]{Nandkishore2015}%
  \BibitemOpen
  \bibfield  {author} {\bibinfo {author} {\bibfnamefont {Rahul}\ \bibnamefont
  {Nandkishore}}\ and\ \bibinfo {author} {\bibfnamefont {David~A.}\
  \bibnamefont {Huse}},\ }\bibfield  {title} {\enquote {\bibinfo {title}
  {Many-body localization and thermalization in quantum statistical
  mechanics},}\ }\href
  {https://doi.org/10.1146/annurev-conmatphys-031214-014726} {\bibfield
  {journal} {\bibinfo  {journal} {Annu. Rev. Cond. Matt. Phys.}\ }\textbf
  {\bibinfo {volume} {6}},\ \bibinfo {pages} {201} (\bibinfo {year}
  {2015})}\BibitemShut {NoStop}%
\bibitem [{\citenamefont {Moessner}\ and\ \citenamefont
  {Sondhi}(2017)}]{Moessner2017}%
  \BibitemOpen
  \bibfield  {author} {\bibinfo {author} {\bibfnamefont {R.}~\bibnamefont
  {Moessner}}\ and\ \bibinfo {author} {\bibfnamefont {S.~L.}\ \bibnamefont
  {Sondhi}},\ }\bibfield  {title} {\enquote {\bibinfo {title} {Equilibration
  and order in quantum floquet matter},}\ }\href
  {http://dx.doi.org/10.1038/nphys4106} {\bibfield  {journal} {\bibinfo
  {journal} {Nat. Phys.}\ }\textbf {\bibinfo {volume} {13}},\ \bibinfo {pages}
  {424} (\bibinfo {year} {2017})}\BibitemShut {NoStop}%
\bibitem [{\citenamefont {Haake}(2010)}]{Haake2010}%
  \BibitemOpen
  \bibfield  {author} {\bibinfo {author} {\bibfnamefont {Fritz}\ \bibnamefont
  {Haake}},\ }\href@noop {} {\emph {\bibinfo {title} {Quantum Signatures of
  Chaos}}}\ (\bibinfo  {publisher} {Springer, Berlin},\ \bibinfo {year}
  {2010})\BibitemShut {NoStop}%
\bibitem [{\citenamefont {Khemani}\ \emph {et~al.}(2016)\citenamefont
  {Khemani}, \citenamefont {Lazarides}, \citenamefont {Moessner},\ and\
  \citenamefont {Sondhi}}]{Khemani2016}%
  \BibitemOpen
  \bibfield  {author} {\bibinfo {author} {\bibfnamefont {Vedika}\ \bibnamefont
  {Khemani}}, \bibinfo {author} {\bibfnamefont {Achilleas}\ \bibnamefont
  {Lazarides}}, \bibinfo {author} {\bibfnamefont {Roderich}\ \bibnamefont
  {Moessner}}, \ and\ \bibinfo {author} {\bibfnamefont {S.~L.}\ \bibnamefont
  {Sondhi}},\ }\bibfield  {title} {\enquote {\bibinfo {title} {Phase structure
  of driven quantum systems},}\ }\href {\doibase
  10.1103/PhysRevLett.116.250401} {\bibfield  {journal} {\bibinfo  {journal}
  {Phys. Rev. Lett.}\ }\textbf {\bibinfo {volume} {116}},\ \bibinfo {pages}
  {250401} (\bibinfo {year} {2016})}\BibitemShut {NoStop}%
\bibitem [{\citenamefont {von Keyserlingk}\ and\ \citenamefont
  {Sondhi}(2016{\natexlab{b}})}]{Keyserlingk2016b}%
  \BibitemOpen
  \bibfield  {author} {\bibinfo {author} {\bibfnamefont {C.~W.}\ \bibnamefont
  {von Keyserlingk}}\ and\ \bibinfo {author} {\bibfnamefont {S.~L.}\
  \bibnamefont {Sondhi}},\ }\bibfield  {title} {\enquote {\bibinfo {title}
  {Phase structure of one-dimensional interacting floquet systems. ii.
  symmetry-broken phases},}\ }\href {\doibase 10.1103/PhysRevB.93.245146}
  {\bibfield  {journal} {\bibinfo  {journal} {Phys. Rev. B}\ }\textbf {\bibinfo
  {volume} {93}},\ \bibinfo {pages} {245146} (\bibinfo {year}
  {2016}{\natexlab{b}})}\BibitemShut {NoStop}%
\bibitem [{\citenamefont {Else}\ \emph {et~al.}(2016)\citenamefont {Else},
  \citenamefont {Bauer},\ and\ \citenamefont {Nayak}}]{Nayak2016}%
  \BibitemOpen
  \bibfield  {author} {\bibinfo {author} {\bibfnamefont {Dominic~V.}\
  \bibnamefont {Else}}, \bibinfo {author} {\bibfnamefont {Bela}\ \bibnamefont
  {Bauer}}, \ and\ \bibinfo {author} {\bibfnamefont {Chetan}\ \bibnamefont
  {Nayak}},\ }\bibfield  {title} {\enquote {\bibinfo {title} {Floquet time
  crystals},}\ }\href {\doibase 10.1103/PhysRevLett.117.090402} {\bibfield
  {journal} {\bibinfo  {journal} {Phys. Rev. Lett.}\ }\textbf {\bibinfo
  {volume} {117}},\ \bibinfo {pages} {090402} (\bibinfo {year}
  {2016})}\BibitemShut {NoStop}%
\bibitem [{Note1()}]{Note1}%
  \BibitemOpen
  \bibinfo {note} {In the absence of a particle-hole or chiral symmetry, the
  classification of all the Hermitian Hamiltonians with a given $E_{\protect
  \rm F}$ is equivalent to that with $E_{\protect \rm F}=0$, since we have a
  time-reversal-symmetry-preserved one-to-one map $H\to H-E_{\protect \rm F}$
  between two sets of Hamiltonians. In the presence of a particle-hole or/and
  chiral symmetry, although $E_{\protect \rm F}$ has arbitrariness for a given
  system, the only choice of $E_{\protect \rm F}$ is zero when considering the
  set of all such Hermitian Hamiltonians.}\BibitemShut {Stop}%
\bibitem [{\citenamefont {Kampen}(2007)}]{Kampen2007}%
  \BibitemOpen
  \bibfield  {author} {\bibinfo {author} {\bibfnamefont {N.~G.~Van}\
  \bibnamefont {Kampen}},\ }\href@noop {} {\emph {\bibinfo {title} {Stochastic
  Processes in Physics and Chemistry}}}\ (\bibinfo  {publisher} {Elsevier, New
  York},\ \bibinfo {year} {2007})\BibitemShut {NoStop}%
\bibitem [{\citenamefont {Albert}\ \emph {et~al.}(2016)\citenamefont {Albert},
  \citenamefont {Bradlyn}, \citenamefont {Fraas},\ and\ \citenamefont
  {Jiang}}]{Jiang2016}%
  \BibitemOpen
  \bibfield  {author} {\bibinfo {author} {\bibfnamefont {Victor~V.}\
  \bibnamefont {Albert}}, \bibinfo {author} {\bibfnamefont {Barry}\
  \bibnamefont {Bradlyn}}, \bibinfo {author} {\bibfnamefont {Martin}\
  \bibnamefont {Fraas}}, \ and\ \bibinfo {author} {\bibfnamefont {Liang}\
  \bibnamefont {Jiang}},\ }\bibfield  {title} {\enquote {\bibinfo {title}
  {Geometry and response of lindbladians},}\ }\href
  {https://link.aps.org/doi/10.1103/PhysRevX.6.041031} {\bibfield  {journal}
  {\bibinfo  {journal} {Phys. Rev. X}\ }\textbf {\bibinfo {volume} {6}},\
  \bibinfo {pages} {041031} (\bibinfo {year} {2016})}\BibitemShut {NoStop}%
\bibitem [{\citenamefont {Kim}\ \emph {et~al.}(2016)\citenamefont {Kim},
  \citenamefont {Ken}, \citenamefont {Kawakami},\ and\ \citenamefont
  {Obuse}}]{Obuse2016}%
  \BibitemOpen
  \bibfield  {author} {\bibinfo {author} {\bibfnamefont {Dakyeong}\
  \bibnamefont {Kim}}, \bibinfo {author} {\bibfnamefont {Mochizuki}\
  \bibnamefont {Ken}}, \bibinfo {author} {\bibfnamefont {Norio}\ \bibnamefont
  {Kawakami}}, \ and\ \bibinfo {author} {\bibfnamefont {Hideaki}\ \bibnamefont
  {Obuse}},\ }\href@noop {} {\enquote {\bibinfo {title} {Floquet topological
  phases driven by $\mathcal{PT}$ symmetric nonunitary time evolution},}\ }
  (\bibinfo {year} {2016}),\ \bibinfo {note} {arXiv:1609.09650}\BibitemShut
  {NoStop}%
\bibitem [{\citenamefont {Bardyn}\ \emph {et~al.}(2013)\citenamefont {Bardyn},
  \citenamefont {Baranov}, \citenamefont {Kraus}, \citenamefont {Rico},
  \citenamefont {\.Imamo\v{g}lu}, \citenamefont {Zoller},\ and\ \citenamefont
  {Diehl}}]{Diehl2013}%
  \BibitemOpen
  \bibfield  {author} {\bibinfo {author} {\bibfnamefont {C.-E.}\ \bibnamefont
  {Bardyn}}, \bibinfo {author} {\bibfnamefont {M.~A.}\ \bibnamefont {Baranov}},
  \bibinfo {author} {\bibfnamefont {C.~V.}\ \bibnamefont {Kraus}}, \bibinfo
  {author} {\bibfnamefont {E.}~\bibnamefont {Rico}}, \bibinfo {author}
  {\bibfnamefont {A.}~\bibnamefont {\.Imamo\v{g}lu}}, \bibinfo {author}
  {\bibfnamefont {P.}~\bibnamefont {Zoller}}, \ and\ \bibinfo {author}
  {\bibfnamefont {S.}~\bibnamefont {Diehl}},\ }\bibfield  {title} {\enquote
  {\bibinfo {title} {Topology by dissipation},}\ }\href
  {https://doi.org/10.1088/1367-2630/15/8/085001} {\bibfield  {journal}
  {\bibinfo  {journal} {New J. Phys.}\ }\textbf {\bibinfo {volume} {15}},\
  \bibinfo {pages} {085001} (\bibinfo {year} {2013})}\BibitemShut {NoStop}%
\bibitem [{\citenamefont {Thouless}(1983)}]{Thouless1983}%
  \BibitemOpen
  \bibfield  {author} {\bibinfo {author} {\bibfnamefont {D.~J.}\ \bibnamefont
  {Thouless}},\ }\bibfield  {title} {\enquote {\bibinfo {title} {Quantization
  of particle transport},}\ }\href {\doibase 10.1103/PhysRevB.27.6083}
  {\bibfield  {journal} {\bibinfo  {journal} {Phys. Rev. B}\ }\textbf {\bibinfo
  {volume} {27}},\ \bibinfo {pages} {6083} (\bibinfo {year}
  {1983})}\BibitemShut {NoStop}%
\bibitem [{Note2()}]{Note2}%
  \BibitemOpen
  \bibinfo {note} {The phenomenon alone may be relevant to a single band, like
  the integer or anomalous quantum Hall effect. After all, at least another
  band is necessary to make the present band nontrivial, such as another band
  with the opposite Chern number.}\BibitemShut {Stop}%
\bibitem [{\citenamefont {Asada}\ \emph {et~al.}(2002)\citenamefont {Asada},
  \citenamefont {Slevin},\ and\ \citenamefont {Ohtsuki}}]{Asada2002}%
  \BibitemOpen
  \bibfield  {author} {\bibinfo {author} {\bibfnamefont {Yoichi}\ \bibnamefont
  {Asada}}, \bibinfo {author} {\bibfnamefont {Keith}\ \bibnamefont {Slevin}}, \
  and\ \bibinfo {author} {\bibfnamefont {Tomi}\ \bibnamefont {Ohtsuki}},\
  }\bibfield  {title} {\enquote {\bibinfo {title} {Anderson transition in
  two-dimensional systems with spin-orbit coupling},}\ }\href {\doibase
  10.1103/PhysRevLett.89.256601} {\bibfield  {journal} {\bibinfo  {journal}
  {Phys. Rev. Lett.}\ }\textbf {\bibinfo {volume} {89}},\ \bibinfo {pages}
  {256601} (\bibinfo {year} {2002})}\BibitemShut {NoStop}%
\bibitem [{\citenamefont {Aoki}\ and\ \citenamefont {Ando}(1981)}]{Aoki1981}%
  \BibitemOpen
  \bibfield  {author} {\bibinfo {author} {\bibfnamefont {H.}~\bibnamefont
  {Aoki}}\ and\ \bibinfo {author} {\bibfnamefont {T.}~\bibnamefont {Ando}},\
  }\bibfield  {title} {\enquote {\bibinfo {title} {Effect of localization on
  the hall conductivity in the two-dimensional system in strong magnetic
  fields},}\ }\href {https://doi.org/10.1016/0038-1098(81)90021-1} {\bibfield
  {journal} {\bibinfo  {journal} {Solid State Commun.}\ }\textbf {\bibinfo
  {volume} {38}},\ \bibinfo {pages} {1079} (\bibinfo {year}
  {1981})}\BibitemShut {NoStop}%
\bibitem [{Note3()}]{Note3}%
  \BibitemOpen
  \bibinfo {note} {Here we tacitly assume a finite system size ($L$), and
  therefore the probability is zero for the disordered Hamiltonian to be not
  invertible.}\BibitemShut {Stop}%
\bibitem [{Note4()}]{Note4}%
  \BibitemOpen
  \bibinfo {note} {If $z_1=z_2$, we have $\psi
  _j=c_1z^j_1+c_2jz^{j-1}_1$.}\BibitemShut {Stop}%
\bibitem [{\citenamefont {Xiong}(2018)}]{Xiong2018}%
  \BibitemOpen
  \bibfield  {author} {\bibinfo {author} {\bibfnamefont {Ye}~\bibnamefont
  {Xiong}},\ }\bibfield  {title} {\enquote {\bibinfo {title} {Why does bulk
  boundary correspondence fail in some non-hermitian topological models},}\
  }\href {https://doi.org/10.1088/2399-6528/aab64a} {\bibfield  {journal}
  {\bibinfo  {journal} {J. Phys. Commun.}\ }\textbf {\bibinfo {volume} {2}},\
  \bibinfo {pages} {035043} (\bibinfo {year} {2018})}\BibitemShut {NoStop}%
\bibitem [{\citenamefont {Lieb}\ and\ \citenamefont
  {Robinson}(1972)}]{Lieb1972}%
  \BibitemOpen
  \bibfield  {author} {\bibinfo {author} {\bibfnamefont {Elliott~H.}\
  \bibnamefont {Lieb}}\ and\ \bibinfo {author} {\bibfnamefont {Derek~W.}\
  \bibnamefont {Robinson}},\ }\bibfield  {title} {\enquote {\bibinfo {title}
  {The finite group velocity of quantum spin systems},}\ }\href
  {https://doi.org/10.1007/978-3-662-10018-9_25} {\bibfield  {journal}
  {\bibinfo  {journal} {Commun. Math. Phys.}\ }\textbf {\bibinfo {volume}
  {28}},\ \bibinfo {pages} {251} (\bibinfo {year} {1972})}\BibitemShut
  {NoStop}%
\bibitem [{\citenamefont {Reichel}\ and\ \citenamefont
  {Trefethen}(1992)}]{Reichel1992}%
  \BibitemOpen
  \bibfield  {author} {\bibinfo {author} {\bibfnamefont {L.}~\bibnamefont
  {Reichel}}\ and\ \bibinfo {author} {\bibfnamefont {L.~N.}\ \bibnamefont
  {Trefethen}},\ }\bibfield  {title} {\enquote {\bibinfo {title} {Eigenvalues
  and pseudo-eigenvalues of toeplitz matrices},}\ }\href
  {https://doi.org/10.1016/0024-3795(92)90374-J} {\bibfield  {journal}
  {\bibinfo  {journal} {Linear Algebra Appl.}\ }\textbf {\bibinfo {volume}
  {162-164}},\ \bibinfo {pages} {153} (\bibinfo {year} {1992})}\BibitemShut
  {NoStop}%
\bibitem [{\citenamefont {Xiao}\ \emph {et~al.}(2010)\citenamefont {Xiao},
  \citenamefont {Chang},\ and\ \citenamefont {Niu}}]{Niu2010}%
  \BibitemOpen
  \bibfield  {author} {\bibinfo {author} {\bibfnamefont {Di}~\bibnamefont
  {Xiao}}, \bibinfo {author} {\bibfnamefont {Ming-Che}\ \bibnamefont {Chang}},
  \ and\ \bibinfo {author} {\bibfnamefont {Qian}\ \bibnamefont {Niu}},\
  }\bibfield  {title} {\enquote {\bibinfo {title} {Berry phase effects on
  electronic properties},}\ }\href
  {https://link.aps.org/doi/10.1103/RevModPhys.82.1959} {\bibfield  {journal}
  {\bibinfo  {journal} {Rev. Mod. Phys.}\ }\textbf {\bibinfo {volume} {82}},\
  \bibinfo {pages} {1959} (\bibinfo {year} {2010})}\BibitemShut {NoStop}%
\bibitem [{\citenamefont {Abanin}\ \emph {et~al.}(2013)\citenamefont {Abanin},
  \citenamefont {Kitagawa}, \citenamefont {Bloch},\ and\ \citenamefont
  {Demler}}]{Abanin2013}%
  \BibitemOpen
  \bibfield  {author} {\bibinfo {author} {\bibfnamefont {Dmitry~A.}\
  \bibnamefont {Abanin}}, \bibinfo {author} {\bibfnamefont {Takuya}\
  \bibnamefont {Kitagawa}}, \bibinfo {author} {\bibfnamefont {Immanuel}\
  \bibnamefont {Bloch}}, \ and\ \bibinfo {author} {\bibfnamefont {Eugene}\
  \bibnamefont {Demler}},\ }\bibfield  {title} {\enquote {\bibinfo {title}
  {Interferometric approach to measuring band topology in 2d optical
  lattices},}\ }\href {\doibase 10.1103/PhysRevLett.110.165304} {\bibfield
  {journal} {\bibinfo  {journal} {Phys. Rev. Lett.}\ }\textbf {\bibinfo
  {volume} {110}},\ \bibinfo {pages} {165304} (\bibinfo {year}
  {2013})}\BibitemShut {NoStop}%
\bibitem [{\citenamefont {Longhi}(2009)}]{Longhi2009}%
  \BibitemOpen
  \bibfield  {author} {\bibinfo {author} {\bibfnamefont {S.}~\bibnamefont
  {Longhi}},\ }\bibfield  {title} {\enquote {\bibinfo {title} {Bloch
  oscillations in complex crystals with $\mathcal{P}\mathcal{T}$ symmetry},}\
  }\href {\doibase 10.1103/PhysRevLett.103.123601} {\bibfield  {journal}
  {\bibinfo  {journal} {Phys. Rev. Lett.}\ }\textbf {\bibinfo {volume} {103}},\
  \bibinfo {pages} {123601} (\bibinfo {year} {2009})}\BibitemShut {NoStop}%
\bibitem [{\citenamefont {Longhi}(2015)}]{Longhi2015c}%
  \BibitemOpen
  \bibfield  {author} {\bibinfo {author} {\bibfnamefont {Stefano}\ \bibnamefont
  {Longhi}},\ }\bibfield  {title} {\enquote {\bibinfo {title} {Bloch
  oscillations in non-hermitian lattices with trajectories in the complex
  plane},}\ }\href {\doibase 10.1103/PhysRevA.92.042116} {\bibfield  {journal}
  {\bibinfo  {journal} {Phys. Rev. A}\ }\textbf {\bibinfo {volume} {92}},\
  \bibinfo {pages} {042116} (\bibinfo {year} {2015})}\BibitemShut {NoStop}%
\bibitem [{\citenamefont {Wimmer}\ \emph {et~al.}(2017)\citenamefont {Wimmer},
  \citenamefont {Price}, \citenamefont {Carusotto},\ and\ \citenamefont
  {Peschel}}]{Carusotto2017}%
  \BibitemOpen
  \bibfield  {author} {\bibinfo {author} {\bibfnamefont {Martin}\ \bibnamefont
  {Wimmer}}, \bibinfo {author} {\bibfnamefont {Hannah~M.}\ \bibnamefont
  {Price}}, \bibinfo {author} {\bibfnamefont {Iacopo}\ \bibnamefont
  {Carusotto}}, \ and\ \bibinfo {author} {\bibfnamefont {Ulf}\ \bibnamefont
  {Peschel}},\ }\bibfield  {title} {\enquote {\bibinfo {title} {Experimental
  measurement of the berry curvature from anomalous transport},}\ }\href
  {http://dx.doi.org/10.1038/nphys4050} {\bibfield  {journal} {\bibinfo
  {journal} {Nat. Phys.}\ }\textbf {\bibinfo {volume} {13}},\ \bibinfo {pages}
  {545} (\bibinfo {year} {2017})}\BibitemShut {NoStop}%
\bibitem [{\citenamefont {Longhi}\ \emph
  {et~al.}(2015{\natexlab{a}})\citenamefont {Longhi}, \citenamefont {Gatti},\
  and\ \citenamefont {Della~Valle}}]{Longhi2015a}%
  \BibitemOpen
  \bibfield  {author} {\bibinfo {author} {\bibfnamefont {Stefano}\ \bibnamefont
  {Longhi}}, \bibinfo {author} {\bibfnamefont {Davide}\ \bibnamefont {Gatti}},
  \ and\ \bibinfo {author} {\bibfnamefont {Giuseppe}\ \bibnamefont
  {Della~Valle}},\ }\bibfield  {title} {\enquote {\bibinfo {title} {Robust
  light transport in non-hermitian photonic lattices},}\ }\href
  {http://dx.doi.org/10.1038/srep13376} {\bibfield  {journal} {\bibinfo
  {journal} {Sci. Rep.}\ }\textbf {\bibinfo {volume} {5}},\ \bibinfo {pages}
  {13376} (\bibinfo {year} {2015}{\natexlab{a}})}\BibitemShut {NoStop}%
\bibitem [{\citenamefont {Longhi}\ \emph
  {et~al.}(2015{\natexlab{b}})\citenamefont {Longhi}, \citenamefont {Gatti},\
  and\ \citenamefont {Della~Valle}}]{Longhi2015b}%
  \BibitemOpen
  \bibfield  {author} {\bibinfo {author} {\bibfnamefont {Stefano}\ \bibnamefont
  {Longhi}}, \bibinfo {author} {\bibfnamefont {Davide}\ \bibnamefont {Gatti}},
  \ and\ \bibinfo {author} {\bibfnamefont {Giuseppe}\ \bibnamefont
  {Della~Valle}},\ }\bibfield  {title} {\enquote {\bibinfo {title}
  {Non-hermitian transparency and one-way transport in low-dimensional lattices
  by an imaginary gauge field},}\ }\href {\doibase 10.1103/PhysRevB.92.094204}
  {\bibfield  {journal} {\bibinfo  {journal} {Phys. Rev. B}\ }\textbf {\bibinfo
  {volume} {92}},\ \bibinfo {pages} {094204} (\bibinfo {year}
  {2015}{\natexlab{b}})}\BibitemShut {NoStop}%
\bibitem [{\citenamefont {Bordia}\ \emph {et~al.}(2016)\citenamefont {Bordia},
  \citenamefont {L\"uschen}, \citenamefont {Hodgman}, \citenamefont
  {Schreiber}, \citenamefont {Bloch},\ and\ \citenamefont
  {Schneider}}]{Schneider2016}%
  \BibitemOpen
  \bibfield  {author} {\bibinfo {author} {\bibfnamefont {Pranjal}\ \bibnamefont
  {Bordia}}, \bibinfo {author} {\bibfnamefont {Henrik~P.}\ \bibnamefont
  {L\"uschen}}, \bibinfo {author} {\bibfnamefont {Sean~S.}\ \bibnamefont
  {Hodgman}}, \bibinfo {author} {\bibfnamefont {Michael}\ \bibnamefont
  {Schreiber}}, \bibinfo {author} {\bibfnamefont {Immanuel}\ \bibnamefont
  {Bloch}}, \ and\ \bibinfo {author} {\bibfnamefont {Ulrich}\ \bibnamefont
  {Schneider}},\ }\bibfield  {title} {\enquote {\bibinfo {title} {Coupling
  identical one-dimensional many-body localized systems},}\ }\href {\doibase
  10.1103/PhysRevLett.116.140401} {\bibfield  {journal} {\bibinfo  {journal}
  {Phys. Rev. Lett.}\ }\textbf {\bibinfo {volume} {116}},\ \bibinfo {pages}
  {140401} (\bibinfo {year} {2016})}\BibitemShut {NoStop}%
\bibitem [{\citenamefont {Shiozaki}\ \emph {et~al.}(2017)\citenamefont
  {Shiozaki}, \citenamefont {Sato},\ and\ \citenamefont {Gomi}}]{Shiozaki2017}%
  \BibitemOpen
  \bibfield  {author} {\bibinfo {author} {\bibfnamefont {Ken}\ \bibnamefont
  {Shiozaki}}, \bibinfo {author} {\bibfnamefont {Masatoshi}\ \bibnamefont
  {Sato}}, \ and\ \bibinfo {author} {\bibfnamefont {Kiyonori}\ \bibnamefont
  {Gomi}},\ }\bibfield  {title} {\enquote {\bibinfo {title} {Topological
  crystalline materials: General formulation, module structure, and wallpaper
  groups},}\ }\href {\doibase 10.1103/PhysRevB.95.235425} {\bibfield  {journal}
  {\bibinfo  {journal} {Phys. Rev. B}\ }\textbf {\bibinfo {volume} {95}},\
  \bibinfo {pages} {235425} (\bibinfo {year} {2017})}\BibitemShut {NoStop}%
\bibitem [{\citenamefont {Moore}\ \emph {et~al.}(2008)\citenamefont {Moore},
  \citenamefont {Ran},\ and\ \citenamefont {Wen}}]{Moore2008}%
  \BibitemOpen
  \bibfield  {author} {\bibinfo {author} {\bibfnamefont {Joel~E.}\ \bibnamefont
  {Moore}}, \bibinfo {author} {\bibfnamefont {Ying}\ \bibnamefont {Ran}}, \
  and\ \bibinfo {author} {\bibfnamefont {Xiao-Gang}\ \bibnamefont {Wen}},\
  }\bibfield  {title} {\enquote {\bibinfo {title} {Topological surface states
  in three-dimensional magnetic insulators},}\ }\href
  {https://link.aps.org/doi/10.1103/PhysRevLett.101.186805} {\bibfield
  {journal} {\bibinfo  {journal} {Phys. Rev. Lett.}\ }\textbf {\bibinfo
  {volume} {101}},\ \bibinfo {pages} {186805} (\bibinfo {year}
  {2008})}\BibitemShut {NoStop}%
\bibitem [{\citenamefont {Kawabata}\ \emph
  {et~al.}(2018{\natexlab{b}})\citenamefont {Kawabata}, \citenamefont
  {Higashikawa}, \citenamefont {Gong}, \citenamefont {Ashida},\ and\
  \citenamefont {Ueda}}]{Kawabata2018b}%
  \BibitemOpen
  \bibfield  {author} {\bibinfo {author} {\bibfnamefont {Kohei}\ \bibnamefont
  {Kawabata}}, \bibinfo {author} {\bibfnamefont {Sho}\ \bibnamefont
  {Higashikawa}}, \bibinfo {author} {\bibfnamefont {Zongping}\ \bibnamefont
  {Gong}}, \bibinfo {author} {\bibfnamefont {Yuto}\ \bibnamefont {Ashida}}, \
  and\ \bibinfo {author} {\bibfnamefont {Masahito}\ \bibnamefont {Ueda}},\
  }\href@noop {} {\enquote {\bibinfo {title} {Topological unification of
  time-reversal and particle-hole symmetries in non-hermitian physics},}\ }
  (\bibinfo {year} {2018}{\natexlab{b}}),\ \bibinfo {note}
  {arXiv:1804.04676}\BibitemShut {NoStop}%
\bibitem [{Note5()}]{Note5}%
  \BibitemOpen
  \bibinfo {note} {Another reason is that the full information of $U(\protect
  \boldsymbol {k},t)=\protect \mathcal {T}e^{-i\DOTSI \intop \ilimits@
  ^t_0dt'H(t')}$ from $t=0$ to $t=T$ is important in a Floquet system. A good
  illustration is the anomalous edge states \cite {Lindner2013}, which exist in
  spite of a trivial $U(\protect \boldsymbol {k},T)=1$. In contrast, we focus
  on time-independent non-Hermitian Hamiltonians, so that the base manifold for
  classification only contains $\protect \boldsymbol {k}$ but not
  $t$.}\BibitemShut {Stop}%
\bibitem [{\citenamefont {Wigner}(1960)}]{Wigner1960}%
  \BibitemOpen
  \bibfield  {author} {\bibinfo {author} {\bibfnamefont {Eugene~P.}\
  \bibnamefont {Wigner}},\ }\bibfield  {title} {\enquote {\bibinfo {title}
  {Normal form of antiunitary operators},}\ }\href {\doibase 10.1063/1.1703672}
  {\bibfield  {journal} {\bibinfo  {journal} {J. Math. Phys.}\ }\textbf
  {\bibinfo {volume} {1}},\ \bibinfo {pages} {409} (\bibinfo {year}
  {1960})}\BibitemShut {NoStop}%
\bibitem [{\citenamefont {Kraus}(1972)}]{Kraus1971}%
  \BibitemOpen
  \bibfield  {author} {\bibinfo {author} {\bibfnamefont {K.}~\bibnamefont
  {Kraus}},\ }\bibfield  {title} {\enquote {\bibinfo {title} {General state
  changes in quantum theory},}\ }\href
  {https://doi.org/10.1016/0003-4916(71)90108-4} {\bibfield  {journal}
  {\bibinfo  {journal} {Ann. Phys.}\ }\textbf {\bibinfo {volume} {64}},\
  \bibinfo {pages} {311} (\bibinfo {year} {1972})}\BibitemShut {NoStop}%
\bibitem [{Note6()}]{Note6}%
  \BibitemOpen
  \bibinfo {note} {This should be understood with respect to the
  Hilbert-Schmidt inner product $(A,B)\equiv {\protect \rm Tr}[A^\protect \dag
  B]$. We can check that $(\protect \mathcal {K}A,\protect \mathcal
  {K}B)={\protect \rm Tr}[AB^\protect \dag ]={\protect \rm Tr}[B^\protect \dag
  A]=(B,A)$.}\BibitemShut {Stop}%
\bibitem [{\citenamefont {Wolf}\ and\ \citenamefont {Cirac}(2008)}]{Wolf2008}%
  \BibitemOpen
  \bibfield  {author} {\bibinfo {author} {\bibfnamefont {Michael~M.}\
  \bibnamefont {Wolf}}\ and\ \bibinfo {author} {\bibfnamefont {J.~Ignacio}\
  \bibnamefont {Cirac}},\ }\bibfield  {title} {\enquote {\bibinfo {title}
  {Dividing quantum channels},}\ }\href
  {https://doi.org/10.1007/s00220-008-0411-y} {\bibfield  {journal} {\bibinfo
  {journal} {Commun. Math. Phys.}\ }\textbf {\bibinfo {volume} {279}},\
  \bibinfo {pages} {147} (\bibinfo {year} {2008})}\BibitemShut {NoStop}%
\bibitem [{\citenamefont {Breuer}\ \emph {et~al.}(2016)\citenamefont {Breuer},
  \citenamefont {Laine}, \citenamefont {Piilo},\ and\ \citenamefont
  {Vacchini}}]{Breuer2016}%
  \BibitemOpen
  \bibfield  {author} {\bibinfo {author} {\bibfnamefont {Heinz-Peter}\
  \bibnamefont {Breuer}}, \bibinfo {author} {\bibfnamefont {Elsi-Mari}\
  \bibnamefont {Laine}}, \bibinfo {author} {\bibfnamefont {Jyrki}\ \bibnamefont
  {Piilo}}, \ and\ \bibinfo {author} {\bibfnamefont {Bassano}\ \bibnamefont
  {Vacchini}},\ }\bibfield  {title} {\enquote {\bibinfo {title} {Colloquium:
  Non-markovian dynamics in open quantum systems},}\ }\href
  {https://link.aps.org/doi/10.1103/RevModPhys.88.021002} {\bibfield  {journal}
  {\bibinfo  {journal} {Rev. Mod. Phys.}\ }\textbf {\bibinfo {volume} {88}},\
  \bibinfo {pages} {021002} (\bibinfo {year} {2016})}\BibitemShut {NoStop}%
\bibitem [{\citenamefont {Bennett}\ \emph {et~al.}(1996)\citenamefont
  {Bennett}, \citenamefont {Brassard}, \citenamefont {Popescu}, \citenamefont
  {Schumacher}, \citenamefont {Smolin},\ and\ \citenamefont
  {Wootters}}]{Bennett1996}%
  \BibitemOpen
  \bibfield  {author} {\bibinfo {author} {\bibfnamefont {Charles~H.}\
  \bibnamefont {Bennett}}, \bibinfo {author} {\bibfnamefont {Gilles}\
  \bibnamefont {Brassard}}, \bibinfo {author} {\bibfnamefont {Sandu}\
  \bibnamefont {Popescu}}, \bibinfo {author} {\bibfnamefont {Benjamin}\
  \bibnamefont {Schumacher}}, \bibinfo {author} {\bibfnamefont {John~A.}\
  \bibnamefont {Smolin}}, \ and\ \bibinfo {author} {\bibfnamefont {William~K.}\
  \bibnamefont {Wootters}},\ }\bibfield  {title} {\enquote {\bibinfo {title}
  {Purification of noisy entanglement and faithful teleportation via noisy
  channels},}\ }\href {\doibase 10.1103/PhysRevLett.76.722} {\bibfield
  {journal} {\bibinfo  {journal} {Phys. Rev. Lett.}\ }\textbf {\bibinfo
  {volume} {76}},\ \bibinfo {pages} {722} (\bibinfo {year} {1996})}\BibitemShut
  {NoStop}%
\bibitem [{\citenamefont {Gong}\ \emph {et~al.}(2018)\citenamefont {Gong},
  \citenamefont {Hamazaki},\ and\ \citenamefont {Ueda}}]{Gong2018}%
  \BibitemOpen
  \bibfield  {author} {\bibinfo {author} {\bibfnamefont {Zongping}\
  \bibnamefont {Gong}}, \bibinfo {author} {\bibfnamefont {Ryusuke}\
  \bibnamefont {Hamazaki}}, \ and\ \bibinfo {author} {\bibfnamefont {Masahito}\
  \bibnamefont {Ueda}},\ }\bibfield  {title} {\enquote {\bibinfo {title}
  {Discrete time-crystalline order in cavity and circuit qed systems},}\ }\href
  {\doibase 10.1103/PhysRevLett.120.040404} {\bibfield  {journal} {\bibinfo
  {journal} {Phys. Rev. Lett.}\ }\textbf {\bibinfo {volume} {120}},\ \bibinfo
  {pages} {040404} (\bibinfo {year} {2018})}\BibitemShut {NoStop}%
\bibitem [{\citenamefont {Macieszczak}\ \emph {et~al.}(2016)\citenamefont
  {Macieszczak}, \citenamefont {Gu\ifmmode \mbox{\c{t}}\else
  \c{t}\fi{}\ifmmode~\u{a}\else \u{a}\fi{}}, \citenamefont {Lesanovsky},\ and\
  \citenamefont {Garrahan}}]{Lesanovsky2016}%
  \BibitemOpen
  \bibfield  {author} {\bibinfo {author} {\bibfnamefont {Katarzyna}\
  \bibnamefont {Macieszczak}}, \bibinfo {author} {\bibfnamefont {M\ifmmode
  \u{a}\else \u{a}\fi{}d\ifmmode \u{a}\else~\u{a}\fi{}lin}\ \bibnamefont
  {Gu\ifmmode \mbox{\c{t}}\else \c{t}\fi{}\ifmmode~\u{a}\else \u{a}\fi{}}},
  \bibinfo {author} {\bibfnamefont {Igor}\ \bibnamefont {Lesanovsky}}, \ and\
  \bibinfo {author} {\bibfnamefont {Juan~P.}\ \bibnamefont {Garrahan}},\
  }\bibfield  {title} {\enquote {\bibinfo {title} {Towards a theory of
  metastability in open quantum dynamics},}\ }\href {\doibase
  10.1103/PhysRevLett.116.240404} {\bibfield  {journal} {\bibinfo  {journal}
  {Phys. Rev. Lett.}\ }\textbf {\bibinfo {volume} {116}},\ \bibinfo {pages}
  {240404} (\bibinfo {year} {2016})}\BibitemShut {NoStop}%
\bibitem [{\citenamefont {Yao}\ \emph {et~al.}(2017)\citenamefont {Yao},
  \citenamefont {Potter}, \citenamefont {Potirniche},\ and\ \citenamefont
  {Vishwanath}}]{Yao2017}%
  \BibitemOpen
  \bibfield  {author} {\bibinfo {author} {\bibfnamefont {N.~Y.}\ \bibnamefont
  {Yao}}, \bibinfo {author} {\bibfnamefont {A.~C.}\ \bibnamefont {Potter}},
  \bibinfo {author} {\bibfnamefont {I.-D.}\ \bibnamefont {Potirniche}}, \ and\
  \bibinfo {author} {\bibfnamefont {A.}~\bibnamefont {Vishwanath}},\ }\bibfield
   {title} {\enquote {\bibinfo {title} {Discrete time crystals: Rigidity,
  criticality, and realizations},}\ }\href {\doibase
  10.1103/PhysRevLett.118.030401} {\bibfield  {journal} {\bibinfo  {journal}
  {Phys. Rev. Lett.}\ }\textbf {\bibinfo {volume} {118}},\ \bibinfo {pages}
  {030401} (\bibinfo {year} {2017})}\BibitemShut {NoStop}%
\bibitem [{\citenamefont {Zhang}\ \emph {et~al.}(2017)\citenamefont {Zhang},
  \citenamefont {Hess}, \citenamefont {Kyprianidis}, \citenamefont {Becker},
  \citenamefont {Lee}, \citenamefont {Smith}, \citenamefont {Pagano},
  \citenamefont {Potirniche}, \citenamefont {Potter}, \citenamefont
  {Vishwanath}, \citenamefont {Yao},\ and\ \citenamefont {Monroe}}]{Zhang2017}%
  \BibitemOpen
  \bibfield  {author} {\bibinfo {author} {\bibfnamefont {J.}~\bibnamefont
  {Zhang}}, \bibinfo {author} {\bibfnamefont {P.~W.}\ \bibnamefont {Hess}},
  \bibinfo {author} {\bibfnamefont {A.}~\bibnamefont {Kyprianidis}}, \bibinfo
  {author} {\bibfnamefont {P.}~\bibnamefont {Becker}}, \bibinfo {author}
  {\bibfnamefont {A.}~\bibnamefont {Lee}}, \bibinfo {author} {\bibfnamefont
  {J.}~\bibnamefont {Smith}}, \bibinfo {author} {\bibfnamefont
  {G.}~\bibnamefont {Pagano}}, \bibinfo {author} {\bibfnamefont {I.-D.}\
  \bibnamefont {Potirniche}}, \bibinfo {author} {\bibfnamefont {A.~C.}\
  \bibnamefont {Potter}}, \bibinfo {author} {\bibfnamefont {A.}~\bibnamefont
  {Vishwanath}}, \bibinfo {author} {\bibfnamefont {N.~Y.}\ \bibnamefont {Yao}},
  \ and\ \bibinfo {author} {\bibfnamefont {C.}~\bibnamefont {Monroe}},\
  }\bibfield  {title} {\enquote {\bibinfo {title} {Observation of a discrete
  time crystal},}\ }\href {https://doi.org/10.1007/s00220-008-0411-y}
  {\bibfield  {journal} {\bibinfo  {journal} {Nature}\ }\textbf {\bibinfo
  {volume} {543}},\ \bibinfo {pages} {217} (\bibinfo {year}
  {2017})}\BibitemShut {NoStop}%
\bibitem [{\citenamefont {Choi}\ \emph {et~al.}(2017)\citenamefont {Choi},
  \citenamefont {Choi}, \citenamefont {Landig}, \citenamefont {Kucsko},
  \citenamefont {Zhou}, \citenamefont {Isoya}, \citenamefont {Jelezko},
  \citenamefont {Onoda}, \citenamefont {Sumiya}, \citenamefont {Khemani},
  \citenamefont {von Keyserlingk}, \citenamefont {Yao}, \citenamefont
  {Demler},\ and\ \citenamefont {Lukin}}]{Choi2017}%
  \BibitemOpen
  \bibfield  {author} {\bibinfo {author} {\bibfnamefont {Soonwon}\ \bibnamefont
  {Choi}}, \bibinfo {author} {\bibfnamefont {Joonhee}\ \bibnamefont {Choi}},
  \bibinfo {author} {\bibfnamefont {Renate}\ \bibnamefont {Landig}}, \bibinfo
  {author} {\bibfnamefont {Georg}\ \bibnamefont {Kucsko}}, \bibinfo {author}
  {\bibfnamefont {Hengyun}\ \bibnamefont {Zhou}}, \bibinfo {author}
  {\bibfnamefont {Junichi}\ \bibnamefont {Isoya}}, \bibinfo {author}
  {\bibfnamefont {Fedor}\ \bibnamefont {Jelezko}}, \bibinfo {author}
  {\bibfnamefont {Shinobu}\ \bibnamefont {Onoda}}, \bibinfo {author}
  {\bibfnamefont {Hitoshi}\ \bibnamefont {Sumiya}}, \bibinfo {author}
  {\bibfnamefont {Vedika}\ \bibnamefont {Khemani}}, \bibinfo {author}
  {\bibfnamefont {Curt}\ \bibnamefont {von Keyserlingk}}, \bibinfo {author}
  {\bibfnamefont {Norman~Y.}\ \bibnamefont {Yao}}, \bibinfo {author}
  {\bibfnamefont {Eugene}\ \bibnamefont {Demler}}, \ and\ \bibinfo {author}
  {\bibfnamefont {Mikhail~D.}\ \bibnamefont {Lukin}},\ }\bibfield  {title}
  {\enquote {\bibinfo {title} {Observation of discrete time-crystalline order
  in a disordered dipolar many-body system},}\ }\href
  {http://dx.doi.org/10.1038/nature21426} {\bibfield  {journal} {\bibinfo
  {journal} {Nature}\ }\textbf {\bibinfo {volume} {543}},\ \bibinfo {pages}
  {221} (\bibinfo {year} {2017})}\BibitemShut {NoStop}%
\bibitem [{\citenamefont {von Keyserlingk}\ \emph {et~al.}(2016)\citenamefont
  {von Keyserlingk}, \citenamefont {Khemani},\ and\ \citenamefont
  {Sondhi}}]{Keyserlingk2016}%
  \BibitemOpen
  \bibfield  {author} {\bibinfo {author} {\bibfnamefont {C.~W.}\ \bibnamefont
  {von Keyserlingk}}, \bibinfo {author} {\bibfnamefont {Vedika}\ \bibnamefont
  {Khemani}}, \ and\ \bibinfo {author} {\bibfnamefont {S.~L.}\ \bibnamefont
  {Sondhi}},\ }\bibfield  {title} {\enquote {\bibinfo {title} {Absolute
  stability and spatiotemporal long-range order in floquet systems},}\ }\href
  {\doibase 10.1103/PhysRevB.94.085112} {\bibfield  {journal} {\bibinfo
  {journal} {Phys. Rev. B}\ }\textbf {\bibinfo {volume} {94}},\ \bibinfo
  {pages} {085112} (\bibinfo {year} {2016})}\BibitemShut {NoStop}%
\bibitem [{\citenamefont {Zilberberg}\ \emph {et~al.}(2018)\citenamefont
  {Zilberberg}, \citenamefont {Huang}, \citenamefont {Guglielmon},
  \citenamefont {Wang}, \citenamefont {Chen}, \citenamefont {Kraus},\ and\
  \citenamefont {Rechtsman}}]{Zilberberg2018}%
  \BibitemOpen
  \bibfield  {author} {\bibinfo {author} {\bibfnamefont {Oded}\ \bibnamefont
  {Zilberberg}}, \bibinfo {author} {\bibfnamefont {Sheng}\ \bibnamefont
  {Huang}}, \bibinfo {author} {\bibfnamefont {Jonathan}\ \bibnamefont
  {Guglielmon}}, \bibinfo {author} {\bibfnamefont {Mohan}\ \bibnamefont
  {Wang}}, \bibinfo {author} {\bibfnamefont {Kevin~P.}\ \bibnamefont {Chen}},
  \bibinfo {author} {\bibfnamefont {Yaacov~E.}\ \bibnamefont {Kraus}}, \ and\
  \bibinfo {author} {\bibfnamefont {Mikael~C.}\ \bibnamefont {Rechtsman}},\
  }\bibfield  {title} {\enquote {\bibinfo {title} {Photonic topological
  boundary pumping as a probe of 4d quantum hall physics},}\ }\href
  {http://dx.doi.org/10.1038/nature25011} {\bibfield  {journal} {\bibinfo
  {journal} {Nature}\ }\textbf {\bibinfo {volume} {553}},\ \bibinfo {pages}
  {59} (\bibinfo {year} {2018})}\BibitemShut {NoStop}%
\bibitem [{\citenamefont {Lohse}\ \emph {et~al.}(2018)\citenamefont {Lohse},
  \citenamefont {Schweizer}, \citenamefont {Price}, \citenamefont
  {Zilberberg},\ and\ \citenamefont {Bloch}}]{Bloch2018}%
  \BibitemOpen
  \bibfield  {author} {\bibinfo {author} {\bibfnamefont {Michael}\ \bibnamefont
  {Lohse}}, \bibinfo {author} {\bibfnamefont {Christian}\ \bibnamefont
  {Schweizer}}, \bibinfo {author} {\bibfnamefont {Hannah~M.}\ \bibnamefont
  {Price}}, \bibinfo {author} {\bibfnamefont {Oded}\ \bibnamefont
  {Zilberberg}}, \ and\ \bibinfo {author} {\bibfnamefont {Immanuel}\
  \bibnamefont {Bloch}},\ }\bibfield  {title} {\enquote {\bibinfo {title}
  {Exploring 4d quantum hall physics with a 2d topological charge pump},}\
  }\href {http://dx.doi.org/10.1038/nature25000} {\bibfield  {journal}
  {\bibinfo  {journal} {Nature}\ }\textbf {\bibinfo {volume} {553}},\ \bibinfo
  {pages} {55} (\bibinfo {year} {2018})}\BibitemShut {NoStop}%
\bibitem [{\citenamefont {Price}\ \emph {et~al.}(2015)\citenamefont {Price},
  \citenamefont {Zilberberg}, \citenamefont {Ozawa}, \citenamefont
  {Carusotto},\ and\ \citenamefont {Goldman}}]{Goldman2015}%
  \BibitemOpen
  \bibfield  {author} {\bibinfo {author} {\bibfnamefont {H.~M.}\ \bibnamefont
  {Price}}, \bibinfo {author} {\bibfnamefont {O.}~\bibnamefont {Zilberberg}},
  \bibinfo {author} {\bibfnamefont {T.}~\bibnamefont {Ozawa}}, \bibinfo
  {author} {\bibfnamefont {I.}~\bibnamefont {Carusotto}}, \ and\ \bibinfo
  {author} {\bibfnamefont {N.}~\bibnamefont {Goldman}},\ }\bibfield  {title}
  {\enquote {\bibinfo {title} {Four-dimensional quantum hall effect with
  ultracold atoms},}\ }\href {\doibase 10.1103/PhysRevLett.115.195303}
  {\bibfield  {journal} {\bibinfo  {journal} {Phys. Rev. Lett.}\ }\textbf
  {\bibinfo {volume} {115}},\ \bibinfo {pages} {195303} (\bibinfo {year}
  {2015})}\BibitemShut {NoStop}%
\bibitem [{\citenamefont {M\"uller}\ \emph {et~al.}(2012)\citenamefont
  {M\"uller}, \citenamefont {Diehl}, \citenamefont {Pupillo},\ and\
  \citenamefont {Zoller}}]{Muller2012}%
  \BibitemOpen
  \bibfield  {author} {\bibinfo {author} {\bibfnamefont {Markus}\ \bibnamefont
  {M\"uller}}, \bibinfo {author} {\bibfnamefont {Sebastian}\ \bibnamefont
  {Diehl}}, \bibinfo {author} {\bibfnamefont {Guido}\ \bibnamefont {Pupillo}},
  \ and\ \bibinfo {author} {\bibfnamefont {Peter}\ \bibnamefont {Zoller}},\
  }\bibfield  {title} {\enquote {\bibinfo {title} {Engineered open systems and
  quantum simulations with atoms and ions},}\ }\href
  {https://doi.org/10.1016/B978-0-12-396482-3.00001-6} {\bibfield  {journal}
  {\bibinfo  {journal} {Adv. At. Mol. Opt. Phys.}\ }\textbf {\bibinfo {volume}
  {61}},\ \bibinfo {pages} {1} (\bibinfo {year} {2012})}\BibitemShut {NoStop}%
\bibitem [{\citenamefont {Feinberg}\ and\ \citenamefont {Zee}(1999)}]{Zee1999}%
  \BibitemOpen
  \bibfield  {author} {\bibinfo {author} {\bibfnamefont {Joshua}\ \bibnamefont
  {Feinberg}}\ and\ \bibinfo {author} {\bibfnamefont {A.}~\bibnamefont {Zee}},\
  }\bibfield  {title} {\enquote {\bibinfo {title} {Non-hermitian localization
  and delocalization},}\ }\href {\doibase 10.1103/PhysRevE.59.6433} {\bibfield
  {journal} {\bibinfo  {journal} {Phys. Rev. E}\ }\textbf {\bibinfo {volume}
  {59}},\ \bibinfo {pages} {6433} (\bibinfo {year} {1999})}\BibitemShut
  {NoStop}%
\bibitem [{\citenamefont {Meyer}(2000)}]{Meyer2000}%
  \BibitemOpen
  \bibfield  {author} {\bibinfo {author} {\bibfnamefont {C.~D.}\ \bibnamefont
  {Meyer}},\ }\href@noop {} {\emph {\bibinfo {title} {Matrix Analysis and
  Applied Linear Algebra}}}\ (\bibinfo  {publisher} {Society for Industrial and
  Applied Mathematics, Philadelphia},\ \bibinfo {year} {2000})\BibitemShut
  {NoStop}%
\bibitem [{\citenamefont {Callias}(1978)}]{Callias1978}%
  \BibitemOpen
  \bibfield  {author} {\bibinfo {author} {\bibfnamefont {Constantine}\
  \bibnamefont {Callias}},\ }\bibfield  {title} {\enquote {\bibinfo {title}
  {Axial anomalies and index theorems on open spaces},}\ }\href
  {https://doi.org/10.1007/BF01202525} {\bibfield  {journal} {\bibinfo
  {journal} {Commun. Math. Phys.}\ }\textbf {\bibinfo {volume} {62}},\ \bibinfo
  {pages} {213} (\bibinfo {year} {1978})}\BibitemShut {NoStop}%
\bibitem [{\citenamefont {Ashida}\ and\ \citenamefont
  {Ueda}(2018)}]{Ashida2018}%
  \BibitemOpen
  \bibfield  {author} {\bibinfo {author} {\bibfnamefont {Yuto}\ \bibnamefont
  {Ashida}}\ and\ \bibinfo {author} {\bibfnamefont {Masahito}\ \bibnamefont
  {Ueda}},\ }\bibfield  {title} {\enquote {\bibinfo {title} {Full-counting
  many-particle dynamics: Nonlocal and chiral propagation of correlations},}\
  }\href {\doibase 10.1103/PhysRevLett.120.185301} {\bibfield  {journal}
  {\bibinfo  {journal} {Phys. Rev. Lett.}\ }\textbf {\bibinfo {volume} {120}},\
  \bibinfo {pages} {185301} (\bibinfo {year} {2018})}\BibitemShut {NoStop}%
\bibitem [{\citenamefont {Graefe}\ \emph {et~al.}(2016)\citenamefont {Graefe},
  \citenamefont {Korsch},\ and\ \citenamefont {Rush}}]{Rush2016}%
  \BibitemOpen
  \bibfield  {author} {\bibinfo {author} {\bibfnamefont {E.~M.}\ \bibnamefont
  {Graefe}}, \bibinfo {author} {\bibfnamefont {H.~J.}\ \bibnamefont {Korsch}},
  \ and\ \bibinfo {author} {\bibfnamefont {A.}~\bibnamefont {Rush}},\
  }\bibfield  {title} {\enquote {\bibinfo {title} {Quasiclassical analysis of
  bloch oscillations in non-hermitian tight-binding lattices},}\ }\href
  {https://doi.org/10.1088/1367-2630/18/7/075009} {\bibfield  {journal}
  {\bibinfo  {journal} {New J. Phys.}\ }\textbf {\bibinfo {volume} {18}},\
  \bibinfo {pages} {075009} (\bibinfo {year} {2016})}\BibitemShut {NoStop}%
\bibitem [{\citenamefont {Poyatos}\ \emph {et~al.}(1996)\citenamefont
  {Poyatos}, \citenamefont {Cirac},\ and\ \citenamefont {Zoller}}]{Zoller1996}%
  \BibitemOpen
  \bibfield  {author} {\bibinfo {author} {\bibfnamefont {J.~F.}\ \bibnamefont
  {Poyatos}}, \bibinfo {author} {\bibfnamefont {J.~I.}\ \bibnamefont {Cirac}},
  \ and\ \bibinfo {author} {\bibfnamefont {P.}~\bibnamefont {Zoller}},\
  }\bibfield  {title} {\enquote {\bibinfo {title} {Quantum reservoir
  engineering with laser cooled trapped ions},}\ }\href
  {https://link.aps.org/doi/10.1103/PhysRevLett.77.4728} {\bibfield  {journal}
  {\bibinfo  {journal} {Phys. Rev. Lett.}\ }\textbf {\bibinfo {volume} {77}},\
  \bibinfo {pages} {4728} (\bibinfo {year} {1996})}\BibitemShut {NoStop}%
\bibitem [{\citenamefont {Kraus}\ \emph {et~al.}(2008)\citenamefont {Kraus},
  \citenamefont {B\"uchler}, \citenamefont {Diehl}, \citenamefont {Kantian},
  \citenamefont {Micheli},\ and\ \citenamefont {Zoller}}]{Buchler2008}%
  \BibitemOpen
  \bibfield  {author} {\bibinfo {author} {\bibfnamefont {B.}~\bibnamefont
  {Kraus}}, \bibinfo {author} {\bibfnamefont {H.~P.}\ \bibnamefont
  {B\"uchler}}, \bibinfo {author} {\bibfnamefont {S.}~\bibnamefont {Diehl}},
  \bibinfo {author} {\bibfnamefont {A.}~\bibnamefont {Kantian}}, \bibinfo
  {author} {\bibfnamefont {A.}~\bibnamefont {Micheli}}, \ and\ \bibinfo
  {author} {\bibfnamefont {P.}~\bibnamefont {Zoller}},\ }\bibfield  {title}
  {\enquote {\bibinfo {title} {Preparation of entangled states by quantum
  markov process},}\ }\href
  {https://link.aps.org/doi/10.1103/PhysRevA.78.042307} {\bibfield  {journal}
  {\bibinfo  {journal} {Phys. Rev. A}\ }\textbf {\bibinfo {volume} {78}},\
  \bibinfo {pages} {042307} (\bibinfo {year} {2008})}\BibitemShut {NoStop}%
\bibitem [{\citenamefont {Diehl}\ \emph {et~al.}(2008)\citenamefont {Diehl},
  \citenamefont {Micheli}, \citenamefont {Kantian}, \citenamefont {Kraus},
  \citenamefont {B\"uchler},\ and\ \citenamefont {Zoller}}]{Diehl2008}%
  \BibitemOpen
  \bibfield  {author} {\bibinfo {author} {\bibfnamefont {S.}~\bibnamefont
  {Diehl}}, \bibinfo {author} {\bibfnamefont {A.}~\bibnamefont {Micheli}},
  \bibinfo {author} {\bibfnamefont {A.}~\bibnamefont {Kantian}}, \bibinfo
  {author} {\bibfnamefont {B.}~\bibnamefont {Kraus}}, \bibinfo {author}
  {\bibfnamefont {H.~P.}\ \bibnamefont {B\"uchler}}, \ and\ \bibinfo {author}
  {\bibfnamefont {P.}~\bibnamefont {Zoller}},\ }\bibfield  {title} {\enquote
  {\bibinfo {title} {Quantum states and phases in driven open quantum systems
  with cold atoms},}\ }\href {http://dx.doi.org/10.1038/nphys1073} {\bibfield
  {journal} {\bibinfo  {journal} {Nat. Phys.}\ }\textbf {\bibinfo {volume}
  {4}},\ \bibinfo {pages} {878} (\bibinfo {year} {2008})}\BibitemShut {NoStop}%
\bibitem [{\citenamefont {Diehl}\ \emph {et~al.}(2011)\citenamefont {Diehl},
  \citenamefont {Rico}, \citenamefont {Baranov},\ and\ \citenamefont
  {Zoller}}]{Diehl2011}%
  \BibitemOpen
  \bibfield  {author} {\bibinfo {author} {\bibfnamefont {Sebastian}\
  \bibnamefont {Diehl}}, \bibinfo {author} {\bibfnamefont {Enrique}\
  \bibnamefont {Rico}}, \bibinfo {author} {\bibfnamefont {Mikhail~A.}\
  \bibnamefont {Baranov}}, \ and\ \bibinfo {author} {\bibfnamefont {Peter}\
  \bibnamefont {Zoller}},\ }\bibfield  {title} {\enquote {\bibinfo {title}
  {Topology by dissipation in atomic quantum wires},}\ }\href
  {http://dx.doi.org/10.1038/nphys2106} {\bibfield  {journal} {\bibinfo
  {journal} {Nat. Phys.}\ }\textbf {\bibinfo {volume} {7}},\ \bibinfo {pages}
  {971} (\bibinfo {year} {2011})}\BibitemShut {NoStop}%
\bibitem [{\citenamefont {Lindblad}(1976)}]{Lindblad1976}%
  \BibitemOpen
  \bibfield  {author} {\bibinfo {author} {\bibfnamefont {G.}~\bibnamefont
  {Lindblad}},\ }\bibfield  {title} {\enquote {\bibinfo {title} {On the
  generators of quantum dynamical semigroups},}\ }\href
  {https://doi.org/10.1007/BF01608499} {\bibfield  {journal} {\bibinfo
  {journal} {Commun. Math. Phys.}\ }\textbf {\bibinfo {volume} {48}},\ \bibinfo
  {pages} {119} (\bibinfo {year} {1976})}\BibitemShut {NoStop}%
\bibitem [{\citenamefont {Reiter}\ and\ \citenamefont
  {S{\o}rensen}(2012)}]{Sorensen2012}%
  \BibitemOpen
  \bibfield  {author} {\bibinfo {author} {\bibfnamefont {F.}~\bibnamefont
  {Reiter}}\ and\ \bibinfo {author} {\bibfnamefont {A.~S.}\ \bibnamefont
  {S{\o}rensen}},\ }\bibfield  {title} {\enquote {\bibinfo {title} {Effective
  operator formalism for open quantum systems},}\ }\href
  {https://link.aps.org/doi/10.1103/PhysRevA.85.032111} {\bibfield  {journal}
  {\bibinfo  {journal} {Phys. Rev. A}\ }\textbf {\bibinfo {volume} {85}},\
  \bibinfo {pages} {032111} (\bibinfo {year} {2012})}\BibitemShut {NoStop}%
\bibitem [{\citenamefont {Gerbier}\ and\ \citenamefont
  {Dalibard}(2010)}]{Dalibard2010}%
  \BibitemOpen
  \bibfield  {author} {\bibinfo {author} {\bibfnamefont {Fabrice}\ \bibnamefont
  {Gerbier}}\ and\ \bibinfo {author} {\bibfnamefont {Jean}\ \bibnamefont
  {Dalibard}},\ }\bibfield  {title} {\enquote {\bibinfo {title} {Gauge fields
  for ultracold atoms in optical superlattices},}\ }\href
  {https://doi.org/10.1088/1367-2630/12/3/033007} {\bibfield  {journal}
  {\bibinfo  {journal} {New J. Phys.}\ }\textbf {\bibinfo {volume} {12}},\
  \bibinfo {pages} {033007} (\bibinfo {year} {2010})}\BibitemShut {NoStop}%
\bibitem [{\citenamefont {Bloch}\ \emph {et~al.}(2008)\citenamefont {Bloch},
  \citenamefont {Dalibard},\ and\ \citenamefont {Zwerger}}]{Bloch2008}%
  \BibitemOpen
  \bibfield  {author} {\bibinfo {author} {\bibfnamefont {Immanuel}\
  \bibnamefont {Bloch}}, \bibinfo {author} {\bibfnamefont {Jean}\ \bibnamefont
  {Dalibard}}, \ and\ \bibinfo {author} {\bibfnamefont {Wilhelm}\ \bibnamefont
  {Zwerger}},\ }\bibfield  {title} {\enquote {\bibinfo {title} {Many-body
  physics with ultracold gases},}\ }\href
  {https://link.aps.org/doi/10.1103/RevModPhys.80.885} {\bibfield  {journal}
  {\bibinfo  {journal} {Rev. Mod. Phys.}\ }\textbf {\bibinfo {volume} {80}},\
  \bibinfo {pages} {885} (\bibinfo {year} {2008})}\BibitemShut {NoStop}%
\bibitem [{\citenamefont {Barber}\ \emph {et~al.}(2006)\citenamefont {Barber},
  \citenamefont {Hoyt}, \citenamefont {Oates}, \citenamefont {Hollberg},
  \citenamefont {Taichenachev},\ and\ \citenamefont {Yudin}}]{Barber2006}%
  \BibitemOpen
  \bibfield  {author} {\bibinfo {author} {\bibfnamefont {Z.~W.}\ \bibnamefont
  {Barber}}, \bibinfo {author} {\bibfnamefont {C.~W.}\ \bibnamefont {Hoyt}},
  \bibinfo {author} {\bibfnamefont {C.~W.}\ \bibnamefont {Oates}}, \bibinfo
  {author} {\bibfnamefont {L.}~\bibnamefont {Hollberg}}, \bibinfo {author}
  {\bibfnamefont {A.~V.}\ \bibnamefont {Taichenachev}}, \ and\ \bibinfo
  {author} {\bibfnamefont {V.~I.}\ \bibnamefont {Yudin}},\ }\bibfield  {title}
  {\enquote {\bibinfo {title} {Direct excitation of the forbidden clock
  transition in neutral $^{174}\mathrm{Yb}$ atoms confined to an optical
  lattice},}\ }\href {\doibase 10.1103/PhysRevLett.96.083002} {\bibfield
  {journal} {\bibinfo  {journal} {Phys. Rev. Lett.}\ }\textbf {\bibinfo
  {volume} {96}},\ \bibinfo {pages} {083002} (\bibinfo {year}
  {2006})}\BibitemShut {NoStop}%
\bibitem [{\citenamefont {Miranda}\ \emph {et~al.}(2015)\citenamefont
  {Miranda}, \citenamefont {Inoue}, \citenamefont {Okuyama}, \citenamefont
  {Nakamoto},\ and\ \citenamefont {Kozuma}}]{Kozuma2015}%
  \BibitemOpen
  \bibfield  {author} {\bibinfo {author} {\bibfnamefont {Martin}\ \bibnamefont
  {Miranda}}, \bibinfo {author} {\bibfnamefont {Ryotaro}\ \bibnamefont
  {Inoue}}, \bibinfo {author} {\bibfnamefont {Yuki}\ \bibnamefont {Okuyama}},
  \bibinfo {author} {\bibfnamefont {Akimasa}\ \bibnamefont {Nakamoto}}, \ and\
  \bibinfo {author} {\bibfnamefont {Mikio}\ \bibnamefont {Kozuma}},\ }\bibfield
   {title} {\enquote {\bibinfo {title} {Site-resolved imaging of ytterbium
  atoms in a two-dimensional optical lattice},}\ }\href {\doibase
  10.1103/PhysRevA.91.063414} {\bibfield  {journal} {\bibinfo  {journal} {Phys.
  Rev. A}\ }\textbf {\bibinfo {volume} {91}},\ \bibinfo {pages} {063414}
  (\bibinfo {year} {2015})}\BibitemShut {NoStop}%
\end{thebibliography}%
\end{document}